\documentstyle[11pt,epsfig,astrobib,aas_macros,myown,twoside]{report}

\def\i{\item}

\def\ni{\noindent}
\def\beb{\begin{thebibliography}{999}}
\def\eeb{\end{thebibliography}}
\def\bei{\begin{itemize}}
\def\eei{\end{itemize}}
\def\bef{\begin{figure}}
\def\eef{\end{figure}}
\def\ben{\begin{enumerate}}
\def\een{\end{enumerate}}
\def\beq{\begin{equation}}
\def\eeq{\end{equation}}
\def\ber{\begin{eqnarray}}
\def\eer{\end{eqnarray}}
\def\edo{\end{document}}
\newcommand{\ibd}{inverse $\beta$-decay }
\newcommand{\dmdt}{{\mbox{{\rm M}$_{\odot}$}} {\rm yr}$^{-1}$}
\newcommand{\en}{e^{-}}
\newcommand{\gcc}{{\rm g} \, {\rm cm}^{-3}}

\newcommand{\msun}{\mbox{{\rm M}$_{\odot}$}}
\newcommand{\mdot}{\mbox{$\dot{M}$}}

\newcommand{\lsim}{\raisebox{-0.3ex}{\mbox{$\stackrel{<}{_\sim} \,$}}}
\newcommand{\gsim}{\raisebox{-0.3ex}{\mbox{$\stackrel{>}{_\sim} \,$}}}

\begin{document}
\thispagestyle{empty}
\begin{center}
{\LARGE \bf Evolution of the Magnetic Field in Accreting Neutron Stars}

\vspace{2.0cm}

{\large A thesis submitted for the degree of \\ {\bf Doctor of Philosophy}\\
in the Faculty of Science}

\vspace{1.0cm}

{\large by}\\
{\Large \bf Sushan Konar}

\vspace{1.0cm}
{\large under the supervision of }\\
{\Large \bf Dipankar Bhattacharya}

\vspace{2.0cm}

\centerline{\epsfxsize=3cm\epsfbox{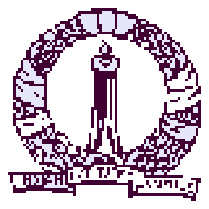}}

\vskip0.5cm

{\large Department of Physics\\Indian Institute of Science \\ Bangalore 
560 012 INDIA \\*[0.75cm] November 1997}

\end{center}

%
%
%
%
%
%
%


\setcounter{page}{0}

\tableofcontents
\listoffigures

\chapter{introduction}
\label{cintro}

Thirty years  of active  research in Pulsars  have made  it abundantly
clear that  these objects are  veritable laboratories for  testing out
theories  for  exotic physics  stretching  far  beyond  the limits  of
present day terrestrial experiments. Indeed, the 1967 discovery of the
first pulsar by Hewish and his  group in Cambridge has been one of the
major    events   in    recent    astronomy~\cite{hews68}.    Pulsars,
characterized by the regular pulses of radiation observed to be coming
from  them, are  actually strongly  magnetized neutron  stars rotating
very  rapidly.  The  concept of  neutron  stars, of  course, has  been
around for about  thirty years before this discovery,  almost from the
day the neutron  was detected for the first time. It  is said that the
day  the  news  of  the  discovery  of  neutrons  reached  Landau,  he
hypothesized on  the possible existence  of stars made up  entirely of
neutrons.   Barely  two  years  after this,  \citeN{baad34}  in  their
seminal paper propounded  the theory of the possible  birth of neutron
stars in the most violent and spectacular event of stellar death, that
of a supernova explosion.

There has  always been a  great interest in  the ultimate fate  of the
stars.   Before   the  advent   of  Fermi-Dirac  statistics,   it  was
inconceivable as to how a star  could escape the final collapse at the
hands of gravity when it exhausts  its nuclear fuel - a view expounded
by  Sir Arthur  Eddington.   But the  work  of \citeN{chan31c}  proved
conclusively  that the  final gravitational  collapse could  be halted
when  the stellar  material  becomes Fermi-degenerate  due to  extreme
compression so that the  degeneracy pressure is sufficient to withhold
gravity.  In  this work  the case of  electron-degeneracy and  the end
state of  stars that we  know of as  white dwarfs has  been discussed.
Soon  afterwards the  companion of  Sirius was  identified as  a white
dwarf which vindicated the existence  of such degenerate end stages of
stars.  The  logical  extension  of Chandrasekhar's  argument  is,  of
course, the state when the neutrons become degenerate. And that is the
state found in neutron stars.

Even before  the discovery  of pulsars, \citeN{paci67}  suggested that
the Crab Nebula must be powered  by a rotating neutron star.  The Crab
Nebula is the remnant of the  supernova of A.D.  1054, recorded by the
Chinese.  The later identification of the Crab Pulsar with the neutron
star  associated with  the supernova  gave the  first direct  proof of
Baade \& Zwicky's hypothesis.  Lately, the attention has been focussed
on  the  most  spectacular  supernova  of  recent  times,  SN1987A,  a
supernova  that  went off  in  the  Large  Magellanic Cloud,  a  small
companion galaxy of the Milky Way,  in February 1987. But even after a
decade of intense search the neutron star that is supposed to be there
has remained elusive.  This,  in all probability, means a modification
of the  theory of  neutron star formation  in supernovae.   Of course,
supernovae   may    not   be   the   only   way    of   neutron   star
formation. Recently, particularly in connection with the possible ways
of  millisecond  pulsar  formation,  the theory  of  accretion-induced
collapse of a white dwarf into a neutron star has been advocated. Even
if that  does happen in  certain systems, supernovae would  still most
likely be the major route through which the neutron stars are born. As
a star  of main-sequence mass $\gsim  8 \msun$ explodes at  the end of
its life, it throws away most of  its mass and the object that is left
behind is  a neutron  star of  mass of about  $\sim 1.4~\msun$  with a
radius   of  ten   kilometers.   Evidently   the  formation   of  this
ultra-compact  object is  accompanied  by a  release  of a  tremendous
amount of energy  ($\sim 10^{53}$ erg) equal to  the binding energy of
it. And  this is the energy  that powers the fantastic  fireworks of a
supernova.

The initial interest in neutron  stars, surprisingly, arose due to the
discovery  of certain  intense radio  emitters.  The  large  values of
red-shift  associated with  these objects  made one  think  that these
could be  neutron stars producing highly red-shifted  radiation due to
the large value  of their surface gravity. That  idea died its natural
death  when  quasars  were  discovered.   Then in  1967  pulsars  were
discovered almost serendipitously, by Jocelyn Bell, a graduate student
working   with   Anthony   Hewish  on   interplanetary   scintillation
\cite{hews68}.   Understandably, the  first signals  from  the neutron
star, due to their extreme regular periodicity, were suspected to have
been  signatures of  the  'little  green men'.   But  all such  highly
speculative theories and their more sober counterparts quickly settled
to   an   identification   of   these   objects   as   neutron   stars
\cite{gold68,gold69a}.  The signals were understood to have originated
due to  the fast rotation of  the neutron stars  possessing very large
magnetic fields (canonical  value of the field being  B $\sim 10^{12}$
Gauss).   Extremely  compact  objects  were required  to  explain  the
rapidity of  the pulses ($P \sim  s$) and the  obvious candidates were
the neutron stars and the white dwarfs. Discovery of fast pulsars like
Crab and  Vela (with  rotation periods of  33 and 89  ms respectively)
excluded the  possibility of these  being white dwarfs (as  they would
not be gravitationally stable at  such high rates of rotation) and the
identification  of   pulsars  with  neutron   stars  was  conclusively
established \cite{gold68,gold69a,gunn69,paci68}.

Pulsars  are characterized by  their pulsed  emission and  the precise
periodicity of these pulses, though the shape and the amplitude of the
pulse profiles vary widely. The  mechanism of the radio emission still
remains  one of  the important  unsolved problems  of  pulsar physics,
though   the  basic   ideas  were   laid  down   quite  early   on  by
\citeN{ostr69}.   In  this  model  of  {\em  magnetic  dipole}  pulsar
emission  is derived  from the  kinetic energy  of a  rotating neutron
star.  It is  assumed that a neutron star  rotates uniformly in vacuum
at a  frequent $\omega$ and  has a dipole  moment {\bf m} at  an angle
$\theta$ to  the axis of rotation.   The dipole field  at the magnetic
pole of the star, $B_p$, is related to {\bf m} by
\beq
|{\bf m}| = \frac{1}{2} B_p R^3,
\eeq
where  $R$ is  the  radius of  the  star. This  {\em oblique  rotator}
configuration has  a time-varying dipole moment as  seen from infinity
and therefore radiates energy at a rate
\ber
\dot{E} &=& - \frac{2}{3c^3}|\ddot{\bf m}|^2 \nonumber \\
&=& -  \frac{1}{6c^3} B_p^2  R^6 \omega^4 \sin^2  \theta \label{edEdt}
\eer
This  energy  carried away  changes  the  kinetic  energy producing  a
slow-down torque on the star given by
\beq
N = I \dot{\omega}, \label{etorque}
\eeq
where  $I$  is the  moment  of inertia  of  the  star. Using  equation
[\ref{edEdt}]  and  [\ref{etorque}] one  obtains  an  estimate of  the
magnetic field  in terms of the period  and the rate of  change of the
period of the  star, which are measurable with  high accuracy from the
timing of  the arriving pulses. Thus,  the magnetic field  is given by
the expression \cite{manc}
\beq
B \sim (\frac{3 I c^3 P \dot P}{8 \pi^2 R^6})^{1/2}, \nonumber
\eeq
where, $P,  \dot P$ are the  period and the period  derivative, $I$ is
the moment  of inertia and $R$ is  the radius of the  star.  There are
problems with  this simple model. Firstly,  as far as  the emission is
concerned,  this  mechanism  does  not  work in  case  of  an  aligned
rotator. Refined theories dealing with  the problem of emission for an
aligned rotator have come into  existence since then.  And this simple
estimate  of the  magnetic  field  provides a  measure  of the  dipole
component only, without  any handle on the total  field.  Yet, this is
still  the most  widely  used and  in  most cases  the  only means  of
estimating  the  magnetic  field.    There  have  been  a  few  direct
measurements of the field,  like that from the cyclotron line-strength
of Her X-1,  which gave the field strength of this  neutron star to be
$\sim 4 - 6 \times 10^{12}$ Gauss~\cite{true78}. But the scope of such
direct measurement is limited only  to the case of X-ray binaries. For
radio pulsars no method for a direct measurement of the field exists.

\bef
\hspace{2.0cm}
\epsfig{file=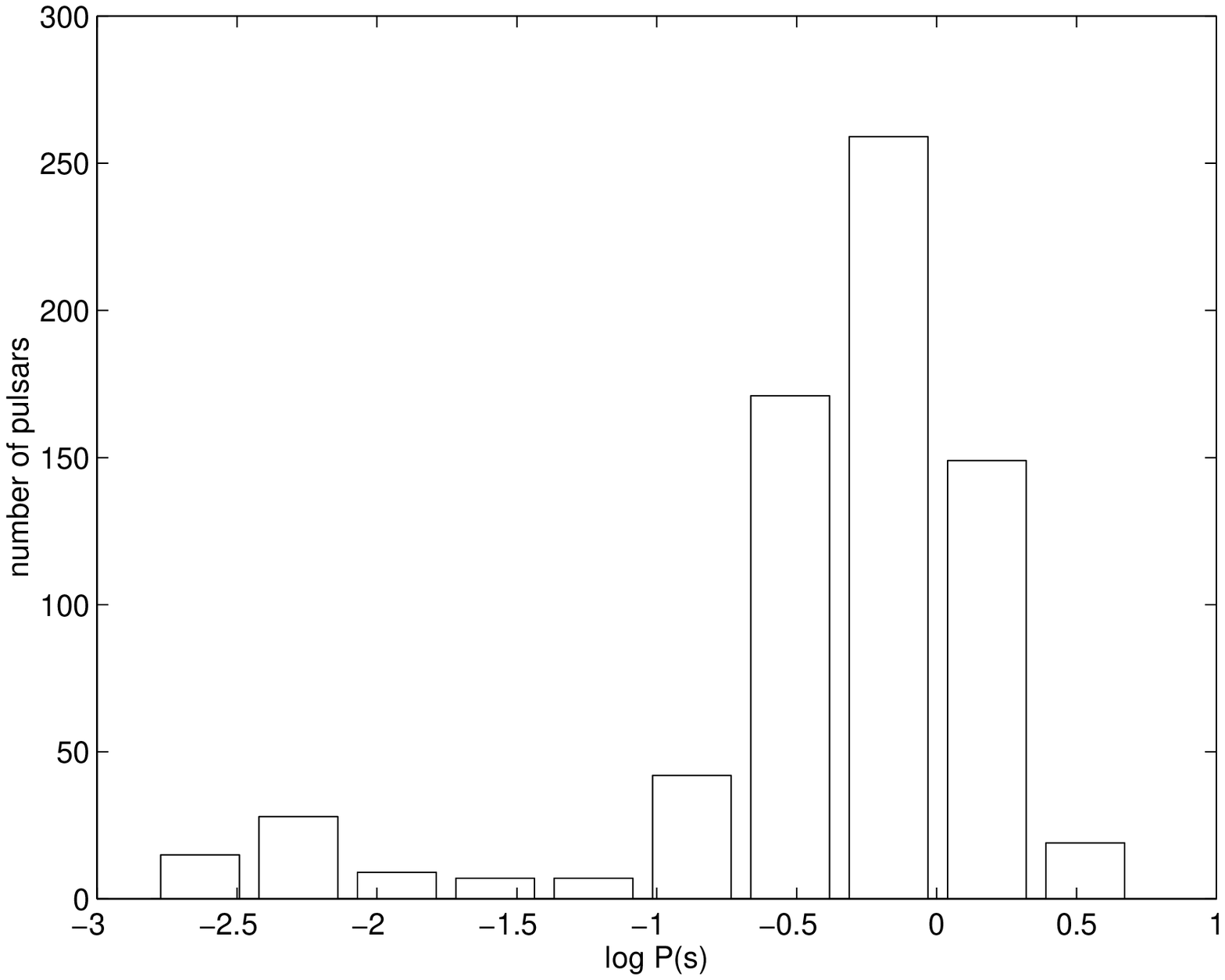,width=300pt}
\caption[period distribution  of radio pulsars]{Histogram  showing the
period distribution of the radio pulsars.}
\label{fperiod}
\eef

With  the discovery  of  a new  variety  of pulsar  by  Backer et  al.
\citeyear{back82} another  horizon in  the pulsar research  opened up.
From now on  the radio pulsars were divided  into two distinct classes
with very  different physical  characteristics.  The new  variety were
named  {\em millisecond  pulsars} as  these have  very  small rotation
periods,  in  the range  of  milliseconds.   The  first one  observed,
PSR1937+21,  had a  period of  1.6~ms. Also  they were  found  to have
extremely  short  (compared   to  the  earlier-observed  {\em  normal}
pulsars) magnetic fields  in the range of $10^8 -  10^9$ Gauss. In all
some  seven  hundred  pulsars  have  been observed  to  date.   Figure
[\ref{fperiod}]  shows  a  histogram  for  the periods  of  all  these
pulsars.  The period distribution is clearly bimodal, with most of the
occupants of the peak at  short periods being millisecond pulsars. One
must admit here that the definition of millisecond pulsars is somewhat
ad hoc, defined as the ones  with spin periods less than $20ms$. Still
the division serves quite well due to the fact that these two sets, as
far  as  present  understanding  goes,  do have  very  different  past
histories. We make a crude  comparison of these two classes of pulsars
here in order to highlight the differences.

\bef
\hspace{2.5cm}
\epsfig{file=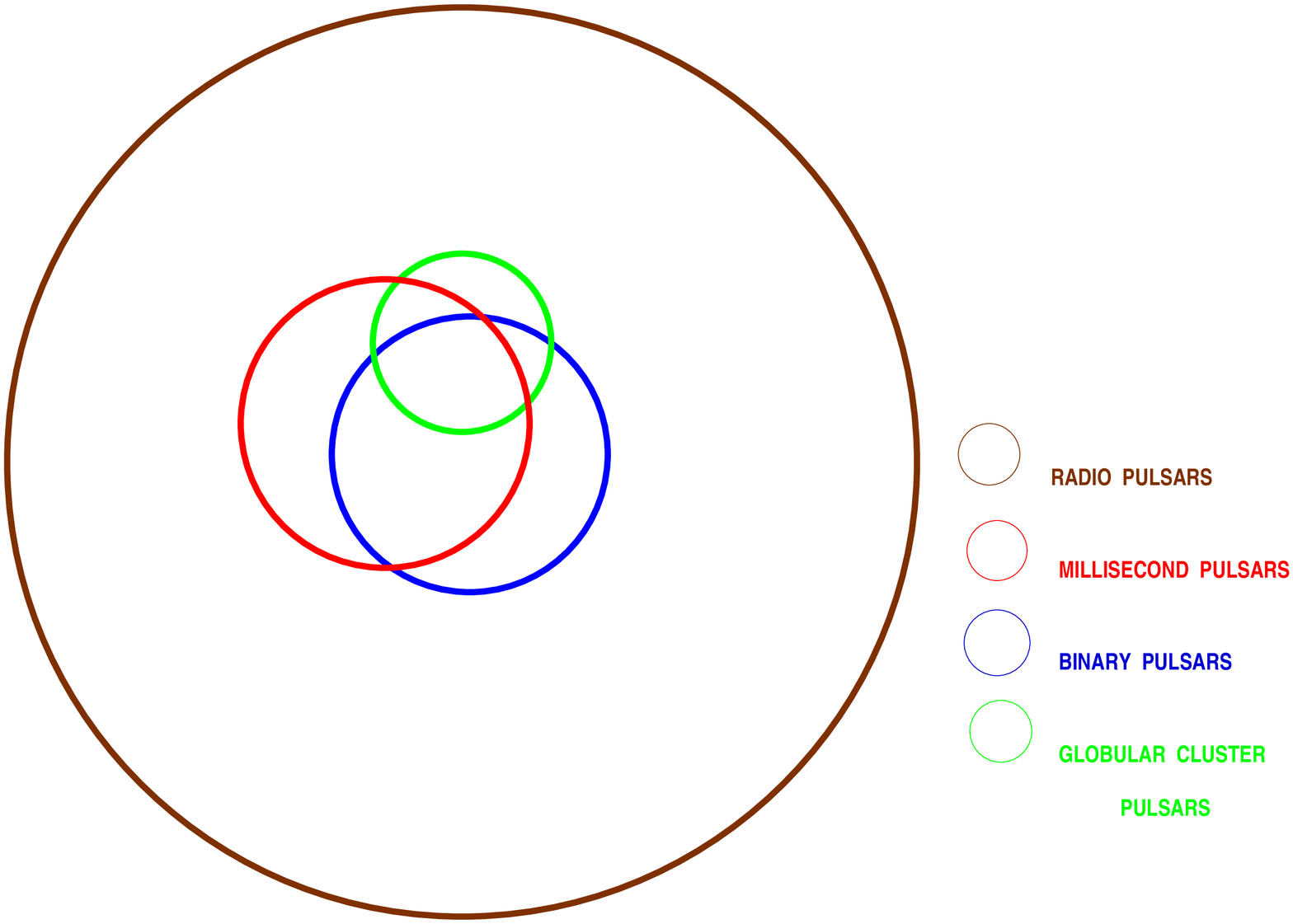,width=215pt,angle=-90}
\caption[overlapping classes of radio pulsars]{A schematic description
of how  the various special types  of radio pulsars  overlap with each
other.}
\vspace{-0.5cm}
\label{fvenn}
\eef

\vspace{0.5cm}
\ni
\hspace{2.5cm}
\begin{tabular}{|l|l|r|} \hline
&& \\
properties & {\bf normal pulsars} & {\bf millisecond pulsars} \\ 
&& \\ \cline{1-3}
&& \\
spin period & $ P \gsim 20$~ms & $P \lsim 20$~ms  \\ 
&& \\ \cline{1-3}
&& \\
magnetic field & $10^{11} - 10^{13}$~Gauss & $10^{8} - 10^{9}$~Gauss  \\
&& \\ \cline{1-3}
&& \\
age & $\gsim 10^6 - 10^7$~yrs & $\sim 10^9$~yrs \\ 
&& \\ \cline{1-3}
&& \\
binarity & mostly isolated & mostly in binaries \\
&& \\ \cline{1-3}
\end{tabular} \\

The most striking difference is,  of course, the fact that whereas the
normal pulsars are  mostly isolated, some 90\% of  the disk population
and about 50\%  of the Globular Cluster population  of the millisecond
pulsars  are  in binaries.   The  age  determination  of some  of  the
millisecond  pulsars are  possible also  due to  that fact.   From the
surface temperature of the white  dwarf companion it has been possible
to put  a lower limit  to the age  of a few millisecond  pulsars which
lies in the range of  $10^9$ years \cite{call89,kulk91}.  On the other
hand, in the case of normal  pulsars it is basically the spin-down age
estimated from the rate of change of the period which turn out to be a
few   million   years   at   most.   To   emphasize   the   remarkable
binary-millisecond pulsar  association we draw here  a Venn-diagram in
figure [\ref{fvenn}]  showing the nature of the  pulsars, their binary
association  and the population  (disk or  globular cluster)  to which
they belong.

\bef
\hspace{2.0cm}
\epsfig{file=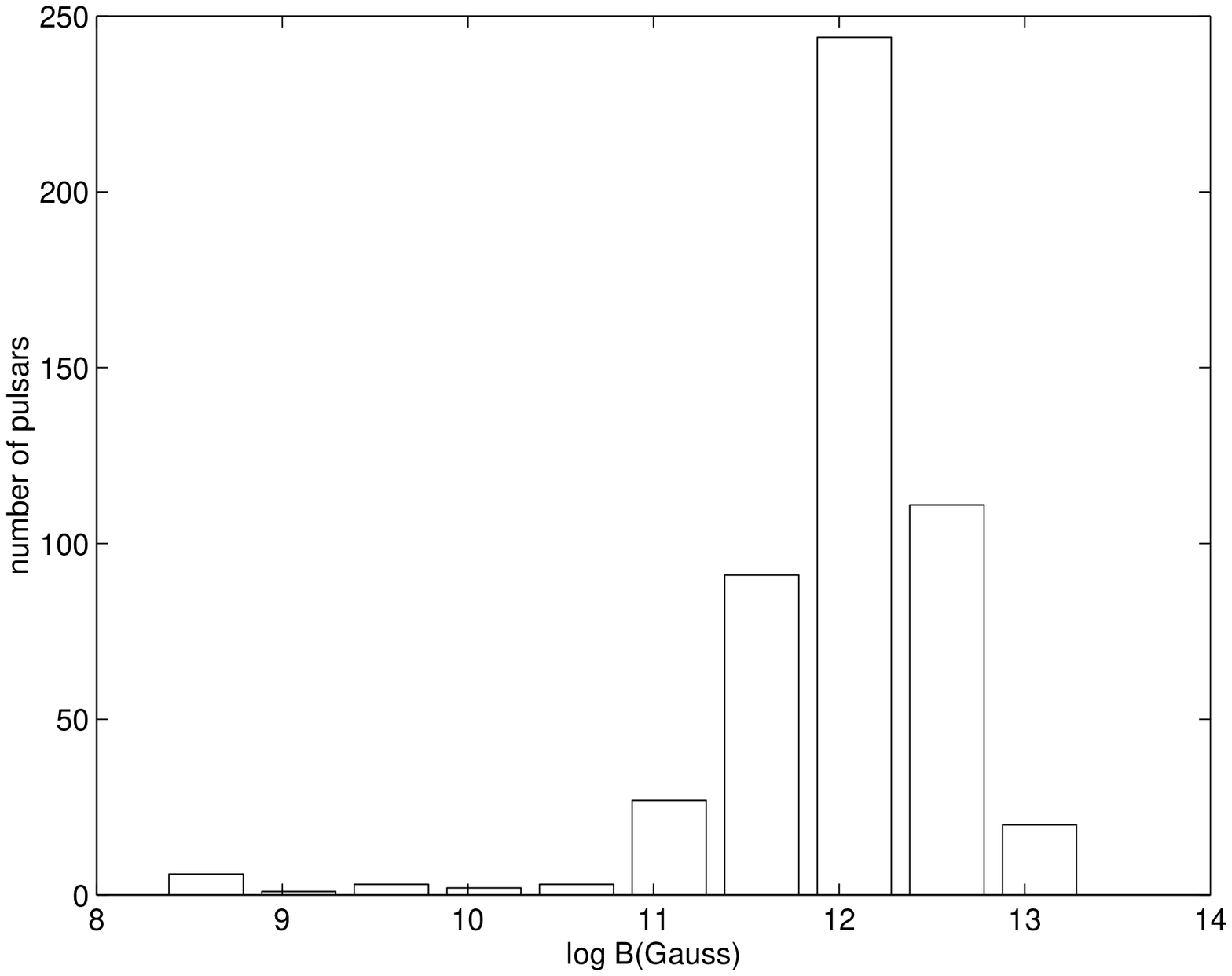,width=300pt}
\caption[magnetic fields of  isolated radio pulsars]{Histogram showing
the distribution of the magnetic field in solitary radio pulsars.}
\label{ffield_solitary}
\eef

\bef
\hspace{2.0cm}
\epsfig{file=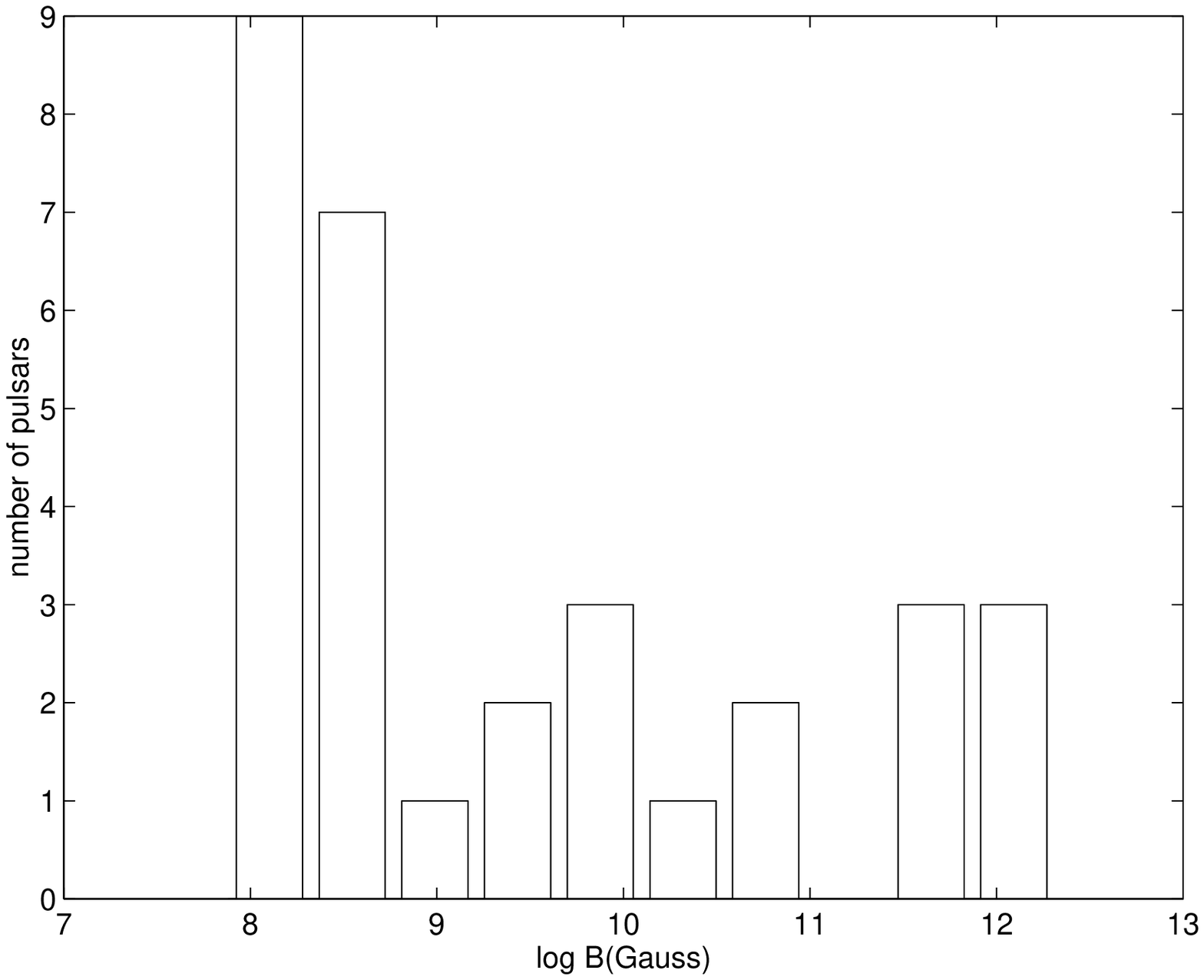,width=300pt}
\caption[magnetic  fields of  binary radio  pulsars]{Histogram showing
the distribution of the magnetic field in binary radio pulsars.}
\label{ffield_binary}
\eef

It has been  observed that, in general, the  binary pulsars have lower
fields than the  isolated pulsars.  In figures [\ref{ffield_solitary}]
and [\ref{ffield_binary}] we have  plotted the histograms of the field
strengths of in these  two categories separately.This fact was evident
even  in  the  first  binary  pulsar  to  be  discovered,  the  famous
PSR1913+16  discovered by \citeN{huls75}.   It was  \citeN{bisn74} who
suggested for  the first  time a connection  between the decay  of the
magnetic  field  in  a   neutron  star  and  its  binary  association.
\citeN{taam86} provided  observational support  to this idea.   In the
early eighties  the idea  of {\em recycled  pulsars} was  forwarded in
this   connection  \cite{srini82,rad82,alpr82}.   According   to  this
scenario an  otherwise {\em normal} pulsar,  at the end  of its normal
lifetime, could  be resurrected  with a reduced  magnetic field  and a
spun-up period  if it is  processed in a  binary. One of  the problems
faced by  the theory of {\em  recycled pulsars} is  that of explaining
the  isolated  low-field   and  millisecond  pulsars.   \citeN{rudr89}
suggested that the  companion could perhaps be ablated  by the intense
radiation falling  on it from the pulsar.   Soon afterwards PSR1957+20
(another  1.6~ms pulsar)  was  caught  in the  act  of vaporising  its
companion~\cite{fruc88}.    So   that   immediately   confirmed   this
conjecture, although  later on doubts  have been raised  regarding the
efficacy of this method to  completely destroy the companion. Yet, the
connection  between a  reduction of  the field  strength and  a binary
history has remained well  endorsed by observations. Though the theory
of  `recycling' has  its  problems, to  date  this has  been the  most
successful  in   explaining  the  population  of   the  low-field  and
millisecond  pulsars by  integrating  them with  the  class of  normal
pulsars through their binary history.

In order to  develop a proper theory of the  evolution of the magnetic
field it is essential to understand the nature of the internal current
configuration supporting  the observed  field.  This also  requires an
accurate  knowledge of  the internal  structure of  the  neutron stars
which determines the long-term behaviour of the embedded current loops
and  hence the  time-evolution  of the  magnetic  field. Roughly,  the
neutron star has two physically  different regions - the crust and the
core. The crust is the outer  shell, about a kilometer thick, which is
a crystalline solid made up of neutron rich nuclei. In this region the
density  changes  by  some  eight  orders  of  magnitude,  going  from
$10^6~\gcc$ at  the very surface, to $10^{14.5}~\gcc$  at the boundary
of the  crust and the core~\cite{pand76}.  Underneath  this crust lies
the region with  an average density of nuclear  density or more, which
is  believed to  have superfluid  neutrons along  with superconducting
protons     and    extremely     relativistic,     Fermi    degenerate
electrons~\cite{latt92,pine92}.  There  is a lot  of controversy about
the exact composition of the  core, opinions ranging from normal n-p-e
plasma                            to                            exotic
quark-condensates~\cite{hana92,latt91,peth92,prak94,tats92},  but  the
above-mentioned picture is what is generally accepted at present.

Regarding the generation  of the magnetic field in  the neutron stars,
there are  two possibilities. The field  could be a  fossil field. The
original magnetic field of the progenitor of the neutron star could be
enhanced to large values  by flux conservation during compactification
of the large star to neutron star dimensions. As one believes the core
to  become  superfluid soon  after  formation  and  in particular  the
protons form a type II superconductor, the field would be supported by
the  proton superconducting  flux tubes  in this  case. But  there are
problems with  this scenario. Firstly, there are  no good measurements
for the core field of the  massive stars, so it is not certain whether
the field strength required to  enhance it to the pulsars field values
really  obtain  in the  progenitor  core. On  the  other  hand, in  an
extremely  violent  process like  a  supernova  explosion, whether  an
adiabatic  process  like flux  conservation  would  hold  good is  not
clearly understood. The other possibility is that of the generation of
the field after the birth of the neutron star.  \citeN{blnd83} pointed
out that it is possible to generate  a field in the crust of a cooling
neutron star due to thermo-magnetic instabilities as the heat flows in
presence of a seed field. This mechanism, again, suffers from the fact
that a seed field of the order of $10^8$ Gauss is required in order to
produce the canonical field values that are observed in neutron stars.
Though none of the theories are  free from hitches, for want of better
alternatives,  these are accepted  as the  best possibilities  for the
origin of the field in neutron stars.

It is, of course, obvious that the evolution of the field would itself
depend on  the generation  mechanism, which determines  the underlying
configuration of  currents. Initially, observational  data appeared to
indicate that the pulsar magnetic fields decay with a time constant of
$\sim 10^6$ years~\cite{gunn70,lyne82} On the other hand, it was shown
by \citeN{baym69c}  that given the state  of matter in  a neutron star
the electrical conductivity  is expected to be extremely  high and the
ohmic dissipation time-scale should be larger than the Hubble time. It
was  borne out by  some recent  statistical work~\cite{bhat92,hart97},
where  they showed  that the  magnetic field  of the  isolated pulsars
indeed do  not show  any significant decay.   An investigation  of the
association of the field decay  and a binary history therefore becomes
extremely pertinent.  In the past  few years a considerable  amount of
effort has  been spent in trying  to find the answer  to this question
(for details  see \citeNP{bhat95}  and references therein).  The basic
understanding in  this regard could  be divided into two  classes. The
underlying physics of  the field evolution in a binary  is that of the
ohmic  dissipation  of  the  current  loops  in  the  accretion-heated
crust.  Therefore, for  the believers  in crustal  field,  the current
dissipates due  to an increase in  temperature as mass  is accreted by
the neutron star from its companion in the course of binary evolution.
When  one assumes  an initial  core  field configuration,  a phase  of
spin-down driven flux-expulsion is necessary prior to the phase of the
ohmic dissipation in  the crust.  Both these ideas  have been explored
in  detail   by  a  multitude   of  researchers  (for  a   review  see
\citeNP{db95a}) yet by no means have all the questions been answered.

Therefore,  fifteen  years  after  the discovery  of  the  millisecond
pulsars  there still remains  a lot  of uncertainties  regarding their
possible past history. On a broader perspective the scenarios for both
the generation and  the evolution of magnetic fields  of neutron stars
lack a consensus. A coherent  theoretical picture is yet to emerge. In
such a  situation, the observational data  is the only  guide. In this
thesis, therefore, we have tried to address a few questions related to
the evolution of the magnetic  field of neutron stars that are members
of binary systems. We try  to make connection with the overall picture
of  the  field  evolution  as  indicated by  observational  data.   In
particular we look  into the problem of the  generation of millisecond
pulsar from  particular kind of binary  systems.  To this  end we have
looked at four related problems as described below :
\bei
\i the effect of diamagnetic screening on the final field of a neutron
star accreting material from its binary companion;
\i evolution  of magnetic  flux located in  the crust of  an accreting
neutron star;
\i  application of  the above-mentioned  model to  real systems  and a
comparison with observations;
\i  an investigation  into  the consequences  of  magnetic flux  being
initially  located in  the  core  of the  star  and its  observational
implications.
\eei

Here then is a brief resume of the problems mentioned above.
\ben
\i The  effect of diamagnetic screening  - The basic idea  is that the
magnetic field is screened due to the stream of the accreting material
that arrives at the polar  cap, being channeled by the strong magnetic
field of the star.  A screening  of the external dipole field, in this
case, is  achieved by the  production of horizontal components  at the
cost of poloidal ones.  If fluid-interchange instabilities are ignored
then  the field  lines  are frozen  to  the material  (since the  flow
time-scales are very much  smaller than the diffusive time-scales) and
as the accreted  material spreads over the surface  it drags the field
lines along.  The  field lines would then reconnect  on the equatorial
plane and  get buried. Before this  field can diffuse  out more matter
will come and spread  on top of it, and push the  field to even deeper
regions.  Finally  the field may  even reach the  superconducting core
from where it will not diffuse  out. But whether such burial is at all
possible depends on the relative magnitude of the time-scales of flow,
diffusion and interchange instability.  This question has already been
addressed before, and  the calculations made by us  re-assert the fact
that the fluid-interchange time-scales are too small for the burial of
the field to  be effective since any stretching of  the field lines is
quickly restored over this overturn time-scale. Therefore the cause of
the low field  observed in some neutron stars in  X-ray binaries or in
millisecond  pulsars can not  be due  to a  simple screening  of their
magnetic field by the accreted matter.
\i  Evolution  of a  crustal  magnetic  flux  under accretion  -  This
investigation  is  carried   out  assuming  the  underlying  currents,
supporting the observed field, to be entirely confined to the crust of
the neutron  star to start  with.  The main mechanism  responsible for
the field decay here is the  ohmic dissipation of the current loops in
the accretion-heated  crust. The evolution is investigated  for a wide
range of values of the  relevant physical parameters, such as the rate
of accretion, the  depth of current carrying layers  etc. We find that
within  a reasonable  range of  parameter values  final fields  in the
correct range for millisecond  pulsars are produced.  A most important
feature has been seen to arise  due to the inward material movement of
the crustal layers because  of accreted overburden.  The current loops
reach the  region of very high  conductivity in the  deeper and denser
regions of the star by such  material movement and this puts a stop to
further  field decay.   This {\em  freezing in}  behaviour  that comes
naturally out  of the input  physics, is very important  in explaining
the limiting field values  observed in binary and millisecond pulsars.
Therefore,  within  the  limits  of uncertainty  this  model,  besides
providing for an effective mechanism  for field reduction by the right
order of magnitude,  also gives an explanation for  the lower bound of
the field observed in millisecond pulsars.
\i Comparison with observations - Here we investigate the evolution of
the magnetic field of neutron stars  in its entirety -- in case of the
isolated  pulsars as  well as  in different  kinds of  binary systems,
assuming  the  field  to  be  originally confined  in  the  crust.   A
comparison of our results for  the field evolution in isolated neutron
stars  with  observational  data   helps  us  constrain  the  physical
parameters of the crust. We also model the full evolution of a neutron
star    in   a    binary    system   through    several   stages    of
interaction. Initially there is  no interaction between the components
of the binary and the evolution of the neutron star is similar to that
of an  isolated one. It  then interacts with  the stellar wind  of the
companion and  finally a phase  of heavy mass transfer  ensues through
Roche-lobe overflow.   We model the field evolution  through all these
stages  and  compare the  resulting  final  field  strength with  that
observed in neutron  stars in various types of  binary systems. One of
the  interesting  aspects of  our  result  is  a positive  correlation
between the rate  of accretion and the final  field strength. Recently
there has been observational  indications for such a correlation.  Our
results also match with the  overall picture of the field evolution in
neutron stars.   In particular, the generation  of millisecond pulsars
from low-mass binaries arises as  a natural consequence of the general
framework.
\i Lastly,  we look  at the outcome  of spindown-induced  expulsion of
magnetic  flux originally  confined to  the  core, in  which case  the
expelled  flux undergoes  ohmic decay.   We  model this  decay of  the
expelled flux. Once  again we look into the  nature of field evolution
-- both for neutron stars that  are isolated and are members of binary
systems.   This scenario  of field  evolution could  also  explain the
observed  field strength  of neutron  stars  but only  if the  crustal
lattice  contains a  large amount  of  impurity. At  present both  the
scenarios (assuming  an original crustal  field and an  expelled flux)
appear  to be  consistent with  the observations  though  they require
rather different assumptions regarding the  state of the matter in the
crusts of the neutron stars. Also, the detailed predictions in the two
scenarios are  different. Therefore  future observations, able  to pin
down these details, should distinguish  between the two models. On the
other  hand, with  an unambiguous  determination of  the state  of the
matter in the  neutron star crust, at some future  date, it will again
be possible to  arrive at a definitive conclusion  regarding the model
of field evolution that actually prevails in neutron stars.
\een

Improved observational  techniques have produced  a wealth of  data in
the recent past which  requires an accurate and detailed understanding
of  pulsar physics. Unfortunately,  the regime  in which  the physical
theories are  lacking are precisely  the regimes in which  the neutron
stars are the  only available systems. And the  data, though enormous,
still    remain    inadequate    for    answering    such    questions
unambiguously. The  handicap is many-faceted, like  the uncertainty in
the form of a  nucleon-nucleon interaction potential or the inadequacy
of  the quantum  many body  techniques to  handle the  nuclear density
systems. Even  though the basic  picture of field reduction  via ohmic
dissipation is  on secure grounds there are  still many uncertainties,
for example due to uncertainties in :
\bei
\i the crustal  structure, in particular regarding the  exact forms of
the nuclei,
\i the transport properties arising  due to a lack of proper knowledge
of the  impurity concentration or  the dislocations that exist  in the
crust,
\i  the  change  in  the  composition  due  to  accretion,  since  the
newly-formed accreted crust do  not contain cold-catalysed matter like
the original crust, or
\i the thermal behaviour in both isolated and accreting neutron star.
\eei
As for the generation of the millisecond pulsars, quite apart from the
birthrate  problem,  all  the  model  calculations  also  suffer  from
uncertainties in the binary evolution.

If the  physics of these are  understood a lot of  accepted wisdom may
change. Refined many-body calculations of proton superconductivity has
already  cast doubts  on the  magnetic  field being  supported by  the
fluxoids  threading the  core~\cite{ains89}.  Therefore  to understand
the basic problems at least  within the standard premises one needs to
have  a  second  look  at  the  problems  incorporating  all  the  new
refinements  that have  been coming  in.  That  seems to  be  the next
logical step.  On a different  level, new and exotic physics is making
inroads like  the Strange stars  being put forward as  possible pulsar
candidates~\cite{chen97}.  Those  probably would start the  new era of
pulsar research.

This thesis has been organized as follows. In chapter 2, we review the
basics of  neutron star physics, aspects  that we have  needed for our
calculations.   Chapter  3 discusses  the  standard  scenario for  the
generation and evolution of  neutron star magnetic fields. In chapters
4 to  7 we describe  the details of  the four problems that  have been
worked on.   Finally in chapter 8  we have made  our conclusions along
with   a   discussion  of   the   uncertainties   inherent  in   these
investigations and the possible  future directions of work along these
lines.

\chapter{microphysics of neutron stars} 
\label{cnsphys}

\section{equation of state of dense matter}
\label{seos}

The conditions in the interior  of Neutron Stars are more extreme than
any   encountered  terrestrially.    The  gravitational   pressure  is
supported mainly by  the pressure of the repulsive  interaction of the
nucleons.  To  a first  approximation a neutron  star is like  a giant
nucleus made  of $10^{57}$ nucleons (mostly neutrons)  with an average
baryon density close to the nuclear density. The star also has a solid
{\em crust}  roughly one  kilometer thick, compositionally  similar to
terrestrial crystalline  solids with highly  neutron-rich nuclei.  The
{\em core} beneath  the crust is essentially a sea  of neutrons with a
mere  ten  percent  sprinkling  of  protons and  an  equal  number  of
electrons  to maintain  charge neutrality.  Besides having  an average
density of  about $\sim 10^{15}~\gcc$ a  neutron star also  has a huge
neutron excess. When a neutron star forms in a supernova explosion the
temperature attained is higher than the characteristic temperatures of
all  the equilibrating  chemical reactions.  Consequently, all  of the
neutron  star  material  is  $\beta$-equilibrated where  most  of  the
protons  have been converted  to neutrons  due to  enhanced \ibd  in a
dense environment. As the electron  Fermi sea is filled up the reverse
process, i.e., the decay of a  neutron to a proton, an electron and an
anti-neutrino,  is  progressively  blocked  resulting in  the  neutron
excess.

Except near  the surface  the neutron star  behaves like  an effective
zero-temperature system, the actual  temperature ($\sim 10^6$K or less
in the crust  and $\lsim 10^8$K in the core  after about $10^4$ years)
being  much smaller  than the  characteristic temperatures  (the Fermi
temperature  of the electrons  or the  neutrons or  the energy  of the
nucleon-nucleon interaction).  Therefore almost whole of  the star can
be described as  a degenerate, free Fermi system  (electrons being the
dominant component near the surface  and neutrons in the interior). We
shall not  discuss here the  superfluid states of neutrons  or protons
believed to exist in the core. Density-wise the neutron star has three
characteristically  different  regions.  The  thin  outer  crust  with
densities  ranging from  $10^6~\gcc$  at the  surface  to $4.0  \times
10^{11}~\gcc$  ({\em neutron  drip  density}) at  which free  neutrons
start dripping out  of the nuclei. Next is the  {\em inner crust} with
densities in-between the neutron-drip  density and the nuclear density
($2.8  \times 10^{14}~\gcc$).  Beyond the  nuclear density  the nuclei
dissolve to produce a soup of nucleons.

\subsection{outer crust : $7.86~\gcc < \rho < 4.0 \times 10^{11}~\gcc$}

This is  the best  understood density regime  of all. The  pressure is
primarily due  to that of the degenerate  electrons, charge neutrality
being maintained by  an ionic crystal. For $\rho  \gsim 10^6~\gcc$ the
electrons become relativistic. As  density increases beyond this value
the electron  Fermi energy approaches  the MeV range where  it becomes
energetically favourable  for the protons to undergo  \ibd and convert
themselves to neutrons  giving rise to the neutron-rich  nuclei in the
crust.  The  equilibrium nuclide  for a given  density is  obtained by
minimizing the free energy of  the system with respect to a particular
combination of ($Z,A$) keeping the baryon number density constant. The
first such  calculation was done by  \citeN{baym71a}, reproduced here
in  table  [2.1], based  on  Bethe-Weiz\"{s}acker semi-empirical  mass
formula   with   parameters   obtained   from   fits   to   laboratory
nuclei.  Recently,  \citeN{haen89b}  have redone  these  calculations
using more  refined methods, though  their results do not  differ very
much from the  earlier ones.  Among the factors  important in deciding
the equilibrium nuclide at a  given density are the neutron and proton
(dominant just below the neutron  drip) shell effects and the strength
of the  spin-orbit interaction which  depends on the three  and higher
body  nucleon-nucleon   interactions  (defining  the   energy  of  the
individual nuclei).

\begin{table}
{\small{\begin{tabular}{|l|l|l|r|} \hline
\multicolumn{4}{|c|}{} \\ 
\multicolumn{4}{|c|}{TABLE 2.1} \\ 
\multicolumn{4}{|c|}{} \\ 
\multicolumn{4}{|c|}{\emph{DATA FROM BAYM, PETHICK AND SUTHERLAND (1971)}} \\ 
\multicolumn{4}{|c|}{} \\ \hline
&&&\\
mass density & baryon number density & mass number & atomic number \\
&& of equilibrium nuclide & of equilibrium nuclide \\
&&&\\ \cline{1-4}
&&&\\
$\rho$ ($\gcc$) & $n_b$ (cm$^{-3}$) & $Z$ & $A$ \\
&&&\\ \cline{1-4}
&&&\\
7.86E0    &  4.73E24   &  26   &  56  \\ 7.90E0    &  4.76E24   &  26   &  56 \\ 8.15E0    &  4.91E24   &  26   &  56 \\
1.16E01   &  6.99E24   &  26   &  56 \\ 1.64E01   &  9.90E24   &  26   &  56 \\ 4.51E01   &  2.72E25   &  26   &  56 \\
2.12E02   &  1.27E26   &  26   &  56 \\ 1.150E03  &  6.93E26   &  26   &  56 \\ 1.044E04  &  6.295E27  &  26   &  56 \\
2.622E04  &  1.581E28  &  26   &  56 \\ 6.587E04  &  3.972E28  &  26   &  56 \\ 1.654E05  &  9.976E28  &  26   &  56 \\
4.156E05  &  2.506E29  &  26   &  56 \\ 1.044E06  &  6.294E29  &  26   &  56 \\ 2.622E06  &  1.581E30  &  26   &  56 \\
6.588E06  &  3.972E30  &  26   &  56 \\ 8.293E06  &  5.000E30  &  28   &  62 \\ 1.655E07  &  9.976E30  &  28   &  62 \\
3.302E07  &  1.990E31  &  28   &  62 \\ 6.589E07  &  3.972E31  &  28   &  62 \\ 1.315E08  &  7.924E31  &  28   &  62 \\
2.624E08  &  1.581E32  &  28   &  62 \\ 3.304E08  &  1.990E32  &  28   &  64 \\ 5.237E08  &  3.155E32  &  28   &  64 \\
8.301E08  &  5.000E32  &  28   &  64 \\ 1.045E09  &  6.294E32  &  28   &  64 \\ 1.316E09  &  7.924E32  &  34   &  84 \\
1.657E09  &  9.976E32  &  34   &  84 \\ 2.626E09  &  1.581E33  &  34   &  84 \\ 4.164E09  &  2.506E33  &  34   &  84 \\
6.601E09  &  3.972E33  &  34   &  84 \\ 8.312E09  &  5.000E33  &  32   &  82 \\ 1.046E10  &  6.294E33  &  32   &  82 \\
1.318E10  &  7.924E33  &  32   &  82 \\ 1.659E10  &  9.976E33  &  32   &  82 \\ 2.090E10  &  1.256E34  &  32   &  82 \\
2.631E10  &  1.581E34  &  30   &  80 \\ 3.313E10  &  1.990E34  &  30   &  80 \\ 4.172E10  &  2.506E34  &  30   &  80 \\
5.254E10  &  3.155E34  &  28   &  78 \\ \cline{1-4}
\end{tabular}}} \\
\end{table}

\begin{table}
{\small{\begin{tabular}{|l|l|l|r|} \hline
\multicolumn{4}{|c|}{} \\ 
\multicolumn{4}{|c|}{TABLE 2.1 (\em{continuted})} \\ 
\multicolumn{4}{|c|}{} \\ 
\multicolumn{4}{|c|}{\em{DATA FROM BAYM, PETHICK AND SUTHERLAND (1971)}} \\ 
\multicolumn{4}{|c|}{} \\ \hline
&&&\\
mass density & baryon number density & mass number & atomic number \\
&& of equilibrium nuclide & of equilibrium nuclide \\
&&&\\ \cline{1-4}
&&&\\
$\rho$ ($\gcc$) & $n_b$ (cm$^{-3}$) & $Z$ & $A$ \\ 
&&&\\ \cline{1-4}
&&&\\
6.617E10  &  3.972E34  &  28   &  78 \\ 8.332E10  &  5.000E34  &  28   &  78 \\ 1.049E11  &  6.294E34  &  28   &  78 \\ 
1.322E11  &  7.924E34  &  28   &  78 \\ 1.664E11  &  9.976E34  &  26   &  76 \\ 1.844E11  &  1.105E35  &  42   &  124 \\ 
2.096E11  &  1.256E35  &  40   &  122 \\ 2.640E11  &  1.581E35  &  40   &  122 \\ 3.325E11  &  1.990E35  &  38   &  120 \\ 
4.188E11  &  2.506E35  &  36   &  118 \\ 4.299E11  &  2.572E35  &  36   &  118 \\ 4.460E11  &  2.670E35  &  40   &  126 \\ 
5.228E11  &  3.126E35  &  40   &  128 \\ 6.610E11  &  3.951E35  &  40   &  130 \\ 7.964E11  &  4.759E35  &  41   &  132 \\ 
9.728E11  &  5.812E35  &  41   &  135 \\ 1.196E12  &  7.143E35  &  42   &  137 \\ 1.471E12  &  8.786E35  &  42   &  140 \\ 
1.805E12  &  1.077E36  &  43   &  142 \\ 2.202E12  &  1.314E36  &  43   &  146 \\ 2.930E12  &  1.748E36  &  44   &  151 \\ 
3.833E12  &  2.287E36  &  45   &  156 \\ 4.933E12  &  2.942E36  &  46   &  163 \\ 6.482E12  &  3.726E36  &  48   &  170 \\ 
7.801E12  &  4.650E36  &  49   &  178 \\ 9.611E12  &  5.728E36  &  50   &  186 \\ 1.246E13  &  7.424E36  &  52   &  200 \\ 
1.496E13  &  8.907E36  &  54   &  211 \\ 1.778E13  &  1.059E37  &  56   &  223 \\ 2.210E13  &  1.315E37  &  58   &  241 \\ 
2.988E13  &  1.777E37  &  63   &  275 \\ 3.767E13  &  2.239E37  &  67   &  311 \\ 5.081E13  &  3.017E37  &  74   &  375 \\ 
6.193E13  &  3.675E37  &  79   &  435 \\ 7.732E13  &  4.585E37  &  88   &  529 \\ 9.826E13  &  5.821E37  &  100   &  683 \\ 
1.262E14  &  7.468E37  &  117   &  947 \\ 1.586E14  &  9.371E37  &  143   &  1390 \\ 2.004E14  &  1.182E38  &  201   &  2500 \\ 
2.004E14  &  1.182E38  &  201   &  2500 \\ \cline{1-4}
\end{tabular}}} \\
\end{table}

\begin{table}
\hspace{2.5cm}
{\small{\begin{tabular}{|l|l|l|r|r|} \hline
\multicolumn{5}{|c|}{} \\ 
\multicolumn{5}{|c|}{TABLE 2.2} \\
\multicolumn{5}{|c|}{} \\ 
\multicolumn{5}{|c|}{\emph{EQUATION OF STATE}} \\
\multicolumn{5}{|c|}{\emph{FROM BAYM, PETHICK AND SUTHERLAND (1971)}} \\ 
\multicolumn{5}{|c|}{} \\ \hline
&&&&\\
mass density & pressure && mass density & pressure \\
&&&&\\ \cline{1-5}
&&&&\\
$\rho$ ($\gcc$) & $P$ (dyne~cm$^{-2}$) && $\rho$ ($\gcc$) & $P$ (dyne~cm$^{-2}$) \\ 
&&&&\\ \cline{1-5}
&&&&\\
7.86E0     &  1.01E09  & & 1.316E09   &  5.036E26   \\ 
7.90E0     &  1.01E10  & & 1.657E09   &  6.860E26   \\ 
8.15E0     &  1.01E11  & & 2.626E09   &  1.272E27   \\ 
1.16E01    &  1.21E12  & & 4.164E09   &  2.356E27   \\ 
1.64E01    &  1.40E13  & & 6.601E09   &  4.362E27   \\ 
4.51E01    &  1.70E14  & & 1.046E10   &  7.702E27   \\ 
2.12E02    &  5.82E15  & & 8.312E09   &  5.662E27   \\ 
1.150E03   &  1.90E17  & & 1.318E10   &  1.048E28   \\ 
1.044E04   &  9.744E18  & & 1.659E10   &  1.425E28   \\ 
2.622E04   &  4.968E19   & & 2.090E10   &  1.938E28   \\ 
6.587E04   &  2.431E20   & & 2.631E10   &  2.503E28   \\ 
1.654E05   &  1.151E21   & & 3.313E10   &  3.404E28   \\ 
4.156E05   &  5.266E21   & & 4.172E10   &  4.628E28   \\ 
1.044E06   &  2.318E22   & & 5.254E10   &  5.949E28   \\ 
2.622E06   &  9.755E22   & & 6.617E10   &  8.089E28   \\ 
6.588E06   &  3.911E23   & & 8.332E10   &  1.100E29   \\ 
8.293E06   &  5.259E23   & & 1.049E11   &  1.495E29   \\ 
1.655E07   &  1.435E24   & & 1.322E11   &  2.033E29   \\ 
3.302E07   &  3.833E24   & & 1.664E11   &  2.597E29   \\ 
6.589E07   &  1.006E25   & & 1.844E11   &  2.892E29   \\ 
1.315E08   &  2.604E25   & & 2.096E11   &  3.290E29   \\ 
2.624E08   &  6.676E25   & & 2.640E11   &  4.473E29   \\ 
3.304E08   &  8.738E25   & & 3.325E11   &  5.816E29   \\ 
5.237E08   &  1.629E26   & & 4.188E11   &  7.538E29   \\ 
4.299E11   &  7.805E29   & & 8.301E08   &  3.029E26   \\ 
1.045E09   &  4.129E26   & & &\\ \cline{1-5}
\end{tabular}}}
\end{table}

The  pressure of a  free, Fermi  degenerate electron  gas in  the zero
temperature phase is given by :
\ber P_e &=& \frac{m_e c^2}{\lambda_e^3} \phi(x), \\
\phi(x)   &=&   \frac{1}{8    \pi}\{x(1+x^2)^{1/2}(2x^2/3   -   1)   +
ln[x+(1+x^2)^{1/2}]\} \nonumber
\eer
where  $x$ ($\frac{p_F}{m_e  c}$)  is the  relativistic parameter  and
$\lambda_e$   ($\frac{\hbar}{m_e   c}$)   is  the   electron   Compton
wavelength.  But the mass  density is  given by  the rest-mass  of the
ions,
\beq
\rho = \mu_e m_u n_e = \frac{3}{\pi^2 \lambda_e^3} \mu_e m_u x^3,
\eeq
where $\mu_e$ is  the mean molecular weight, $m_u$  is the atomic mass
unit  and $n_e$  ($\frac{x^3}{3 \pi^2  \lambda_e^3}$) is  the electron
number  density.  To  obtain the  correct equation  of  state, several
corrections  have  to be  incorporated  in  the  above expression  for
pressure.  Firstly, the  electrostatic correction  arises  because the
positively charged ions are not uniformly distributed, but arranged in
a crystal  lattice with lattice sites  having a charge  $Z$ each. This
decreases the energy and the  pressure of the ambient electrons as the
distance between the repelling electrons  is on an average larger than
the  mean  distance  between  nuclei  and  electrons.  Therefore,  the
repulsion  is weaker  than attraction.  In a  non-degenerate  gas, the
ratio between this Coulomb correction to the thermal energy is
\beq
\frac{E_c}{k_BT}  \simeq \frac{Ze^2/ \left<  r \right>  }{k_BT} \simeq
{Ze^2 n_e^{1/3}}{k_BT}
\eeq
and in a degenerate gas when Coulomb energy is comparable to the Fermi
energy we have,
\beq
\frac{E_c}{E_F} \simeq c{Ze^2/ \left< r \right> }{p_F^2/2 mc} \simeq 2
(\frac{1}{3 \pi^3})^{2/3} \frac{Z}{a_0} \frac{1}{n_e^{1/3}}
\eeq
where $a_0  = \frac{\hbar^2}{m_e e^2}$  is the Bohr radius.  When this
correction is taken  into consideration it is found  that the pressure
is modified as $P  = P_e - P_{\rm Coulomb}$, with $P  = 0$ for $\rho =
7.86~\gcc$.  Therefore,  this   is  the  minimum  equilibrium  density
obtained at the very surface  of the neutron star. At higher densities
the most  important correction is due  to the \ibd.  The condition for
the \ibd ($\en + p \rightarrow n + \nu$) is that the kinetic energy of
the electrons be  larger than 1.24~MeV, the mass  difference between a
neutron and a  proton. The $\beta$-decay of a  neutron ($n \rightarrow
\en + p +  \nu$) is blocked when the density is  so large that all the
electron levels  in the Fermi sea are  filled up to the  energy of the
emitted electron.

The  pressure is  obtained by  the thermodynamic  relation $P  = n_B^2
\frac{\partial (\epsilon/n_B)}{\partial n_B}$, where $\epsilon$ is the
total free-energy  density including the rest-mass of  the baryons and
$n_B$  is the  baryon  number  density. When  one  species of  nuclide
changes to another  as $n_B$ changes there is  a phase transition with
an  accompanying  discontinuity  in  $n_B$.   Since there  can  be  no
discontinuity in the  pressure and the temperature inside  the star to
obtain the  equilibrium composition and  the equation of  state Gibbs'
free energy  should be  minimized. In this  density range  usually the
equation  of  state obtained  by  \citeN{baym71a}, incorporating  the
results of \citeN{feyn49} in the range $7.9~\gcc < \rho < 10^4~\gcc$,
is  used.  In table  [2.2] the  equation of  state (pressure  vs. mass
density) as calculated by them is shown.

\subsection{inner crust : $4.0 \times 10^{11}~\gcc < \rho < 2.8 \times 10^{14}~\gcc$}

\bef
\begin{center}{\mbox{\epsfig{file=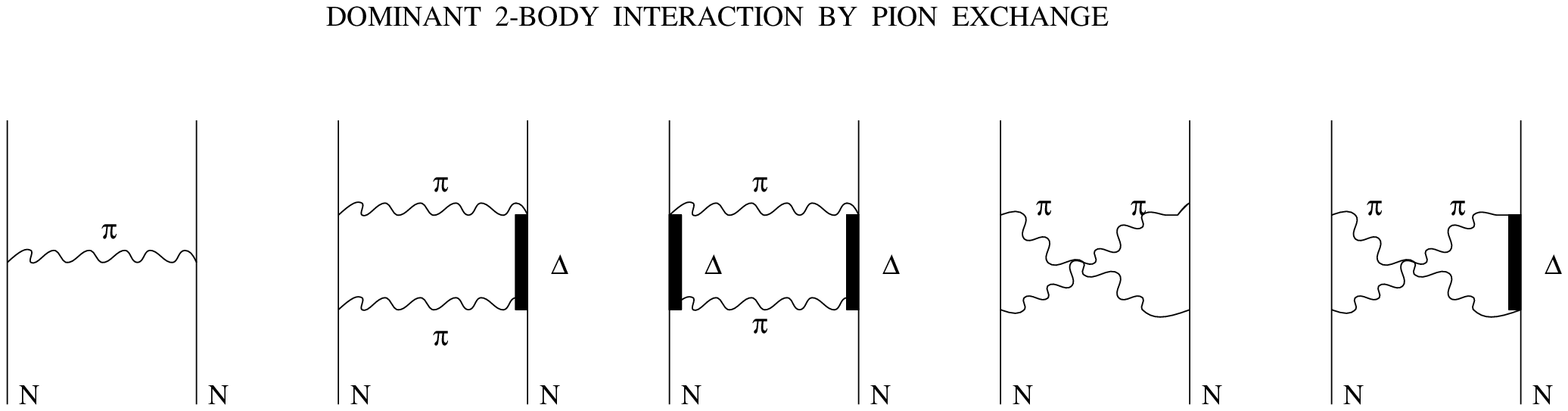,width=100pt,angle=-90}}}\end{center}
\caption[two-nucleon     interactions]{Feynman     diagrams    showing
two-nucleon interaction  by exchange of pions.  The  single lines, the
wavy  lines and  the thick  lines  stand for  the world  lines of  the
nucleons, the pions and the virtual $\Delta$ particles respectively.}
\label{fnnscat}
\eef

At the lower edge of this regime, the neutron energy levels within the
nuclei  merge into  a continuum  and they  drip out  of the  nuclei to
comprise a  free neutron gas  co-existing with the crystal  lattice of
the  neutron-rich  nuclei.  The  problem  of  calculating an  accurate
equation of  state here is that the  correct nucleon-nucleon potential
is  not  known  to any  degree  of  certainty,  and that  the  quantum
many-body   techniques   are  not   quite   adequate   to  solve   the
Schr\"{o}dinger equation given the potential. In this regime, with the
proton-to-neutron ratio ranging from  0.1 to 0.3, extrapolations based
on   semi-empirical  mass   formula  is   used.  The   work   done  by
\citeN{baym71b} took  care of the  fact that the neutrons  inside and
outside the  nuclei behave in  a similar fashion. The  nuclear surface
energy is  modified by the free  neutron gas outside. By  using a {\em
compressible  liquid drop model}  of nuclei  they minimized  the total
energy,  for  a  fixed value  of  the  baryon  density $n_B$,  for  an
equilibrium configuration. Free neutrons supply an increasingly larger
fraction of the pressure as the density increases.

But these  earlier works did not  take the nuclear  shell effects into
account, as  was later done  by \citeN{nege73}.  The main  feature of
this work has been the  modeling of the nucleon-nucleon interaction by
taking  into  consideration   the  two-body  interactions  only.   The
dominant  two-body  interaction,  by  exchange  of  pions,  come  from
processes  like the ones  in figure  [\ref{fnnscat}]. The  equation of
state in the  above mentioned density range is  given by the following
interpolation formula :
\ber
E_t &=& m_n + \Sigma^7_{i=0} c_i x^{i-1} \\
\rho &=& \frac{n_b E_T}{c^2} \\
P &=& n_b^2 \frac{\partial E_T}{\partial n_b}
\eer
where  $m_n$  is the  mass  of  the neutron  and  $x  = ln(n_b  \times
10^{-35})$,  $n_b$  being the  baryon  number  density. The  constants
$c_i$s are given in table [2.3].

Another important fact is that  at these densities the solid state and
the nuclear energies are comparable.  Hence they require to be treated
on  equal footing.   This leads  to  the possibility  of existence  of
non-spherical nuclei.   It has been  shown by \citeN{lorn93}  that at
sub-nuclear  densities nuclei  with rod  or disc  shape are  likely to
exist. If  they indeed do, that  will introduce a  modification in the
equation of state  in these density ranges and may  in turn affect the
structure   and  other   physical  properties   (like   the  transport
coefficients or the thermal evolution) of a neutron star.

\begin{table}
\hspace{2.5cm}
{\small{\begin{tabular}{|l|r|} \hline
\multicolumn{2}{|c|}{} \\ 
\multicolumn{2}{|c|}{TABLE 2.3} \\ 
\multicolumn{2}{|c|}{} \\ 
\multicolumn{2}{|c|}{\em{COEFFICIENTS FOR CALCULATION OF}} \\
\multicolumn{2}{|c|}{\em{THE EQUATION OF STATE}} \\
\multicolumn{2}{|c|}{\em{FROM NEGELE AND VAUTHERIN (1973)}} \\ 
\multicolumn{2}{|c|}{} \\ \hline
&\\
{\bf i} & {\bf c$_i$} (ground state) \\ 
&\\ \cline{1-2}
&\\
0 & $- 4.0$ \\ 
1 & $2.8822899 \times 10^{-1}$ \\
2 & $5.9150523 \times 10^{-1}$ \\
3 & $9.0185940 \times 10^{-2}$ \\
4 & $- 1.1025614 \times 10^{-1}$ \\
5 & $2.9377479 \times 10^{-2}$ \\
6 & $- 3.2618465 \times 10^{-3}$ \\
7 & $1.3543555 \times 10^{-4}$ \\ 
&\\ \cline{1-2}
\end{tabular}}} 
\end{table}

\subsection{the core : $\rho > 2.8 \times 10^{14}~\gcc$ }

The  theories  at  these  densities  are  faced  with  a  plethora  of
problems. There is a lack of understanding of the correct form for the
nucleon-nucleon  potential  added  to   the  fact  that  there  is  no
laboratory data available  to test the theory against.  As the density
increases the effects of  relativity becomes important. Also at higher
densities it is essential  to incorporate the non-nucleonic degrees of
freedom as mesons and higher  mass baryons make appearance. At extreme
high  densities there  may probably  occur a  phase transition  to the
quark phase and then quark  and gluonic degrees of freedom should also
have  to  be  taken  into  account. And  even  at  nuclear  saturation
densities the predictions regarding the possible phase transition to a
superfluid/superconducting phase are not without uncertainties. One of
the  major problems  in  trying to  understand  the nuclear  phenomena
inside a neutron star is due to the huge neutron excess. The parameter
$\delta = (N-Z)/(N+Z)$, used to denote the neutron excess is about 1/4
in terrestrial nuclei.  In neutron stars, starting from  that value at
the surface $\delta$ becomes as large as unity deep in the interior of
the star.  Any extrapolation,  that requires going  up by a  factor of
four, is bound to be unreliable.

\bef
\begin{center}{\mbox{\epsfig{file=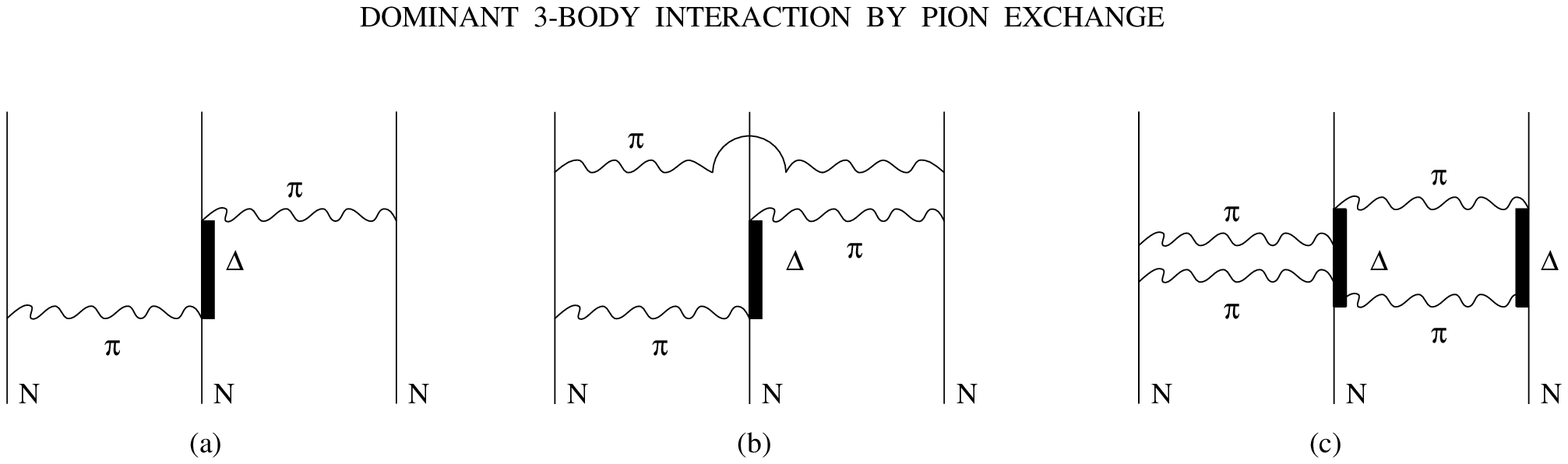,width=120pt,angle=-90}}}\end{center}
\caption[three-nucleon    interactions]{Feynman    diagrams    showing
three-nucleon interaction by exchange of pions.  The single lines, the
wavy  lines and  the thick  lines  stand for  the world  lines of  the
nucleons, the pions and the virtual $\Delta$ particles respectively.}
\label{fnnnscat}
\eef

Nevertheless,  we have  reasonable estimates  for  the nucleon-nucleon
interaction  based   on  the  scattering  data   from  the  laboratory
experiments. But  these provide information only  about the long-range
behaviour  of the  potential. There  is no  handle on  the short-range
behaviour,  which   is  likely  to   dominate  at  the   neutron  star
densities. From  the data  on the binding  energy of light  nuclei the
microscopic Hamiltonian  is modelled. But aspects  of interaction that
are relatively  unimportant for  such light nuclei  (deuterium, He$^3$
etc.)  may play  significant roles  in a  neutron star.  Of particular
importance are the three and  higher body interactions. At long range,
the most important three-body interaction  is that due to the exchange
of pions,  where one of the  nucleons becomes converted  to a $\Delta$
and  then de-excites  back by  exchanging  another pion  with a  third
nucleon  (figure [\ref{fnnnscat}a]).  At  short-range other  processes
like   those  in   figures  [\ref{fnnnscat}b]   and  [\ref{fnnnscat}c]
dominate.

To  summarize,   we  mention  the  three  equations   of  state  which
incorporate some of the recent developments, following \citeN{wiri88}
(though more  recent calculations  for the equation  of state  in this
density range has been  performed, see for example \citeN{prak92}. In
this  paper, they  compare the  equations of  state obtained  by using
different types  of two-body and three-body potentials  as against the
equation  of  state  for  a  pure, free  neutron  gas.   The  two-body
potentials used  by them  are AV14 (Argonne  14) and UV14  (Urbana 14)
both  of  which fit  the  scattering data  well  but  differ in  their
short-range  behaviour.   These   are  modified  with  the  three-body
interaction  UVII which  is adjusted  to fit  the binding  energies of
He$^3$  and He$^4$.   The  other three-body  interaction  TNI is  less
complete  in  taking  into  account  all  aspects  of  the  three-body
interaction. It  is observed  that, at $\rho  \sim 3-4  \rho_s$, where
$\rho_s$  is the  saturation  nuclear density,  the  total energy  per
particle  differs by  an  amount small  compared  to the  mass of  the
neutrons from that  obtained by using a free-neutron  gas.  It is also
seen  that  the energy  per  particle depends  on  the  choice of  the
two-body  as well as  the three-body  interaction. Lastly,  though the
energy does not  change much, the pressure, given by  the slope of the
energy curve ($P(\rho)  \sim \frac{\partial E(\rho)}{\partial \rho}$),
is very different  for different equations of state.   In table [2.4],
is the data from \citeN{wiri88} for the three equations of state.

\begin{table}
{\small{\begin{tabular}{|l|l|l|l|l|l|r|} \hline
\multicolumn{7}{|c|}{} \\ 
\multicolumn{7}{|c|}{TABLE 2.4} \\ 
\multicolumn{7}{|c|}{} \\ 
\multicolumn{7}{|c|}{\em{DATA FROM WIRINGA, FIKS AND FABROCINI (1971)}} \\ 
\multicolumn{7}{|c|}{} \\ \hline
& \multicolumn{2}{|c|}{} & \multicolumn{2}{|c|}{} & \multicolumn{2}{|c|}{} \\
& \multicolumn{2}{|c|}{AV14 + UVII} & \multicolumn{2}{|c|}{UV14+UVII} & \multicolumn{2}{|c|}{UV14+TNI} \\
& \multicolumn{2}{|c|}{} & \multicolumn{2}{|c|}{} & \multicolumn{2}{|c|}{} \\ \hline
&&&&&& \\
mass & proton & energy & proton & energy & proton & energy \\
density & fraction & density & fraction & density & fraction & density \\
&&&&&& \\
$\rho$ & $x(\rho)$ & $E(\rho,x)$ & $x(\rho)$ & $E(\rho,x)$ & $x(\rho)$ & $E(\rho,x)$ \\ 
&&&&&&\\ \cline{1-7}
&&&&&& \\
$fm^{-3}$ & & Mev/nucleon & & Mev/nucleon & & Mev/nucleon \\ 
&&&&&&\\ \cline{1-7}
&&&&&& \\
0.07 & 0.017 & 7.35 & 0.019 & 8.13 & 0.026 & 5.95 \\
0.08 & 0.019 & 7.94 & 0.021 & 8.66 & 0.029 & 6.06 \\
0.10 & 0.023 & 8.97 & 0.025 & 9.79 & 0.033 & 6.40 \\
0.125 & 0.027 & 10.18 & 0.030 & 11.06 & 0.037 & 7.17 \\
0.15 & 0.031 & 11.43 & 0.035 & 12.59 & 0.042 & 8.27 \\
0.175 & 0.036 & 12.74 & 0.042 & 14.18 & 0.047 & 9.70 \\
0.20 & 0.044 & 14.12 & 0.052 & 15.92 & 0.051 & 11.55 \\
0.25 & 0.051 & 16.96 & 0.069 & 20.25 & 0.057 & 16.29 \\
0.30 & 0.051 & 20.48 & 0.079 & 25.78 & 0.059 & 22.19 \\
0.35 & 0.052 & 24.98 & 0.087 & 32.60 & 0.060 & 28.94 \\
0.40 & 0.055 & 30.44 & 0.097 & 40.72 & 0.060 & 36.60 \\
0.50 & 0.060 & 45.15 & 0.116 & 61.95 & 0.051 & 56.00 \\
0.60 & 0.077 & 66.40 & 0.132 & 90.20 & 0.039 & 79.20 \\
0.70 & 0.099 & 93.60 & 0.155 & 126.20 & 0.023 & 106.10 \\
0.80 & 0.101 & 132.10 & 0.172 & 170.50 & 0.005 & 135.50 \\
1.00 & 0.094 & 233.00 & 0.177 & 291.10 & 0.0009 & 200.9 \\
1.25 & 0.066 & 410.00 & 0.122 & 501.00 & 0.00 & 294.00 \\
1.50 & 0.014 & 635.00 & 0.026 & 753.00 & 0.00 & 393.00 \\ 
&&&&&&\\ \cline{1-7}
\end{tabular}}} 
\end{table}

A  combination of  the Baym,  Pethick \&  Sutherland (BPS),  Negele \&
Vautherin (NV) and Wiringa, Fiks \& Fabrocini (WFF) equations of state
in  the respective  density  ranges  seem to  be  the most  acceptable
considering all  the uncertainties mentioned above.  In our subsequent
calculation  of the  structure of  a neutron  star we  shall  use this
combination  as our starting  point.  Amongst  the three  equations of
state  given by  Wiringa  et al.  we  have used  only  the second  one
mentioned as UV14+UVII in the discussion above.

\section{mass and density profile of a neutron star}
\label{sprofile}

In this thesis  we investigate the temporal behaviour  of the magnetic
fields  assuming  a  crustal   current.   This  requires  an  accurate
knowledge  of  the various  transport  coefficients (most  importantly
thermal and electrical conductivity)  in the crust. Therefore, we need
an accurate  density profile, particularly in the  low density crustal
regions,   to   obtain  the   radial   behaviour   of  the   transport
coefficients.  The  mass  and  density profiles  for  a  non-rotating,
self-gravitating  object are obtained  by integrating  the hydrostatic
pressure balance equation
\beq
\frac{dP(r)}{dr} = - \frac{GM(r)\rho(r)}{r^2}, \label{edPdr1}
\eeq
along with the equation of mass distribution,
\beq
\frac{dM(r)}{dr} = 4 \pi r^2 \rho(r), \label{edMdr}
\eeq
where $P(r), M(r) and \rho(r)$ are the pressure, mass and density at a
given     radius    $r$     and    $G$     is     the    gravitational
constant. Equation[\ref{edPdr1}] is  modified, when effects of general
relativity is incorporated, to :
\beq
\frac{dP(r)}{dr} =  - \frac{G(M(r)  + 4 \pi  r^3 P(r)  /c^2)(\rho(r) +
P(r)/c^2)}{r^2(1 - \frac{2G M(r)}{r^2 c^2})}, \label{edPdr}
\eeq
where $c$  is the speed  of light. This  is known as the  TOV equation
after Tolman, Oppenheimer and Volkoff~\cite{oppn39}.  A measure of the
importance of  general relativity is  given by the  quantity $\epsilon
\sim \frac{GM}{Rc^2}$ for a self-gravitating body of rest mass $M$ and
total radius $R$. For $\epsilon << 1$, the effect of relativity can be
neglected. Putting in the typical numbers for a neutron star we obtain
$\epsilon$ to  be close  to 1. Therefore,  to obtain  the mass-density
profile of  a neutron star  it is required  to solve the  TOV equation
instead  of the  Newtonian hydrostatic  equation.  We  solve equations
[\ref{edMdr}] and [\ref{edPdr}] numerically.  The equation of state we
use for  this structure  calculation is of  moderate stiffness  and is
given by \citeN{baym71a} in the  density range $10^6~\gcc < \rho < 4.0
\times  10^{11}~\gcc$,  by \citeN{nege73}  in  the  range $4.0  \times
10^{11}~\gcc < \rho <  2.8 \times 10^{14}~\gcc$, and by \citeN{wiri88}
in  the   range  $2.8  \times   10^{14}~\gcc  <  \rho$.    In  figures
[\ref{feos_bps}], [\ref{feos_nv}], [\ref{feos_wff}] - the pressure vs.
density as obtained in these three ranges have been plotted.

Though the  structure calculations have been performed  by many people
(see for  example~\citeNP{wiri88}) an accurate density  profile in the
low  density regime  of the  crust, has  been lacking.   Therefore, we
undertook  the task  of obtaining  the density  profile for  a typical
neutron star,  by integrating the TOV  equation, using above-mentioned
equations of state.   It must be mentioned here that  in a recent work
\citeN{datt95}  have performed  detailed calculations  of  the crustal
density profile of  neutron stars for a number  of equations of state.
One ought  to note that the  equations of state  for different density
regimes  are not  exactly  matched at  the  boundaries.  So  we use  a
smoothing procedure  ensuring the continuity  of the pressure  and the
pressure gradient at each  boundary.  This smoothed composite equation
of state is plotted in figure [\ref{feos_all}].

\bef
\begin{center}{\mbox{\epsfig{file=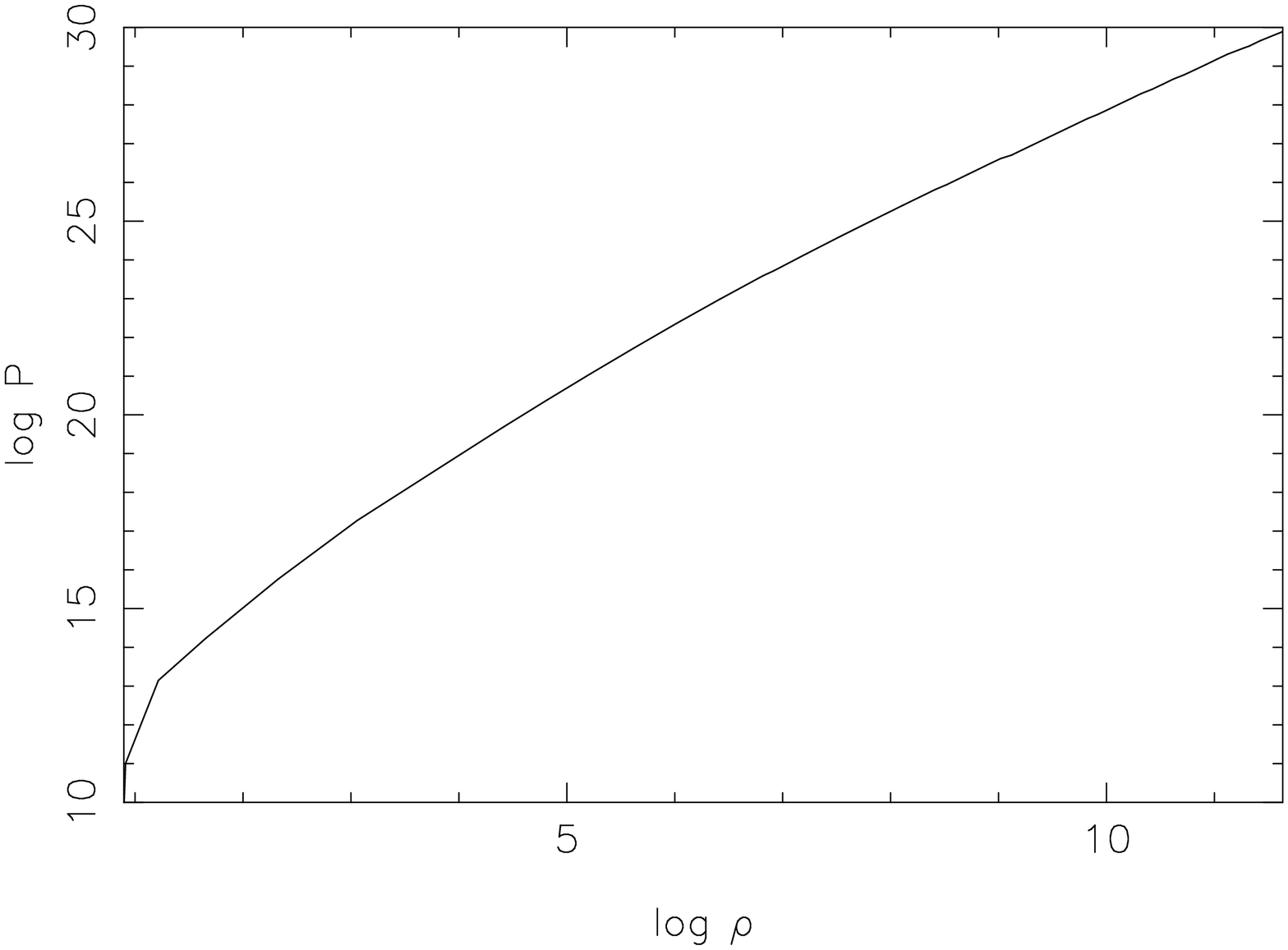,width=235pt,angle=-90}}}\end{center}
\caption[Baym-Pethick-Sutherland     equation    of    state]{Pressure
vs. Density from \citeN{baym71a}.}
\label{feos_bps}
\eef
\bef
\begin{center}{\mbox{\epsfig{file=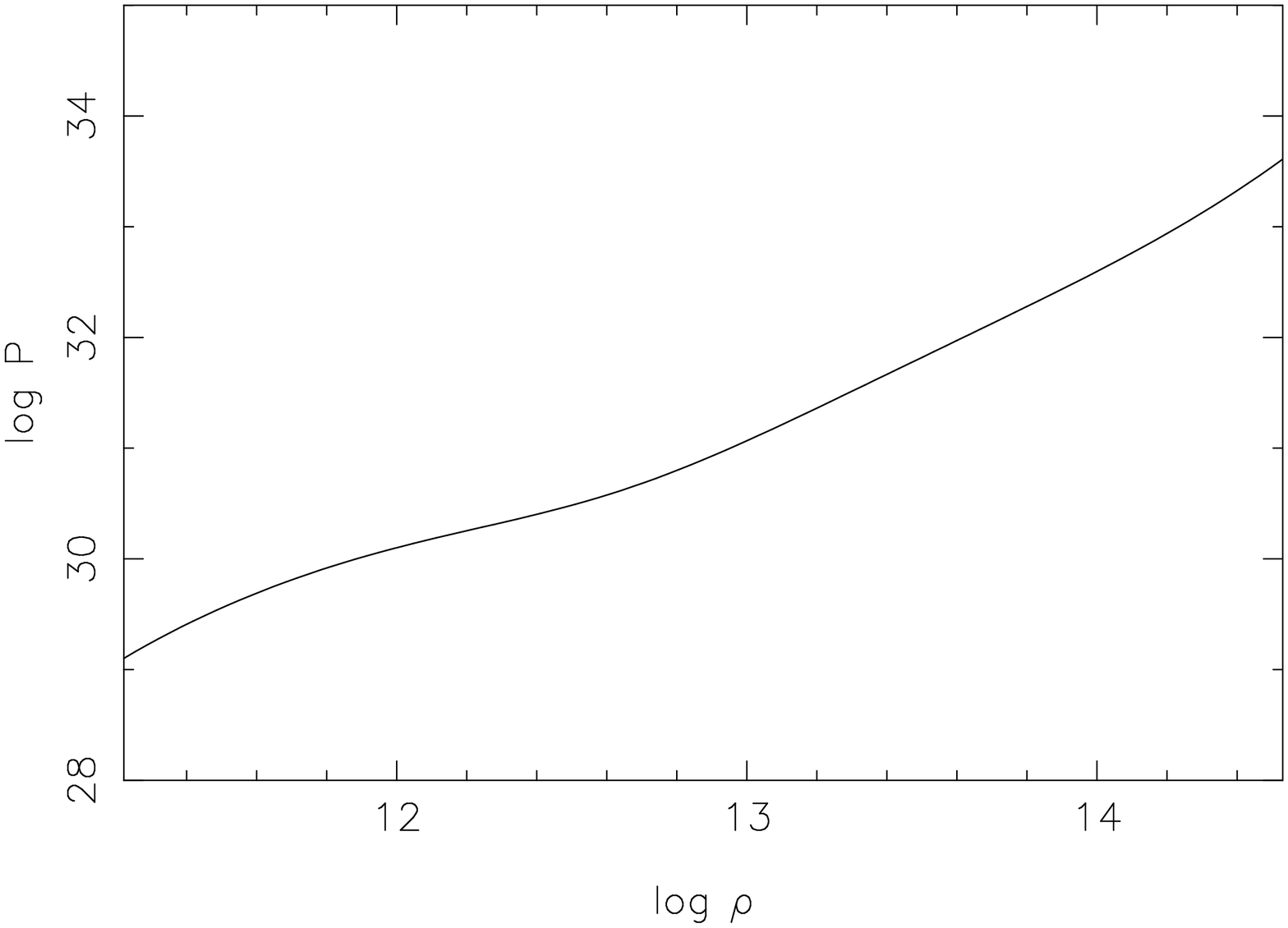,width=235pt,angle=-90}}}\end{center}
\caption[Negele-Vautherin equation of state]{Pressure vs. Density from
\citeN{nege73}.}
\label{feos_nv}
\eef
\bef
\begin{center}{\mbox{\epsfig{file=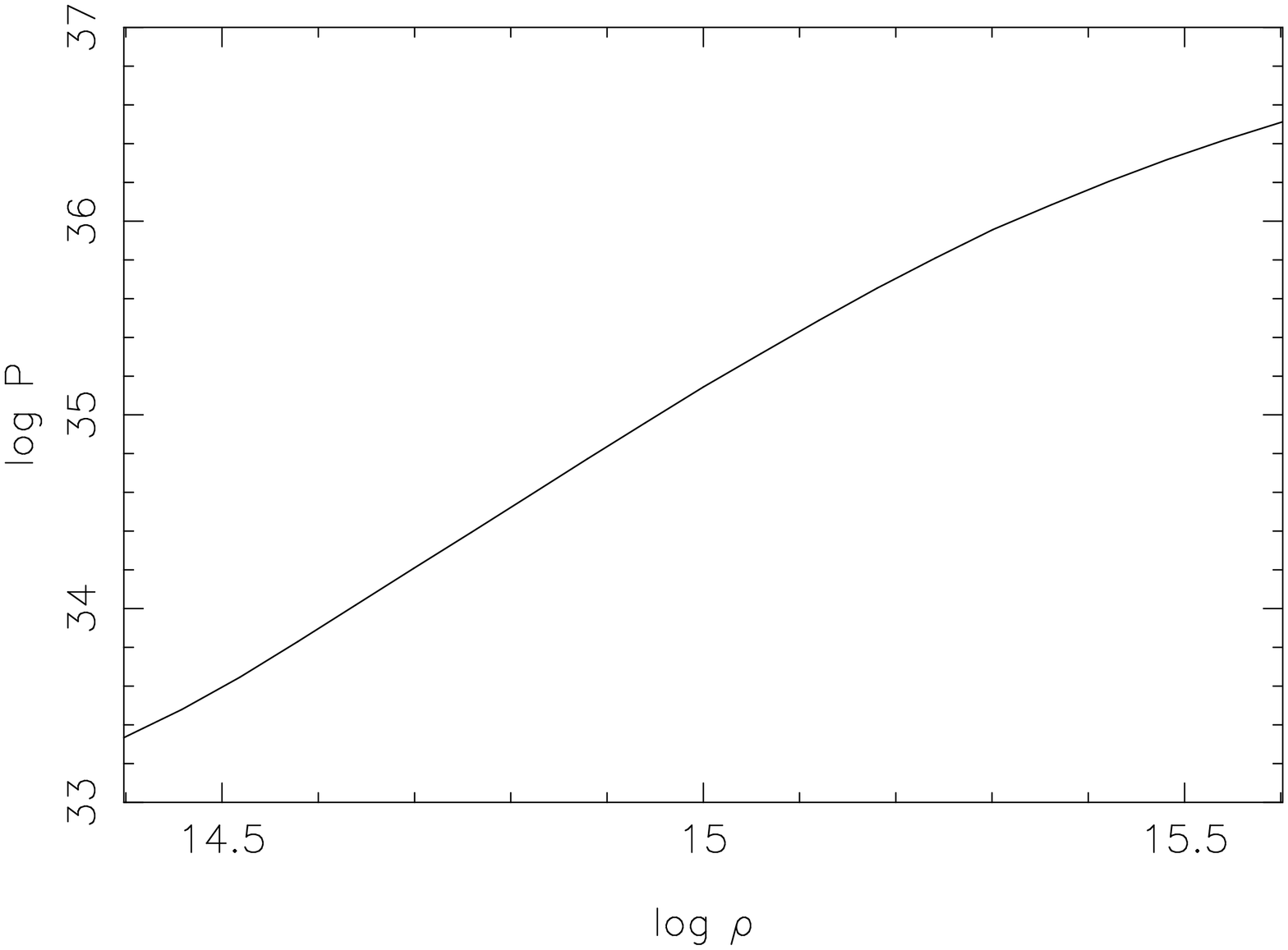,width=235pt,angle=-90}}}\end{center}
\caption[Wiringa-Fiks-Fabrocini     equation     of    state]{Pressure
vs. Density from \citeN{wiri88}.}
\label{feos_wff}
\eef
\bef
\begin{center}{\mbox{\epsfig{file=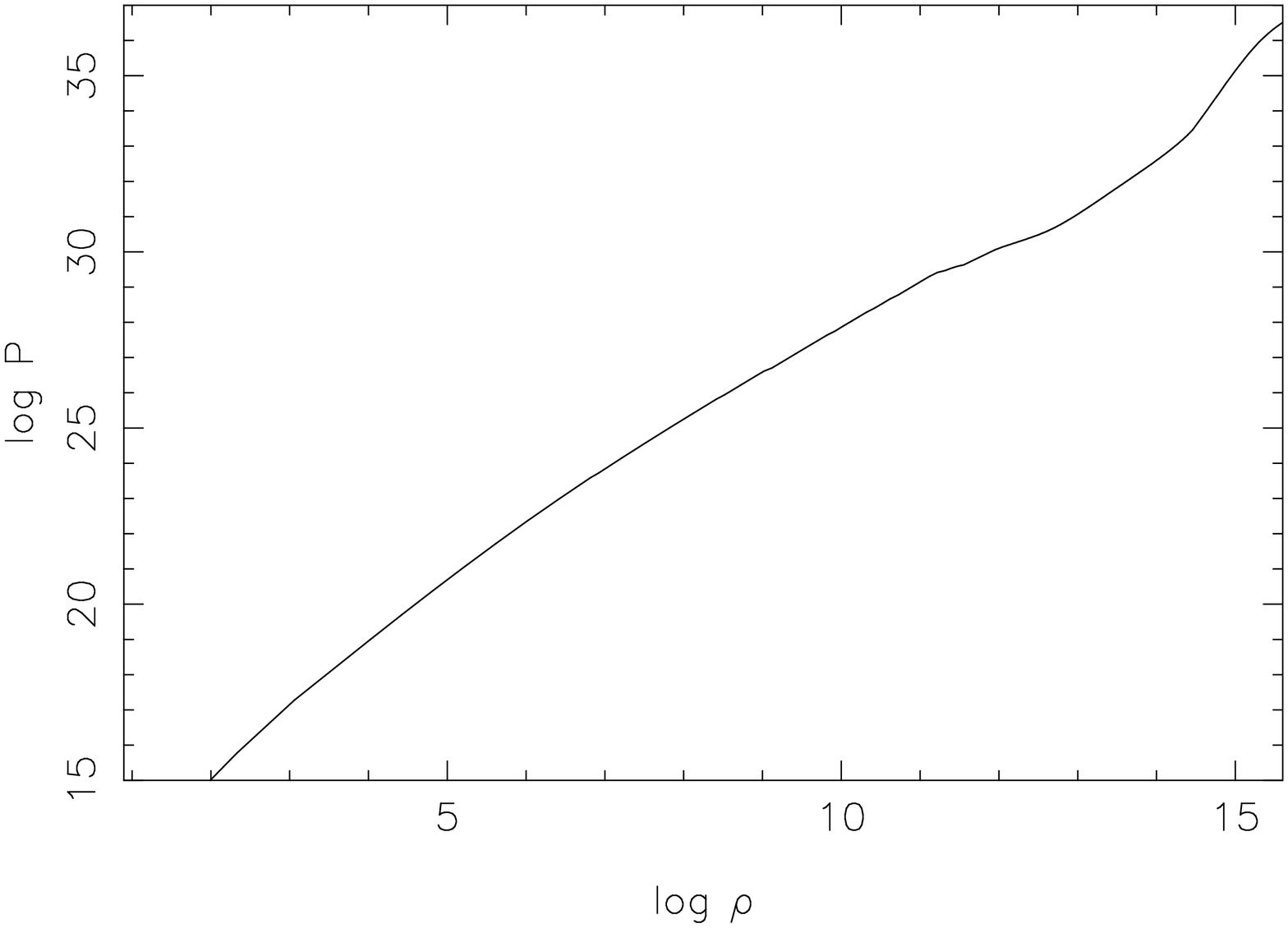,width=235pt,angle=-90}}}\end{center}
\caption[smoothed  out  equation  of  state]{Pressure vs.   Density  :
Smoothed out over the whole density range.}
\label{feos_all}
\eef

We  integrate the  TOV  equation starting  from  a particular  central
density and  corresponding central pressure at zero  radius. The other
boundary condition  at the  centre is  that of zero  mass. The  set of
coupled second order ordinary  differential equations are solved using
a fourth order  Runge-Kutta scheme of differencing.  We  have used the
ordinary differential equation solver programs by \citeN{pres92}.  for
the  Runge-Kutta  driver  with  an  adaptive  step-size  control.  The
adaptive   step-size   control  is   essential   in  integrating   the
mass-density  profile since  both the  functions show  extremely steep
behaviour  near  the surface,  at  the  low  density regime.   In  our
computation  the surface corresponds  to a  density of  7.86~$\gcc$ as
that is the minimum density  obtained in the neutron star. The density
and the mass profiles for a  neutron star of total mass 1.4$\msun$ are
plotted in figures [\ref{frho_r}] and [\ref{fm_r}] respectively.

\bef
\begin{center}{\mbox{\epsfig{file=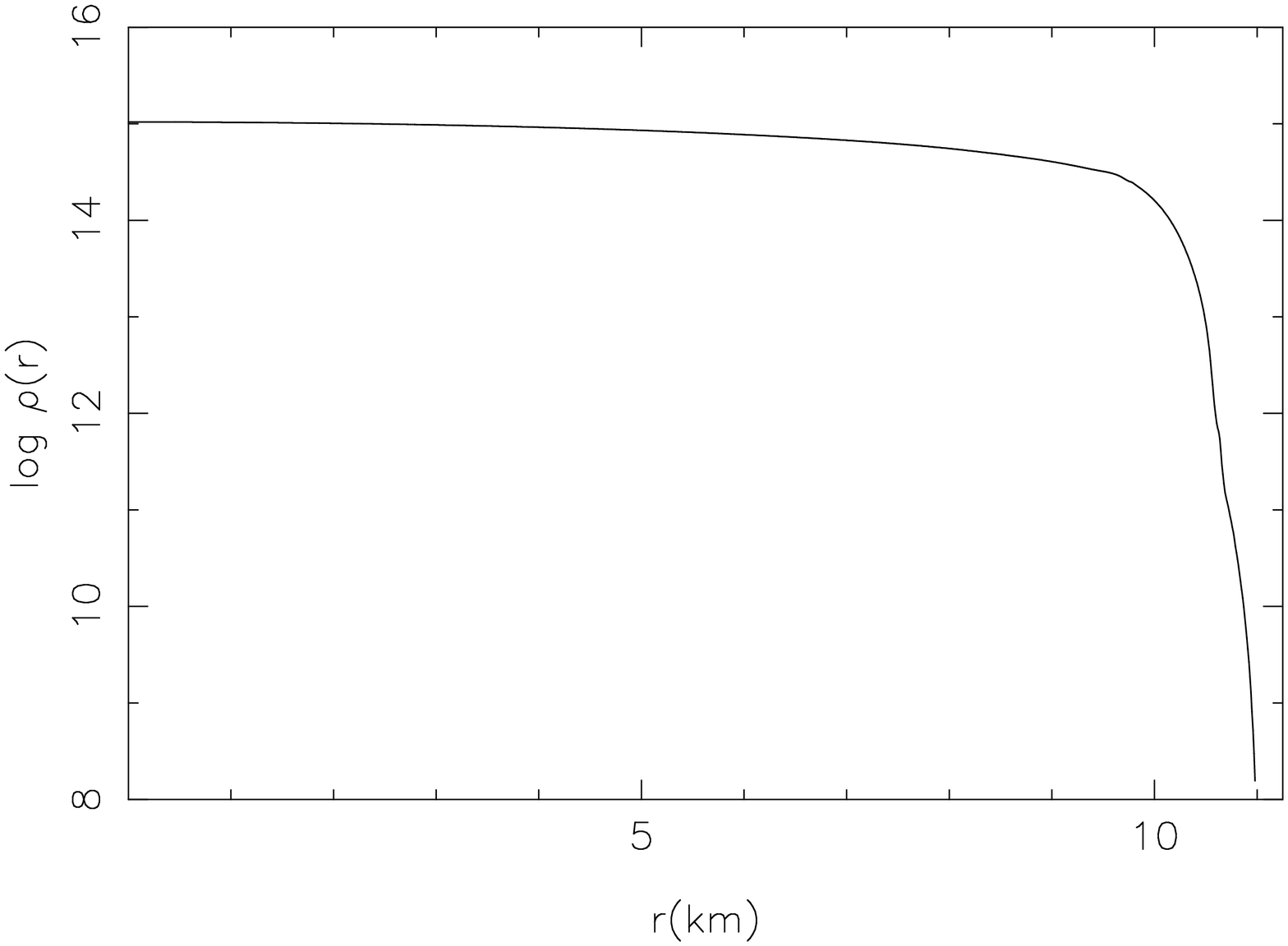,width=235pt,angle=-90}}}\end{center}
\caption[neutron star  density profile]{Density  vs. Radius for  a 1.4
$\msun$ neutron star.}
\label{frho_r}
\eef
\bef
\begin{center}{\mbox{\epsfig{file=m_r.ps,width=235pt,angle=-90}}}\end{center}
\caption[neutron star mass profile]{Mass  vs. Radius for a 1.4 $\msun$
neutron star.}
\label{fm_r}
\eef
\bef
\begin{center}{\mbox{\epsfig{file=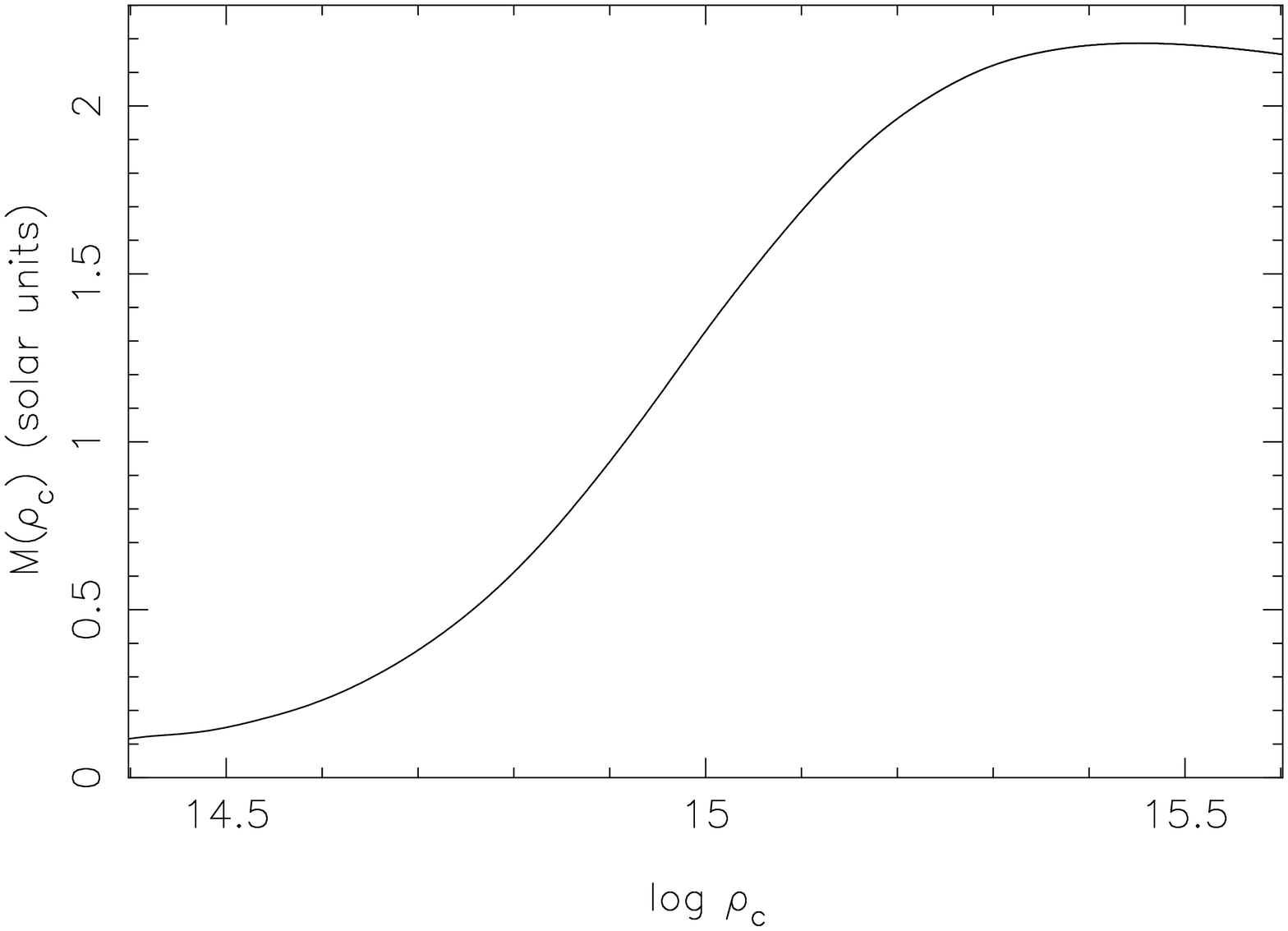,width=235pt,angle=-90}}}\end{center}
\caption[total  mass  as a  function  of  central density]{Total  Mass
vs. Central Density}
\label{fm_rhoc}
\eef
\bef
\begin{center}{\mbox{\epsfig{file=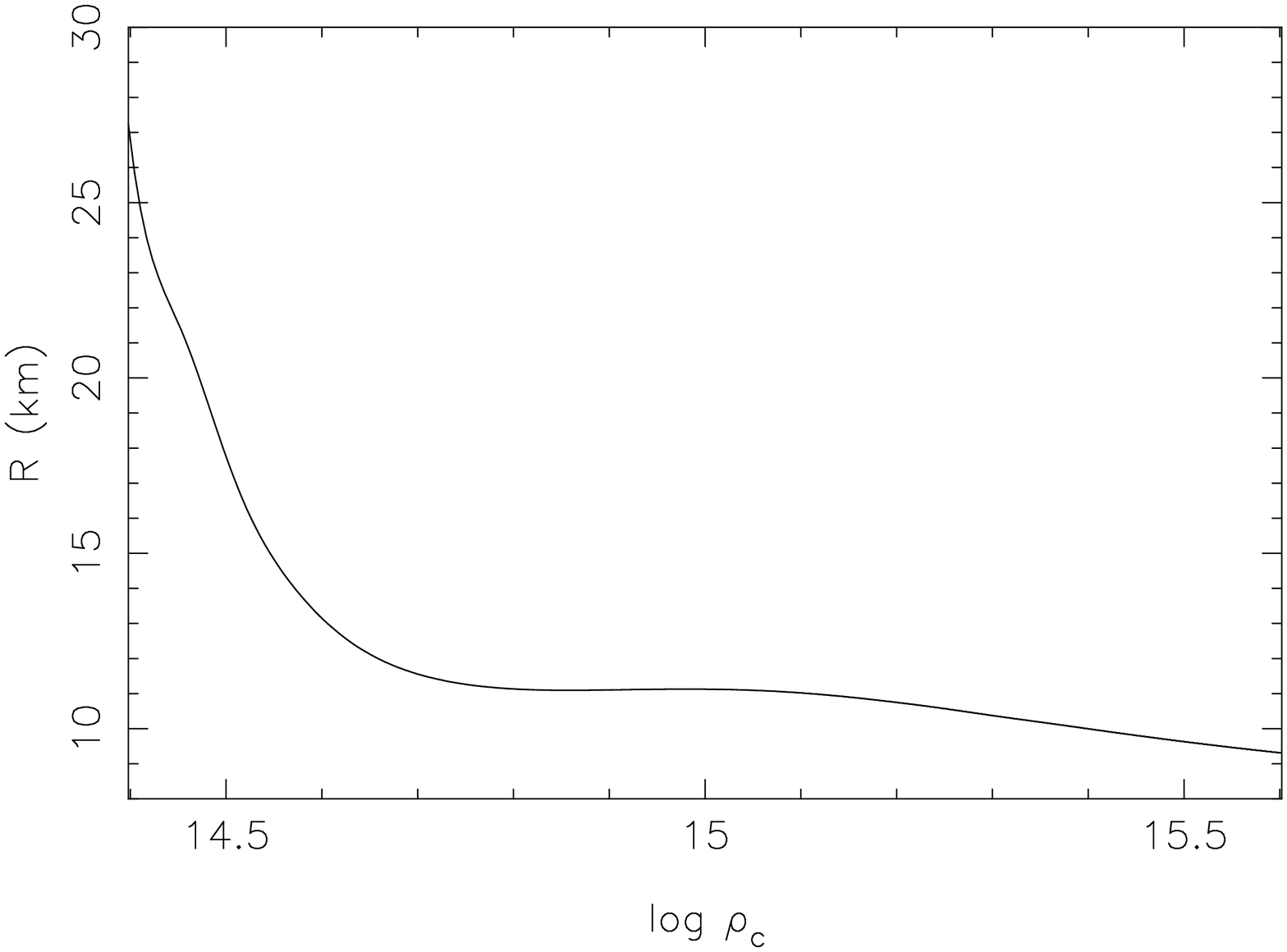,width=235pt,angle=-90}}}\end{center}
\caption[radius  as  a  function  of central  density]{Stellar  Radius
vs. Central Density}
\label{fr_rhoc}
\eef
\bef
\begin{center}{\mbox{\epsfig{file=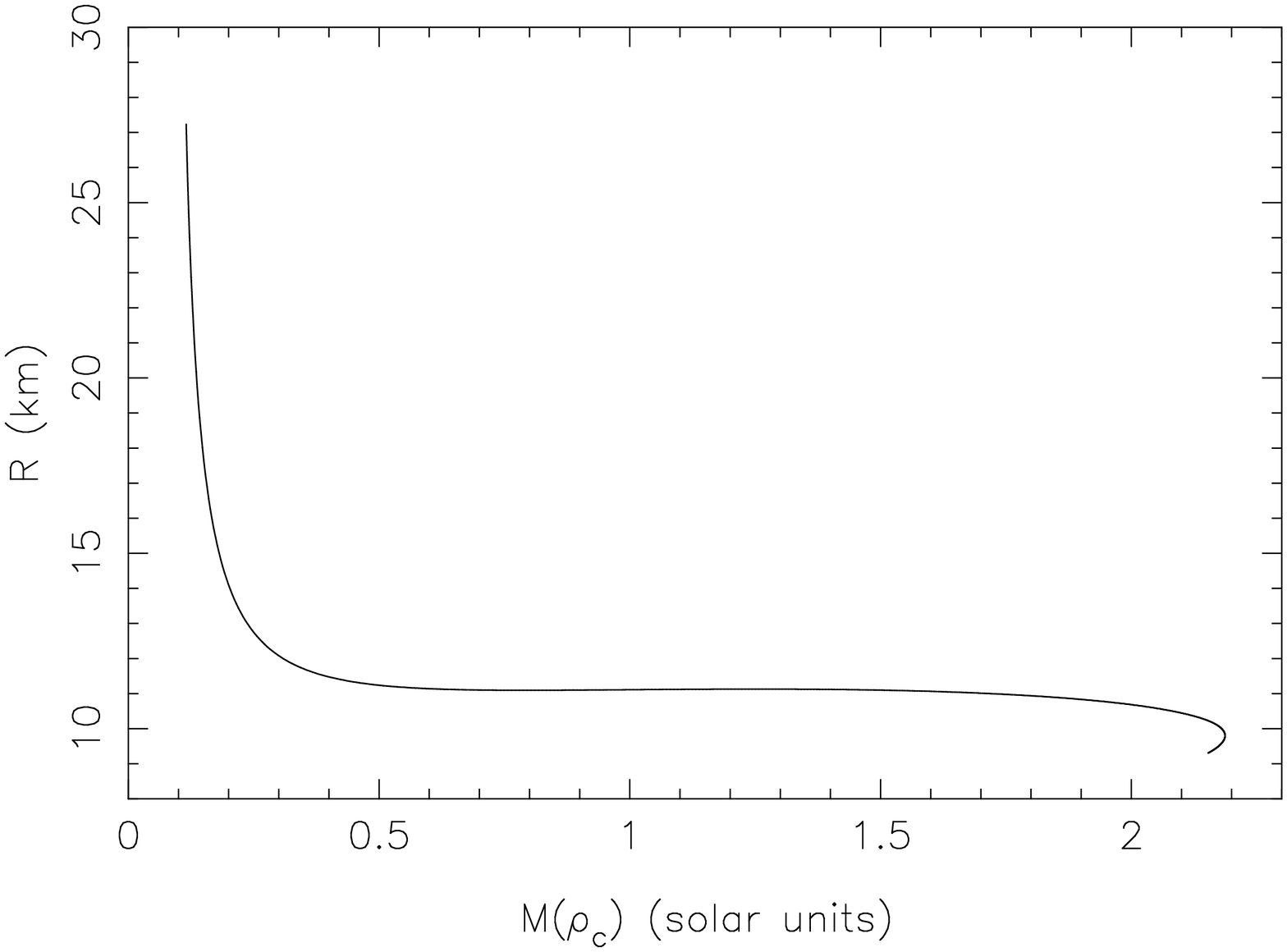,width=235pt,angle=-90}}}\end{center}
\caption[radius as a function  of total mass]{Stellar Radius vs. Total
Mass}
\label{fr_mstar}
\eef

For different central densities, the  total mass and the radius of the
star  differ quite  a lot.  The variation  of the  total mass  and the
radius   with   central  density   have   been   plotted  in   figures
[\ref{fm_rhoc}] and [\ref{fr_rhoc}].  And the mass-radius relation for
a   set    of   neutron   stars    state   is   plotted    in   figure
[\ref{fr_mstar}]. This clearly shows  the existence of a maximum mass,
which  could also  be seen  (albeit  with some  difficulty) in  figure
[\ref{fm_rhoc}].  This  maximum mass of about  2.2 $\msun$ corresponds
to a central density of $\sim 2.5 \times 10^{15}~\gcc$ and a radius of
10 km.

\bef
\begin{center}{\mbox{\epsfig{file=dm_z.ps,width=235pt,angle=-90}}}\end{center}
\caption[mass  in the crust]{Mass  (in solar  units) of  the overlying
layers vs.   Depth (from the  surface) of the  layer in a  1.4 $\msun$
neutron star.}
\label{fdm_z}
\eef
\bef
\begin{center}{\mbox{\epsfig{file=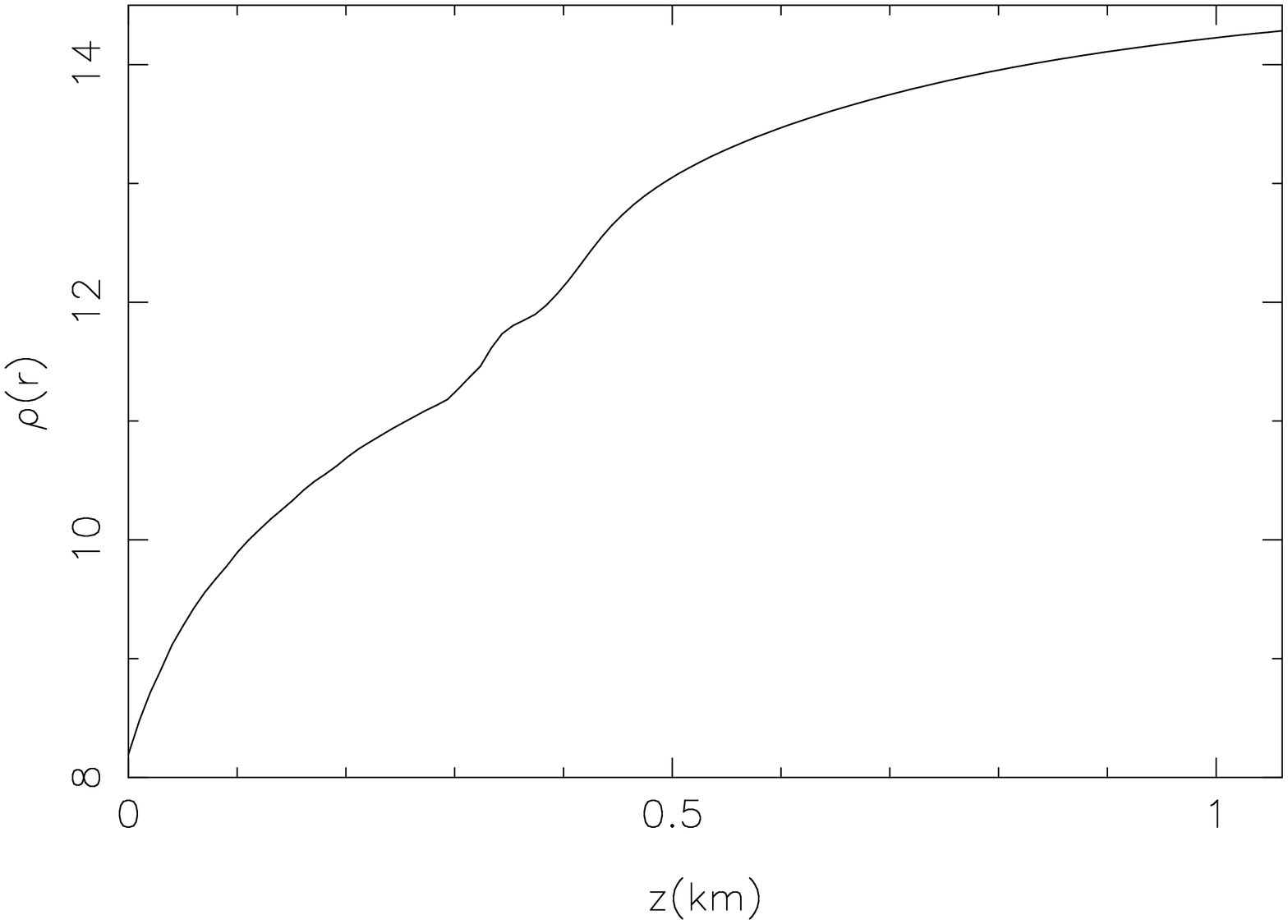,width=235pt,angle=-90}}}\end{center}
\caption[density in the crust]{Density  vs. Depth (from the surface) a
  1.4 $\msun$ neutron star.}
\label{frho_z}
\eef
\bef
\begin{center}{\mbox{\epsfig{file=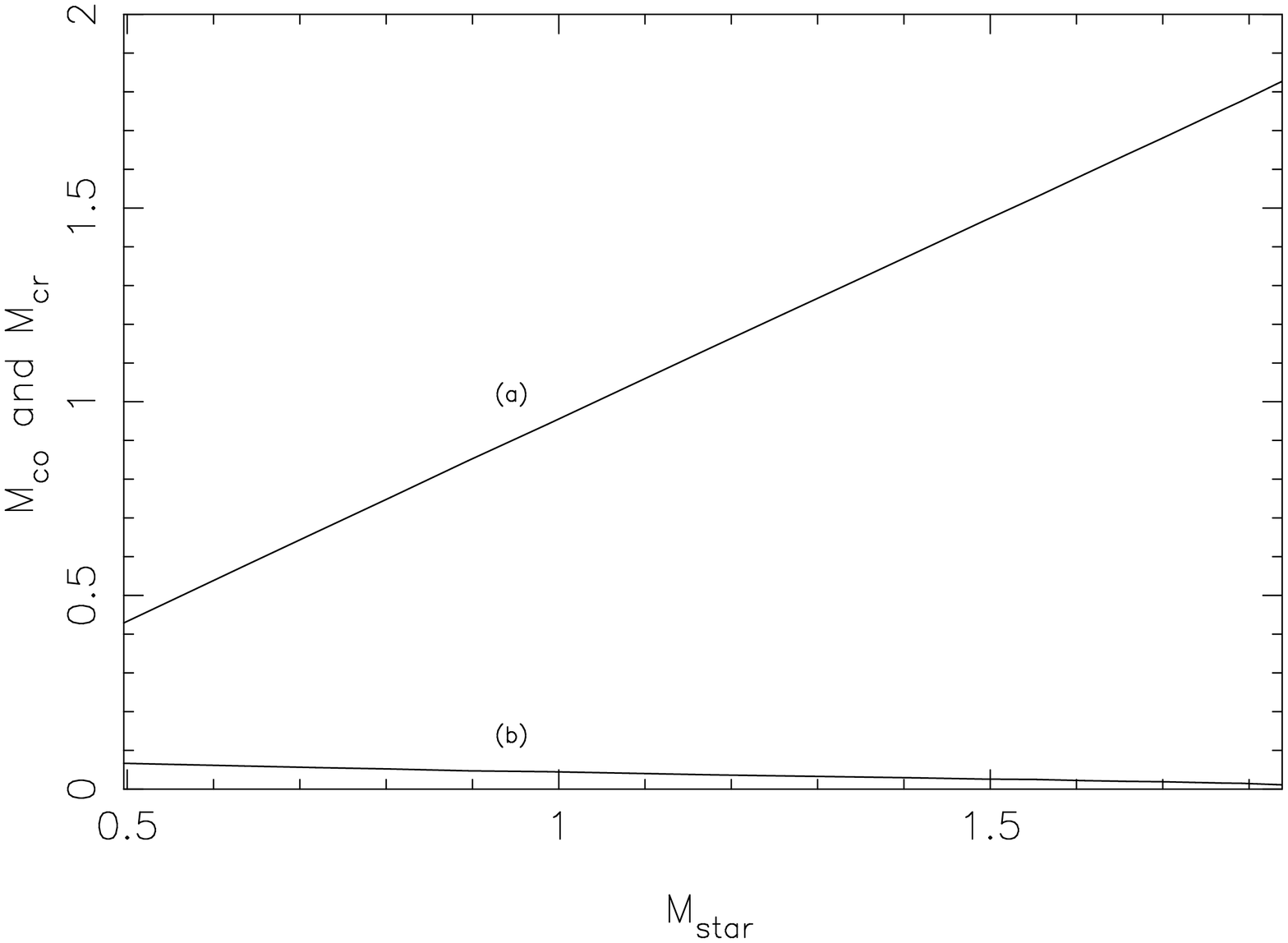,width=235pt,angle=-90}}}\end{center}
\caption[mass in the core \& crust]{Variation of the core-mass and the
crust-mass with the total mass of  a neutron star.  Curves (a) and (b)
refer to the core and crust mass respectively.}
\label{fm_ms}
\eef
\bef
\begin{center}{\mbox{\epsfig{file=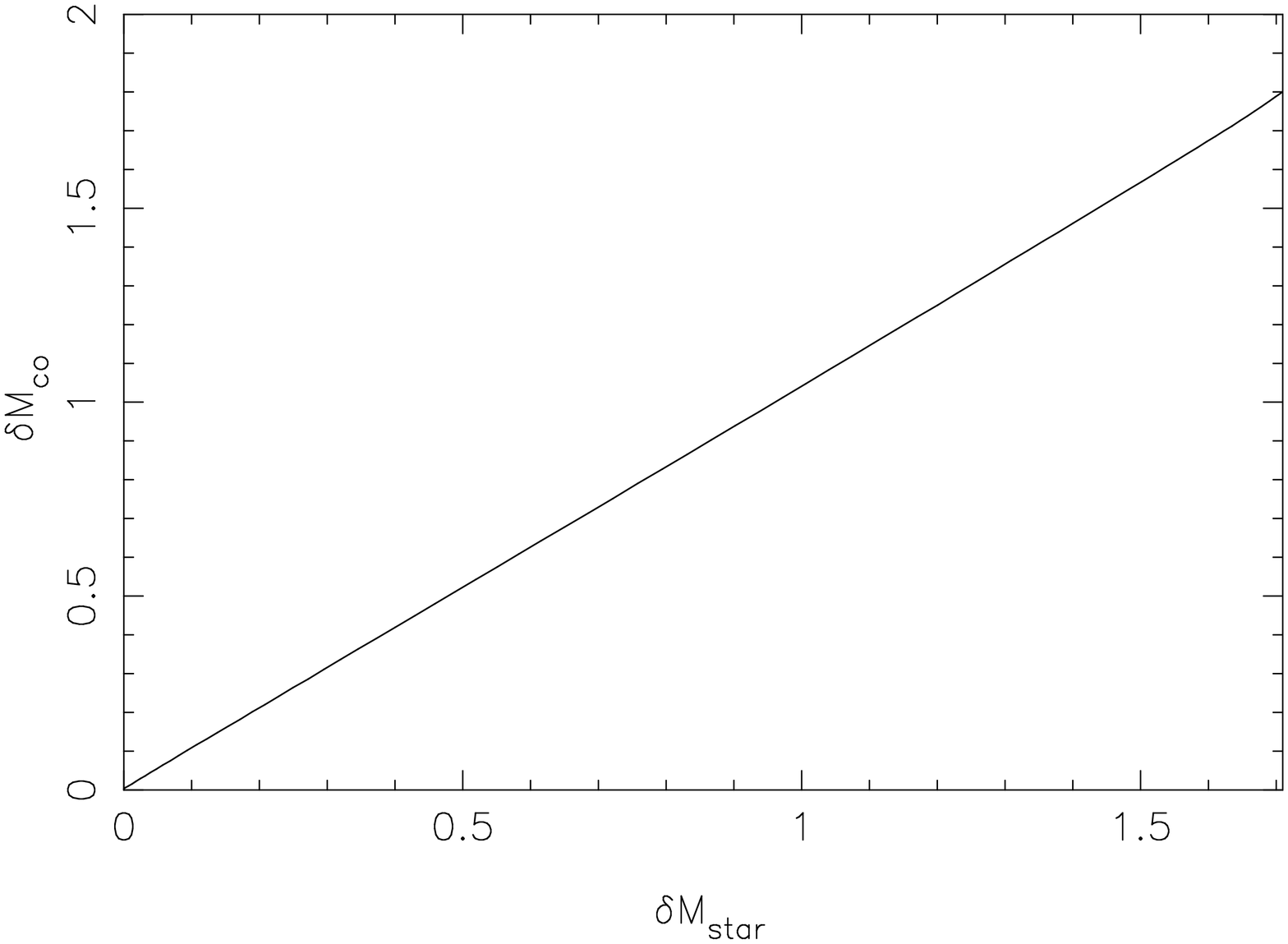,width=235pt,angle=-90}}}\end{center}
\caption[mass of  the core]{Variation of  the change in  the core-mass
with a change in the total mass of a neutron star.}
\label{fdmco_dms}
\eef
\bef
\begin{center}{\mbox{\epsfig{file=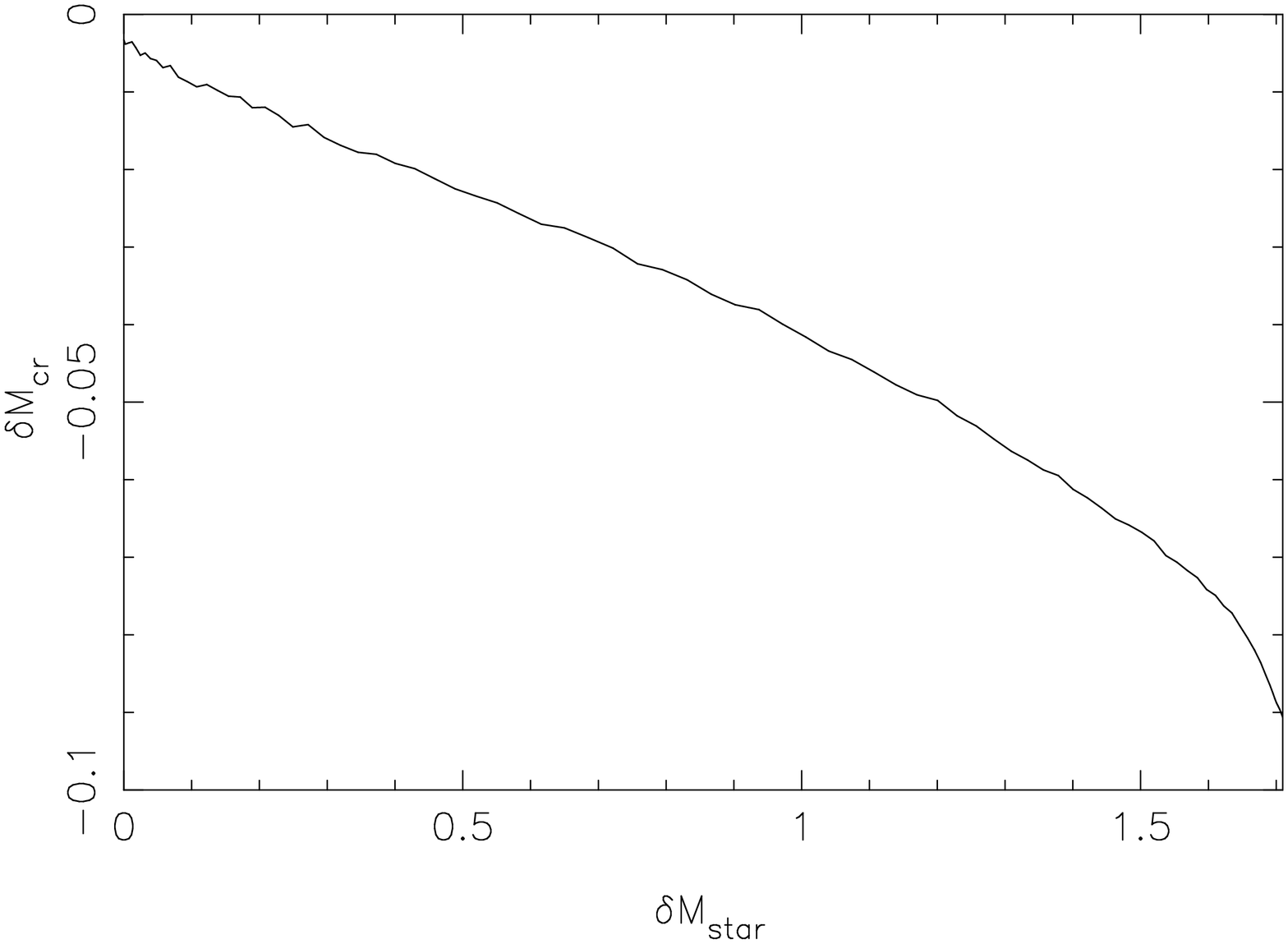,width=235pt,angle=-90}}}\end{center}
\caption[mass of the crust]{Variation  of the change in the crust-mass
with a change in the total mass of a neutron star.}
\label{fdmcr_dms}
\eef

We plot the variation of the  mass of the overlying layers and density
with  the  depth  from   the  surface  in  figures  [\ref{fdm_z}]  and
[\ref{frho_z}]. It  should be noted  that the density  changes sharply
with  depth whereas  the mass  remains  almost constant  close to  the
surface and then shows a sharp  increase. This is due to the fact that
the mass  in the outer  layers of the  neutron star is very  small. In
figure [\ref{fm_ms}] we have plotted the mass of the core and the mass
of the crust as  functions of the total mass. It is  seen that with an
increase in  the total mass the  mass of the core  increases almost by
the  same amount.  Whereas, the  change in  the mass  of the  crust is
minimal. Figures  [\ref{fdmco_dms}] and [\ref{fdmcr_dms}]  in which we
plot the  change in crustal  and core mass  vs. a change in  the total
mass brings this fact out more dramatically.

\section{thermal evolution of neutron stars}
\label{sthermal}

\subsection{isolated neutron star}

Thermal evolution of a system is determined by the processes of energy
loss and those of heat generation.  In the case of a neutron star heat
loss  is mainly  by emission  of neutrinos  from the  interior  and by
emission of photons  from the surface of the  star.  There are various
mechanisms for internal heat  generation, for example, friction due to
differential  rotation  of  crustal  neutron  superfluid,  dissipative
processes  due to  the  core proton  superconductor,  heat release  by
chemical change in  the crust induced by spin-down  of the star, ohmic
dissipation of  current loops (supporting  the magnetic field)  due to
the  finite conductivity  in  the  crust or  crust  cracking etc  (for
details     of     neutron      star     thermal     evolution     see 
\citeNP{latt91,peth92,page98} and references therein).

The  dominant mechanism  of cooling  in  the early  phases of  thermal
evolution is that of neutrino emission.  Different regions of the star
produce neutrinos by different  mechanisms, namely, by URCA process in
the  core   and  neutrino  pair  bremsstrahlung  in   the  crust.  The
comparability of the  two processes depends on the  presence of exotic
phases in the core and whether  direct URCA process can proceed in the
core after it  has cooled down below $\sim  10^{11}$K. It also depends
on the band-structure of the electrons in the crust of the star, which
may suppress  the neutrino  pair bremsstrahlung considerably.   In the
core, if the  matter is a normal n-p-e plasma  and the proton fraction
is  not  too  high  then  neutrinos  are  emitted  via  modified  URCA
process.  Through this  process the  star cools  with a  time-scale of
$T^{-8}$.  In presence  of exotic  phases  like quark  matter or  Bose
condensates of kaons or pions  direct URCA process can proceed. With a
$T^{-6}$  dependence  on temperature  this  process  results in  rapid
cooling. Since the  state of the matter in the core  of a neutron star
is not known  with any certainty, there is a  lot of controversy about
whether direct or modified URCA processes control neutron star thermal
evolution. Moreover, there is  uncertainty in the rate calculation for
the modified  URCA process due to  medium effects etc  and therefore a
comparison with observation does not yet provide a definite answer.

All of  the above discussion assumes  the matter to be  normal and the
spectrum of  elementary excitations smooth near the  Fermi surfaces of
the particles. In presence  of superfluidity or superconductivity gaps
would open up near the Fermi surfaces suppressing neutrino emission at
temperatures  less than  the gap  energy. Under  these  conditions the
neutrino pair  bremsstrahlung is the dominant  cooling process. Recent
work by \citeN{peth94}  has shown that this crust  cooling process may
get suppressed due to the  creation of the band structure as electrons
move  in  the  periodic  lattice  potential, below  a  temperature  of
$10^{10}$K.   Recently,  the  effect   of  Cooper  pair  breaking  and
formation  has  also  been   incorporated  in  the  thermal  evolution
calculations~\cite{scha97}.

\bef
\begin{center}{\mbox{\epsfig{file=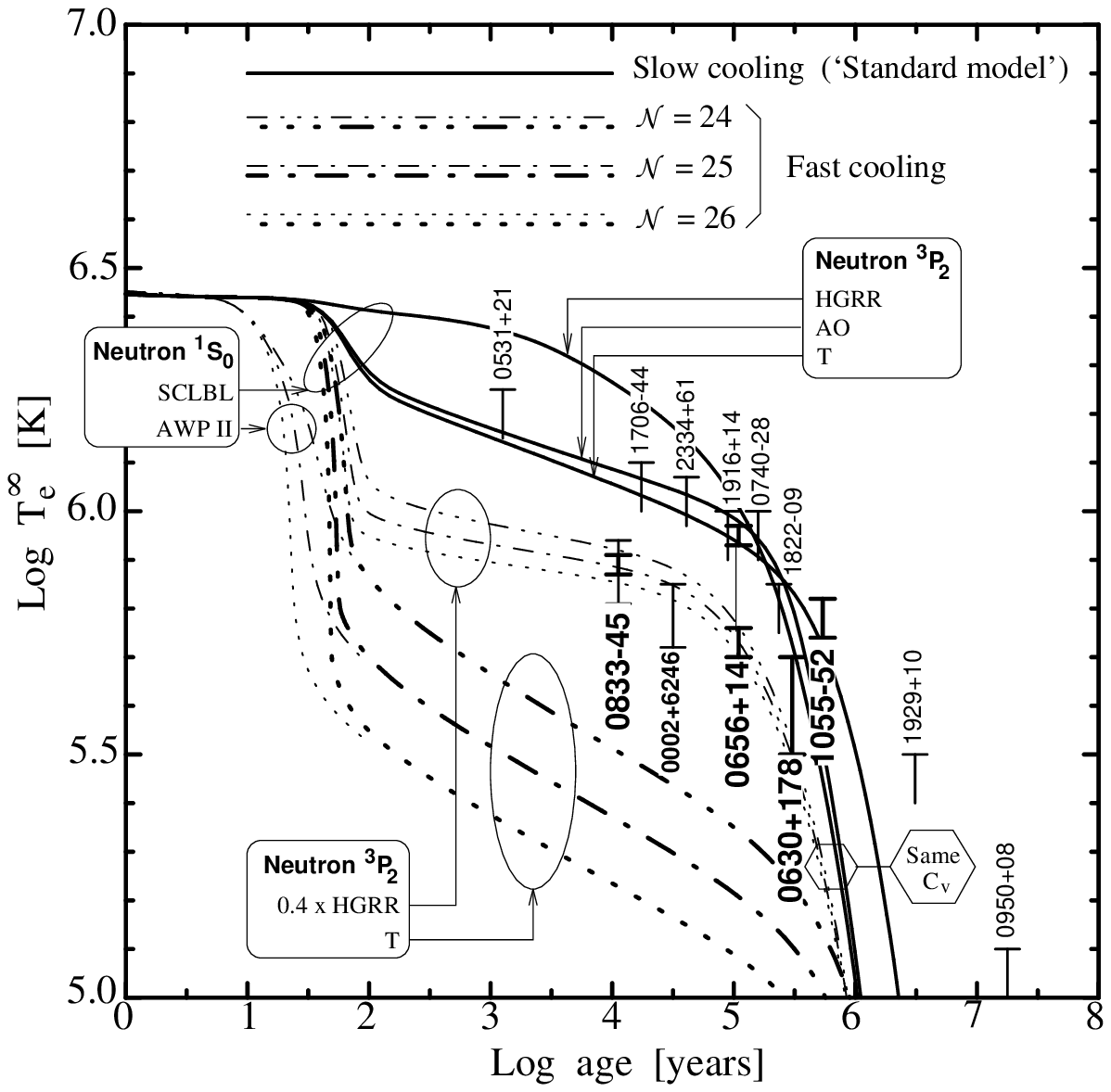,width=415pt}}}\end{center}
\caption[neutron  star   cooling]  {{\bf  Typical   behavior  of  slow
(`standard') and fast cooling scenarios.} 1.4~\msun neutron stars with
the Friedman \& Pandharipande (1981) equation of state. The cases with
$\cal N$ = 24, 25, and  26 correspond approximately to the effect of a
kaon condensate,  pion condensate, and  the direct URCA  process (with
hyperons or  nucleons), respectively. The various  curves, within each
scenario, show the  effect of various assumptions about  pairing : all
models use the  proton $^1$S$_0$ $T_c$ `T', and  the neutron $^1$S$_0$
and $^3$P$_2$  $T_c$'s are as labeled. All  models have non-magnetized
iron envelopes.   Neither Cooper pair breaking  and formation neutrino
emission nor heating  are included. The main effect  of pairing in the
crust  (neutron $^1$S$_0$)  is  to  shorten the  length  of the  early
plateau. Core pairing suppresses  the neutrino emission, which results
in a higher $T_e$ during the neutrino cooling era (age from $\sim$ 100
to $\sim$ 10$^5$  yr), and the specific heat,  which results in faster
cooling during  the photon cooling  era (age above $\sim$  10$^5$ yr).
The reduction  of the  specific heat during  the neutrino  cooling era
does not show up  as much as during the photon cooling  era due to the
small slope of  the curves at this phase. All  points are really upper
limits (in several  cases based on a non-detection  of the pulsar) but
for the  radio pulsars 0833$-$45 (Vela),  0656+14, 0630+178 (Geminga),
1055$-$52, and the neutron star 0002+6246, there is good evidence that
the observed X-rays are  from surface thermal emission. Uncertainty on
the temperature  estimate is  illustrated in the  case of  PSR 0656+14
where two values are reported~\cite{page98}.}
\label{ft_page}
\eef

In  a recent  work \citeN{iwam95}  have shown  that a  finite magnetic
moment of neutrino would significantly modify the cooling history of a
neutron star in the very  early phases. This makes the crustal cooling
compete  with the  core cooling  within  the typical  time scale  that
conduction  takes to  transport thermal  energy from  the core  to the
surface.

It appears that the present data  is compatible with both the slow and
fast cooling processes (modified and direct URCA) as there is a lot of
uncertainty in all the mechanisms involved in the thermal evolution of
a neutron  star.  In figure  [\ref{ft_page}] taken from~\citeN{page98}
different theoretical  scenarios could be seen and  how these theories
compare with the observational values of surface temperatures measured
for various pulsars.  There are  other factors that may be responsible
for a discrepancy  between the theory of the  thermal evolution of the
neutron stars  and the observed  values for the  surface temperatures.
For example the temperature  is usually estimated assuming the neutron
star  to behave  like a  perfect black-body,  but the  pressure  of an
atmosphere   and  the  effects   of  a   strong  magnetic   field  may
significantly modify this result~\cite{pavl96,shib96}.

\subsection{thermal structure of an isolated neutron star}

Temperature  fluctuations in  the  interior of  the  neutron star  are
smoothed  out very  fast due  to  its large  thermal conductivity  and
effectively the whole  of the star behaves like  an isothermal system,
except at  the layers close to the  surface~\cite{gudm82}.  Though the
temperature of the entire region beyond a density of $10^{10}~\gcc$ is
practically the  same, it drops by  almost two orders  of magnitude at
the outermost layers of the star. The work of \citeN{gudm82,gudm83} on
the  envelopes   of  non-magnetic   neutron  stars  showed   that  the
temperature  of the isothermal  interior, $T_b$,  depends only  on the
surface temperature and the surface gravity of the star :
\beq
T_b = 1.288 \times 10^8 K \left(\frac{(T_s/10^6 K)^4}{g_s/10^{14} {\rm
cm} {\rm s}^{-2}}\right)^{0.455}. \nonumber
\eeq
where  $T_s$ is  the  surface  temperature and  $g_s$  is the  surface
gravity. These  authors also present the variation  of the temperature
with  density between  the  surface and  the  isothermal interior.  To
obtain temperature as a function  of density in these outer regions of
the crust we use the following fitting formula to their plots :
\beq
T(\rho) =  (\frac{\rho}{\rho_{\rm boundary}})^{1/4} T_b,  \mbox{ $\rho
\le \rho_{\rm boundary}$}
\eeq
where  $\rho_{\rm   boundary}$  is   the  density  beyond   which  the
temperature stays effectively constant.

\subsection{accreting neutron star}

The thermal  history of  an accreting neutron  star is  very different
from that of an isolated one.  The cooling of an isolated neutron star
brings  the surface  temperature down  to $\sim  10^{4.5}$~K  in about
$10^7$~yr  with  an  attendant  interior  temperature  of  the  nearly
isothermal  core of the  order of  $10^7$~K~\cite{ripe91a}.  Therefore
when mass  accretion starts  this cold  star is heated  up due  to the
entropy inflow of the accreted  matter.  The temperature rise might be
enough  to  start nuclear  burning  at  the  surface and  one  expects
pycnonuclear shell burning of hydrogen and helium. Within a short time
($\sim  10^5$ yr)  almost the  entire crust  is heated  to  a constant
temperature  of the  order of  $10^{7.5}  - 10^{8.5}$~K~\cite{mira90}.
This is  ignoring an  initial short  phase in which  both the  rate of
accretion and the  temperature of the crust show  time evolution.  The
rate of  accretion stabilizes  in a few  thousand years~\cite{savo78}.
The temperature that the crust will finally attain in the steady phase
has been  computed by  Fujimoto et \citeN{fuji84},  \citeN{mira90} and
\citeN{zdun92}.  However, these computations are restricted to limited
range  of  mass accretion  and  also do  not  yield  the same  crustal
temperature  under  similar   conditions.   The  results  obtained  by
\citeN{zdun92} for the crustal temperatures for a given accretion rate
in the range $10^{-15}$~\msun/yr$< \mdot < 2 \times 10^{-10}$~\msun/yr
could be fitted to the following formula:
\beq
\log T = 0.397 \log \mdot + 12.35.  \label{etdmdt}
\eeq
But  extrapolation of  this fit  to  higher rates  of accretion  gives
extremely  high temperatures which  would not  be sustainable  for any
reasonable  period  due  to   rapid  cooling  by  neutrinos  at  those
temperatures.  For the purpose  of our  calculations, we  use equation
[\ref{etdmdt}] as long as the temperature of the crust is smaller than
$10^{8.5}$~K.  Beyond that  we freeze  the temperature  at  that upper
limit. In figure [\ref{ft_dmdt}] we  have plotted the variation of the
crustal  temperature   with  accretion  rate   according  to  equation
[\ref{etdmdt}]. The thermal state of  the core depends strongly on the
neutrino emissivity  whereas the crust remains  largely indifferent to
that.  The core  stays relatively  cool  if there  is pion  condensate
inside  which induces  enhanced neutrino  cooling, otherwise  the core
temperature may also be raised to a large extent by mass accretion.

\bef
\begin{center}{\mbox{\epsfig{file=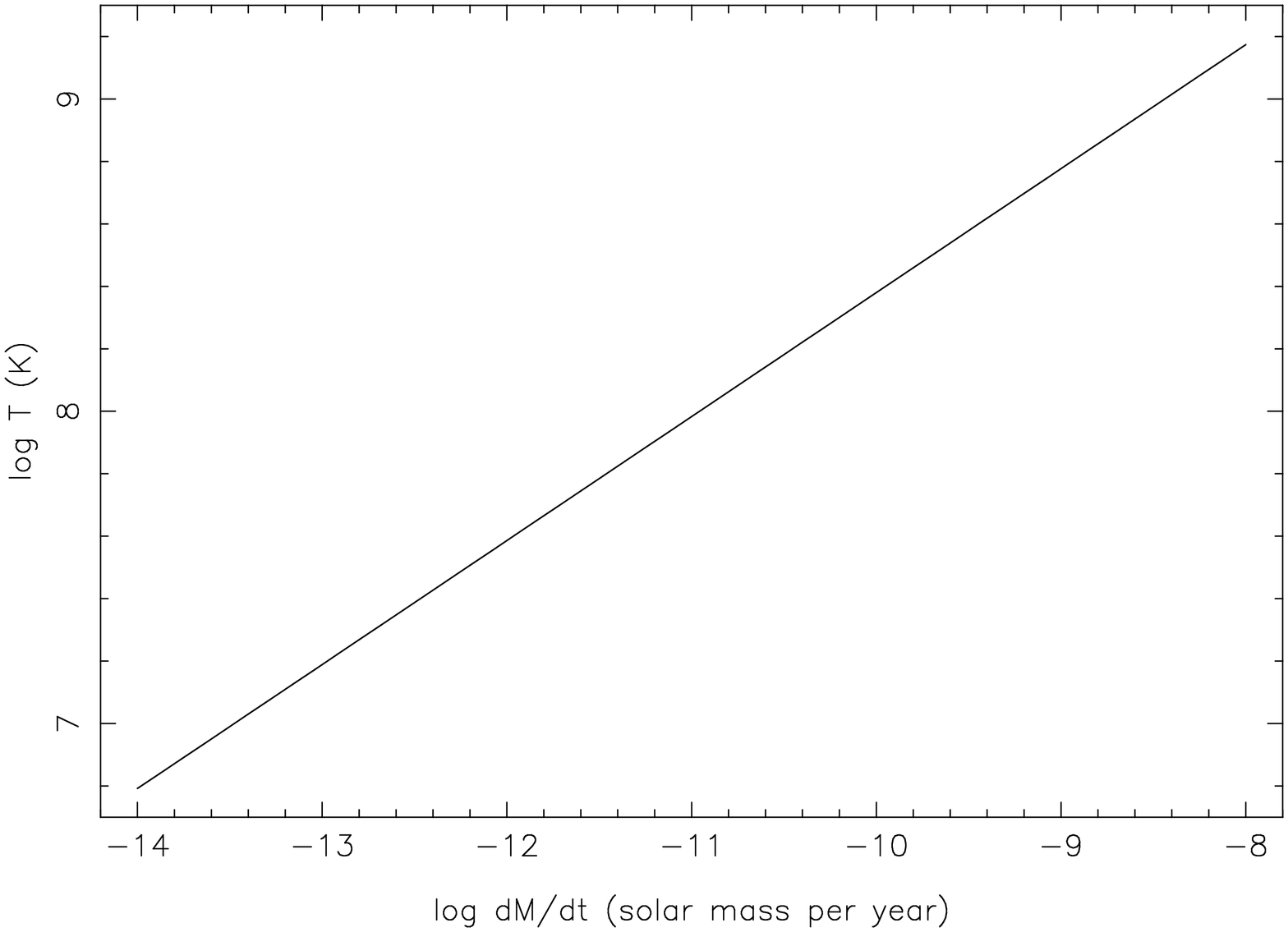,width=235pt,angle=-90}}}\end{center}
\caption[crustal  temperature]{Variation  of  the constant  isothermal
temperature of the crust of a neutron star with accretion rate.}
\label{ft_dmdt}
\eef

The  above discussion does  not take  into account  the fact  that the
composition of the  accreted layers could be very  different from that
of the original  cold catalysed composition.  In a  recent work it has
been  shown  that the  presence  of  light  elements in  the  accreted
envelope enhances the emission processes in the photon cooling era and
hence      ultimately      a      faster     cooling      rate      is
achieved~\cite{chab97,pote97a}.  Such  effects show drastic difference
in  the surface  temperature  (see \citeNP{page98}  for a  discussion)
already within ten thousand years.  If incorporated, this might change
the evolution of the magnetic field considerably.

\section{transport properties in the crust of neutron stars} 
\label{sns-transport}

The investigations  of the transport properties  of ultra-dense matter
arise out  of the interest in  the evolution of the  thermal state and
the  magnetic field  in  white  dwarfs and  neutron  star crusts.  See
\citeN{itoh94}  and  references  therein  for  a good  review  on  the
transport  properties of  neutron  star crust.   It  has already  been
mentioned in  section [\ref{seos}]  that the crust  of a  neutron star
consists of a relativistic,  Fermi-degenerate free electron gas plus a
non-relativistic, non-degenerate liquid/crystal of ions. It is assumed
that the material is completely pressure-ionized. The density at which
this happens is given by the condition
\ber
\rho \ge 0.378 A Z^2~\gcc,
\eer
which turns out to be  $\sim 10^4~\gcc$ for Fe$^{56}$ ions. Therefore,
the lower boundary for which the transport properties have been worked
out is this particular density. Though, recently, \citeN{itoh93b} have
investigated the entire density range below this value.

The thermal and electrical conduction  is basically carried out by the
electrons. The  electrical conductivity  is given by  following simple
Drude formula~\cite{ashc}
\ber
\sigma = \frac{n_e e^2 \tau}{m_{\ast}},
\eer
where $n_e$ is  the number density of electrons  and $m_{\ast}$ is the
effective  mass  of  the  electron  in  the  crystal.  $\tau$  is  the
time-scale  of the  collision of  electrons with  the ions  (in liquid
phase) or phonons/impurities (in case of a crystalline solid). It must
be mentioned here that  although the importance of quantum corrections
have been realized in the  present context, not much progress has been
made in that direction

In the crust of a neutron  star both density and temperature vary with
radius. Whereas the  uppermost layers close to the  surface are likely
to be in  a liquid state, the inner crust is  a crystalline solid. The
condition  for melting/crystallization  of  a classical  one-component
plasma  is given by  Lindeman criterion.  According to  this 
criterion~\cite{slat82},
\beq
\Gamma =  \frac{\rm Coulomb \;  Energy \; of  \; the \;  Crystal} {\rm
Thermal \; Energy \; of \; the \; Lattice \; Ions},
\eeq
equals  172 at  the melting  point. For  a crystal  composed  of ionic
species of charge $Z$ and  lattice spacing $a$, the Coulomb Energy per
ion  is  $\frac {(Ze)^2}{a}$  and  the thermal  energy  of  an ion  is
approximately   $k_B  T$  where   $T$  is   the  temperature   of  the
crystal. Therefore,
\beq
\Gamma = \frac {(Ze)^2}{a k_B T}.
\eeq
The inter atomic spacing $a$, in terms of density is,
\beq
a =  (\frac {4 \pi}{3})^{-  1/3} (\frac {\rho}{A})^{-  1/3} m_p^{1/3},
\label{elatconst}
\eeq
$m$ and  $A$ being  the proton mass  and the  mass number of  the ion,
respectively. Then the melting temperature is
\beq
T_{m} = 0.22692 \times 10^8 \frac {Z^2 (\frac {\rho_6}{A})^{1/3}}{171}
K, \label{etmelt}
\eeq
where   $\rho_6$  is   the  density   in  $10^6   gm/cc$.   In  figure
[\ref{ftm_rho}]  the  melting  temperature  has  been  plotted  versus
density in the crust of a 1.4 $\msun$ neutron star.

\bef
\begin{center}{\mbox{\epsfig{file=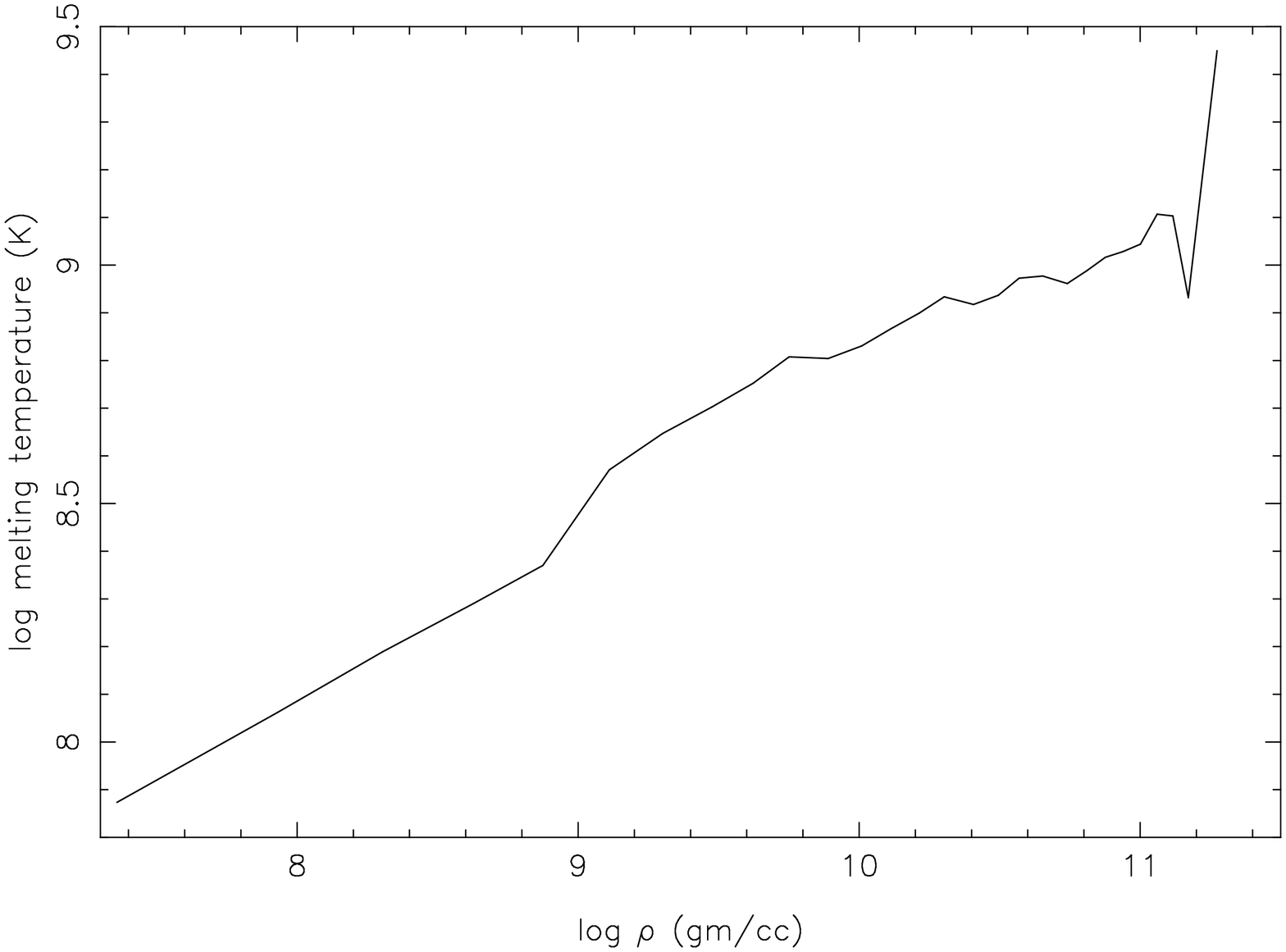,width=235pt,angle=-90}}}\end{center}
\caption[melting  temperature  in   the  crust]{Variation  of  melting
temperature with density in the crust on a neutron star. }
\label{ftm_rho}
\eef

Densities  for  which the  actual  temperature  is  above the  melting
temperature,  the  material  is  in  a  liquid  state.  The  transport
properties  in  such  a   state  is  determined  by  the  electron-ion
collisions and  by electron-phonon collisions in the  solid phase. The
three  factors   important  factors  in   calculating  electron-phonon
collision  time-scale are  - the  dielectric screening  of  the phonon
spectrum   by  the   relativistic,  Fermi-degenerate   electrons,  the
Debye-Waller factor for  the pure Coulomb, bcc crystal  and the atomic
form  factor. The Debye-Waller  factor changes  the conductivity  by a
factor  of  two to  four  at the  melting  temperature.  And when  the
electron de-Broglie wavelength becomes  comparable to the nuclear size
the third correction becomes  rather important. Unlike the terrestrial
situation,  in  the  crust  of  a neutron  star  the  Umklapp  process
dominates.  For  lower temperatures, the  dominant process is  that of
the collision  of electrons with the impurity  atoms. These collisions
are similar to the electron-ion  collision in the liquid phase, except
that here the effective charge is the difference between the charge of
the  impurity  atom  and  the  charge of  the  dominant  species.  The
temperature  or density  of the  cross-over from  phonon  dominated to
impurity dominated process depends on the impurity strength $Q$, given
by,
\beq
Q = \frac{1}{n} \sum_{i}{{n_{i}}(Z - Z_{i})^2} \nonumber
\eeq
where $n$ is  the total ion density, $n_i$ is  the density of impurity
species $i$ with charge $Z_i$, and $Z$ is the ionic charge in the pure
lattice~\cite{yako80}.

For our work, we have taken the expression for electrical conductivity
of  the liquid and  due to  impurity concentration  in the  solid from
\citeN{yako80}.   For the  pure  crystalline phase  we  have used  the
results of \citeN{itoh84}. The conductivity in the liquid is given by,
\beq
\sigma_{\rm  liquid} = 8.53  \times 10^{21}  \frac{x^3}{Z \Lambda_{\rm
Coulomb} (1 + x^2)},
\eeq
where $x$ is defined by the relation
\beq
x = (Z/ \rho_{6})^{1/3},
\eeq
and $\Lambda_{\rm  Coulomb}$ is the  Coulomb logarithm. In  the solid,
the  conductivity  has contributions  from  both  the  phonon and  the
impurity processes. Therefore, the conductivity is given by,
\beq
\sigma_{\rm solid}  = \frac{1}{\sigma_{\rm phonon}^{-1}  + \sigma_{\rm
impurity}^{-1}}, \label{esigma}
\eeq
where
\ber
\sigma_{\rm impurity} &=& 8.53 \times 10^{21} x Z/Q /s\\
\sigma_{\rm  phonon}  &=& 1.24  \times  10^{20} \frac{x^{4}}{u  T_{8}}
\frac{(u^{2} + 0.0174)^{1/2}}{(1 + 1.018 x^{2}) I_{\sigma}},
\eer
with,
\ber
u &=& \frac{2 \pi}{9} (log \rho - 3) \nonumber \\
T_{8} &=& \mbox{temperature in units of $10^8$ K} \nonumber \\
\rho_{6} &=& \mbox{density in units of $10^6~\gcc$} \nonumber \\
I_{\sigma} &=& \mbox{a function of  density, $Z$, $A$ given by Itoh et
al. 1984}. \nonumber
\eer

In the following diagrams  we have plotted the electrical conductivity
in  the  crust  of a  neutron  star,  as  a  function of  density  and
emphasizing   the  dependence   on  various   parameters.   In  figure
[\ref{fsigma_cool}], the plot is  for different values of the impurity
concentration $Q$ for a given surface temperature. Notice that in this
case we assume a temperature  variation with density as is expected in
a  cool,  isolated neutron  star  (section~\ref{sthermal}). In  figure
[\ref{fsigma_t}], on the other  hand, we have plotted the conductivity
for different  values of  the temperature which  is constant  over the
whole of  the crust.   In figure [\ref{fsigma_q}],  we have  shown the
change in conductivity with different values of $Q$, assuming the same
constant crustal temperature in each case.

\bef
\begin{center}{\mbox{\epsfig{file=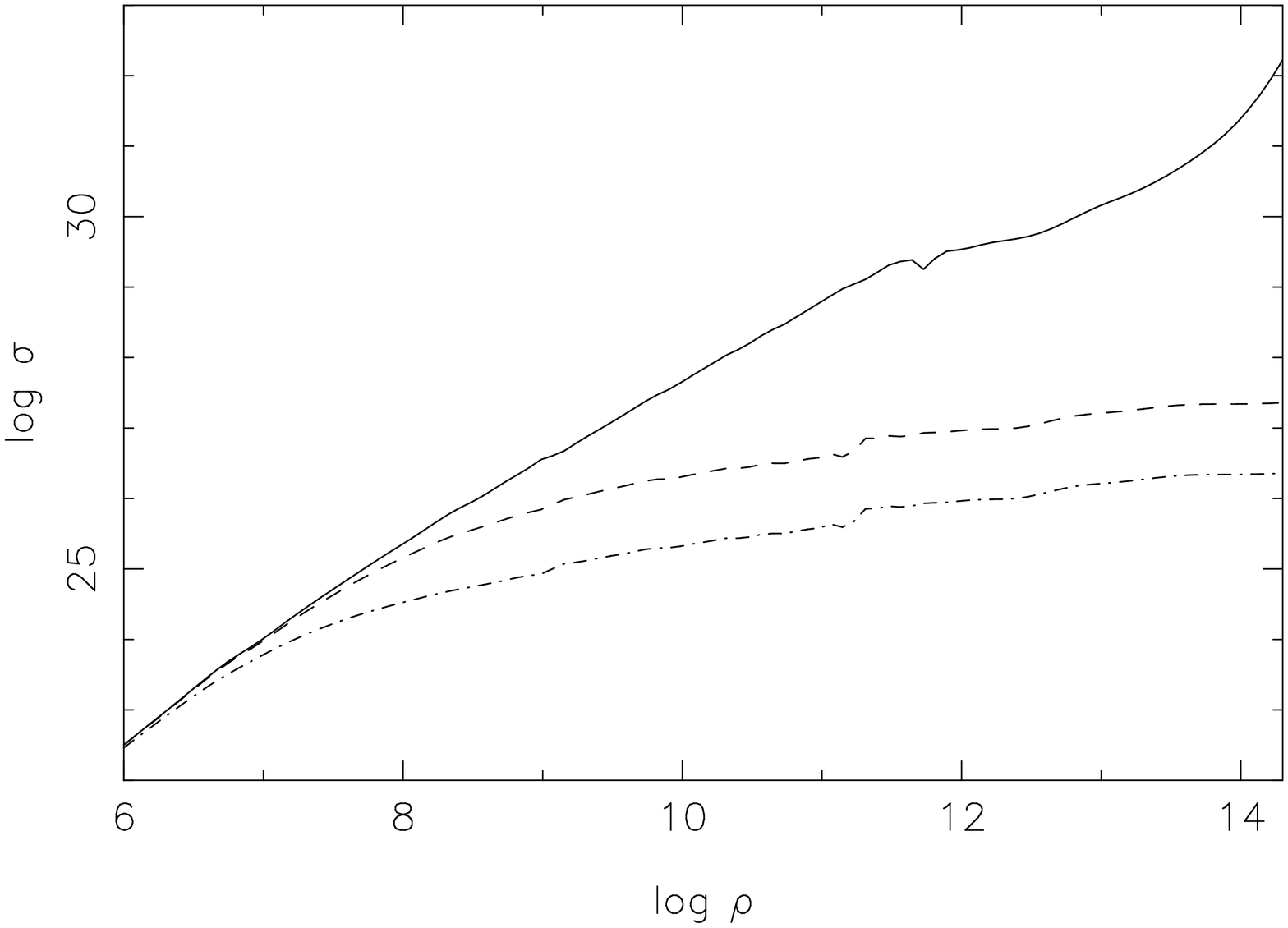,width=235pt,angle=-90}}}\end{center}
\caption[electrical  conductivity  in the  crust  I]{Variation of  the
electrical conductivity  with density in  the crust of a  cool neutron
star. The solid, dashed and dash-dotted curves correspond to $Q = 0.0,
0.01, 0.1$. For  all curves the surface temperature  has been taken to
be equal to $10^{4.5}$~K. }
\label{fsigma_cool}
\eef
\bef
\begin{center}{\mbox{\epsfig{file=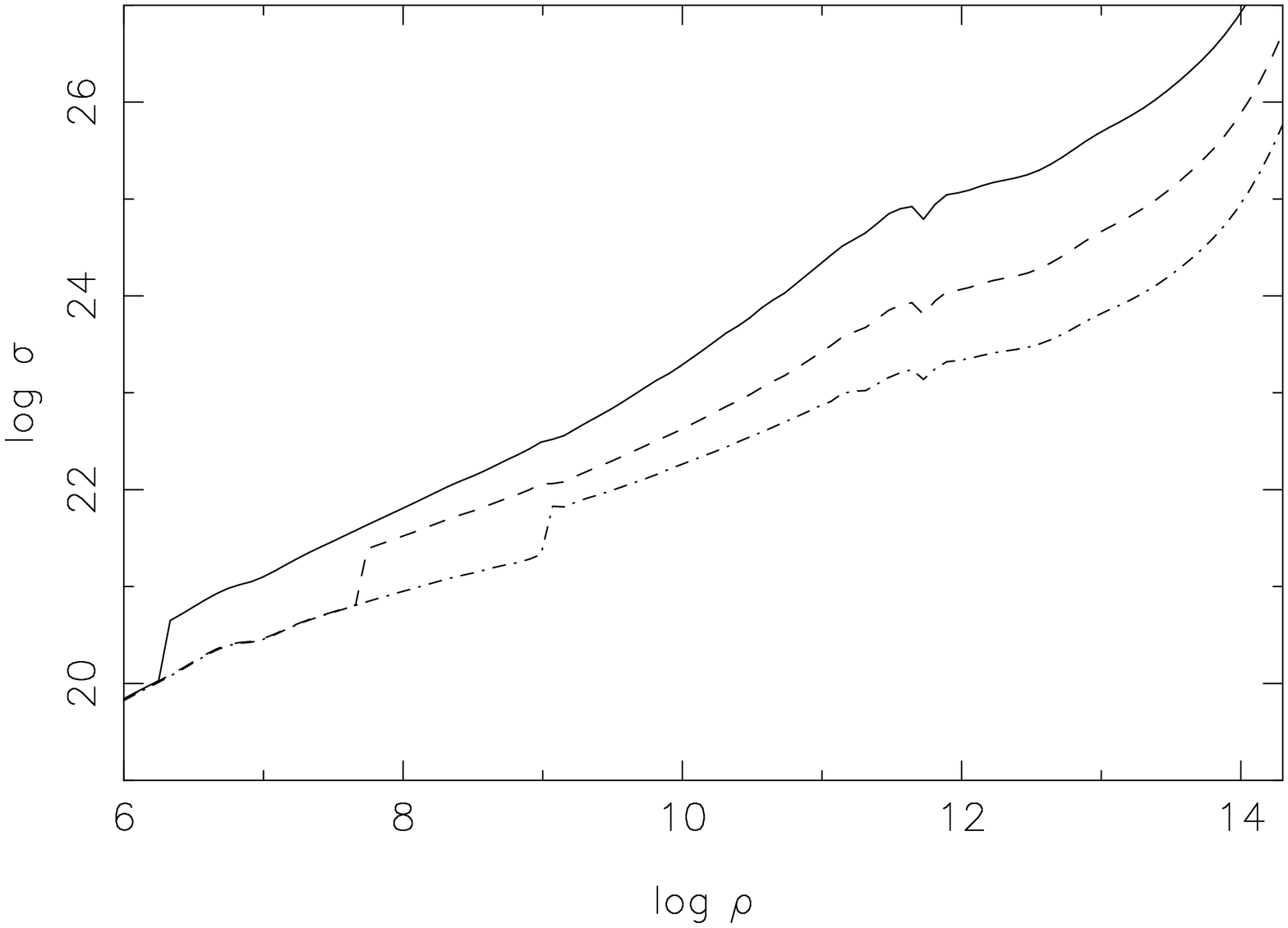,width=235pt,angle=-90}}}\end{center}
\caption[electrical  conductivity in  the crust  II]{Variation  of the
electrical conductivity with  density in the crust of  a neutron star.
The  solid, dashed and  dash-dotted curves  correspond to  the crustal
temperatures  of  $10^{7.5}, 10^8,  10^{8.5}$~K.   In  all the  curves
$Q=0$. }
\label{fsigma_t}
\eef
\bef
\begin{center}{\mbox{\epsfig{file=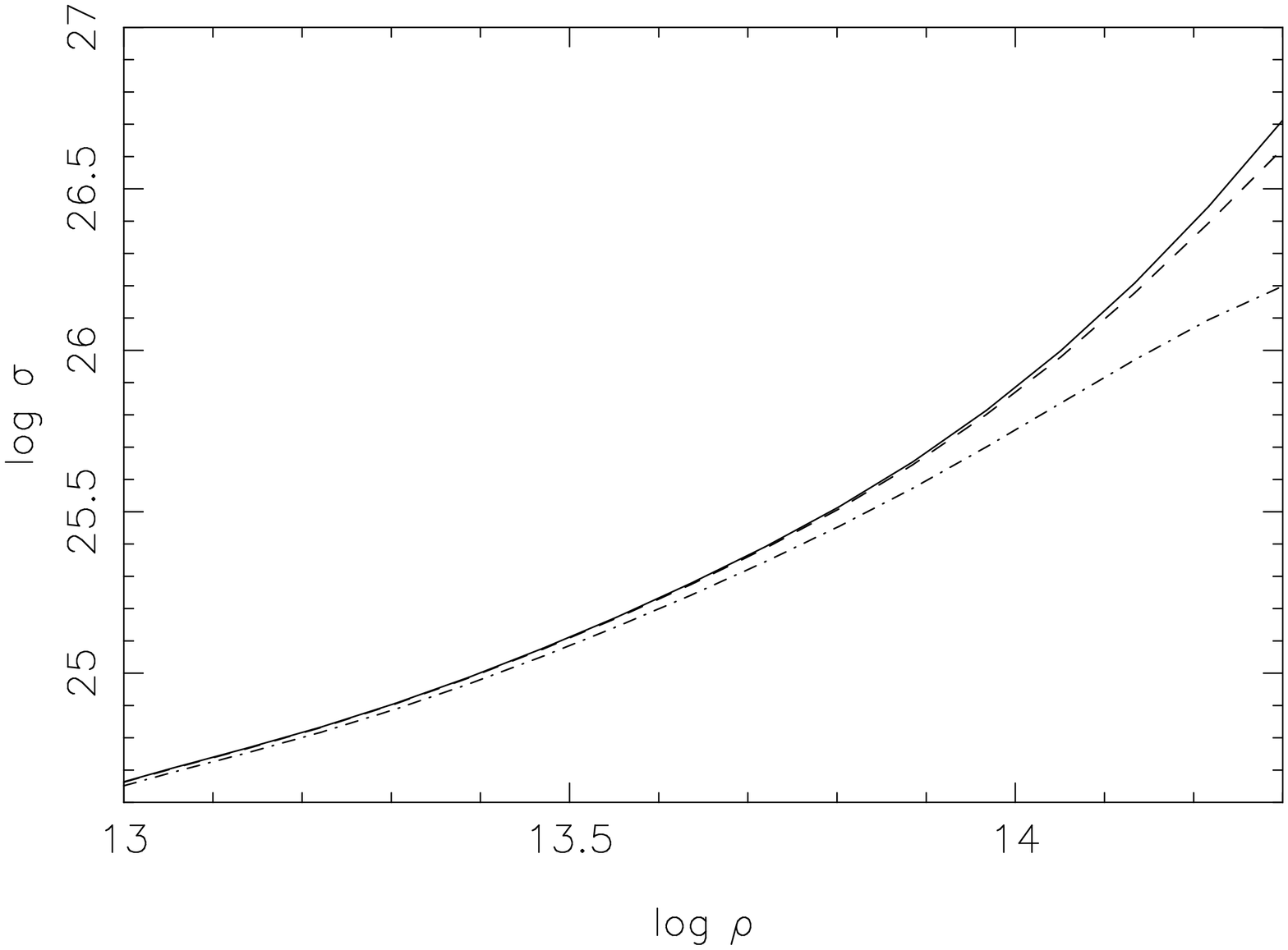,width=235pt,angle=-90}}}\end{center}
\caption[electrical  conductivity in the  crust III]{Variation  of the
electrical  conductivity  with  density  in  the crust  of  a  neutron
star. The solid, dashed and dash-dotted curves correspond to $Q = 0.0,
0.01, 0.1$. For all curves $T = 10^8$~K. }
\label{fsigma_q}
\eef

In figures  [\ref{fsigma_t}] and [\ref{fsigma_q}] we  have plotted the
conductivity assuming  the temperature to be constant  over the entire
crust. That  is the  case for  a star with  an accretion  heated crust
after the temperature has stabilized. For an isolated star with a very
low  surface temperature and  a non-zero  temperature gradient  in the
outermost layers  (as described in  section[\ref{sthermal}] above) the
variation of  conductivity with  density looks somewhat  different. In
figure [\ref{fsigma_cool}] we plot the conductivity profile for such a
cool star.  Note  that the impurity strength $Q$  becomes important in
this case.

It should be  mentioned here that the above  discussion does not refer
to the  fact that the transport  properties in the crust  of a neutron
star must also  take into account the presence  of magnetic fields. As
early as in 1980, Urpin \&  Yakovlev had looked into this problem. And
recently,  very   refined  results   have  been  available   in  which
conductivity    calculations   have    been    made   with    magnetic
field~\cite{pote97a}.  Also, all the above calculations have been made
assuming   a  bcc   lattice.   Recently,   \citeN{baik95}   have  also
investigated the  case of  fcc lattice.  But  for our  calculations we
have not made use of these refined results.

\chapter{magnetic fields of neutron stars : a general introduction}
\label{cfield}

\section{overview}

In the  cosmic scheme of things  the major players behind  most of the
interesting  phenomena  are rotation  and  magnetic  field.  A  unique
combination of very rapid rotation  and a large magnetic field is what
makes a neutron star act as  a pulsar. The rotation period of a pulsar
can  be  as small  as  1.6  millisecond~\cite{back82}, and  even  the
smallest  field observed  in pulsars  could be  about three  orders of
magnitude higher than the maximum field so far achieved in terrestrial
laboratories.  The  typical values of magnetic field  in pulsars range
from $\sim 10^8$  Gauss to $10^{13.5}$ Gauss. It  is the ultra-compact
nature of  a neutron star  that allows the  extremes in both  the spin
rate and  the magnetic field  strength. Because of the  compactness it
can  support a  fast rotation  against the  centrifugal forces  and at
close to  nuclear densities  even such high  fields do not  affect the
state of  the matter significantly because the  energy associated with
the  magnetic field is  insignificant compared  to the  other relevant
energy scales~\cite{shap}.

It should  also be noted that  the neutron star  material is something
like the  ultimate high-$T_{\rm  c}$ super-conductor. Even  though the
temperature  in the interior  of a  newly born  neutron star  could be
about  $10^9$K, it  is  still small  compared  to the  superconducting
transition temperature, believed to be a few times $10^9$K. Hence, the
material inside a neutron  star quickly settles into a superconducting
state     soon     after      its     birth     in     a     supernova
explosion~\cite{alpr91,pine91}.  We shall see later that this plays an
important role in shaping the magnetic history of the star.

Unfortunately, there is  as yet no satisfactory theory  for either the
generation  of  the neutron  star  magnetic  field  or its  subsequent
evolution~\cite{db95a}.  There  is a major uncertainty  even about the
possible  location of  the field  in the  interior of  the  star. This
question is in fact related to  the problem of the epoch and mechanism
of    field   generation,   as    we   shall    see   in    the   next
section~\cite{srini95}.    It    is   obvious   that,    under   these
circumstances, there can be no consensus regarding the theory of field
evolution as any  such scenario will have to  ultimately depend on the
nature of the  underlying structure and location of  the current loops
that support the observable field.

Nevertheless, a compilation of current observational facts provides us
with the  nature of the  questions that need  to be looked  into. They
also  give an  indication  of  the range  of  possible answers.  These
observational  facts  strongly suggest  that  the  field evolution  is
intricately related to  the binary history of a  neutron star.  In the
rest of this chapter we shall discuss the observational status and the
current  theoretical attempts  to  understand the  generation and  the
subsequent  evolution of  the magnetic  field in  neutron  stars. This
provides the background  for the problems addressed by  us in chapters
[4], [5], [6] and [7]. 

\section{origin}
\label{sorigin}

There  are two  main  possibilities regarding  the  generation of  the
magnetic  field in neutron  stars (for  a review  see \citeNP{bhat95},
\citeNP{srini95} and  references therein).  The field can  either be a
fossil remnant  from the  progenitor star, or  be generated  after the
formation  of  the  neutron  star.  Uncertainties  surround  both  the
scenarios and observations  are yet to be able  to distinguish between
the two.  This has led  to a large  variety of field  evolution models
that we shall discuss in section [\ref{sevolution}].

\subsection{fossil field}
\label{ssfossil}

Originally suggested by  \citeN{ginz64} and \citeN{wolt64} long before
the discovery of  pulsars, the idea of the  fossil field is considered
to be the  most promising. The magnetic field existing  in the core of
the  progenitor  star gets  enhanced  when  the  core collapses  in  a
supernova, conserving the magnetic  flux. Flux conservation demands an
increase in the field strength by a factor $(R_{\rm progenitor}/R_{\rm
NS})^2$ which  is of  the order of  $10^{10}$. This can,  depending on
fields in the  cores of the progenitor stars,  produce fields as large
as $10^{14} - 10^{16}$ Gauss.   The field observed on neutron stars is
mainly the dipole component of the surface field.  It is possible that
the subsurface/interior field is much higher than this value.

In the core of a neutron star, the proton fraction is small (a maximum
of  10\% when  presence  of  exotic states  like  Kaon condensates  is
considered~\cite{pera92}.  Nevertheless, the  protons are  believed to
exist in a superconducting  state.  Calculations indicate that this is
a  type II superconductor  with a  lower critical  field in  excess of
$10^{15}$ Gauss.  Evidently, the observed field values  fall far short
of this  critical field and one  expects a complete  flux expulsion in
accordance with Meissner effect.  But unlike in a laboratory situation
the  flux  expulsion  encounters   a  problem  because  the  electrons
coexisting with the  protons do not form a  condensate state. The time
scale in which the flux can be expelled is therefore dictated by ohmic
diffusion  through the  electron  component.  Even  at extremely  high
temperatures at the  time of the birth of the  neutron star this ohmic
diffusion  time turns  out to  be  larger than  $10^8$ years.   Hence,
superconducting   transition   occurs    with   the   field   embedded
in~\cite{giki64}.   The magnetic  field is  carried in  quantized flux
tubes called  Abrikosov fluxoids, each  fluxoid carrying a  quantum of
magnetic    flux,     $\phi    =    \frac{hc}{2e}     =    2    \times
10^{-7}$~\mbox{G-cm$^2$}.   Therefore,  some  $10^{31}$ such  fluxoids
would be present  in a neutron star with a  typical field of $10^{12}$
Gauss.

One major  problem with the concept  of a fossil field  is that strong
surface  fields are  not  observed  in massive  stars  except in  some
dynamically peculiar ones.   \citeN{rudr73} suggest that the so-called
fossil magnetic field may not be a relic of the main sequence phase of
the  star  but can  be  generated in  the  core  during the  turbulent
Carbon-burning phase.  The strong field can therefore be hidden in the
core, but given  the short duration of evolution  during and after the
Carbon-burning phase,  it is unclear  whether this field  can organize
itself  into large-scale  poloidal components.  A new  input  into the
physics  of the  fossil  field  has come  from  the refined  many-body
calculations  performed   recently  on  the  behaviour   of  the  core
superconductor.  These  new results hint that  the core superconductor
is likely to be of  type I~\cite{ains89}. This would mean a completely
different structure  for the  field and would  require a  redressal of
some  of  the existing  theories  of  field  evolution involving  flux
expulsion  associated  with the  spin-down  of  pulsars (discussed  in
subsection [\ref{ssmodel}]).

\subsection{post-formation field generation}
\label{sspost-formation}

Almost all of the existing field evolution models ultimately depend on
ohmic  dissipation of the  currents in  the crust  of a  neutron star,
where  the   electrical  conductivity,   though  very  large   by  any
terrestrial   standard,  is   still   finite  (in   contrast  to   the
superconducting  interior which  can be  thought to  have  an infinite
conductivity).  Therefore, any model that makes generation of currents
in  the crust  possible  is  a very  attractive  proposition. We  have
mentioned earlier that the interior  of a neutron star quickly settles
into a superconducting  state immediately after birth, as  soon as its
temperature  falls  below  a  few times  $10^{9}$~K.  This  transition
happens within  a day or so after  the formation of the  star, and the
observed field values are much smaller than the critical field of this
superconductor. Hence, any field that is generated after the formation
of the star has to be embedded  in the region in which the protons are
in a  normal state. So post-formation generation  mechanisms give rise
to fields  that are completely  crustal and in particular  confined to
densities  below neutron  drip~\cite{blnd83,roma90,urpn92a}.

Following   an    idea   originally   proposed    by   \citeN{yako80},
\citeN{blnd83}   worked   out   a   possible   mechanism   for   field
generation. They suggested that magnetic field arises as a consequence
of thermal effects occurring in the outer crust in the early phases of
the thermal evolution of the star immediately after it is formed.  The
investigation  is  confined  to   the  degenerate  surface  layers  of
non-rotating neutron stars, assuming that the plasma is in hydrostatic
equilibrium.   The density range  considered is  $\sim 10^7  - 10^{11}
\gcc$,  where  the  mechanism  for field  enhancement  functions  most
effectively.

The field  can grow either in  the liquid phase and  then be convected
into the solid regions, or it could grow in the solid crust itself. In
the solid, the heat flux is carried by the degenerate electrons giving
rise to thermoelectric instabilities  that in turn make the horizontal
components  of the magnetic  field grow  exponentially with  time. The
field grows with  a time-scale of $\sim 10^5$  years. Such instability
can  also  develop if  the  liquid phase  that  lies  above the  solid
contains a horizontal  magnetic field. The coexistence of  a heat flux
and a  seed magnetic field, in  excess of $10^8$ Gauss,  in the liquid
will cause the  fluid to circulate which may  lead to effective dynamo
action. If that  is the case then the field will  grow rapidly, with a
time  scale of  about  a $10^2$  years,  and supply  the  flux to  the
solid. Either of these two instabilities will soon saturate to produce
a  field strength  of $\sim  10^{12}$ Gauss,  where  the instabilities
become non-linear.   Further growth will be prevented  when either the
magnetic stress  exceeds the lattice  yield stress or  the temperature
perturbations become non-linear, both  of which happen at a subsurface
field  of $\sim  10^{14}$ Gauss.  The corresponding  surface  field is
$\sim 10^{12}$ Gauss.  

The above mentioned mechanism  is most effective for large temperature
gradients.  And the  instabilities grow  only if  the  ohmic diffusion
time-scales are such that the  increase in the field is not dissipated
away  at a faster  rate than  the growth.  Unfortunately, for  that to
happen in  the solid the conductivity  should be at least  a factor of
three higher  than the present estimates.  Or if the  surface layer is
made of helium the instabilities can  grow to large values. It must be
noted here that the mechanism of field-growth mentioned here generates
small  loops only. The  growth time-scales  in the  solid is  too long
which makes the field generation  in the liquid layers the only viable
option. The most  effective way is to generate the  flux in the liquid
and then quickly  anchor it in the solid crust  either by advection of
the flux or by freezing the  liquid layer itself due to cooling of the
star. 

This  work was  extended for  the case  of rotating  neutron  stars by
\citeN{urpn86}.  They studied  the thermo-magnetic instability in less
deep  and hence  less dense  liquid layers.   In  non-rotating neutron
stars, the smearing of the  instability due to hydrodynamic motions is
an impediment to  fast growth of field in  the liquid.  Fast rotations
($P  \lsim 1$~s),  however  suppresses these  hydrodynamic motions  by
Coriolis force.   This mechanism is  most effective when  the electron
temperature is $\gsim  3 \times 10^6$~K and leads to  the growth of an
azimuthally symmetric  toroidal magnetic field.  Typical time-scale of
field growth  is of  the order  of a year  and the  typical horizontal
wavelength is  about 100 metres.  The field is  created at a  depth of
about 50 metres at density of the order of $10^7 \gcc$.

The post-formation field generation  mechanism is besieged by a number
of problems,  which were recognized  by the early  workers themselves.
First  and foremost  is the  fact that  this mechanism  is  capable of
generating only  toroidal modes.  And  the scale-size of the  field is
confined to  the melting depth of the  crust which is of  the order of
hundred    meters.    In    a   series    of   papers,    Geppert   \&
Wiebicke~\citeyear{gepp91,gepp95}        and        Wiebicke        \&
Geppert~\citeyear{wieb91,wieb92,wieb95,wieb96}  have  investigated the
problem of field  generation in the crust of  neutron stars in detail.
Their work  reiterates the fact  that the temperature  gradient driven
thermo-magnetic  instability can  only  give rise  to toroidal  field,
albeit  strong.  They  have  shown  that it  is  possible to  generate
toroidal modes of very high polarities ($n \sim 1000$) in a short time
($\sim$  years) and that  the magnitude  of such  field could  also be
quite large. But  how such field could get  restructured to a generate
large-scale polar fields,  the only kind that is  actually observed in
neutron stars, is still a point of much speculation.

As yet, there is no observational evidence for distinguishing one kind
of field  generation mechanism from the  other. The only  hope lies in
the nature of the evolution  of the field itself. Since the generation
mechanism decides  the question  of underlying current  structure, one
expects  that the  predicted field  evolution would  be  different for
these  two  cases,  and  observations  would be  able  to  probe  this
difference.  In chapter  [8] we  shall  make some  comments about  the
conclusions we  have drawn regarding  this problem from the  nature of
the field  evolution and the evolutionary link  between the population
of normal pulsars and their millisecond counterparts. 

\section{evolution}
\label{sevolution}

The  evolution of  neutron star  magnetic  field has  been of  abiding
interest both because  it is a challenging problem  in itself and also
due to  the fact that many  other aspects of  pulsar physics crucially
depend on the nature of field  evolution. Even though over the years a
somewhat coherent  picture has emerged,  some of the key  issues still
remain  unresolved.   In  particular,  the generation  of  millisecond
pulsars  and  their evolutionary  link  to  the  population of  normal
pulsars through a  processing in binaries has become  one of the major
challenges  that faces the  researchers at  present.  There  have been
several excellent reviews detailing the current worries and the status
of  the  theoretical  endeavors~\cite{db95a,db95b,bhat95,rudr95}.   In
this chapter, we shall  recapitulate, in brief, the present situation,
the observational indications and the theoretical models.

From observational facts and from the recent statistical analyses made
on  the pulsar  population,  the following  conclusions  can be  drawn
regarding the field evolution.
\bei
\i  Isolated pulsars  with high  magnetic fields  ($ \sim  10^{11}$ --
$10^{13}$ G) do  not undergo any significant field  decay during their
lifetimes~\cite{bhat92,waka92,lorm94,hart97}.
\i The  fact that binary pulsars  as well as  millisecond and globular
cluster pulsars which  almost always have a history  of being a member
of  a  binary,  possess  much  lower field  strengths,  suggests  that
significant field decay occurs only  as a result of the interaction of
a neutron star with its binary companion~\cite{bail96}.
\i The age  determination of pulsars with white  dwarf companions show
that most of the millisecond pulsars are extremely long-lived. The old
age ($\sim  10^9$ years) of  the low-field pulsars implies  that their
field is  stable over  long time-scales. That  is after  the recycling
process  in the  binaries  is over,  the  field does  not undergo  any
further decay~\cite{dbgs86,kulk86,heuv86,vwb90}.
\i  The evolutionary  link  between millisecond  pulsars and  low-mass
X-ray binaries seem to be  borne out both from binary evolution models
and  from  the  comparative  study  of the  kinematics  of  these  two
populations. To spin a neutron star up to millisecond periods at least
an amount $\sim 0.1 \msun$ needs  to be accreted.  Such huge amount of
mass transfer is  possible only in low mass  X-ray binaries, where the
duration of mass  transfer could be as long as  $10^9$ years. And even
though  the present  estimates for  the  birthrate of  low mass  X-ray
binaries     fall    short    of     that    of     the    millisecond
pulsars~\cite{kulk88,lorm95},                                 kinematic
studies~\cite{cord90,wols94,nice95,nica95,bhat96a,rama97,cord97},
indicate that  the two  populations are most  likely to be  related as
both have very similar kinematic properties.
\eei

\subsection{models of field evolution}
\label{ssmodel}

In order  to understand  the above facts,  attempts have been  made to
relate the  field decay  to the star's  binary history. There  are two
classes of  models which have been  explored in this  context one that
relates the magnetic field evolution to the spin evolution of the star
and the  other attributing  the field evolution  to direct  effects of
mass  accretion (see \citeNP{rudr95}  and \citeNP{db95a}  for detailed
reviews).  It should also be mentioned here that almost all the models
are  built  upon  two   themes,  namely,  a  large  scale  macroscopic
restructuring of the fields in  the interior of the star (for example,
the spin-down  and expulsion of flux  from the superfluid  core of the
star)  and a  microscopic  mechanism (like  ohmic  dissipation of  the
currents in the crust) acting to actually kill the underlying currents
supporting  the  observable  field.   The different  class  of  models
usually assume different kind  of initial field configuration.  Models
depending  on  spin-down  assume   a  core-flux  supported  by  proton
superconductor flux tubes.  Whereas, models invoking ohmic dissipation
usually assume  an initial crustal  configuration. It should  be noted
here that a whole host of models exist that discuss field evolution of
isolated neutron stars (for a  list of such models see \citeN{db96b}).
But we shall confine our discussion to accretion-induced models alone.

The former class of models involves the inter-pinning of the Abrikosov
fluxoids  (of  the superconducting  protons)  and the  Onsager-Feynman
vortices     (of      the     superfluid     neutrons)      in     the
core~\cite{srini90,rudr91a}).   This class  also  includes the  models
involving    the    plate    tectonics    of    the    neutron    star
crust~\cite{rudr91b,rudr91c}.    \citeN{srini90}   pointed  out   that
neutron stars interacting with the companion's wind would experience a
major  spin-down,  causing  the  superconducting  core  to  expel  the
magnetic  flux, which  would then  undergo ohmic  decay in  the crust.
This coupled  evolution of  spin and magnetic  field has  been modeled
both for the  cases of wide low mass  X-ray binaries~\cite{miri94} and
high mass X-ray binaries~\cite{miri96}.  They have assumed ohmic decay
of the expelled  field in the crust with a  constant decay constant of
$10^8  - 10^9$ years.   An investigation  of the  nature of  the ohmic
decay of  such expelled  field in an  accretion heated crust  has also
been attempted~\cite{bhat96},  and shows  that the final  field values
could be quite consistent with those observed for millisecond pulsars.
Ruderman~\citeyear{rudr91b,rudr91c}  on  the  other hand,  suggests  a
coupling between the  spin and the magnetic evolution  of the star via
crustal plate  tectonics -  torques acting on  the star  cause crustal
plates, and the magnetic poles anchored in them, to migrate, resulting
in major changes of the effective dipole moment.

The other class of models  attribute the field decay to direct effects
of accretion. The  work done in this thesis  comes under this category
of models.   Previous work  by \citeN{bisn74} and  \citeN{taam86} have
suggested that  accreted matter  might screen the  pre-existing field.
Computations  by \citeN{roma90} indicate  that hydrodynamic  flows may
bury the pre-existing field reducing  the strength at the surface.  We
shall  see  in chapter[4]  that  strong Rayleigh-Taylor  instabilities
prevent such hydrodynamic flows to create horizontal components of the
field at  the expense of the  dipolar component, and  therefore such a
screening mechanism  does not provide  for a viable scenario  of field
evolution in an accreting neutron star.  The mechanism of ohmic decay,
being unique  to the  crustal currents, is  also used in  models where
spin-down is invoked for  flux expulsion, for a subsequent dissipation
of such flux in the crust~\cite{miri94,bhat96}.

The most important microscopic mechanism invoked for accretion-induced
field decay is that of  fast ohmic decay. In an accretion-heated crust
the decay takes place principally  as a result of rapid dissipation of
currents due to the decrease  in the electrical conductivity and hence
a        reduction        in        the       ohmic        dissipation
time-scale~\cite{gepp94,urpn95,urpn96,kb97}.    The   crustal   field
undergoes ohmic diffusion due to the finite electrical conductivity of
the crustal  lattice, but  the time-scale of  such decay is  very long
under   ordinary   conditions~\cite{sang87,urpn92b}.   The   situation
changes significantly when accretion is  turned on. The heating of the
crust  reduces  the  electrical  conductivity  by  several  orders  of
magnitude, thereby reducing the ohmic decay time-scale.

There is also  an additional effect that acts  towards stabilizing the
field.  As  the mass increases, a  neutron star becomes  more and more
compact  and the  mass  of the  crust  actually decreases  by a  small
amount.  So  the newly  accreted  material  forms  the crust  and  the
original  crustal  material  gets  continually  assimilated  into  the
superconducting core  below. The original current  carrying layers are
thus  pushed   into  deeper  and  more  dense   regions  as  accretion
proceeds.  The  higher  conductivity   of  the  denser  regions  would
progressively  slow  down  the  decay,  till  the  current  loops  are
completely inside  the superconducting region where  any further decay
is prevented~\cite{kb97}.

\subsection{the millisecond pulsar question}
\label{sfield-msp}

A  summary  of  the  current  status  of  the  important  question  of
generation  of  millisecond pulsars  has  recently  been presented  by
\citeN{bhat96a}  in the  proceedings of  the IAU  colloquium  160. The
debate  that   followed  the   presentation  (recorded  in   the  same
proceedings) gives a  fair indication as to how  bad the situation is.
Here we shall  just record the facts relevant  to the field evolution,
ignoring such questions that  are associated with binary evolution and
other factors.

In the fifteen years following  the discovery of the first millisecond
pulsar more  than fifty pulsars  have been discovered which  belong to
this particular  category. In an  attempt to understand the  origin of
millisecond pulsars,  it was suggested  that these are  {\em recycled}
pulsars,  pulsars that  have  evolved to  the characteristic  magnetic
field and spin-period by virtue of their binary history. 

The major  problem regarding the generation of  millisecond pulsars in
the binaries  is the question  of progenitors.  What kind  of binaries
would give rise to millisecond pulsars? We have already mentioned that
there is a  problem of birthrate mismatch if one  assumes that all the
millisecond  pulsars come  from low  mass X-ray  binaries  we normally
observe.  There is no satisfactory  explanation for the origin of {\em
isolated} millisecond  pulsars either.  Any model  for field evolution
has to be consistent with the nature of binary evolution that produced
these  objects. The  field evolution  scheme must  also provide  for a
limiting  minimum field  strength---the so-called  `flooring'  seen at
$\sim 10^8$~G should  arise out of the evolution  itself. Except for a
few  of the models~\cite{roma95,miri94,kb97}  an explanation  for this
has not been attempted so far.  And the models of field evolution must
also   match  the   spin-evolution  of   the  neutron   star   in  the
binaries. Since  there is  no consensus on  either the  internal field
structure or the  evolution of the field, one  important check on such
models  would be  to look  for a  consistency of  the  field evolution
models with  that of the  binary evolution scenario. In  chapter[6] we
attempt to look  for such a consistency in general  and from the point
of view of millisecond pulsar generation in particular.

\section{general introduction to accretion-induced field evolution}

In  the previous  section we  have seen  that observational  facts and
statistical analyses of the existing data clearly relate the evolution
of magnetic  field of neutron  stars with binary interaction.   In the
next  chapter  we  shall  address  three  different  aspects  of  such
binary-induced field evolution. We investigate the underlying physical
mechanisms that take  place in the interior of  the star, changing the
currents that are  flowing to maintain the observable  field. We would
interest ourselves with  a neutron star that is  in active interaction
with its  binary companion.  This  companion could be a  main sequence
star affecting  the neutron star through its  wind.  Alternatively, it
could be a  giant or a super giant, in which  case heavy mass transfer
may take  place through Roche-lobe  overflow~\cite{bhat91,king95}.  In
either of these  cases, there are some basic  changes that the neutron
star  undergoes.  These  changes  in various  physical parameters  are
responsible for  a change in the  magnetic field. So, here  is a brief
resume of how the process of accretion affects a neutron star.

\bei
\i The  process of  accretion mainly changes  three parameters  of the
star,  the  total mass,  the  total  angular  momentum and  the  total
energy. The change in energy  comes about when the potential energy of
the accreting  particles get converted  into random kinetic  energy of
the system and is then manifested as an increase in temperature.
\i Models of  field evolution have been proposed  using the changes in
each of these quantities, namely relating the change in the observable
field $\delta B$ to $\delta M$ (change in mass), $\delta J$ (change in
angular  momentum) and  $\delta T$  (change in  temperature).  In this
connection  it should  be mentioned  that earlier  attempts to  show a
proportionality between  $\delta B$ and $\delta M$  have recently been
challenged~\cite{wije97}.
\i The change  in angular momentum has been  exploited in models which
assume an  initial field confined  to the superconducting core  of the
star. In particular, a rapid  slow-down ($\delta J \ll 0$) experienced
by the neutron star during the `propeller phase' causes the core field
to be expelled  to the crust, where it  could undergo subsequent ohmic
dissipation.
\i Ultimately most of the accretion-related models depend on the ohmic
dissipation of currents  in the crust for a  permanent decrease in the
field. The rapidity  of such dissipation depends on  a decrease of the
electrical  conductivity, caused  by  an increase  in the  temperature
($\delta T \ge 0$).
\i  Another important factor  influencing the  field evolution  is the
hydrodynamic mass  motion in the  star. Near the surface  the material
flow  is   lateral,  moving   from  polar  (magnetic)   to  equatorial
region. But interchange instabilities  appear to prevent the screening
of the  field due  to such hydrodynamic  motions (chapter [4]  of this
thesis).
\i  Deep inside  the star  the material  movement is  a  simple radial
inflow, due to the compression of  the star as a result of an increase
in  the total  mass.  The  material movement  is, evidently,  a direct
effect  of the  change  in mass  ($\delta  M$), with  the actual  flow
dynamics determined by the rate of such change, i.e., the rate of mass
accretion, $\mdot$,  on the star.  This radially inward motion  in the
deeper layers helps  to advect crustal currents to  more dense layers,
ultimately stabilizing the field in such cases.
\eei

From the above  discussion, we see that processes  of both macroscopic
and  microscopic nature  are active  in the  interior of  an accreting
neutron star.   The change in  angular momentum induces a  large scale
flux movement whereas a change in total mass gives rise to large scale
material movement.  Both these  cause macroscopic restructuring of the
current pattern.  On  the other hand, a change  in temperature induces
the  microscopic process  of ohmic  dissipation, by  which  the energy
stored in the  large scale field is transferred  to the random kinetic
energy of the  particles. In the following sections,  we shall see how
these two  mechanisms complement each other in  the problems addressed
by us.

\chapter{effect of diamagnetic screening}
\label{cds}

\section{introduction}
     
First proposed by \citeN{bisn74}, the  idea of a possible screening of
the magnetic  field of a neutron  star by accreting  material onto it,
has  been a  recurrent  theme in  the  field-evolution scenario.  This
suggestion was  substantiated by the  work of \citeN{taam86}  in which
they indicated  that the accreted matter, which  is completely ionized
plasma and hence diamagnetic  in nature, might screen the pre-existing
field   reducing   the  strength   of   the   surface  field.    Later
\citeN{blnd79}   proposed   a    possible   mechanism   for   such   a
screening. They  suggested that the material accreting  onto the poles
of a magnetized  neutron star will be confined  by the strong magnetic
stresses  near the surface  of the  star. At  low accretion  rates the
material  sinks  below  the  stellar  surface  until  the  hydrostatic
pressure of the stellar material is as large as the magnetic pressure.
The plasma then flows sideways giving rise to horizontal components of
the magnetic  field.  The creation of this  horizontal component comes
at the expense  of the vertical field and may result  in a decrease in
the observed field  strength. This work, however, did  not provide any
quantitative details of the mechanism.

Later  it was  shown  by \citeN{wosl82}  and  \citeN{hame83} that  the
accretion column  is actually like a  small mountain on  the polar cap
rather  than  being  subsurface.   The accreting  material  is  mostly
ionized hydrogen and hydrogen has  a smaller mean molecular weight per
electron, $\mu_e$, than the iron crust  of the star. Since most of the
pressure  in the  accreted layer  comes from  the  electrons, hydrogen
tends  to  float over  the  underlying iron  layer  -  if no  material
displacement  perpendicular  to the  magnetic  field  is possible.   A
hydrogen mountain then  forms at the surface of  the neutron star. The
height  of  this  mountain  depends   on  the  density  at  which  the
transmutation  of  hydrogen (by  electron  capture  on protons)  takes
place.  When  the material pressure  in this accretion  column becomes
much  larger  than the  magnetic  pressure  then  the material  starts
flowing sideways  giving rise to horizontal  field components reducing
the external dipolar field strength.

The  first  quantitative  calculations  of  diamagnetic  screening  of
neutron star  magnetic field were performed by  \citeN{roma90}. It was
shown in this  work that hydrodynamic flows in  the surface layers may
bury  the field  to  deeper layers  effectively  reducing the  surface
strength.   In a  later article~\cite{roma95}  in continuation  to the
earlier work, various time-scales, relevant for an effective screening
of the surface field, were estimated.   It was also shown here how the
depth and  density at which the  field might get buried  depend on the
strength of the initial surface field.

In the  present work, we investigate the  effectiveness of diamagnetic
screening as  a possible  mechanism for a  permanent reduction  in the
surface field strength  of the neutron stars. In  particular, we shall
try to answer the following questions :
\ben
\i  Are  diffusive time-scales,  in  the  layers  where the  field  is
expected  to  be buried,  long  enough to  allow  screening  to be  an
effective mechanism for long-term field reduction ?
\i  Is it  at all  possible  to bury  the field  or create  horizontal
components at the expense  of the vertical one against Rayleigh-Taylor
overturn instability ?
\een

Recently, calculations  of screening have been  performed assuming the
accreting material to  be ferromagnetic~\cite{zhan94,zhan98}.  Such an
assumption of extremely large  magnetic permeability makes it possible
to reduce the surface field by  about four orders of magnitude. In our
work though, we regard the accreting material to be completely ionized
plasma (mainly hydrogen) and therefore to be diamagnetic in nature.

The  layout  of the  chapter  is in  the  following  form. In  section
[\ref{sds-matflow}] we have discussed  the nature of the material flow
in an  accreting neutron star. In section  [\ref{sds-phys}] we discuss
the  actual mechanism  of the  diamagnetic screening  and  the various
relevant  time-scales.  And  finally in  section  [\ref{sds-concl}] we
present our conclusions. 

\section{accretion and material flow}
\label{sds-matflow}

For  a neutron  star with  a strong  magnetic field,  the flow  of the
accreting material  outside the star  is completely determined  by the
field.   The material flows  in along  open field  lines and  hits the
surface within the polar cap region. The area of this polar cap region
is  determined  by the  rate  of accretion  and  the  strength of  the
magnetic  field~\cite{shap}.   In all  our  subsequent discussions  we
shall assume a situation where the magnetic axis and the rotation axis
are aligned. Even though that is hardly ever the case in reality, this
assumption does not affect the general conclusions much.

The  polar cap area,  region constrained  by open  field lines  on the
surface of the star, is given by,
\beq
A_P = 2 \pi R_s^2 [1 - \cos \theta],
\eeq
where, $R_s$ is  the radius of the star and $\theta$  is the angle the
last open  field line  makes with magnetic  axis. This  limiting field
line is  the one which goes  through the Alfv\'{e}n  radius, $R_A$, in
the  equatorial plane. Therefore,  using the  equation for  the dipole
field lines one finds  $\sin \theta = \sqrt{\frac{R_s}{R_A}}$, so that
the polar cap area is given by,
\beq
A_P = 2 \pi R_s^2 \left[1 - \sqrt{1 - \frac{R_s}{R_A}}\right].
\eeq
In the limit $R_A >> R_s$ (which is a reasonable approximation for the
range of  field strength  and accretion rate  we consider),  the above
expression reduces to the following simple form :
\beq
A_P = \pi \frac{R_s^3}{R_A}.
\eeq

The  Alfv\'{e}n radius  of an  accreting system  is determined  by the
condition  that at  this radius  the  ram pressure  of the  in-falling
material equals the  pressure of the magnetic field.  The ram pressure
at the Alfv\'{e}n radius is :
\beq
P_{\rm ram} = \frac{1}{2} \rho(R_A) V^2(R_A),
\eeq
where $\rho(R_A)$ is  the density and $V(R_A)$ is  the velocity of the
accreting  material  at  Alfv\'{e}n  radius.   In  an  accretion  disc
material rotates  with Keplerian velocity at  every radius. Therefore,
the velocity of the material at Alfv\'{e}n radius is given by,
\beq
V(R_A) = (\frac{GM}{R_A})^{1/2}. \label{ealfvel}
\eeq
The density of the accreting material at Alfv\'{e}n radius is :
\beq
\rho(R_A) = \frac{\mdot}{4 \pi V(R_A) R_A^2},
\eeq
where,  the  dependence  on  the  rate  of  mass  accretion  is  quite
obvious. As the field is dipolar  in nature, the field strength at the
Alfv\'{e}n radius is,
\beq
B(R_A) = (\frac{R_s}{R_A})^3 B_s,
\eeq
where, $B_s$  stands for the strength  of the field at  the surface of
the neutron star. Equating the ram pressure and the magnetic pressure,
one obtains  the following expressions  for the Alfv\'{e}n  radius and
the polar cap area respectively,
\ber
R_A &=& (2GM)^{-1/7} R_s^{12/7} B_s^{4/7} {\mdot}^{-2/7} \\ 
\label{ealfrad}
A_P &=& \pi (2G)^{1/7} M^{1/7} R_s^{9/7} B_s^{-4/7} {\mdot}^{2/7}. 
\label{epolcap}
\eer

The density  profile within  the accretion column  is described  by an
`atmosphere' solution~\cite{bild95}.  The pressure in  the column, due
to the relativistic degenerate electrons and the ions, is given by
\beq
P = \frac{1}{4} n_e m_e c^2 x + n_I k_B T, \label{eeos-atmos}
\eeq
where $n_e(n_I)$ is  the number density of electrons  (ions) and $m_e$
is the electron mass. $x$ is the relativity parameter defined as
\ber
x &=& \frac{p_F}{m_e c} \nonumber \\
&=& 1.008 (\frac{Z \rho_6}{A})^{1/3},
\eer
where $p_F$ is the Fermi momentum of the electrons, $(Z,A)$ correspond
to the atomic  number and mass number of the  dominant ion species and
$\rho_6$ is density  in units of $10^6 \gcc$. The  scale height of the
accretion column is given by  $H = \frac{P}{\rho g}$, where $\rho$ and
$P$  are the  density and  pressure respectively  at the  base  of the
accretion column  and $g$  is the acceleration  due to gravity  at the
surface  of the state.  Using the  above equation  of state  the scale
height, for a $1.4~\msun$ star, is approximately found to be :
\beq
H \sim 324 \; \rho_6^{1/3} {\rm cm}.
\eeq

As long as  the pressure in the accretion column  is much smaller than
the magnetic  pressure there can  not be any material  movement across
the field lines so the accreted material remains confined to the polar
cap.   This kind  of polar  cap accretion  causes  transverse pressure
gradients that are balanced by the curvature of the magnetic field. An
order of magnitude estimate  by Hameury et al.~\citeN{hame83} shows
that in  order to create a  significant distortion of  the field lines
due to the overpressure of the accretion column the condition
\beq 
\frac{P_{\rm accretion}}{P_{\rm magnetic}}  \gsim \frac{R_P}{H} 
\label{erh-fl} 
\eeq
should be satisfied. Here,  $P_{\rm accretion}$ and $P_{\rm magnetic}$
stand for pressure due to the material in the accretion column and due
to the magnetic  field respectively. $R_P$ is the  radius of the polar
cap and $H$ is the density scale height of the accretion column. For a
given value  of the crustal field  and a given rate  of accretion, the
density at  which the material starts flowing  sideways, obtained from
the condition  given by equation  [\ref{erh-fl}], is :  
\beq 
\rho_{\rm flow}  =  2.0961 \times  10^8  B_{13}^{36/35} \mdot_{-9}^{3/35},  
\eeq
where $B_{13}$ is  the field strength in units  of $10^{13}$~Gauss and
$\mdot_{-9}$ is  the accretion rate in units  of $10^{-9}$~\dmdt. This
condition  is, of  course, only  valid in  absence of  any interchange
instabilities  about   which  we   shall  discuss  later.   In  figure
[\ref{frh-fl}]  we plot  the variation  of the  flow density  with the
field strength for various values of the accretion rate. 

\bef
\begin{center}{\mbox{\epsfig{file=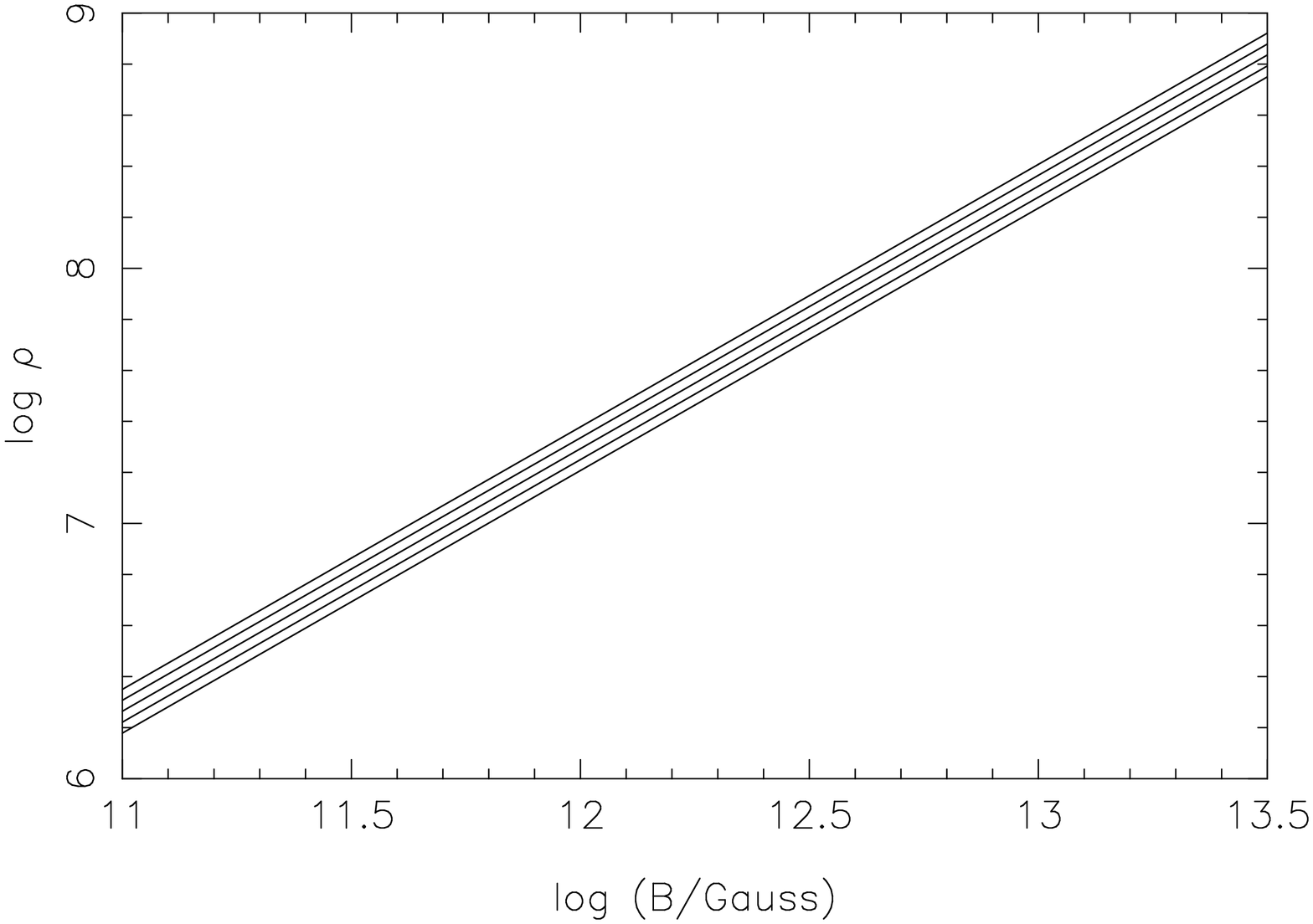,width=235pt,angle=-90}}}\end{center}
\caption[flow density]{Variation of the  flow density with the surface
fields strength. The different  curves (from top to bottom) correspond
to  $\mdot = 10^{-8},  10^{-8.5}, 10^{-9},  10^{-9.5}, 10^{-10}$~\dmdt
respectively.}
\label{frh-fl}
\eef

\section{physics of diamagnetic screening}
\label{sds-phys}

We shall, as has been mentioned  above, work under the assumption of a
{\em  flux-frozen} situation.  This basically  amounts to  assuming an
effectively infinite conductivity.  Admittedly, even though large, the
actual  value of  conductivity  is  finite. But  the  large values  of
conductivity   in   crust  provide   for   extremely  long   diffusive
time-scales.  In this  case, the  assumption of  `flux-freezing'  is a
valid one since the flow time-scales are much smaller compared to this
diffusive time-scale (to be discussed in the next section). 

Therefore  a reduction in  the surface  field is  obtained due  to the
combined effect of  i) flux freezing and ii)  hydrodynamic flow of the
material.  Due to  flux-freezing the  total flux  contained  within an
amount of material remains constant.  And when this material flows and
spreads out  covering a larger area,  the field gets  adjusted to keep
the flux conserved. Say, the total flux contained within the polar cap
at time  $t=0$ is  $B \times A_P$.  Conservation of flux  then implies
that we have,
\beq
\frac {d B}{d t} \times A_P + B \times \frac {d A_P}{d t} = 0.  \label{edBdt}
\eeq

Therefore the change in the field strength is related to the change in
the area of the original polar cap surface due to material flow during
accretion. Assuming  that the  flow of material  is such that  at each
instant the amount  accreted equals the amount flowing  out from under
the accretion  column one obtains  the following relation  between the
change in  the polar cap area  and the rate of  mass accretion $\mdot$
given by,
\beq
\mdot dt = \rho_{\rm flow} h_{\rm flow} dA_P \label{edAdt} 
\eeq
where,  $dA_P$  is   the  change  in  the  polar   cap  area  in  time
$dt$. $\rho_{\rm flow}$ is the  density where the flow takes place and
$h_{\rm flow}$ is the effective  height of the accretion column, which
hereafter  we take roughly  to be  equal to  the density  scale height
$H$. Integrating  equation [\ref{edAdt}]  above the time  variation of
the polar cap area is obtained in the following form -
\beq
A_P(t) = A_P(t=0) + \frac {\mdot t}{\rho_{\rm flow} H},
\eeq
where,  $A_P(t=0)$  is the  original  area  of  the polar  cap.   Then
equation [\ref{edBdt}]  implies the  following time dependence  of the
surface field :
\beq
B(t) = \frac {B_0} {[1 + \frac {t}{\tau_{\rm screen}}]},
\eeq
where we have defined the screening time-scale by the relation
\beq
\tau_{\rm screen} = \frac{\rho_{\rm flow} H A_P}{\mdot}. 
\label{etscreen}
\eeq
This is  the time in which the  original field reduces to  half of its
original value.  It should be noted here  that \citeN{shib89} obtained
an expression  for the decay of  magnetic field due  to accretion very
similar  to   equation  [\ref{edBdt}]  from   purely  phenomenological
considerations. Here we have arrived at the above relation from a more
physical point of view.

To find  the effectiveness  of the screening  we need to  compare this
time-scale  to  the diffusive  time-scale  over  which  the field  may
re-emerge  by  ohmic  diffusion  through the  overlaying  layers.  The
diffusive time-scale is defined by,
\beq
\tau_{\rm diff} = \frac {4 \pi \sigma (\rho_{\rm flow}, T({\mdot})) H^2}{c^2}, 
\label{etdiff}
\eeq
where, $\sigma$ is the electrical  conductivity which is a function of
the density and the crustal  temperature.  We shall show later that if
one  ignores interchange instabilities,  then the  flow time  scale is
much  smaller than  diffusion time  scale, and  therefore  field would
remain  frozen and  be dragged.  Reconnection will  then occur  on the
equatorial plane and that will  bury the field.  Before this field can
diffuse out more  matter will come and spread on top  of it, and drive
the field deeper.  So the eventual re-emergence time  scale is set not
by the  initial depth at  which it was  buried but the final  depth to
which it is driven by continuous  accretion. This can be very deep and
the  field  can  even reach  the  core  and  therefore never  get  out
again. But for the moment, we take the burial depth to be equal to the
scale height  of the  density. For the  screening time-scale  this, of
course, is the actual length-scale. For the diffusive case we find the
time-scale at the depth where the  field gets buried to begin with. We
have not  considered the  case of deep  burial mentioned  above, since
from the subsequent discussion that will prove to be unnecessary.

The  point  to be  noted  here is  that  for  the screening  mechanism
described above the important  factor is the anisotropic material flow
through the polar cap. The area of polar cap increases with increasing
rate  of accretion. Hence,  for low  rates of  accretion, the  flow is
maximally anisotropic.  For high values  of accretion rate,  the polar
cap  could cover a  large fraction  of the  stellar surface  area. The
effect of anisotropic  material flow would not be  very severe in that
case.  Also as  mentioned  earlier, the  temperatures  are higher  for
higher values $\mdot$,  a lower value of conductivity  and therefore a
smaller diffusive  time. Hence, situations with low  values of $\mdot$
has better chance of screening the field. 

In all of the above  discussion, the implicit assumption has been that
it is possible to create a horizontal component of the magnetic field,
at the expense of the vertical component, as a result of material flow
in a `flux-frozen' condition. But this mechanism is not viable against
the  Rayleigh-Taylor  instability.  This  instability  arises  due  to
magnetic buoyancy.  As the  accreting material slowly  builds up  in a
column,  the field lines  tend to  spread out  due to  the diamagnetic
property of  the material.  At the base  of the accretion  column, the
field lines are firmly anchored to the solid crust. Therefore material
flow at the base of the accretion column stretches the field lines out
horizontally. This  creates horizontal components of  the field giving
rise to a screening of the  external dipole. We have seen earlier that
the pressure of the accretion  column is much larger than the pressure
of the magnetic field in this region. Under such a condition the field
lines re-organize  themselves into flux  tubes as that is  the minimum
energy  configuration. These flux  tubes move  upward due  to magnetic
buoyancy  and thereby destroy  the horizontal  component of  the field
(see \citeNP{spru83} and references therein).

In a tube  in pressure equilibrium with its  surroundings the pressure
is
\beq
P_{\rm in} + P_{\rm magnetic} = P_{\rm out}.
\eeq
However at the base of the accretion column $\beta \equiv \frac{P_{\rm
gas}}{P_{\rm  magnetic}} \sim  \frac{R_P}{H}$ and  the  above relation
simplifies to
\beq
P_{\rm in} = (1 - 1/\beta) P_{\rm out}.
\eeq
Therefore, the gas pressure within a flux tube is smaller than the gas
pressure outside.  Then from  equation [\ref{eeos-atmos}] we  find the
following relation between internal and external densities :
\ber
\rho_{\rm in}  &=& (1  - 1/\beta)^{3/4} \rho_{\rm  out} \\  
&{\rm or}& \nonumber \\
\frac{\delta \rho}{\rho_{\rm in}} &=& \frac{3}{4 \beta},
\eer
where $\delta \rho$ is the density deficit within the flux tubes. This
density deficit  makes the  flux tube move  upward under the  force of
buoyancy. The upward velocity $V_F$ of  the flux tubes is given by the
equality of  the buoyancy  force with the  aerodynamic drag  force per
unit area of the flux tube :
\beq
\frac{1}{2} \rho_{\rm in} V_F^2 = g \, r \, \delta \rho,
\eeq
where $r$ is the characteristic radius of the flux tube and $g$ is the
acceleration due to gravity. Therefore, the velocity is :
\beq
V_F = \sqrt{\frac {3 g r}{2 \beta}}
\eeq
and the time-scale required for a flux-tube to traverse the density scale-height is :
\beq
\tau_{\rm RT} = \frac{H}{V_F}.
\eeq
It should  be noted that the  characteristic scale length  of the flux
tubes is again that of the density scale-height, and therefore,
\beq
\tau_{\rm RT} = \sqrt{\frac{2 \beta H}{3 g}}. \label{etRT}
\eeq

The time-scale for  the reconnection of the field  lines, on the other
hand, is typically given by
\ber \tau_{\rm  recon} &=& \frac{l_{\rm  instability}}{V_A}, \mbox{ if
$V_A < V_S$} \nonumber \\
&=& \frac{l_{\rm instability}}{V_S}, \mbox{ if $V_S < V_A$};
\eer
where,   $V_A,  V_S$  are   the  Alfv\'{e}n   and  the   sound  speed,
respectively.  $l_{\rm  instability}$   is  the  length-scale  of  the
instability which is again of the order of the density scale height in
the accretion column.  Within the accretion column the  sound speed is
always much  greater than the  Alfv\'{e}n velocity and  therefore only
the top expressions  of the above equations are  relevant here. At the
flow density, the Alfv\'{e}n velocity is
\beq
V_A = \frac{B}{\sqrt{4 \pi \rho_{\rm flow}}}.
\eeq
Therefore, the expressions for the reconnection time-scale is given by
\ber
\tau_{\rm  recon} &=&  \frac{H  \sqrt{4 \pi  \rho_{\rm flow}}}{B},  \\
\label{etrecon}
\eer

For the  screening to be effective,  the total time taken  to create a
loop of field  lines by turn-over instability and  to destroy the loop
by reconnection  has to be  much larger than the  screening time-scale
defined above. So we define an instability time-scale as
\beq
\tau_{\rm instability} = \tau_{\rm RT} + \tau_{\rm recon} \label{etinst}
\eeq
which should  be compared with  $\tau_{\rm screen}$. It must  be noted
here   that  we   use   the  `pressure   scale-height'  and   `density
scale-height'  interchangeably  in  the  above  discussion  since  for
barotropic (atmospheric)  equation of  state (as is  the case  here) a
pressure gradient implies the same density gradient.

\bef
\begin{center}{\mbox{\epsfig{file=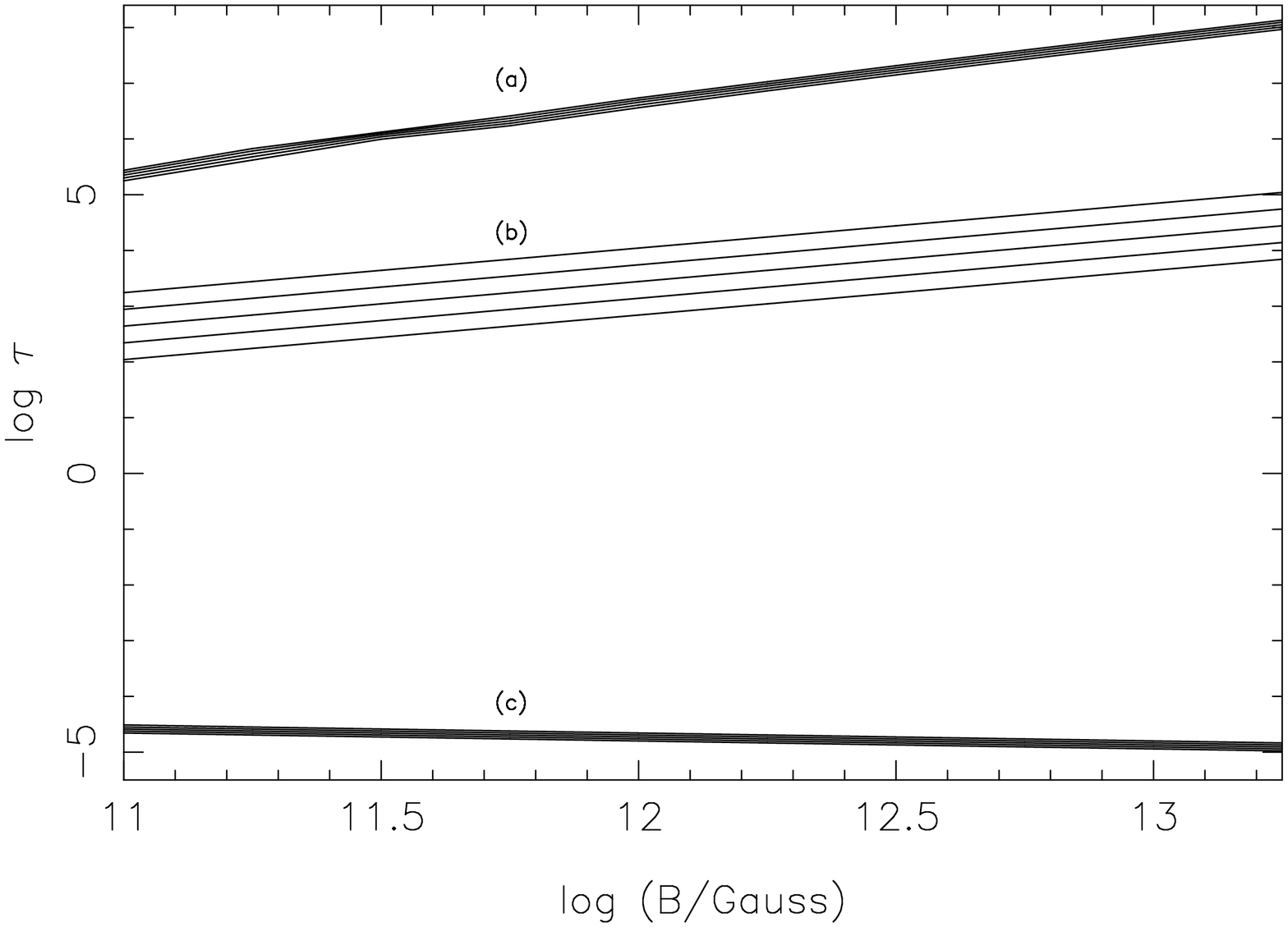,width=235pt,angle=-90}}}\end{center}
\caption[time-scales]{Variation of the time-scales with surface fields
strength.  The  (a), (b),  and  (c)  groups  of curves  correspond  to
diffusive,  screening and  instability  time-scales respectively.  The
curves  (in each group)  correspond to  $\mdot =  10^{-10}, 10^{-9.5},
10^{-9},  10^{-8.5}$~\dmdt  respectively, the  upper  curves being  of
lower values of accretion.}
\label{ftime-scale}
\eef

Therefore the three relevant time-scales of the problem are :
\bei
\i screening time-scale -  given by equation [\ref{etscreen}];
\i diffusive time-scale - given by equation [\ref{etdiff}]; and
\i instability time-scale - given by equation [\ref{etinst}].
\eei
In figure[\ref{ftime-scale}] we plot  all the time-scales as functions
of surface field strength and the accretion rate. 

\section{discussion and conclusions}
\label{sds-concl}

Looking at the  figures plotted above, obtained for  a range of values
for the accretion rate and the surface field strength, we find that,
\ben
\i  The  density of  flow  increases with  an  increase  in the  field
strength. The larger  the field, the deeper and  denser it gets buried
(see figure [\ref{frh-fl}]). The flow density also has a mild positive
dependence  on the rate  of mass  accretion. But  even for  very large
values  of the  surface  field strength  the  flow does  not occur  at
densities beyond  $\sim 10^9~\gcc$. That  means the flow  always takes
place within the liquid layer. And the earlier contentions of a burial
within the solid layer does not really happen.
\i  As the  screening time-scales  are  always much  smaller than  the
diffusive  time-scales  a  condition  of  flux-freezing  prevails  and
material  movement  indeed should  proceed  dragging  the field  lines
along.
\i But, since the instability  time-scale (sum of the overturn and the
re-connection  time-scales) is  much too  smaller than  the  other two
time-scales of  the problem, any  stretching of field line  is quickly
restored   (over  the   instability  time-scale)   and   the  material
effectively flows past the field lines without causing any change.
\een
We find  that the time-scale of  overturn and reconnection  is so much
smaller than any other time-scale of the problem that it is not at all
possible to create horizontal field components at the cost of vertical
ones and effect  a screening. Therefore, it is  not possible to screen
the  magnetic field of  a neutron  star by  the accreting  material in
order to reduce the magnitude of the external dipole observed.

\chapter{evolution of crustal magnetic field in an accreting neutron star}
\label{cmn}

\section{introduction}

It  has been  discussed  in chapter  [\ref{cfield}] that  observations
suggest  a connection  between the  low magnetic  field of  binary and
millisecond  pulsars  and their  being  processed  in binary  systems,
indicating accretion-induced  field decay in such  systems. A possible
mechanism  for  such  decay  is  understood  to  be  the  rapid  ohmic
dissipation of the currents in the accretion heated crust (see section
[\ref{sevolution}] for  details). In this work we  explore models that
primarily  depend on  this microscopic  process of  ohmic decay  for a
permanent reduction  in the observed  field strength, and also  on the
large  scale material  movement  for the  final  stabilization of  the
field. Here we shall investigate  a model based on the assumption that
the  current loops responsible  for the  magnetic field  are initially
confined entirely within the crust. This situation is likely to be the
result  of   a  generation  of   the  field  due   to  thermo-magnetic
instabilities after the birth of the star (see section [\ref{sorigin}]
for   details).  If   the  initial   field  resides   mainly   in  the
superconducting core then our scenario  would apply only after most of
this flux has been expelled into  the crust, a situation which will be
treated in chapter [7]. 

The  crustal  field  undergoes  ohmic  diffusion  due  to  the  finite
electrical conductivity of the  crustal lattice, but the time-scale of
such      decay       is      very      long       under      ordinary
conditions~\cite{sang87,urpn92b}.  The situation changes significantly
when  accretion is turned  on. The  heating of  the crust  reduces the
electrical  conductivity  by  several  orders  of  magnitude,  thereby
reducing the ohmic decay time-scale.  As the mass increases, a neutron
star becomes more and more compact  and the mass of the crust actually
decreases by  a small amount  (see section [\ref{sprofile}]).   So the
newly  accreted material  forms  the crust  and  the original  crustal
material  gets continually assimilated  into the  superconducting core
beneath.  The  original current carrying  layers are thus  pushed into
deeper and more dense  regions as accretion proceeds. This compression
results in  a decrease  of the effective  length scale of  the current
loops which  makes a fast dissipation  possible~\cite{kbu95}.  But the
higher  conductivity of  the denser  regions would  progressively slow
down the  decay winning over  the effect of  a decrease in  the length
scale,   till   the   current   loops  are   completely   inside   the
superconducting region where any further decay is prevented.

The   organization  of  the   chapter  is   as  follows.   In  section
[\ref{smn-phys}] we discuss  the physics of the field  reduction in an
accreting neutron star, computational details are discussed in section
[\ref{smn-comp}].  Finally  in section [\ref{smn-results}]  we present
our results and draw the conclusions in section [\ref{smn-concl}]. 

\section{physics of the mechanism}
\label{smn-phys}

The evolution of the crustal field  as a result of ohmic diffusion and
material    motion   has    been    discussed   by    a   number    of
authors~\cite{wend87,sang87,gepp94}   and  essentially   concerns  the
solution of the following equation~\cite{jack}
\beq
\frac{\partial \vec B}{\partial t} = \vec \nabla \times (\vec V \times
\vec  B)  - \frac{c^2}{4  \pi}  \vec  \nabla \times  (\frac{1}{\sigma}
\times \vec \nabla \times \vec B) \label{emhd}
\eeq
where $\vec  V$ is the velocity  of material movement  and $\sigma$ is
the electrical conductivity of the  medium.  As in previous studies we
solve this equation by introducing the vector potential $\vec A$, such
that
\beq
\vec B = \nabla \times \vec A
\eeq
where,  ${\vec A}  =  (0, 0,  A_{\phi})$.  This choice  of the  vector
potential ensures a poloidal geometry  for $\vec B$. In particular, we
choose
\beq
A_{\phi}  = \frac{g(r,t)  \sin  \theta}{r}, \;  \mbox{$g(r,t)$ is  the
Stokes' function},
\eeq
$(r,\theta,\phi)$ being the spherical polar co-ordinates. Here we work
in the lowest order of multipole, obtaining the following dipolar form
of the magnetic field :
\ber
\vec B(r, \theta) &=& \vec \nabla \times (\frac{g(r,t) \sin \theta}{r}
\; \hat \phi) \nonumber \\
&=&   \frac{2  \cos   \theta  g(r,t)}{r^2}   {\hat  r}   -  \frac{\sin
\theta}{r}\frac{\partial        g(r,t)}{\partial        r}       {\hat
\theta}. \label{eBrtheta}
\eer
The magnitude of the field is given by,
\beq
B(r,   \theta)  =  {\left[\frac{4   \cos^2  \theta   g^2(r,t)}{r^4}  +
\frac{\sin^2    \theta}{r^2}(\frac{\partial   g(r,t)}{\partial   r})^2
\right]}^{1/2}
\eeq
Therefore, at the pole ($\theta = 0$, $r = R$) the magnitude is
\beq
B(R, 0) = \frac{2 g(R,t)}{R^2}, \label{eBmag}
\eeq
which  is simply  proportional to  the value  of the  Stokes' function
there. We shall  use this fact to obtain the  evolution of the surface
field with time.

The underlying  current distribution corresponding to  the above field
is obtained by using Maxwell's equation :
\ber
\vec J &=& \frac{c}{4 \pi} \nabla \times \vec B \nonumber \\
&=& - \frac{c}{4  \pi} \; \frac{\sin \theta}{r} \left[\frac{\partial^2
g(r,t)}{\partial r^2} - \frac{2 \, g(r,t)}{r^2} \right] {\hat \phi}.
\eer

\subsection{ohmic diffusion}

In order  to understand the  ohmic dissipation of the  field strength,
let us  consider equation [\ref{emhd}]  without the first term  on the
right  hand  side.  The  first  term  corresponds  to  the  convective
transport, which  we shall discuss  later. Without this  term equation
[\ref{emhd}] takes the following form,
\beq
\frac{\partial \vec  B}{\partial t} = - \frac{c^2}{4  \pi} \vec \nabla
\times (\frac{1}{\sigma} \times \vec \nabla \times \vec B).
\eeq
For the  moment, let  us consider the  conductivity to be  a constant,
without any  dependence on space.  Then, the above equation  takes the
form of  a pure diffusion  equation (by virtue of  the divergence-free
condition for the magnetic fields), given by
\beq
\frac{\partial \vec B}{\partial t} = \frac{c^2}{4 \pi \sigma} \nabla^2
\vec B.
\eeq
The  diffusion constant for  the above  equation is  $\frac{c^2}{4 \pi
\sigma}$. One can define  a time-scale characteristic of the diffusion
process :
\beq
\tau_{\rm diff} = \frac{4 \pi \sigma L^2}{c^2}, \label{ediffconst}
\eeq
where $L$  is the length-scale associated with  the underlying current
distribution supporting the field.

It is  clear from equation  [\ref{ediffconst}] that the rate  of ohmic
diffusion is  determined mainly by the electrical  conductivity of the
crust which is  a steeply increasing function of  density (see figures
in section [\ref{sns-transport}]). As density in the crust spans eight
orders of magnitude with a very large radial gradient the conductivity
changes sharply as a function  of depth from the neutron star surface.
Thus the deeper  the location of the current  distribution, the slower
is the decay. We have discussed, in detail, the effects of temperature
and  that of  impurity concentration  on the  conductivity  in section
[\ref{sns-transport}], also about the  change in the thermal behaviour
in   presence   of   accretion   in   a  neutron   star   in   section
[\ref{sthermal}]. The  increase in temperature  of an accretion-heated
crust  lowers the conductivity  and therefore  the time-scale  for the
diffusion decreases. In other words, the transfer of the energy of the
systematic motion  of the charge  carriers (the electrons)  within the
current loops into the random  kinetic energy of the electrons proceed
at  a faster  rate at  a  higher temperature.  For the  models we  are
considering  now, it  has  been  assumed in  the  literature that  the
impurity strength  $Q$ lies  in the  range 0.01 -  0.1. The  effect of
impurities  is  most  important   at  lower  temperatures  and  higher
densities        (see        figures       [\ref{fsigma_t}]        and
[\ref{fsigma_q}]).  Therefore the  field evolution  does not  show any
significant dependence on  impurity strength for this range  of $Q$ in
an accretion-heated  crust. We therefore restrict  our computations to
the $Q = 0.0$ case. However,  the impurity strength will still play an
important role in field decay  during the pre-accretion phase when the
crustal temperatures can be quite low.

\subsection{accretion and material transport}
\label{ssaccn}

In a  neutron star,  for a given  equation of  state, the mass  of the
crust  is uniquely  determined  by the  total  mass of  the star  (see
section [\ref{sprofile}]).  And this crustal  mass remains effectively
constant for  the accreted  masses of the  order of 0.1~\msun\  with a
slight   decrease   as  the   total   mass   increases  (see   figures
[\ref{fm_ms}], [\ref{fdmco_dms}]  and [\ref{fdmcr_dms}]). For example,
for the equation of state  that we have used (see section [\ref{seos}]
for details) the  accretion of 0.1~\msun\ on the  star causes a change
in the crustal mass of only 0.004~\msun. As a result, accretion causes
continuous assimilation of material from  the bottom of the crust into
the core. At the same time  the upper layers of the original crust are
pushed to  deeper and  denser regions, leading  to extreme  squeeze of
this material.  This also  causes the current distribution embedded in
this material to be sharpened,  reducing the effective length scale of
the system.

The  change in the  crustal mass  and the  crustal density  profile is
negligible for the amount of  mass accretion we consider. We therefore
take the mass  flux to be the same throughout the  crust, equal to its
value  at  the surface.  Assuming  the  mass  flow to  be  spherically
symmetric  in the  crustal layers  of  interest, one  the obtains  the
following condition  for the  equality of mass  flux at  all densities
within the crust,
\ber
\mdot dt &=& 4\pi r^2 \rho (r) dr, \nonumber \\
\Rightarrow V(r) &=& \frac {\mdot}{4\pi r^2 \rho (r)}, \label{eaccvel}
\eer
where $\mdot$  is the  rate of  mass accretion and  $\rho (r)$  is the
density  as a  function of  radius  $r$.  The  above equation  defines
$V(r)$ - the velocity of radial material flow at a given radius in the
crust.  It  should be  noted here that  the material flow  is radially
inwards, hence writing the flow velocity in its full vectorial form we
have,
\beq
\vec V(r) = - \frac {\mdot}{4\pi r^2 \rho (r)} {\hat r}. \label{evecV}
\eeq

The  result of  accretion on  the magnetic  field  evolution therefore
manifests itself as  a combination of three effects:  transport of the
current  distribution to regions  of higher  density and  hence higher
conductivity, reduction  of conductivity due to heating  and change in
the  effective   length  scale   of  the  current   distribution  (see
\citeNP{db95a} for  a detailed discussion).  We find  that the overall
effect turns  out to be a  rapid initial decay followed  by a leveling
off when  an amount of mass  equivalent to about 10\%  of the original
crust  has been accreted.  By this  time the  current loops  reach the
regions  of very  high  density and  consequently  of extremely  large
conductivity  where the diffusion  time-scales are  much too  long. As
further accretion proceeds the  original crust is assimilated into the
superconducting  interior  freezing   the  currents  there.  Following
\citeN{baym69b}  we  assume  that  the  newly  formed  superconducting
material retains the magnetic flux through it in the form of Abrikosov
fluxoids rather than expelling it through the Meissner effect.

\subsection{the field evolution equation}

We  use   the  form  of   $B(r,  \theta,  \phi)$  given   by  equation
[\ref{eBrtheta}]  and   the  form  of  $\vec  V$   given  by  equation
[\ref{evecV}] to  cast equation [\ref{emhd}]  in terms of  the Stokes'
function.
\ben
\i The left hand side --
\ber
\frac{\partial  \vec B}{\partial t}  &=& \frac{\partial  \nabla \times
\vec A}{\partial t} \nonumber \\
&=&  \nabla \times  \frac{\partial }{\partial  t}  \; \left[\frac{g(r,
\theta) \; \sin \theta}{r}\right] \; {\hat \phi}. \label{elhs}
\eer
\i The first term in the right hand side --
\ber
\nabla \times (\vec V \times \vec B)
&=&  \nabla \times  \left[ \left(  -  V(r) \;  {\hat r}\right)  \times
\left(\frac{2  \cos  \theta  \;  g(r,t)}{r^2} {\hat  r}  -  \frac{\sin
\theta}{r}\;    \frac{\partial    g(r,t)}{\partial    r}   \;    {\hat
\theta}\right) \right] \nonumber \\
&=&   \nabla  \times   \left[   \frac{\sin  \theta   \;  V(r)}{r}   \;
\frac{\partial    g(r,    \theta)}{\partial    r}\right]   \;    {\hat
\phi}. \label{erhs1}
\eer
\i The second term in the right hand side --
\ber
\nabla \times \left[\frac{1}{\sigma} \nabla \times \vec B\right]
&=& \nabla  \times \left[\frac{1}{\sigma} \nabla  \times \left(\frac{2
\cos  \theta \;  g(r,t)}{r^2} \;  {\hat r}  -  \frac{\sin \theta}{r}\;
\frac{\partial  g(r,t)}{\partial  r}  \;  {\hat  \theta}\right)\right]
\nonumber \\
&=& \nabla  \times \left[ - \frac{1}{\sigma}  \; \frac{\sin \theta}{r}
\;    \left(\frac{\partial^2   g(r,t)}{\partial    r^2}    -   \frac{2
g(r,t)}{r^2}\right) \; {\hat \phi}\right].
\label{erhs2}
\eer
\een
Incorporation  of  the  expressions  [\ref{elhs}],  [\ref{erhs1}]  and
[\ref{erhs2}] in the equation[\ref{emhd}] then leads to
\beq
\frac{\partial    g(r,t)}{\partial    t}    =   V(r)    \frac{\partial
g(r,t)}{\partial  r} + \frac{c^2}{4\pi  \sigma} \left(\frac{\partial^2
g(r,t)}{\partial r^2} - \frac{2g(r,t)}{r^2} \right). \label{edgdt}
\eeq
The results  of this chapter will  be based on  numerical solutions of
the equation [\ref{edgdt}].

\section{computations}
\label{smn-comp}

The  aim of  our computations  is to  solve equation  [\ref{edgdt}] to
obtain $g(r,t)$ using the  following boundary conditions valid for all
times (see, e.g., Geppert \& Urpin 1994):
\ber
\frac{\partial g(r,t)}{\partial r}|_{r=R} + \frac {g(R,t)}{R} = 0,\\
g(r_{\rm co},t) = 0
\eer
where $R$  is radius of  the star and  $r_{\rm co}$ is that  radius to
which the original  boundary between the core and  the crust is pushed
to,  due to  accretion, at  any point  of time.   The  first condition
matches the  interior field to  an external dipole  configuration. The
second condition  indicates that as accretion proceeds  along with the
crustal material  the frozen-in  flux moves inside  the core,  but the
field  can not diffuse  through the  original crust-core  boundary. To
simulate an  effectively infinite  conductivity in the  region between
the bottom  of the crust and  the original boundary  between the crust
and  the core,  we  set $\sigma  \sim  10^{50} {\rm  s^{-1}}$ in  this
region. As mentioned before, we take into account the combined effects
of accretion driven material  motion and ohmic diffusion. We construct
the  density  profile  of  a  neutron star  as  described  in  section
[\ref{sprofile}] for an  assumed mass of 1.4 $\msun$.  This star has a
total crustal mass  of 0.044 $\msun$ and we  restrict our evolutionary
calculations to a  maximum net accretion of this  additional amount on
the star, because  an accretion of that amount  of material pushes the
original crust completely  within the core. The change  in the crustal
density   profile    resulting   from   this    additional   mass   is
negligible. Hence  we work with  an invariant crustal  density profile
throughout our calculation.

We assume  that the  matter settling onto  the star does  so uniformly
across the  entire surface.  This allows  us to use  the expression of
$V(r)$ as in equation [\ref{eaccvel}]. When the surface magnetic field
is strong this is a poor  approximation very close to the surface (see
the  discussion  about   anisotropic  material  transport  in  section
[\ref{sds-matflow}]   However   in    deeper   layers   ($\rho   \gsim
10^{10}~\gcc$) which we are mainly concerned with, the material motion
is essentially dictated by the added weight and is going to be more or
less isotropic.

We further assume that the  incoming matter fully threads the existing
magnetic field  before settling  onto the surface.  In other  words we
allow for no  reduction of the external magnetic  field arising out of
diamagnetic screening  by the incoming material. We  adopt this scheme
to ensure that the  effects under investigation here, namely diffusion
and  convection,  do   not  get  masked  by  other   effects  such  as
screening.  In the  light  of  the results  obtained  in the  previous
chapter  regarding diamagnetic  screening, we  are quite  justified in
making such an assumption.

We assume that  during the phase of mass  accretion the temperature in
the crust  is uniform  and remains constant  in time. This  ignores an
initial  short phase  in  which both  the  rate of  accretion and  the
temperature of  the crust show  time evolution. The rate  of accretion
stabilizes in  a few thousand years~\cite{savo78}  and the temperature
within $10^5$~yrs~\cite{mira90}.   Computations by \citeN{urpn96} show
that  the  decay during  this  initial  phase  is insignificant.   The
temperature that the crust will finally attain in the steady phase has
been  computed by  \citeN{fuji84}, \citeN{mira90}  and \citeN{zdun92}.
However, these  computations are restricted  to limited range  of mass
accretion and  also do  not yield the  same crustal  temperature under
similar conditions.   The results  obtained by \citeN{zdun92}  for the
crustal  temperatures  for  a   given  accretion  rate  in  the  range
$10^{-15}$~\dmdt \lsim \mdot \lsim  $2 \times 10^{-10}$~\dmdt could be
fitted to the equation  [\ref{etdmdt}].  But extrapolation of this fit
to higher  rates of accretion gives extremely  high temperatures which
may not be sustainable for  any reasonable period due to rapid cooling
by neutrinos at those  temperatures.  We have therefore restricted our
computations to  a maximum accretion  rate of $10^{-9}$~\dmdt  and for
\mdot~in  the range  $10^{-10}$--$10^{-9}$~\dmdt, we  have  explored a
range   of  constant   crustal  temperatures   between   $10^{8}$  and
$10^{8.75}$~K.

In a  series of papers  \citeN{gepp94}, \citeN{urpn95}, \citeN{urpn96}
and \citeN{gepp96} have considered evolution of crustal magnetic field
for  accretion rates  in the  range $10^{-15}-10^{-9}$~\dmdt.  In this
work we too consider accretion rates covering much of the above range.
The difference between our computations  and those in the above papers
lies in  the range of  total mass accretion  and the treatment  of the
inner boundary  condition. Our computations  proceed to a  maximum net
accretion of $\sim 4 \times 10^{-2}$~\msun whereas the computations by
other authors are restricted to  that of the order of $10^{-3}$~\msun.
Our  choice of  the  range  of accretion  rates  also facilitates  the
comparison of our results with those available in the literature, over
the  range of overlap  in the  net mass  accreted. At  accreted masses
$\gsim 10^{-3} \msun$, the inward convection of currents into the core
becomes significant. As computations  by previous authors do not allow
for this  possibility, our  results begin to  diverge from  theirs for
high values of accreted mass.

\subsection{numerical scheme}

We  solve the equation  [\ref{edgdt}] in  terms of  fractional radius,
$x$, instead of  the radius. Re-writing this equation  in terms of the
fractional radius, we obtain,
\beq
\frac{\partial    g(x,t)}{\partial    t}    =   V(x)    \frac{\partial
g(x,t)}{\partial  x} + S(x,t)  \left(\frac{\partial^2 g(x,t)}{\partial
x^2} - \frac{2g(x,t)}{x^2} \right), \label{edgdtx}
\eeq
where,
\ber
V(x) &=& V(r)/R, \\ \label{eVx}
S(x,t) &=& \frac{c^2}{4 \pi \sigma(r,t) R^2}. \label{eSx}
\eer
To  solve  the  above  equation  we  use  a  modified  Crank-Nicholson
scheme. Since the  conductivity is a function of  density and hence of
radius, it has  been necessary to incorporate the  space dependence of
conductivity    in   the    standard    Crank-Nicholson   scheme    of
differencing.  In addition  we also  allow for  a  pre-accretion phase
where  the neutron star  undergoes normal  cooling. This  introduces a
time-dependence in the temperature and hence in the diffusion-constant
as well.   The overall  effect is to  consider an  explicit space-time
dependence  of  the function  $S(x)$  which  is  a pre-factor  to  the
diffusion term. We introduce  the convection term into the computation
through  upwind  differencing  and  operator  splitting  of  the  full
differential equation~\cite{pres}.

Using  the  method  of  operator  splitting, we  first  find  out  the
differenced form of the equation,
\beq
\frac{\partial    g(x,t)}{\partial    t}    =   V(x)    \frac{\partial
g(x,t)}{\partial x}
\eeq
by {\em upwind differencing}, to obtain,
\beq
g^{n+1}_{j}  =  g^n_j  +  \frac{\delta  t}{\delta x}  V_j  (  g^n_j  -
g^n_{j-1} ), \label{edif-upwind}
\eeq
where, $\delta t$  and $\delta x$ represent the  time interval and the
size  of the  space-grid. The  superscript  $n$ stand  for the  $n$-th
time-step and  the subscript  $j$ stand for  the $j$-th  space-grid of
integration.  Similarly, for the diffusive part, given by,
\beq
\frac{\partial  g(x,t)}{\partial  t}  = S(x,t)  \left(\frac{\partial^2
g(x,t)}{\partial x^2} - \frac{2g(x,t)}{x^2} \right)
\eeq
we use {\em Crank-Nicholson} scheme to obtain,
\beq
g^{n+1}_{j} = g^n_j + \frac{\delta  t \, S^{n+1/2}_j }{2 (\delta x)^2}
\left(g^{n+1}_{j+1} -  2 g^{n+1}_j +  g^{n+1}_{j-1} + g^{n}_{j+1}  - 2
g^{n}_j  +  g^{n}_{j-1} \right)  -  \frac{2  \delta  t \,  S^{n+1/2}_j
g^n_j}{x_j^2},
\label{edif-CR}
\eeq
where  the  various symbols  have  the  same  meaning as  in  equation
[\ref{edif-upwind}].  To incorporate  the time-dependence  of $S(x,t)$
without  making the differencing  scheme too  complicated, we  use the
time-averaged (over two neighbouring  intervals) value of the function
at  each time-step.  This does  not introduce  too much  error  as the
function  is slowly-varying over  the time-intervals  typically chosen
for  our integrations.   In  the end  we  combine the  schemes in  the
following manner,
\ber
g^{n+1}_{j}  &=& g^n_j  +  \frac{\delta  t}{\delta x}  V_j  ( g^n_j  -
g^n_{j-1} ) \nonumber \\
&   &   +   \frac{\delta   t   \,  S^{n+1/2}_j   }{2   (\delta   x)^2}
\left(g^{n+1}_{j+1} -  2 g^{n+1}_j +  g^{n+1}_{j-1} + g^{n}_{j+1}  - 2
g^{n}_j + g^{n}_{j-1} \right) \nonumber \\
& & - \frac{2 \delta t \, S^{n+1/2}_j \, g^n_j}{x_j^2},
\eer
to  obtain the final  differenced form  of the  equation [\ref{edgdt}]
which is used in the numerical code.

In connection with  the numerical code an important  point needs to be
mentioned. It is  imperative, before it is used  to obtain results, to
check  for  the  stability  of  the numerical  scheme  employed.   The
Crank-Nicholson scheme  of differencing is  unconditionally stable for
any size of the temporal or spatial grid. Hence any instability in our
scheme  arises from  the convective  part  for which  the scheme  used
(upwind differencing) is, unlike the  other scheme, not stable for all
combinations of  the spatial and temporal intervals.  The stability of
this scheme depends  on whether the Courant condition  is satisfied or
not. This condition demands that the inequality
\beq
\frac{\delta t}{\delta x} V(x) \lsim 1,
\eeq
be  satisfied at  all grid  points and  at all  points of  time. Using
equations [\ref{eaccvel}]  and [\ref{eVx}] to arrive at  the values of
$V(x)$  for a  given rate  of accretion  and for  a space-grid  of 100
points we find that for an accretion rate of $10^{-8}$~\dmdt (which is
roughly  the Eddington  rate of  accretion for  a 1.4  $\msun$ neutron
star) the maximum time-interval  allowable by the Courant condition is
$10^{-2}$  years.  This  of  course  places  a  severe  constraint  on
computational resources in terms of run-time.

\bef
\begin{center}{\mbox{\epsfig{file=epsilon.ps,width=235pt,angle=-90}}}\end{center}
\caption[$\epsilon$  vs.  radius]{$\epsilon$  as  a  function  of  the
fractional radius $x$.}
\label{fepsilon} 
\eef

In  equation [\ref{edgdt}]  above,  the convective  and the  diffusive
terms  dominate   the  evolution  in  different   density  ranges.  To
illustrate this  fact, similar to the diffusive  time-scale defined in
equation [\ref{ediffconst}], we define a convective time-scale :
\beq
\tau_{\rm conv} = \frac{L}{V(r)}
\eeq
where $L$ is the system  size as before. In figure [\ref{fepsilon}] we
plot the  relative magnitude of  these two time-scales defined  by the
coefficient,
\beq
\epsilon = \frac{\tau_{\rm diff}}{\tau_{\rm conv}}.
\eeq
This plot is made for an assumed accretion rate of $10^{-9}$~\dmdt. It
is clear  from this figure that  both in the low  density regions near
the surface and  in the deep high density regions  it is the diffusive
term  that is  more important.  Only in  the intermediate  regions the
effect of convection becomes significant.

\section{results and discussions}
\label{smn-results}

The results are summarized in a series of figures. 

\bef
\begin{center}{\mbox{\epsfig{file=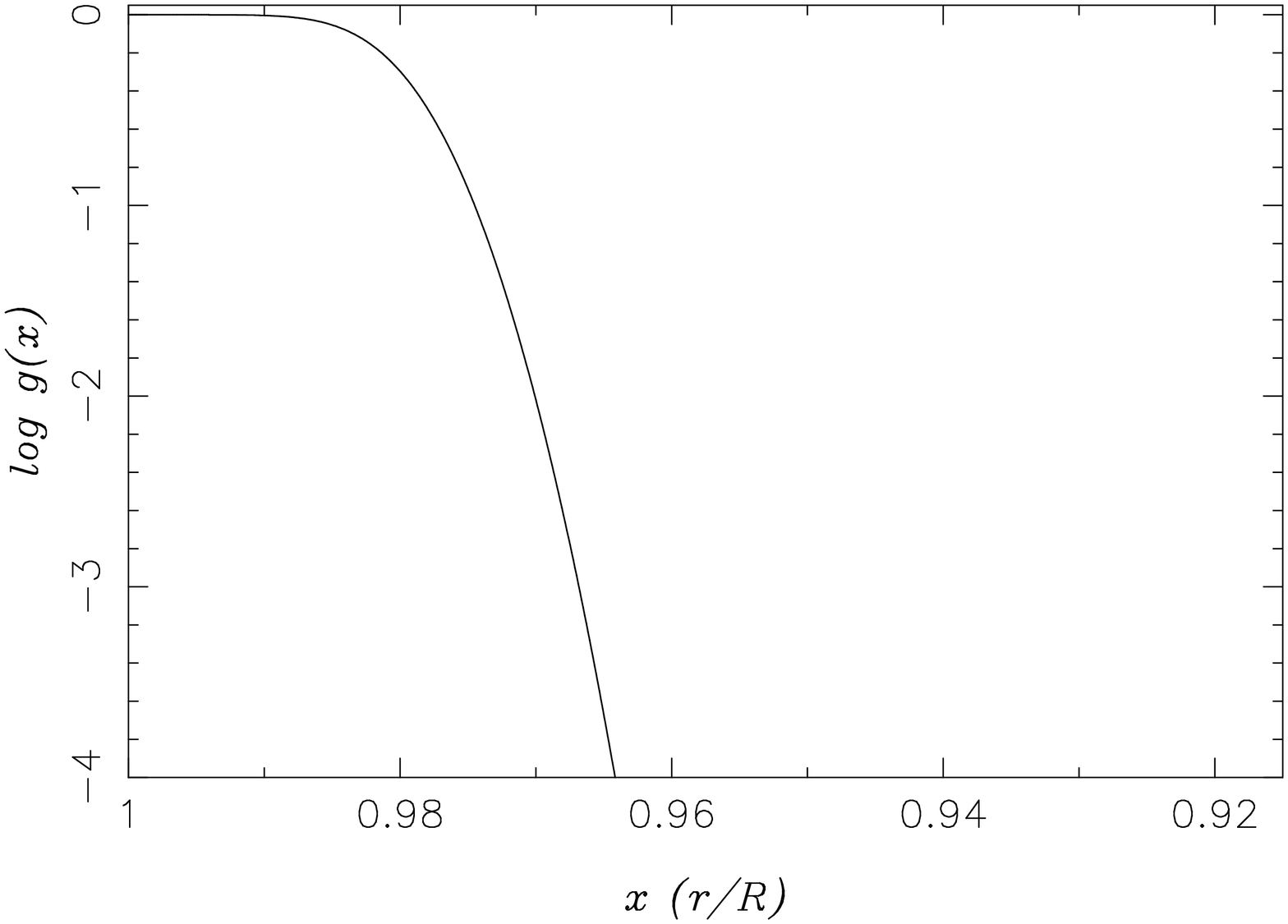,width=235pt,angle=-90}}}\end{center}
\caption[initial  $g$-profile]{The initial  radial  dependence of  the
$g$-profile  centered at  $x$ =  0.98,  which corresponds  to $\rho  =
10^{11}\;\gcc$,  with a width  $\delta x$  = 0.006;  where $x$  is the
fractional radius $r/R$. }
\label{fg_init}
\eef
\bef
\begin{center}{\mbox{\epsfig{file=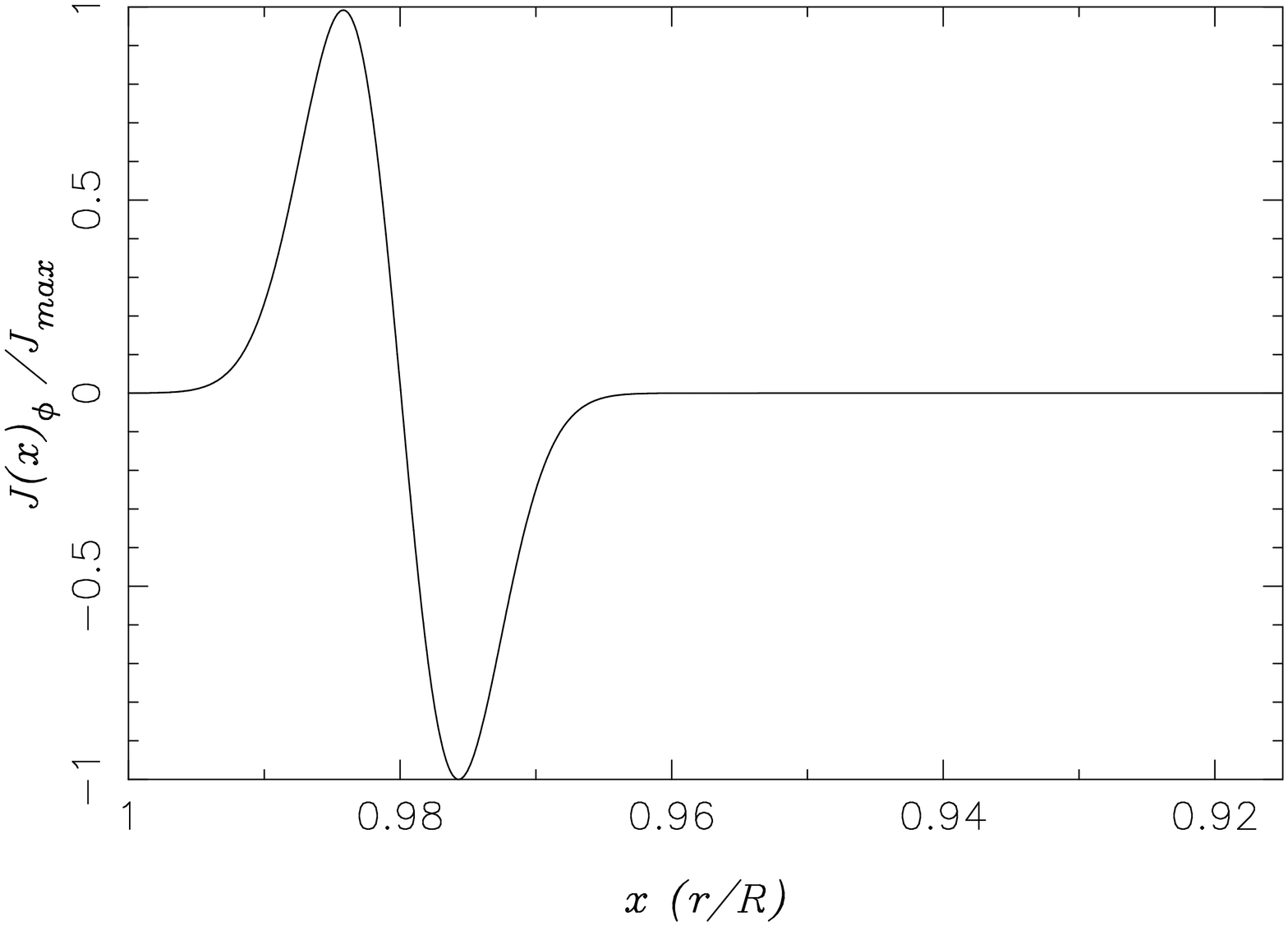,width=235pt,angle=-90}}}\end{center}
\caption[initial  $J$-profile]{The initial  radial  dependence of  the
$\phi$-component of the corresponding current configuration.}
\label{fj_init}
\eef

Figure [\ref{fg_init}] shows the  distribution of the $g$-function and
figure [\ref{fj_init}]  the toroidal currents,  $J_{\phi}$, assumed at
the starting point of the field evolution.

\bef
\begin{center}{\mbox{\epsfig{file=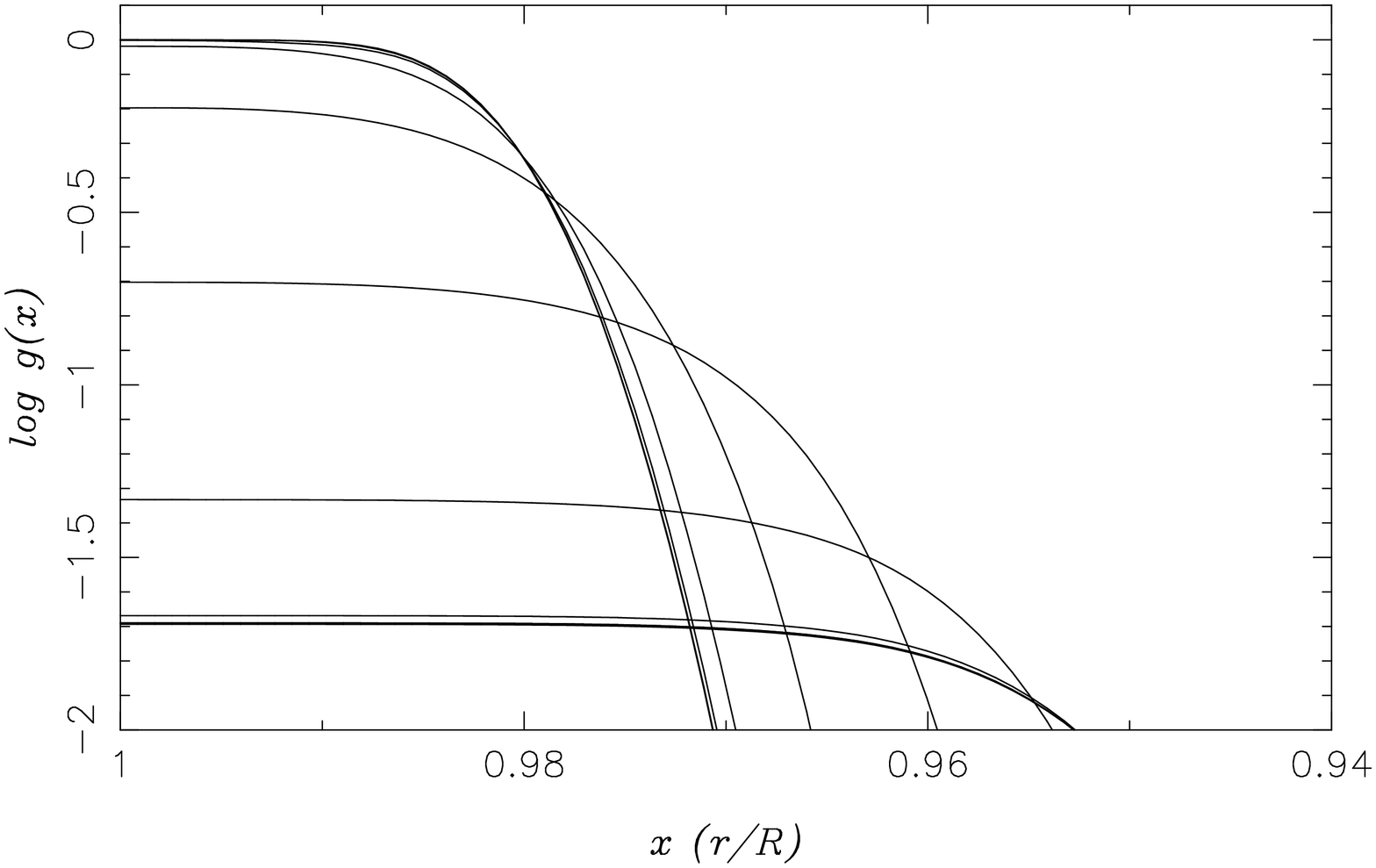,width=235pt,angle=-90}}}\end{center}
\caption[ohmic diffusion  of $g$-profile]{Pure ohmic  diffusion of the
$g$-profile   for  $\tau   \sim  10^9$   yrs,  centred   at   $\rho  =
10^{11}\;\gcc$,  with $Q$  =  0.0,  in a  neutron  star with  standard
cooling.  The  curves shown at  intermediate times correspond to,  t =
$10, 10^2, 10^3,  10^4, 10^5, 10^6, 10^7, 10^8,  10^9$ years (the last
three  are almost  indistinguishable),  respectively, with  decreasing
values at the surface ($x$ = 1). }
\label{fg_diff}
\eef

\bef
\begin{center}{\mbox{\epsfig{file=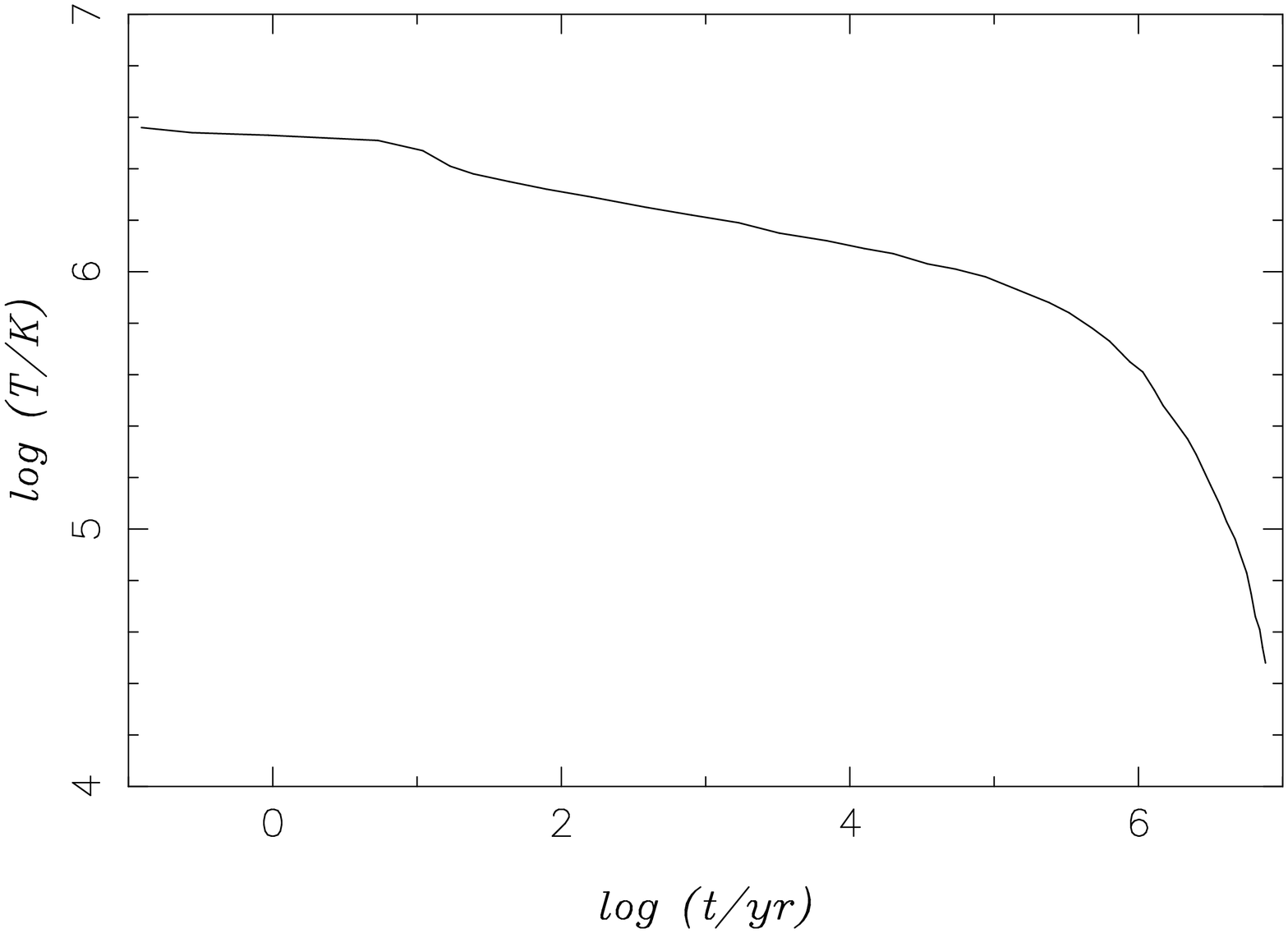,width=235pt,angle=-90}}}\end{center}
\caption[thermal  evolution]{Evolution   of  the  Neutron   Star  core
temperature, after van Riper~\citeyear{ripe91a,ripe91b}.}
\label{fns-thermal}
\eef

Field decay due to pure diffusion in an isolated neutron star is shown
in  figures [\ref{fg_diff}] and  [\ref{fb_diff}].  In  computing this,
neutron   star    cooling   according   to   the    results   of   van
Riper~\citeyear{ripe91a,ripe91b}    (reproduced    here   in    figure
[\ref{fns-thermal}])  for  normal  matter  in 1.4~\msun\  Friedman  \&
Pandharipande~\citeyear{frie81} model star has  been used. It is to be
noted   that  our  adopted   equation  of   state,  namely,   that  of
\citeN{wiri88}  is an  updated  version of  Friedman \&  Pandharipande
equation  of state with  only minor  differences. Among  the published
cooling curves this is the  nearest to that appropriate to our adopted
neutron  star model.   Computations similar  to the  one  displayed in
figures  [\ref{fg_diff}]   and  [\ref{fb_diff}]  have   been  made  by
\citeN{urpn92a} and  our result matches very closely  with theirs.  It
is  evident  from  the  figure  [\ref{fb_diff}]  that  the  net  decay
decreases  with the increasing  density at  which the  initial current
configuration  is  centred  at.  This  is expected  as  the  diffusive
time-scales   are  larger   at  higher   densities  owing   to  larger
conductivities there.

\bef
\begin{center}{\mbox{\epsfig{file=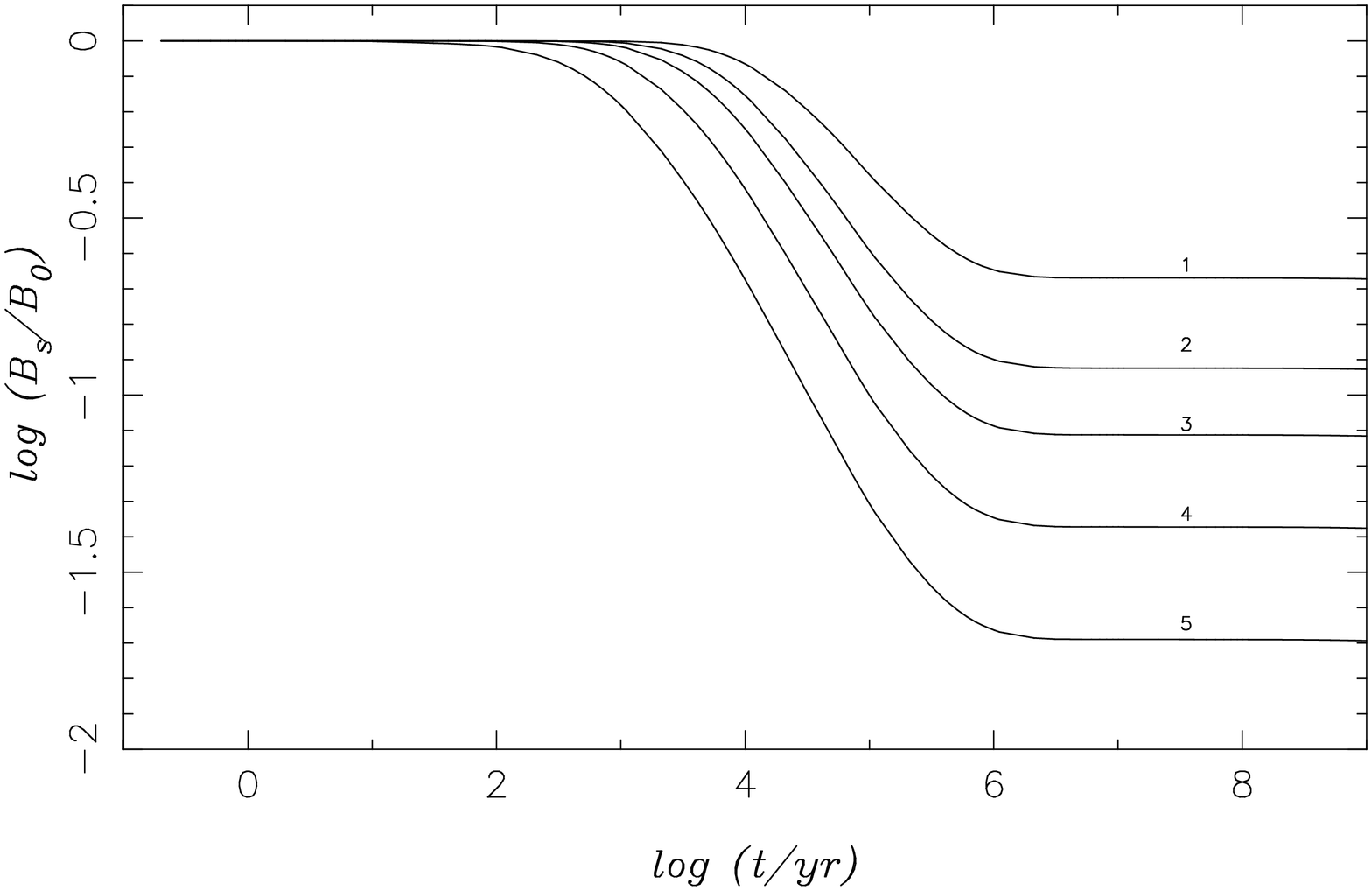,width=235pt,angle=-90}}}\end{center}
\caption[ohmic  diffusion  of  surface  field]{The  evolution  of  the
surface magnetic field due to  pure diffusion.  Curve 5 in this figure
correspond     to     the     $g$-profile    plotted     in     figure
[\ref{fg_diff}].   Curves  1  to   4  correspond   to  $10^{13}~\gcc$,
$10^{12.5}~\gcc$, $10^{12}~\gcc$ and $10^{11.5}~\gcc$ respectively, at
which the $g$-profiles are centred. All curves correspond to $Q$ = 0.}
\label{fb_diff}
\eef

\bef
\begin{center}{\mbox{\epsfig{file=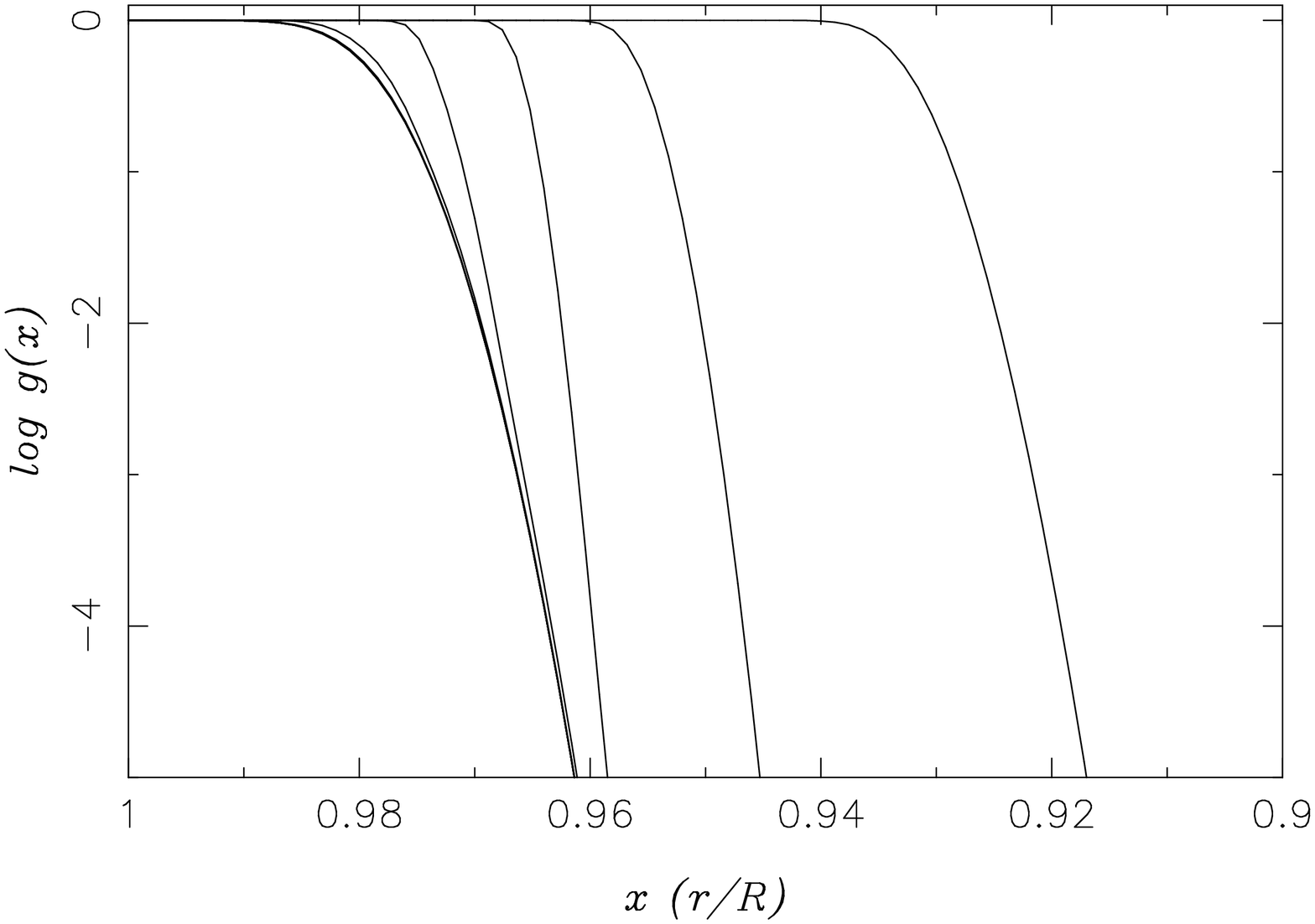,width=235pt,angle=-90}}}\end{center}
\caption[convective  transport of surface  field]{Convective transport
of the  $g$-profile over $10^9$  years with $\mdot  = 10^{-10}$~\dmdt,
surface  field  is  constant   by  assumption.  The  curves  shown  at
intermediate times  correspond to,  $t = 10,  10^2, 10^3,  10^4, 10^5,
10^6,   10^7,  10^8,   10^9$  years   (the  first   four   are  barely
distinguishable), respectively, with the profiles progressively moving
inwards.}
\label{fg_conv}
\eef

Figure [\ref{fg_conv}]  displays the result  of the convection  due to
material movement  alone. We obtain  this by setting  the conductivity
$\sigma$ to  an artificially high  value of $10^{50}$~s$^{-1}$  in our
code.  It shows  the migration  of  $g$-profile to  regions of  higher
density (and consequent  sharpening of the profile). The  field at the
surface  ($B_{\rm  s}  =  2g(R,t)/R^2$) remains  constant  under  pure
convection according to our assumptions.

\bef
\begin{center}{\mbox{\epsfig{file=g_evolv.ps,width=210pt,angle=-90}}}\end{center}
\caption [evolution  of $g$-profile]{Evolution of  the $g$-profile due
to ohmic diffusion  and convective transport, over a  period of $10^6$
years  for $\mdot  = 10^{-9}$~\msun/yr,  $T  = 10^{8.0}$~K  and $Q$  =
0.0. The curves shown at  intermediate times correspond to, t = $10^2,
10^3, 10^4,  10^5, 10^6,  10^7, 4.0 \times  10^7$~years, respectively,
with decreasing values at the surface ($x$ = 1).}
\label{fg_evolv}
\eef

Figure [\ref{fg_evolv}]  shows the results of the  combined effects of
convection and diffusion on the  $g$-function, in other words the full
evolution described by equation  [\ref{edgdt}], for a particular value
of the accretion rate.

\bef
\begin{center}{\mbox{\epsfig{file=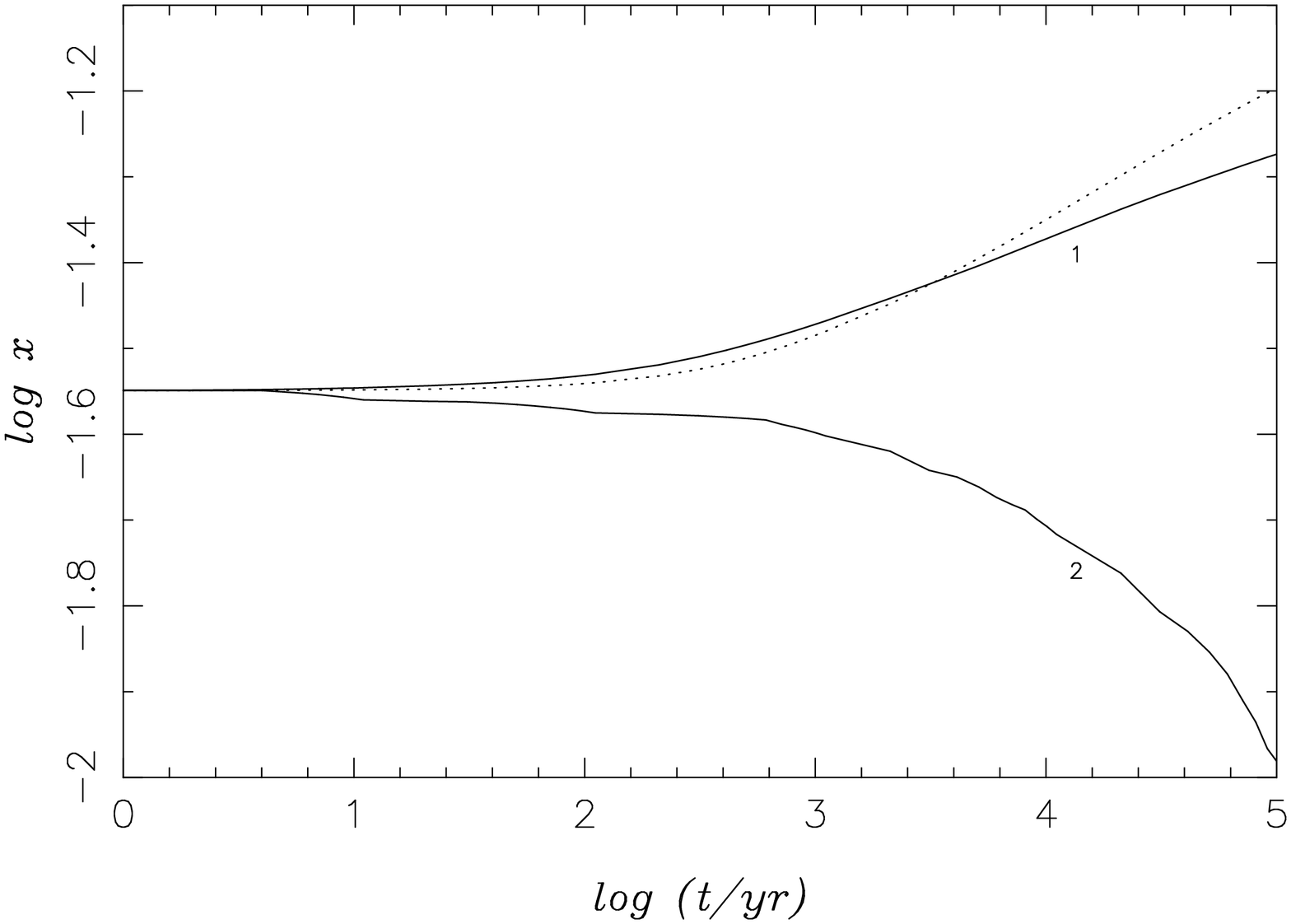,width=210pt,angle=-90}}}\end{center}
\caption [evolution of length-scale]{Evolution of the length scale (in
fractional radius) of  the $g$-profile over a period  of $10^5$ years.
Curve 1 correspond  to pure ohmic diffusion with  standard cooling and
$q$ = 0. Curve 2 correspond  to pure convection with an accretion rate
of $\mdot = 10^{-10}$~\msun/yr and the dotted curve corresponds to the
case of actual accretion with the same $\mdot$.}
\label{fgscale}
\eef

In  section [\ref{ssaccn}]  we have  mentioned that  accretion induces
compression of the crustal layers  and hence effect a reduction of the
effective   length-scale   of   the   current  profile.   In   figures
[\ref{fg_diff}], [\ref{fg_conv}] and [\ref{fg_evolv}] we have seen how
the $g$-profile evolves  due to pure diffusion, pure  convection and a
combination  of both.  Here,  in figure  [\ref{fgscale}]  we plot  the
change of the length-scale of the $g$-profile with time in each of the
above-mentioned cases. As an estimate of the effective length-scale of
the current distribution  we use the width (defined  as the separation
between the peak of the distribution  and a point which has 1\% of the
peak  magnitude)  of the  $g$-profile.  Though  this  is but  a  crude
estimate,  it  nevertheless serves  the  purpose  of illustrating  the
qualitative nature of the change  of the length-scale.  The curves for
pure  diffusion  and  pure  convection show  monotonous  increase  and
decrease respectively.  Even though the  curve for the case  of actual
accretion (where both diffusion and convection are present) the nature
is again that  of a monotonic increase in length  scale similar to the
case of pure diffusion,  it rises comparatively slowly initially. This
is due to the fact that the convective compression slows the diffusive
spreading.  Of course,  in  the end  the  diffusive spreading  becomes
faster than the case of pure diffusion due to heating which lowers the
conductivity.

\bef
\begin{center}{\mbox{\epsfig{file=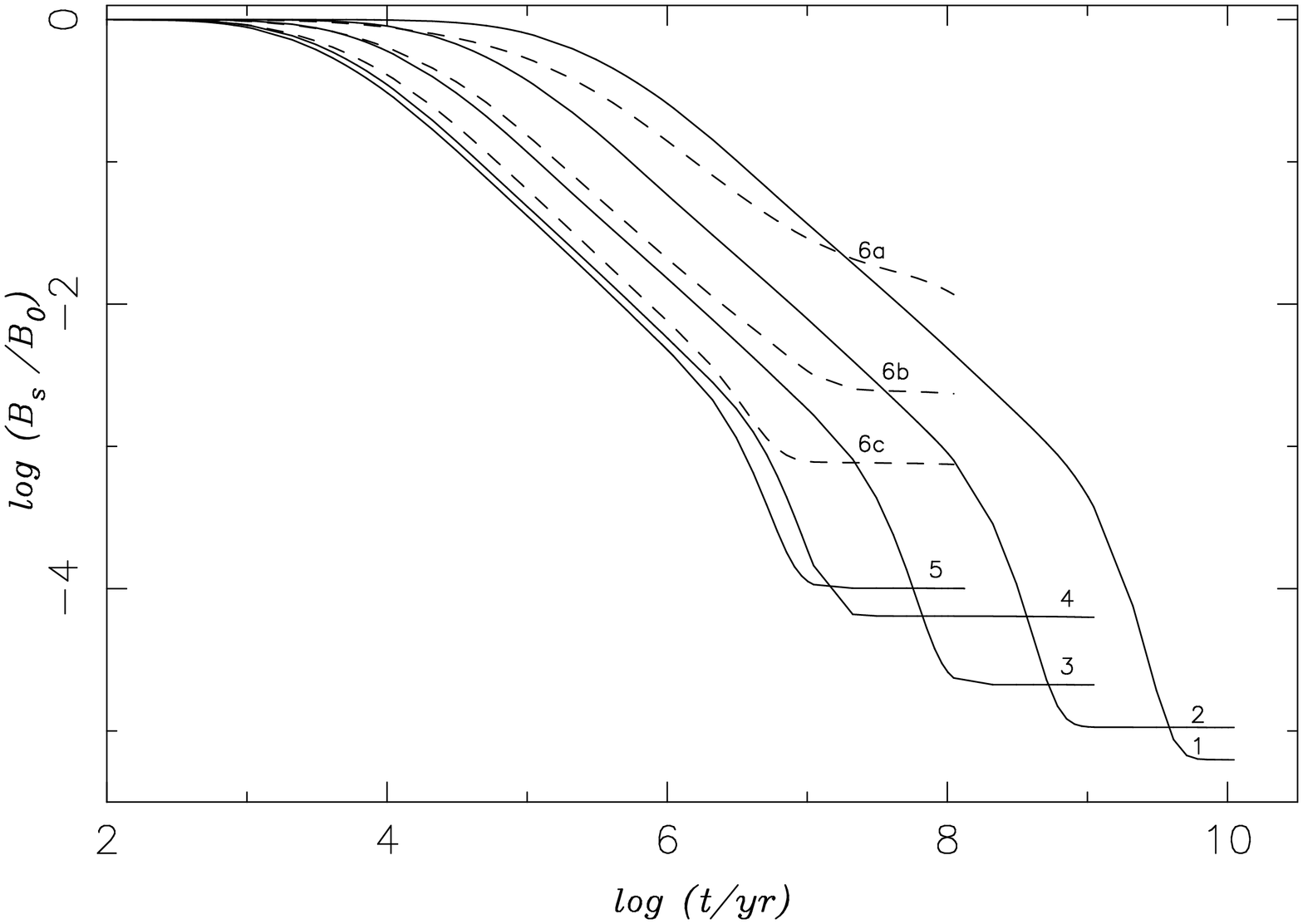,width=235pt,angle=-90}}}\end{center}
\caption[evolution of  the surface field Ia]{Evolution  of the surface
magnetic field  for six values of  accretion rate.  The curves  1 to 5
correspond  to $\mdot  = 10^{-13},  10^{-12}, 10^{-11},  10^{-10}, 2.0
\times  10^{-10}$~\dmdt with  the crustal  temperatures  obtained from
equation [\ref{etdmdt}]. The dashed curves 6a, 6b and 6c correspond to
$T =  10^{8.0}, 10^{8.25},  10^{8.5}$~K respectively for  an accretion
rate of $\mdot  = 10^{-9}$~\dmdt. All curves correspond  to $Q$ = 0.0,
but are insensitive to the value of $Q$.}
\label{fb_evolv1}
\eef

\bef
\begin{center}{\mbox{\epsfig{file=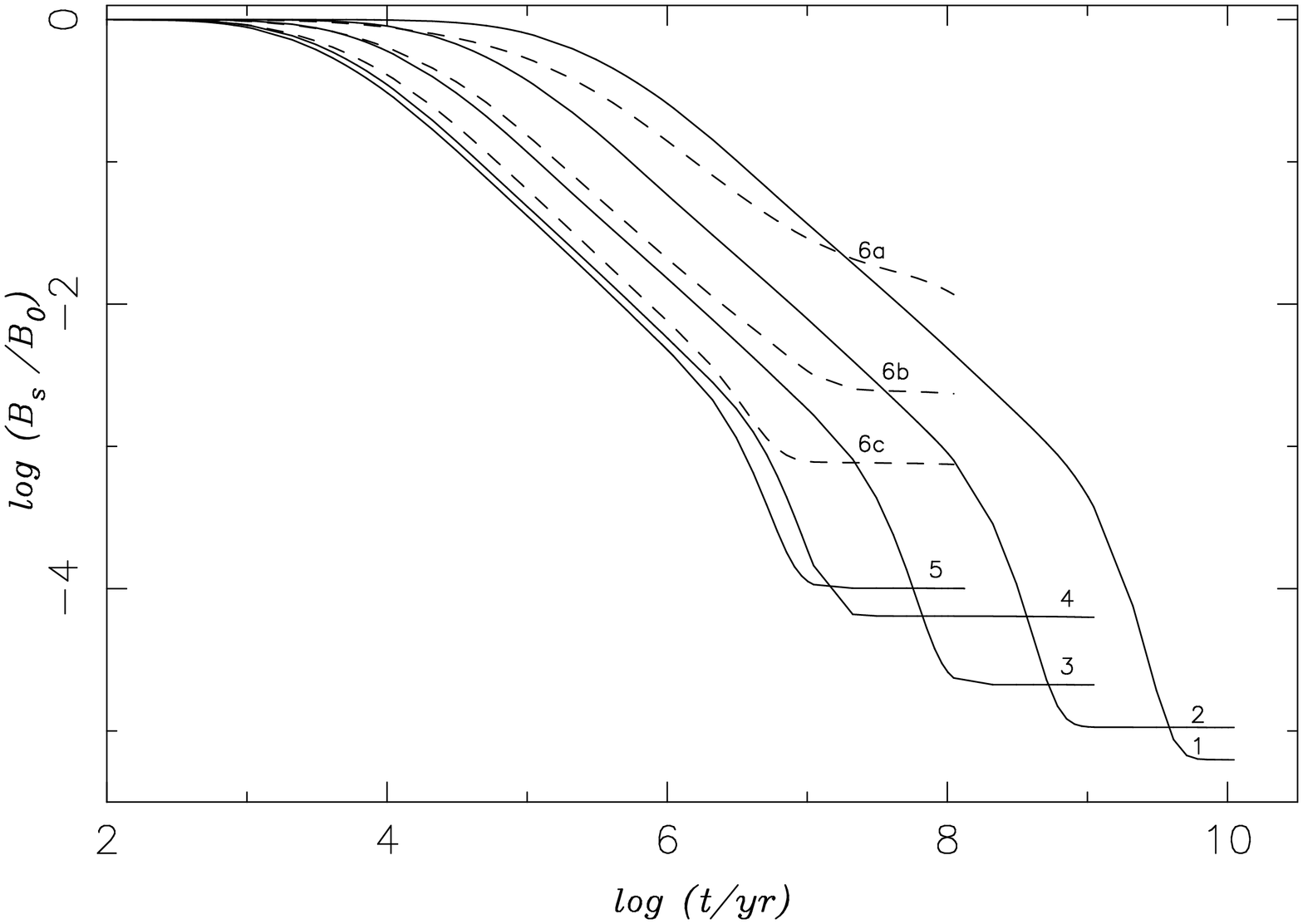,width=235pt,angle=-90}}}\end{center}
\caption[evolution   of   the  surface   field   Ib]{Same  as   figure
[\ref{fb_evolv1}]    with    the    initial   profile    centred    at
$10^{11.5}~\gcc$. }
\label{fb_evolv2}
\eef

\bef
\begin{center}{\mbox{\epsfig{file=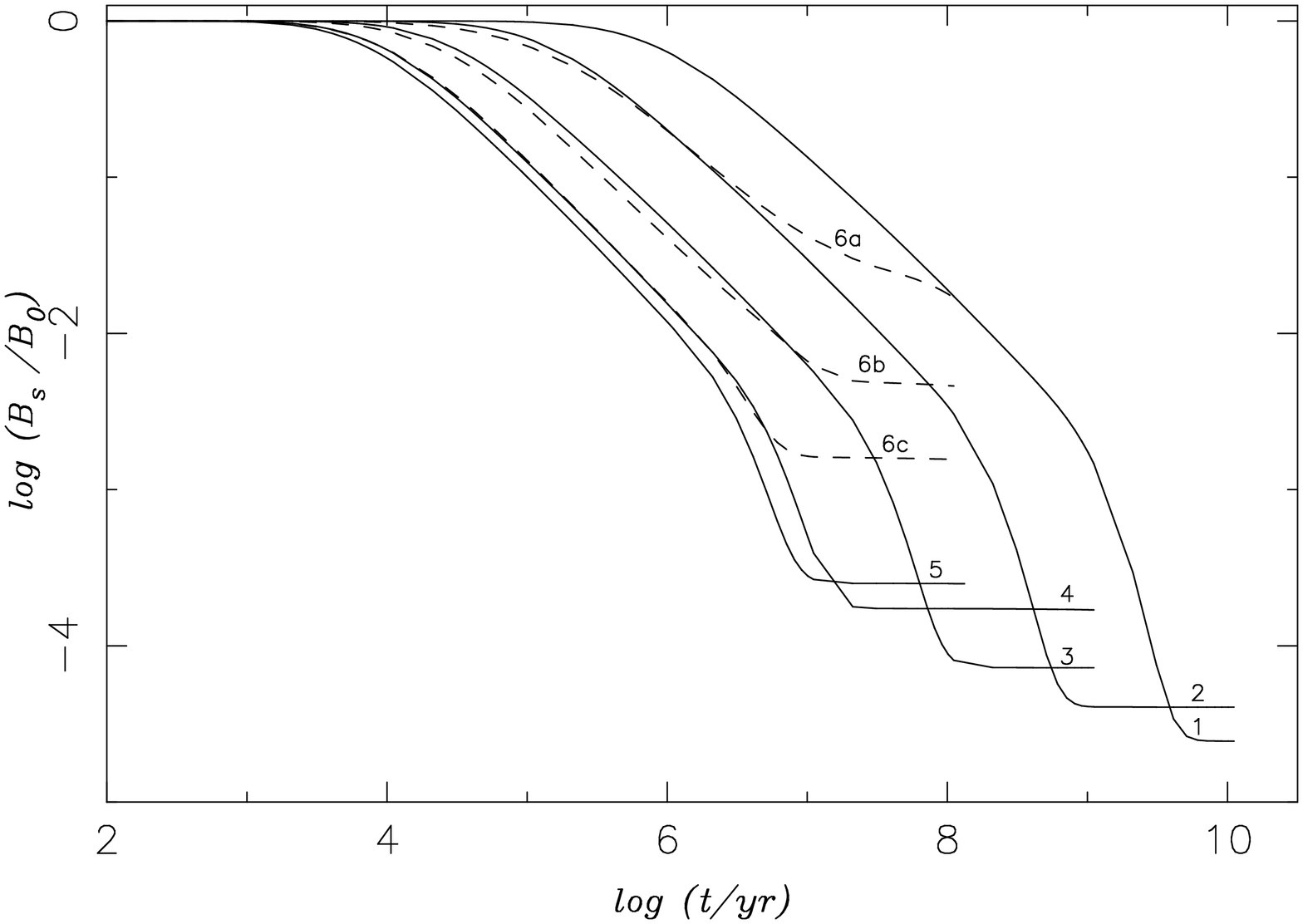,width=235pt,angle=-90}}}\end{center}
\caption[evolution   of   the  surface   field   Ic]{Same  as   figure
[\ref{fb_evolv1}],   with   the   initial   $g$-profile   centred   at
$10^{12}~\gcc$. }
\label{fb_evolv3}
\eef

\bef
\begin{center}{\mbox{\epsfig{file=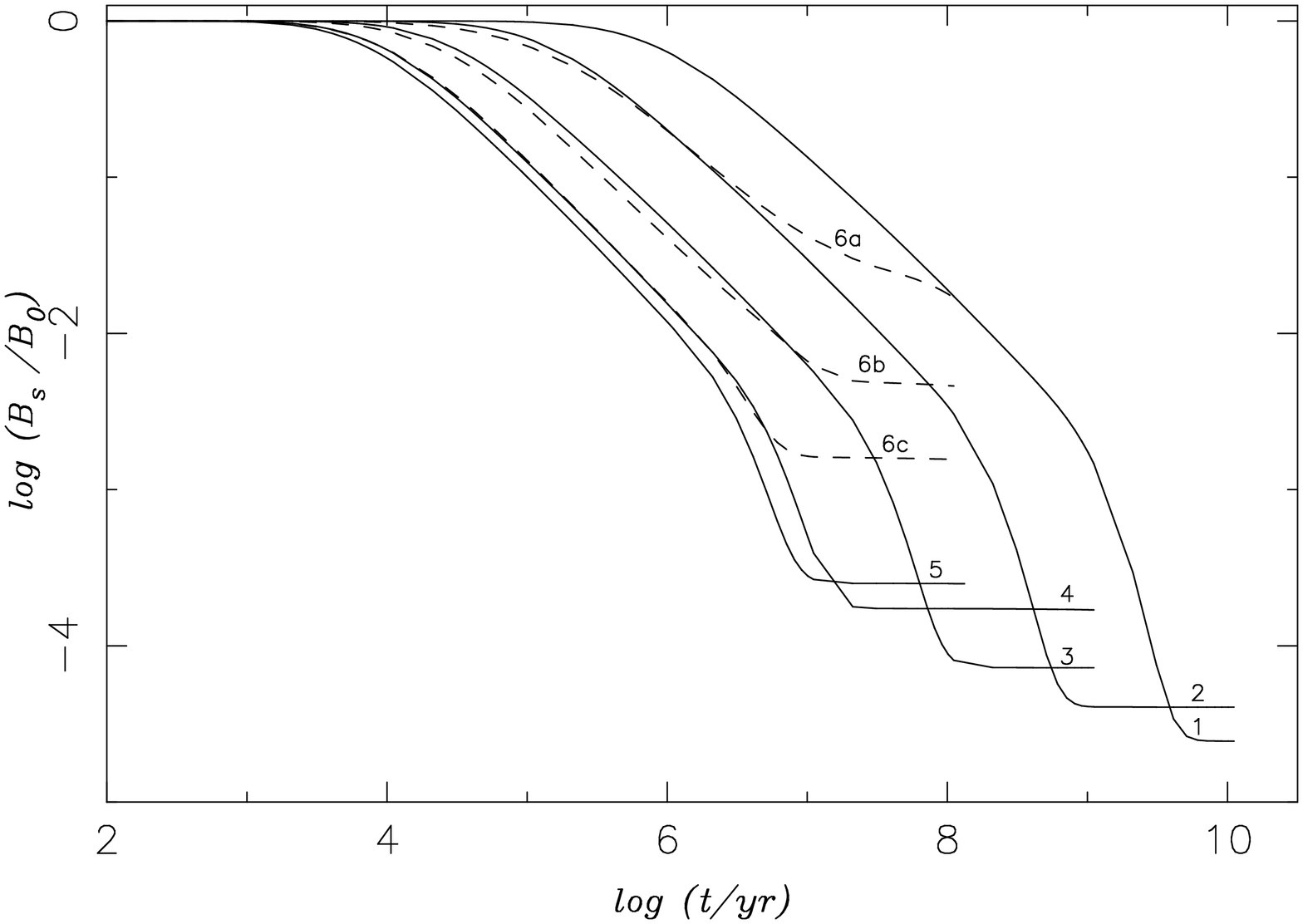,width=235pt,angle=-90}}}\end{center}
\caption[evolution   of   the  surface   field   Id]{Same  as   figure
[\ref{fb_evolv1}],   with   the   initial   $g$-profile   centred   at
$10^{12.5}~\gcc$. }
\label{fb_evolv4}
\eef

\bef
\begin{center}{\mbox{\epsfig{file=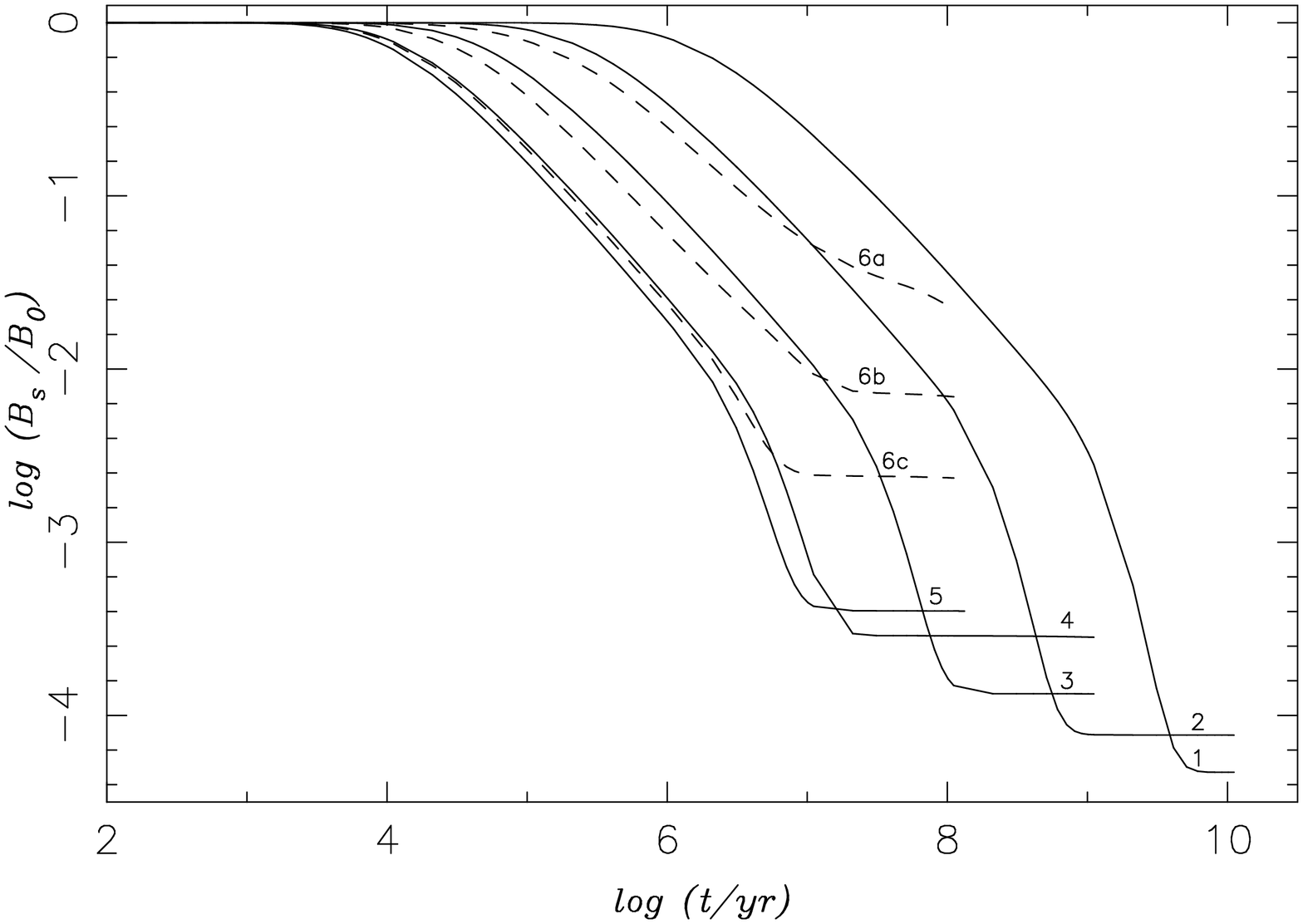,width=235pt,angle=-90}}}\end{center}
\caption[evolution   of   the  surface   field   Ie]{Same  as   figure
[\ref{fb_evolv1}],   with   the   initial   $g$-profile   centred   at
$10^{13}~\gcc$. }
\label{fb_evolv5}
\eef

\bef
\begin{center}{\mbox{\epsfig{file=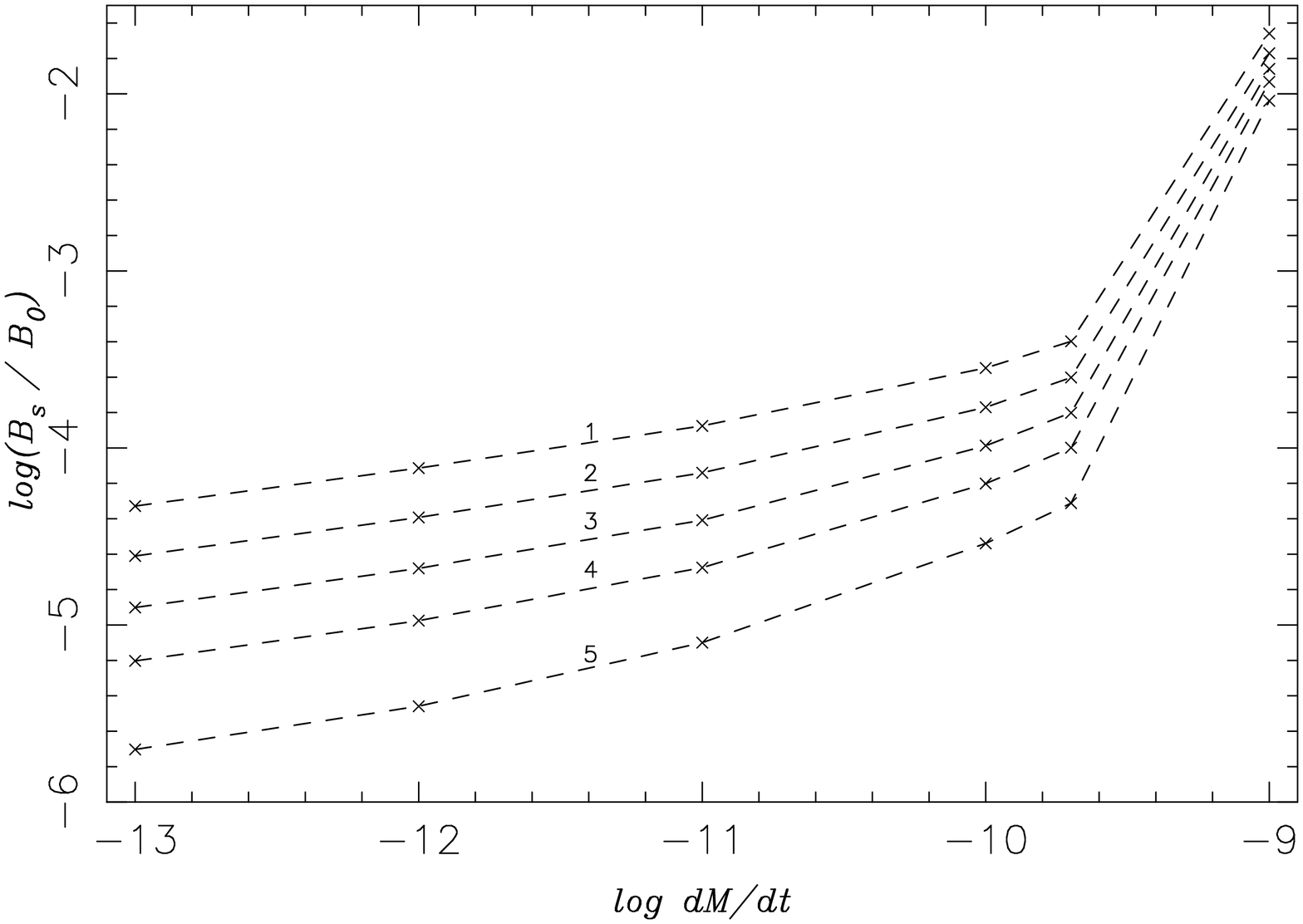,width=220pt,angle=-90}}}\end{center}
\caption[final  surface  field   vs.  accretion]{Final  surface  field
vs.  rate  of accretion.  Curves  1 to  5  correspond  to the  initial
$g$-profile  centering  densities  of  $10^{11},  10^{11.5},  10^{12},
10^{12.5},  10^{13}~\gcc$  respectively.  For  the  accretion rate  of
$\mdot = 10^{-9}$~\dmdt, we have  used the values corresponding to the
crustal temperature of $10^8$~K.}
\label{fb_dmdt_all}
\eef

In figure  [\ref{fb_evolv1}] we display  the evolution of  the surface
field for  different values of \mdot, with  temperatures obtained from
equation  [\ref{etdmdt}]  over  its   validity  range  and  for  three
different assumed temperatures at  the highest accretion rate. All the
curves in this figure refer  to an initial $g$-profile that is centred
at  a  density  of   $10^{11}  \gcc$.  In  figures  [\ref{fb_evolv2}],
[\ref{fb_evolv3}],  [\ref{fb_evolv4}]  and  [\ref{fb_evolv5}] we  plot
similar curves  for different densities  at which the  initial current
profiles  are centred.  In figure  [\ref{fb_dmdt_all}] we  display the
final  surface field  values obtained  as a  function of  the  rate of
accretion  for  different  values  of  initial  current  concentration
densities.

The following  features emerge from  the behaviour displayed  in these
figures:
\ben
\i The  general nature  of the decay  corresponds to an  initial rapid
phase exhibiting a power law behaviour for the most part with an index
ranging from 0.1 to 0.46 (i.e., $B \sim t^{-n}, 0.1 \lsim n \lsim 0.46
$), followed by  a short exponential phase and  then a freezing, which
stabilizes  the surface  field. This  stability is  the result  of the
current  distribution  responsible  for  the field  moving  to  highly
conducting  parts  of  the  star,   much  of  it  migrating  into  the
core. According to  our adopted scenario, the ohmic  time scale in the
core is much  longer than the Hubble time and  hence the surface field
at this  stage will  essentially be stable  forever. We refer  to this
surface field as the {\em `residual field'}.
\i The duration of the exponential phase and consequently the value of
the magnetic  field at which freezing  occurs is a  strong function of
the  accretion rate.   The higher  the accretion  rate the  sooner the
freezing sets in resulting in  a higher value of the `residual field'.
This   effect  can  be   understood  as   follows.  As   explained  by
Bhattacharya~\citeyear{db95a} the  decay behaviour turns  from a power
law  to an  exponential,  once the  diffused $g$-distribution  reaches
nearly  the bottom of  the crust.   The transition  from there  to the
frozen state happens  by further accretion of matter  which pushes the
crustal  material into  the core.  The  time required  for this  final
transition is of course dependent on the accretion rate and the higher
the  accretion rate  the smaller  it is.  For a  rate of  accretion of
$10^{-9}$~\msun/yr, this exponential phase is nearly absent.
\i  The dependence  of  the decay  on  the crustal  temperature is  as
expected, namely, the decay proceeds faster at a higher temperature.
\i  In figure [\ref{fb_dmdt_all}]  we have  plotted the  final 'frozen
field'  for different  rates of  accretion corresponding  to different
densities at which  the initial current profile is  centred at.  It is
clearly seen  from this figure that  the lower the  rate of accretion,
the lower is the final `frozen in' field.
\bef
\begin{center}{\mbox{\epsfig{file=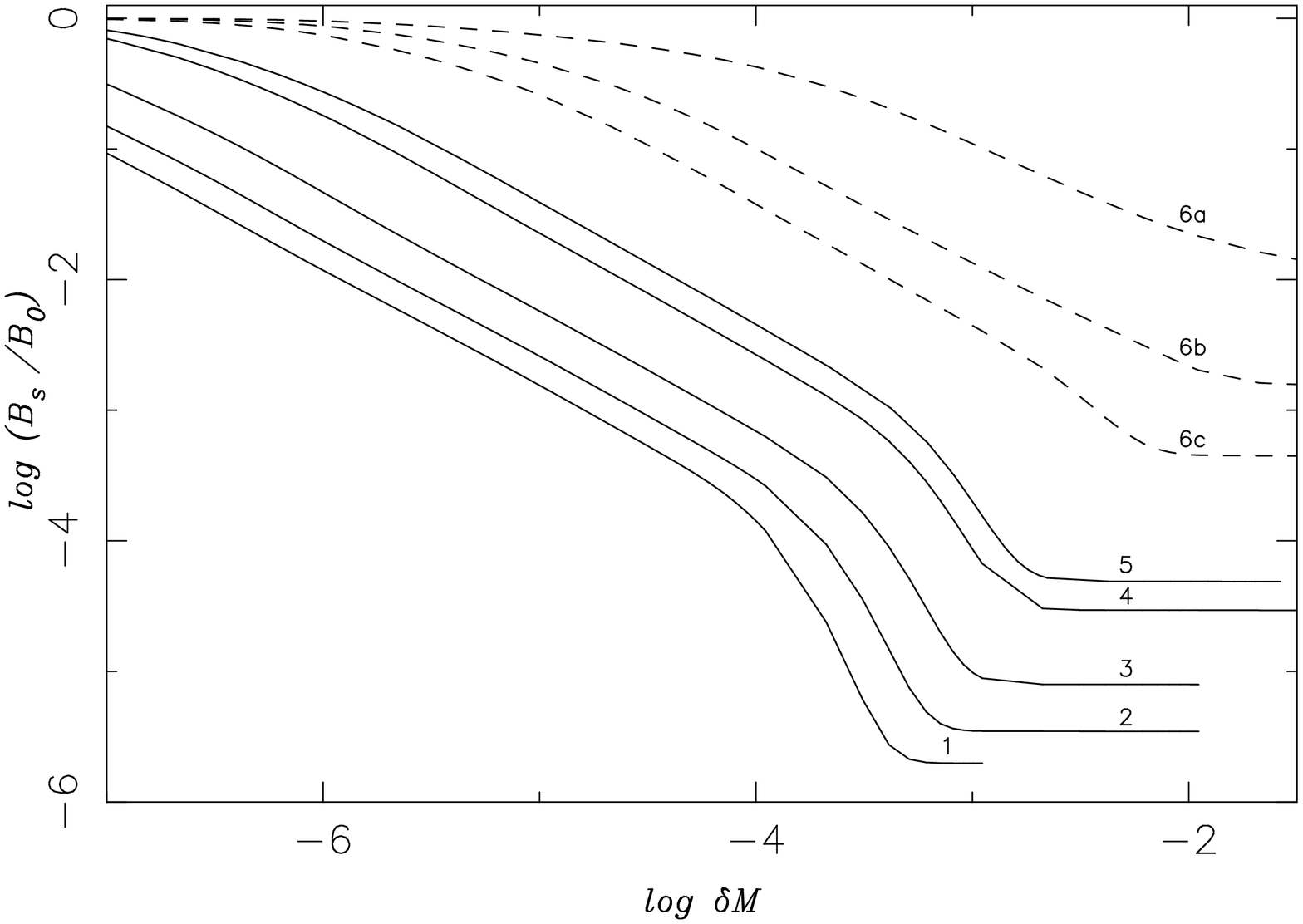,width=235pt,angle=-90}}}\end{center}
\caption[evolution of the surface  field IIa]{Evolution of the surface
magnetic field as a function of  total mass accreted.  The curves 1 to
6 correspond  to $\mdot = 10^{-13}, 10^{-12},  10^{-11}, 10^{-10}, 2.0
\times 10^{-10}, 10^{-9}$~\dmdt, the crustal temperatures are obtained
from  equation  [\ref{etdmdt}].  The  dashed  curves  6a,  6b  and  6c
correspond to  $T = 10^{8.0}, 10^{8.25},  10^{8.5}$~K respectively for
an accretion rate of $\mdot = 10^{-9}$~\dmdt. All curves correspond to
$Q$ = 0.0. and an initial profile centred at $10^{11}~\gcc$.}
\label{fb_dmdt1}
\eef
\bef
\begin{center}{\mbox{\epsfig{file=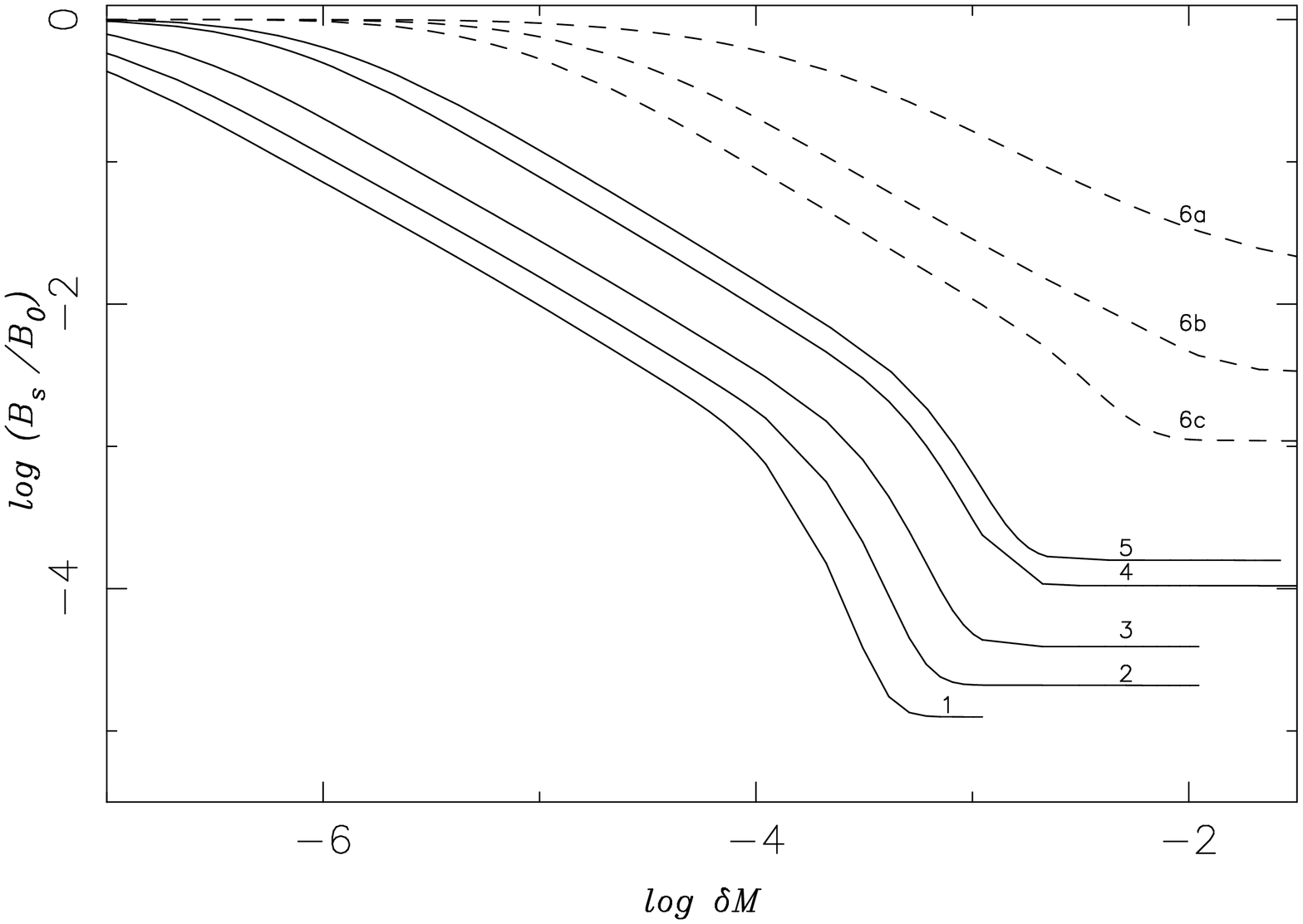,width=235pt,angle=-90}}}\end{center}
\caption[evolution   of  the   surface  field   IIb]{Same   as  figure
[\ref{fb_dmdt1}],   with   the    initial   $g$-profile   centred   at
$10^{12}~\gcc$. }
\label{fb_dmdt2}
\eef
\bef
\begin{center}{\mbox{\epsfig{file=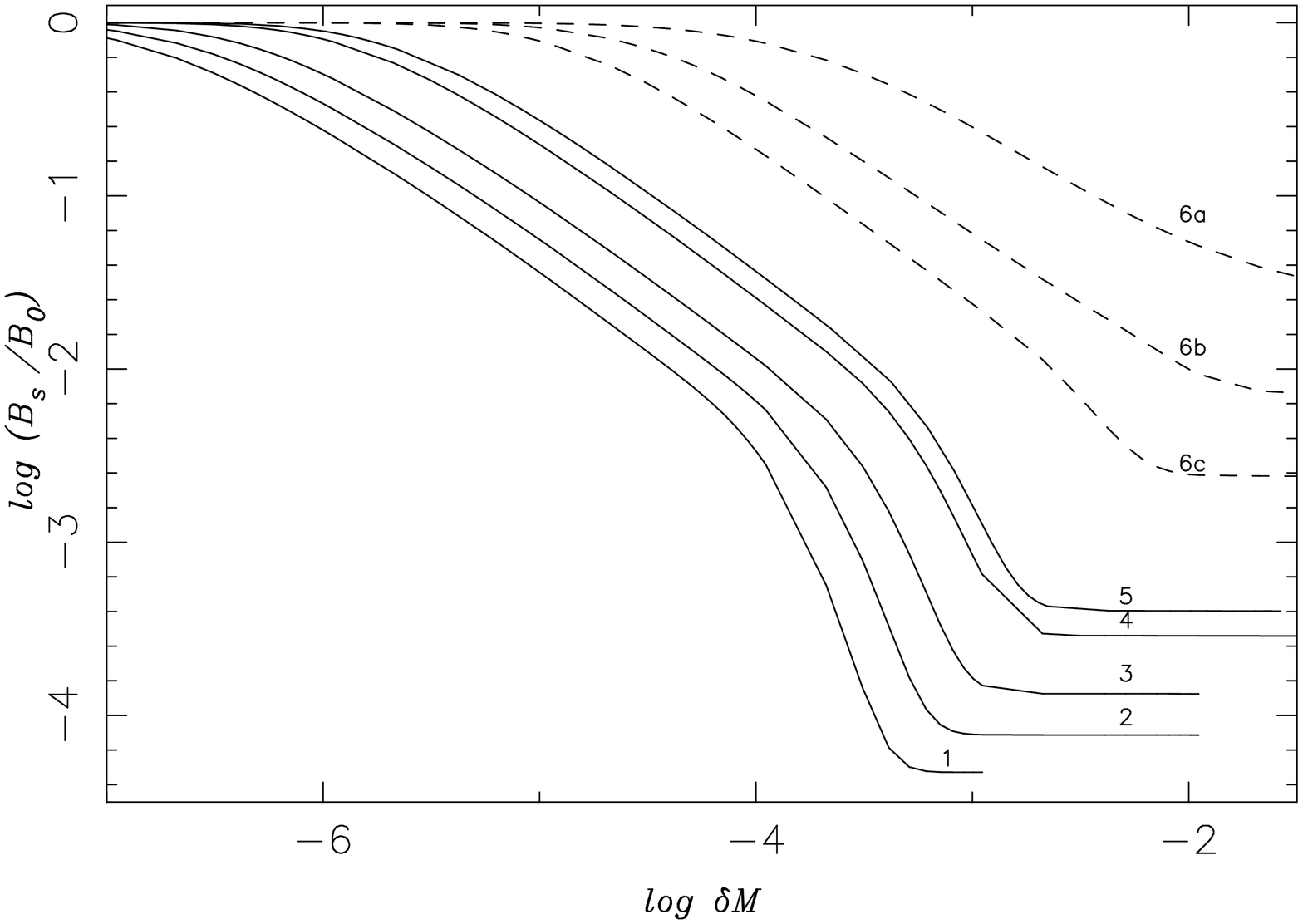,width=235pt,angle=-90}}}\end{center}
\caption[evolution   of  the   surface  field   IIc]{Same   as  figure
[\ref{fb_dmdt1}],   with   the    initial   $g$-profile   centred   at
$10^{12}~\gcc$. }
\label{fb_dmdt3}
\eef
\bef
\begin{center}{\mbox{\epsfig{file=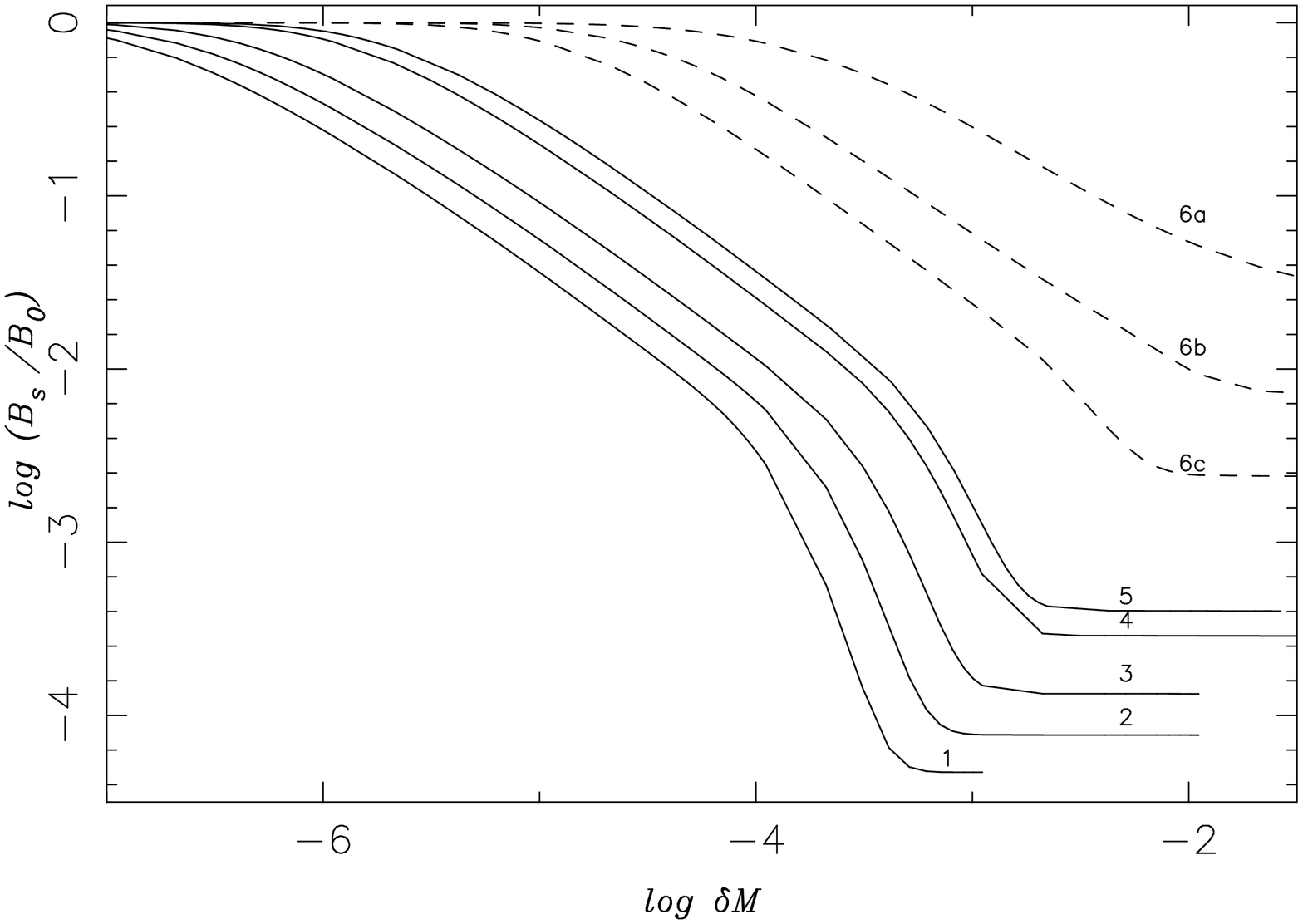,width=235pt,angle=-90}}}\end{center}
\caption[evolution   of  the   surface  field   IId]{Same   as  figure
[\ref{fb_dmdt1}],   with   the    initial   $g$-profile   centred   at
$10^{12.5}~\gcc$. }
\label{fb_dmdt4}
\eef
\bef
\begin{center}{\mbox{\epsfig{file=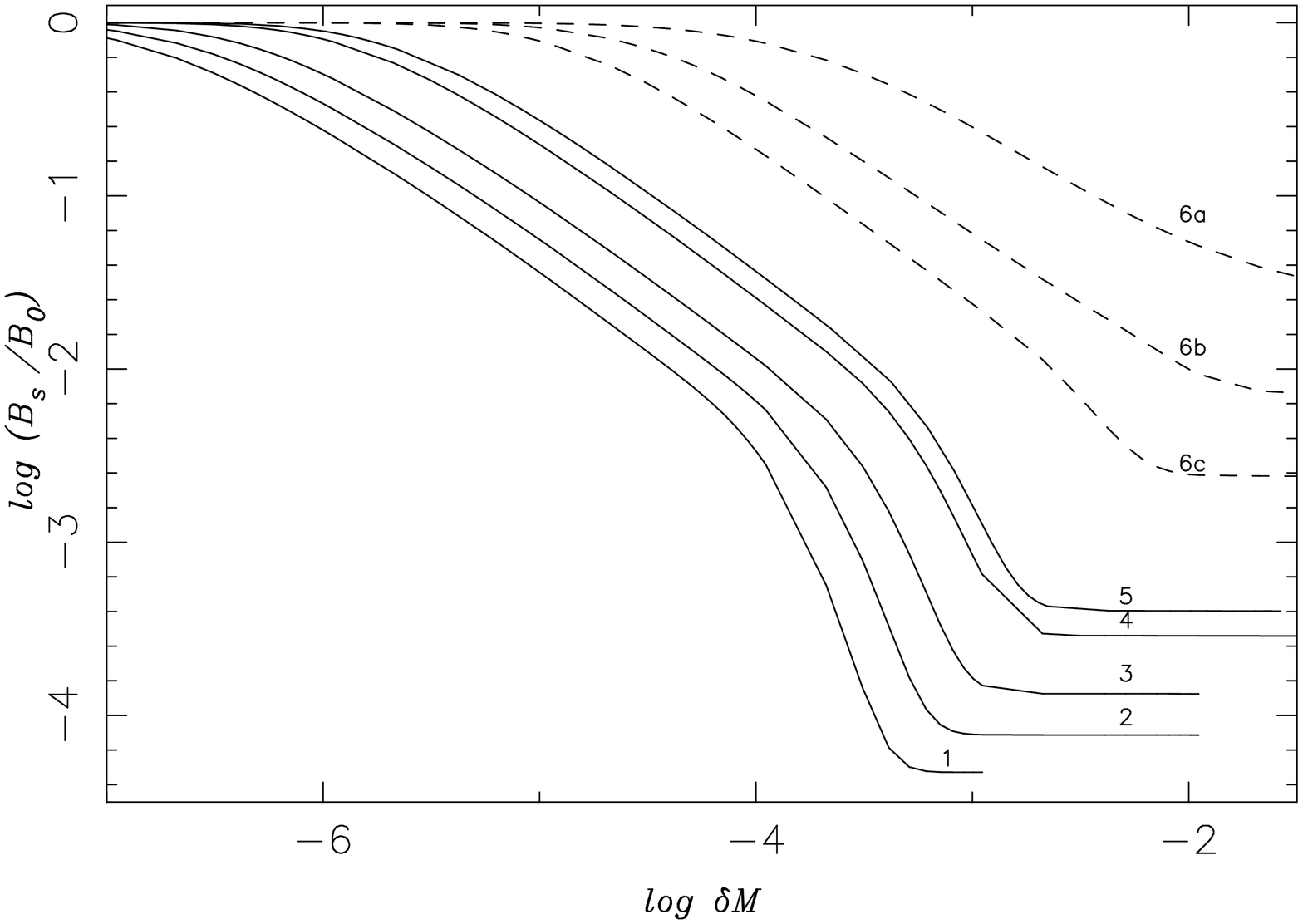,width=235pt,angle=-90}}}\end{center}
\caption[evolution   of  the   surface  field   IIe]{Same   as  figure
[\ref{fb_dmdt1}],   with   the    initial   $g$-profile   centred   at
$10^{13}~\gcc$. }
\label{fb_dmdt5}
\eef
\i In  figure [\ref{fb_dmdt1}] the  evolution of the  surface magnetic
field  as a  function  of total  accreted  mass has  been plotted  for
different rates of accretion.  Corresponding to a given accretion rate
the   crustal   temperature    has   been   obtained   from   equation
[\ref{etdmdt}]. It is observed that the {\em freezing in} of the field
occurs for  larger accreted  mass for higher  rates of  accretion. The
high accretion rate also  ensures quicker material transport to higher
densities,  causing  the field  strength  to  level  off at  a  higher
value. Another  important point  to note here  is that unlike  what is
assumed  in heuristic  evolutionary models~\cite{taam86,shib89,bitz95}
the amount  of field  decay is  dependent not only  on the  total mass
accreted but also on the accretion rate itself. Once the accreted mass
exceeds  $\sim 10^{-3}$~\msun,  the final  magnetic field  strength is
decided only  by the  rate at which  this mass was  accreted.  Whereas
figure [\ref{fb_dmdt1}]  correspond to  an initial profile  centred at
$10^{11}~\gcc$,  in  the  figures [\ref{fb_dmdt2}],  [\ref{fb_dmdt3}],
[\ref{fb_dmdt4}]  and  [\ref{fb_dmdt5}]  we  plot similar  curves  for
different densities.
\bef
\begin{center}{\mbox{\epsfig{file=pre-acc.ps,width=250pt}}}\end{center}
\caption[evolution  in  the   pre-accretion  phase]{Evolution  of  the
surface magnetic field  (1) without any pre-accretion phase  and (2 \&
3) for  such a phase  lasting $\sim 10^9$  yrs. In all cases  $\mdot =
10^{-9}~\msun yr^{-1}$.  and  $T = 10^{8.0}$~K.  $Q$ =  0.0 for curves
1, 2a, 2b and  $Q$ = 0.1 for 3a, 3b.  In 2a  and 3a the actual surface
field has been plotted, whereas in 2b and 3b it has been scaled to the
value of  the field at  the beginning of  the accretion phase.  In the
isolated phase standard cooling van Riper (1991a,b) has been used.}
\label{fpre-acc}
\eef
\i In practice a neutron star will often undergo a non-accreting phase
of   considerable  duration   before  accretion   can  begin   on  its
surface. During this  initial phase its magnetic field  will evolve as
in an isolated  neutron star, namely according to  the evolution shown
in  figure [\ref{fb_diff}]. This  has important  physical consequences
for  the  evolution in  the  accretion  phase.  The diffusion  in  the
pre-accretion phase causes the currents to already penetrate to deeper
and denser  regions of  high conductivity. As  a result the  net decay
achieved   {\em  during   subsequent  accretion}   is   lower.  Figure
[\ref{fpre-acc}] compares  the evolution of the  surface field without
and with  the pre-accretion phase  lasting a billion years.  It should
also  be  noted  that   for  accretion-induced  field  decay  with  an
effectively isolated  pre-accretion phase, the  impurity concentration
plays  an important  role too.  Though the  actual final  field values
obtained  is  lower  for  large  $Q$, the  decay  experienced  in  the
accretion  phase is  significantly less  in such  cases and  the `{\em
freezing}' sets in much faster.
\i As mentioned before, the range of conditions explored by us overlap
with   those   in   the   work  of   \citeN{urpn95},   \citeN{urpn96},
\citeN{gepp96} and goes beyond. We have performed detailed comparisons
of our  results with  theirs in the  overlap range.  The  agreement in
general is found to be  excellent giving us confidence in the validity
of our approach.
\een

To summarize,  we have explored accretion-driven  evolution of crustal
magnetic fields over a range of conditions not previously attempted in
the  literature.  The  new  behaviour revealed  by these  computations
include  a near  exponential  decay  of the  surface  field after  the
initial  power law phase  and most  importantly an  eventual freezing.
The `residual field' corresponding to  this frozen state is a function
of the accretion rate and  the temperature during the evolution. It is
interesting   to  note  that   for  near-Eddington   accretion  rates,
applicable  to  Roche-lobe  overflow   phase  in  real  binaries,  the
`residual  field' lies  between $10^{-2}$--$10^{-4}$  of  the original
value.   So, if  the  neutron  star originally  started  with a  field
strength  of  the  order of  $10^{12}$  G,  this  would mean  a  final
post-accretion field strength  of $\sim 10^8$--$10^{10}$~G, exactly as
observed  in most recycled  pulsars. Unfortunately,  we have  not been
able to treat accretion rates equal to or larger than Eddington in the
present work due  to lack of knowledge about  the crustal temperatures
at  those accretion  rates.   But  judging by  the  dependence of  the
behaviour  on  accretion  rates  (see figure  {\ref{fb_dmdt_all}])  it
appears that even somewhat higher post-accretion field strengths might
be possible under such conditions. Recycled pulsars with the strongest
magnetic fields, namely PSR 0820+02 and PSR 2303+46 have, according to
evolutionary  scenarios~\cite{bhat91}, undergone  super-Eddington mass
transfers. Their field strengths  would therefore be in agreement with
the trend described above.

\section{conclusions}
\label{smn-concl}

In this chapter we have explored the evolution of the crustal magnetic
field of  accreting neutron stars.  The combination  of enhanced ohmic
diffusion due to crustal heating and the transport of current-carrying
layers to higher densities due  to the accreted overburden, causes the
surface field strength to exhibit the following behaviour:
\ben
\i An initial rapid decay (power law behaviour followed by exponential
behaviour) followed by a leveling off (freezing),
\i Faster  onset of freezing at  higher crustal temperatures  and at a
lower final value of the surface field,
\i Lower  final fields for lower  rates of accretion for  the same net
amount of accretion,
\i The  longer the  duration of the  pre-accretion phase the  less the
amount of field decay during the accretion phase, and
\i  The deeper  the initial  current loops  are the  higher  the final
surface field.
\een

\chapter{comparison with observations}
\label{cobs}

\section{introduction}
\label{sobs-intro}

In the previous chapter we have discussed a model for the evolution of
the  magnetic field in  accreting neutron  stars, assuming  an initial
crustal flux. It has been borne out by our calculations that the model
possesses some  essential features required  to explain the  origin of
the observed  low-field pulsars.  But in a  broader perspective  it is
also  necessary for a  model to  explain the  present paradigm  of the
field evolution  in its entirety.  The astronomical  objects that have
so far been unambiguously identified with neutron stars can be divided
into two distinct classes :
\bei
\i the radio pulsars, and
\i the X-ray binaries containing a neutron star.
\eei
Interestingly,  observations suggest  that  a class  of radio  pulsars
descend  from X-ray  binaries,  and neutron  stars  in X-ray  binaries
themselves  represent an  evolutionary state  beyond that  of isolated
radio  pulsars (see  \citeNP{db95b},  \citeNP{db96a}, \citeNP{heuv95},
\citeNP{verb95} and references  therein). Therefore the present belief
is that  this evolutionary  link can be  understood within  an unified
picture of evolution of their spin as well as the magnetic field.  The
spin-evolution  of  the neutron  stars  in  binary  systems have  been
investigated  following the  detailed binary  evolution  assuming some
simple  model for the  field evolution  (see \citeNP{verb95}  (1995) and
references  therein).   In the  present  work  we  concentrate on  the
details of the evolution of the  magnetic field. We apply the model of
field evolution to isolated neutron stars as well as to those that are
members  of binaries.  Comparison of  these results  with observations
allows us to test the validity of the field evolution model.

Radio pulsars can  be classified into two groups,  namely the solitary
pulsars and the  binary pulsars.  The binary pulsars  are again of two
types - the high mass binary  pulsars and the low mass binary pulsars,
the reference to  the masses being to those of  the companions. It has
been suggested by  recent statistical analyses that the  fields of the
isolated    neutron   stars   do    not   undergo    any   significant
decay~\cite{rudr91a,rudr91b,rudr91c,dbgs91a,ding93,miri94,miri96}.  On
the other hand almost all binary pulsars possess field values that are
smaller  than   the  canonical  field  values   observed  in  isolated
pulsars.  Usually   the  high  mass  binary  pulsars,   of  which  the
Hulse-Taylor  pulsar is one  famous example,  have field  strengths in
excess of $10^{10}$~Gauss, whereas the low mass binary pulsars include
both  high-field   pulsars  and   very  low-field  objects   like  the
millisecond pulsars. It must be noted here that when we talk about the
solitary  pulsars  we mean  high-field  solitary  pulsars without  any
history of  binary association.   The millisecond pulsars  and pulsars
with  obvious or  suspected binary  history are  discussed  along with
their binary  counterparts for the  sake of convenience.   The present
belief  regarding the evolutionary  history of  the binary  pulsars is
that the  high mass  binary pulsars come  from systems similar  to the
high  mass  X-ray binaries  whereas  the  low  mass binaries  are  the
progenies of  the low mass X-ray  binaries. In the  following table we
present this evolutionary scenario.

\vspace{0.5cm}

\hspace{0.25cm}
\begin{tabular}{|l|l|r|r|} \hline
&&& \\
systems & {\bf solitary pulsars} & {\bf high mass} & {\bf low mass} \\
& (millisecond pulsars)& {\bf binary pulsars} & {\bf binary pulsars} \\
& excluded)& & \\
&&& \\ \cline{1-4}
&&& \\
companion & & neutron stars, & low mass stars  \\
& & massive ($\gsim 0.6$~\msun) & low mass ($\lsim 0.4$~\msun) \\
& & white dwarfs & white dwarfs  \\
&&& \\ \cline{1-4}
&&& \\
magnetic field & $10^{11} - 10^{13}$~Gauss & usually $\gsim 10^{10}$~Gauss & $10^{8} - 10^{11.5}$~Gauss  \\
&&& \\ \cline{1-4}
&&& \\
progenitors & & high mass & low mass \\
& & X-ray binaries & X-ray binaries \\
&&& \\ \cline{1-4}
\end{tabular} \\

\vspace{0.5cm} Even though the millisecond pulsars fall in the broader
category of low mass binary  pulsars, we shall mention them separately
due  to their  unique characteristic  features.  The  rotation powered
radio  pulsars  separated  out  into  two distinct  classes  with  the
discovery   of  the  1.6   ms  pulsar,   PSR  1937+21~\cite{back82}.In
accordance with  the extremely  small rotation periods  ($\sim$~ms) of
this new variety they came to be known as {\em Millisecond Pulsars} as
opposed to the population of  normal pulsars that have longer rotation
periods. Loosely, the  term millisecond pulsar refers to  the class of
pulsars with rotation periods less than 20 ms. This definition, though
somewhat ad-hoc, actually serves  the purpose of classification rather
well.  In  other words,  all the members  of the class  of millisecond
pulsars show remarkable similarity  in several of their characteristic
physical properties  (listed below),  which also serve  to distinguish
them from the rest of the pulsar population.

The  characteristic  features  of  the millisecond  pulsars  could  be
summarized as follows.
\ben
\i fast rotation -- P $\lsim$ 20~ms;
\i extremely small magnetic fields  (three to four orders of magnitude
smaller than the canonical field values observed in normal pulsars) --
B $\sim 10^{8}-10^{9}$~G;
\i binary association  -- 90\% of the disc population  and 50\% of the
Globular Cluster  population of the millisecond  pulsars have low-mass
binary companions, with most probable mass of the companion $\lsim 0.3
\msun$, in nearly circular orbits;
\i  old age --  age determination  from the  presence in  the Globular
Cluster or from the surface  temperature measurements of a white dwarf
companion   (e.g.,   for  PSR   1855+09)   indicate   a  lifetime   of
$10^{9}$~years or more~\cite{hans97}.
\een

It has been shown that a total mass of $\sim 0.1 \msun$ is required to
be accreted  to achieve  millisecond period through  accretion induced
spin-up. Present  theories of binary  evolution predict that  the only
systems capable of supporting  mass-transfer long enough (about $10^7$
years at the  Eddington rate of accretion) to  allow the neutron stars
to  accrete  $\gsim 0.1~\msun$  are  the  ones  with low-mass  ($\lsim
1.5~\msun$)  donors   (the  Low  Mass  X-ray  Binaries   or  LMXBs  in
short).  Hence, it is  generally believed  that the  evolutionary link
between  a normal  radio  pulsar and  its  millisecond counterpart  is
through  a  phase  of  binary  processing in  LMXBs.  In  general  the
companion of a binary pulsar is indicative of the system from which it
originated.  A   pulsar  with  neutron  star  or   a  massive  ($\gsim
0.6~\msun$) white  dwarf companion is  understood to have come  from a
system  containing a  massive  ($\gsim 5~\msun$)  donor. Similarly,  a
pulsar with a low mass  ($\lsim 0.4~\msun$) white dwarf companion must
originate  in a  low mass  X-ray  binary. All  the binary  millisecond
pulsars have low mass  companions, indicating their origin in low-mass
systems. 

At this stage, it becomes  imperative that the models for evolution of
the   magnetic   field   be   consistent   with   the   spin-evolution
scenario. There  is, as  yet, no consensus  about the models  of field
evolution or  about those  of the origin  and structure of  the field.
Based on  the two theories  of field generation and  the corresponding
internal  structures  (see section  [\ref{sorigin}])  one can  broadly
classify  the models  of  field evolution  into  two categories.   The
original theory  of {\em recycling} assumed a  spontaneous ohmic decay
of magnetic field  with age, which according to  the present evidence,
appears unlikely. For a magnetic  flux confined to the core, the model
of       spindown-induced      flux      expulsion       has      been
explored~\cite{miri94,miri96,bhat96}, We  shall discuss this  model in
the next chapter.

In this chapter, we confront the field evolution model of chapter [5],
namely that of assuming an  initial crustal flux, with observations of
both isolated  neutron stars and  neutron stars in binary  systems. In
particular, we  address the question of  millisecond pulsar generation
purely from the point of view of the field evolution.  Our results are
consistent with  the general view  that millisecond pulsars  come from
low mass x-ray binaries. We also find that the neutron stars processed
in high mass  x-ray binaries would retain fairly  high field strengths
in conformity with what is observed in high mass radio pulsars. 

In  section  [\ref{sobs-binaries}]  we  discuss the  nature  of  field
evolution  of a  neutron star  vis-a-vis the  binary evolution  of the
system which  it is a member of.  Section [\ref{sobs-comp}] elaborates
the  computational   details.  We  present  our   results  in  section
[\ref{sobs-results}]     and     the     conclusions    in     section
[\ref{sobs-concl}]. 

\section{field evolution in solitary neutron stars}
\label{sobs-solitary}

According to modern statistical analyses the isolated neutron stars do
not undergo  any significant field reduction over  their active pulsar
life time~\cite{bhat92,waka92,hart97}. In an isolated neutron star the
field decreases due to pure  ohmic dissipation of the current loops in
the crust. We have already  mentioned that the diffusive time-scale is
dependent on three factors, namely
\bei
\i the density at which the initial current loops are located,
\i the impurity content of the crust, and
\i the temperature of the crust.
\eei
We shall  look at the  effect of all  these factors and  determine the
acceptable  ranges for  these  parameters for  which  the final  field
strengths are within the limits of statistical uncertainty. 

An isolated neutron star cools down after its birth chiefly by copious
neutrino  emission. It  has recently  been shown  (see Page,  1998 and
references therein  for details) that there are  lots of uncertainties
in this  regard. The  data could  be made to  fit both  the `standard'
cooling  and the  `accelerated' cooling  with  appropriate assumptions
regarding the state  of the stellar interior. Therefore,  for the sake
of completeness, we have looked at  both the cases - with standard and
accelerated cooling. The actual behaviour of the system is most likely
to be something in between. We  have used data from Urpin \& van Riper
(private communication) for both  standard and accelerated cooling. In
figure [\ref{fcool_compare}]  the two  cooling curves have  been shown
for a comparison.

\bef
\begin{center}{\mbox{\epsfig{file=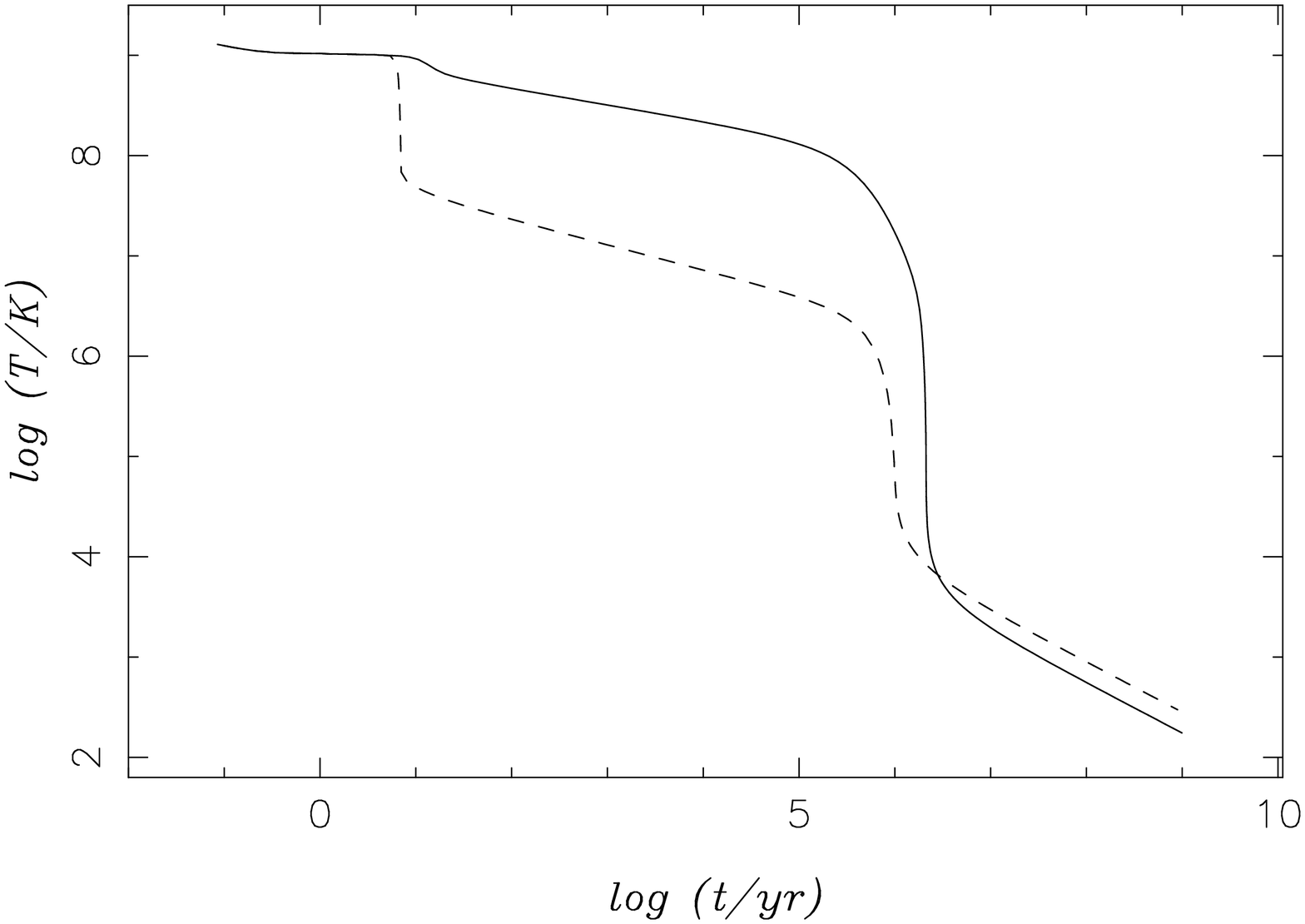,width=250pt,angle=-90}}}\end{center}
\caption[thermal evolution]{Evolution  of the interior  temperature of
an isolated  neutron star. The  solid and dotted curves  correspond to
standard cooling and accelerated cooling, respectively.}
\label{fcool_compare}
\eef

\section{field evolution in neutron stars with binary companions}
\label{sobs-binaries}

\subsection{binary and spin evolution}
\label{ssobs-binary}

From the point  of view of the interaction between  the two stars, the
binaries, in general, go through three distinct phases of evolution as
detailed below.
\ben
\i The  Isolated Phase -  Though the stars are  gravitationally bound,
there is  no mass transfer. Therefore  both the spin  and the magnetic
field evolve as they would in an isolated pulsar. The spin undergoes a
pure dipole slow-down during this phase.
\i The Wind Phase - The interaction is through the stellar wind of the
companion which  is likely to be  in its main-sequence.  In this phase
the interaction  of the  wind material with  the magnetosphere  of the
neutron star proceeds  in two distinct sequences. In  the early stages
of  the  wind phase  the  magnetospheric  interaction  brings about  a
spin-down  of  the  neutron  star  in the  following  manner.  In  the
accretion disc the  matter rotates with a Keplerian  velocity at every
point within the  disc. When the matter arrives  at the magnetospheric
boundary  of the  star (determined  roughly by  the  Alfv\'{e}n radius
where the  magnetic pressure equals  the ram pressure of  the incoming
material)  the magnetic  field  starts controlling  the material  flow
dynamics.  This  point  onwards  the  field  forces  the  material  to
co-rotate with  the star.  If the Keplerian  velocity of  the incoming
material is  smaller than the  co-rotation velocity at  the Alfv\'{e}n
radius then the  material, in being forced to  co-rotate with the star
there,  extracts angular  momentum from  the star  thereby  slowing it
down. This  material then gets  expelled from the  magnetosphere which
has  given  this  situation   the  name  {\em  propeller  phase}.  For
`spin-down induced flux expulsion' models the `propeller phase' is the
most crucial for the field evolution  because it is in this phase that
a significant flux  expulsion is achieved due to  a rapid spin-down of
the star. Once the star  has been spun-down sufficiently the Keplerian
velocity  at Alfv\'{e}n  radius  becomes larger  than the  co-rotation
velocity. In  this case the accreting material  loses angular momentum
as it reaches  Alfv\'{e}n radius, spinning the star  up. This material
eventually reaches the polar cap of the star by moving along the field
lines and gets deposited there.  Therefore, this is the phase in which
there is actual mass accretion  by the star. An equilibrium is reached
when  the  co-rotation  speed   equals  the  Keplerian  speed  at  the
magnetospheric   boundary,  i.e,  at   the  Alfv\'{e}n   radius.  This
equilibrium condition then determines  the maximum spin-up for a given
rate of accretion  and a given strength of  the surface field, through
the following relation~\cite{alpr82,chen93}
\beq
P_{\rm  eq}  =   1.9  \,  {\rm  ms}  \;   B_9^{6/7}  \;  (\frac{M}{1.4
\msun})^{-5/7} \; (\frac{\mdot}{\mdot_{\rm Edd}})^{-3/7} \; R_6^{18/7}
\label{ePeq}
\eeq
where $B_9$ is  the surface field in $10^9$ Gauss, $M$  is the mass of
the star, $\mdot$ is the rate  of accretion and $R_6$ is the radius of
the star in units of 10~km. Therefore, the wind phase has two distinct
sub-phases, namely - the propeller  phase and the phase of actual wind
accretion.  It must  be  noted here  that  the duration  of these  two
sub-phases vary widely  from system to system and  the phase of actual
wind accretion  may not at all  be realized in some  cases. For models
based on  an initial  crustal field configuration,  the phase  of wind
accretion  and  the  subsequent   phase  of  Roche  contact  play  the
all-important role.
\i The  Roche-contact Phase - When  the companion of  the neutron star
fills  its Roche-lobe  a phase  of  heavy mass  transfer ($\mdot  \sim
\mdot_{\rm Edd}$)  ensues. Though short-lived in case  of HMXBs ($\sim
10^4$  years),  this  phase can  last  as  long  as $10^9$  years  for
LMXBs. Consequently, this phase  is very important for field evolution
in LMXBs.
\een

The nature of the binary evolution  is well studied in the case of the
LMXBs where  the mass  transfer proceeds in  a controlled  manner. The
same is true for the wind phase  in the HMXBs. But the exact nature of
mass  transfer  in  the   common-envelope  phase,  due  to  Roche-lobe
overflow, has  not been studied in  any detail yet. Still,  due to the
short-lived  nature this  phase does  not affect  the  field evolution
significantly  and therefore the  lack of  precise knowledge  does not
affect our  calculations much. On  the other hand, not  much attention
has  been paid  to  the  evolution of  the  intermediate systems  with
companion masses in the range $\sim$ 2 - 5 \msun. They are most likely
to have an intermediate nature in that the wind phase is prolonged and
the accretion rates are similar to those in the wind phase of low-mass
systems, whereas,  the Roche-contact phase is perhaps  similar to that
in HMXBs.   In either of the  phases it would be  difficult to observe
these systems, owing  to the low luminosity in the  wind phase and due
to the  short-lived nature of  the Roche-contact phase.   Moreover, in
HMXBs and intermediate mass  binaries Roche-contact would usually lead
to  a common-envelope  evolution. In  this phase  the neutron  star is
engulfed by the common envelope and  the X-ray flux is hidden from the
view.  So  far no  intermediate-mass system has  been observed  in the
X-ray phase. As  for the pulsars processed in  them, there are perhaps
three    examples    PSR   B0655+64,    PSR    J2145-0750   and    PSR
J1022+1001~\cite{cami96}.    But    because   of   the   uncertainties
surrounding  their  mass transfer  history  we  exclude  this kind  of
binaries from the present discussion.

\section{computations}
\label{sobs-comp}

Using  the  methodology  developed  in chapter  [\ref{cmn}]  we  solve
equation [\ref{edgdt}] following  the mass-transfer history on neutron
stars  n  high-mass and  low-mass  binary  systems.  For the  case  of
isolated pulsars we solve the following equation,
\beq
\frac{\partial \vec  B}{\partial t} = - \frac{c^2}{4  \pi} \vec \nabla
\times   (\frac{1}{\sigma}  \times   \vec  \nabla   \times   \vec  B).
\label{ediff}
\eeq
For all of the above cases  we shall assume an initial crustal current
configuration (the kind that has been used in chapter [\ref{cmn}]).

\subsection{binary parameters}

The binary evolution parameters for the LMXBs and the HMXBs used by us
are as follows~\cite{verb90,bhat91,bitz95,king95} -
\ben
\i Low Mass X-ray Binaries -
\ben
\i Isolated  phase - Though binaries  with narrow orbits  may not have
too-long-lived a  phase of completely detached  evolution, long period
binaries (like  the progenitor system  of PSR 0820+02 with  $\sim$ 250
day orbital period~\citeNP{verb95}) may spend longer than $10^9$ years
before contact  is established.  In general, the  isolated phase lasts
between $10^8 - 10^9$ years.
\i Wind phase  - This phase again lasts for about  $10^8 - 10^9$ years
with attendant rates of  accretion ranging from about $10^{-15}$~\dmdt
to $10^{-12}$~\dmdt.
\i Roche-contact phase  - In this phase, the  mass transfer rate could
be as  high as the Eddington  rate ($10^{-8}$~\dmdt for  a 1.4 $\msun$
neutron star),  lasting for  $\lsim 10^8$ years.   But there  has been
recent  indications that the  low-mass binaries  may even  spend $\sim
10^{10}$  years  in  the  Roche-contact  phase  with  a  sub-Eddington
accretion rate~\cite{hans97}. For  wide binaries, however, the contact
phase may  last as little as  $10^7$ years.  We  have investigated the
cases    with     accretion    rates    of     $10^{-10}$~\dmdt    and
$10^{-9}$~\dmdt. With a higher accretion rate the material movement is
faster and therefore the `freezing-in' takes place earlier (see figure
[\ref{fb_evolv1}].  Moreover, the  equation [\ref{etdmdt}], used by us
to find the  crustal temperature for a given  rate of accretion, gives
temperatures   that   are  too   high   for   accretion  rates   above
$10^{-10}$~\dmdt.   The  neutrino cooling  is  likely  to prevent  the
temperature  from reaching such  large values  and therefore  for high
rates  of accretion  probably the  crustal temperature  would  reach a
maximum  saturation value.   In our  calculations we  assume  that the
temperature  that could  be attained  by accretion-induced  heating is
$\sim  10^{8.5}$~K  for an  accretion  rate  of $10^{-9}$~\dmdt.   For
reasons noted in  the previous chapter we are  unable to make detailed
calculations for $\mdot \gsim 10^{-9}$~\dmdt.
\een
\i High Mass X-ray Binaries -
\ben
\i Isolated  phase - This  phase is short  in binaries with  a massive
companion and may last as little as ten thousand years.
\i Wind phase  - This phase is also relatively  short (compared to the
low-mass systems),  lasting not more than $10^7$  years (equivalent to
the main-sequence life time of the massive star), with accretion rates
ranging from $10^{-14}$~\dmdt to $10^{-10}$~\dmdt.
\i Roche-contact phase - A  rapid phase of Roche-lobe overflow follows
the wind phase. The rate of mass shedding by the companion could be as
high as one tenth  of a solar mass per year, of  which a tiny fraction
is  actually  accreted  by  the  neutron star  (the  maximum  rate  of
acceptance being presumably equal to the Eddington rate). The duration
of this phase is $\lsim 10^4$ years.
\een
\een
We must  mention here  about the Globular  Cluster binaries  though we
have  not performed any  explicit calculations  for such  systems. The
duration and the rates of  mass accretion vary widely depending on how
the mass transfer  phase ends in such binaries.  The mass transfer may
end through  binary disruption,  by the neutron  star spiraling  in or
through a slow turn off.

\subsection{thermal behaviour}

The  thermal behaviour  of  isolated pulsars  and  of those  accreting
material  from  their  binary   companions  differs  from  each  other
significantly (for details  see section[\ref{sthermal}]. Since crustal
temperature is one of the major factors controlling field evolution we
need to consider the thermal  behaviour with some care.  For a neutron
star  that is  a member  of a  binary, the  thermal behaviour  will be
similar  to that  of an  isolated neutron  star before  the  advent of
actual mass transfer. Therefore through the isolated and the propeller
phase the neutron  star cools like an isolated  one. In particular, in
the  low-mass systems  the  duration  of the  isolated  phase and  the
propeller phase could be quite  long and therefore is rather important
in affecting the  subsequent evolution of the surface  field.  We have
shown in the  previous chapter that a phase of  field evolution in the
isolated      phase     modifies     the      subsequent     evolution
considerably.  Therefore, it  is necessary  to take  into  account the
proper cooling history of the  neutron star prior to the establishment
of  contact  with  its  binary  companion.  As  mentioned  before,  we
investigate both the cases - with standard and accelerated cooling. Of
course, the isolated phase itself is  of importance only in case of an
initial crustal  field configuration, the  topic of our  discussion in
the   present  chapter.   For  initial   currents  supported   in  the
superconducting  core,  the crustal  physics  of  the  star is  mostly
irrelevant as the field has not  yet been expelled to the crust. When,
in the course of binary evolution, the neutron star actually starts to
accrete mass - the  thermal behaviour changes from that characteristic
of an isolated  phase. The crustal temperature then  settles down to a
steady value  determined by  the accretion rate  as given  in equation
[\ref{etdmdt}].

\subsection{crustal physics}

We have seen in chapter  [\ref{cmn}] that the field evolution stops as
the field  {\em freezes  in} when about  10\% of the  original crustal
mass  is accreted.  This happens  due  to the  fact that  by then  the
current  loops   reach  the  regions  of   extremely  high  electrical
conductivity. The  mass of  the crust of  a 1.4 $\msun$  neutron star,
with our adopted equation of  state, is $\sim 0.044 \msun$. Therefore,
in the present investigation we  stop our evolutionary code when $\sim
0.01 - 0.04~\msun$ is accreted. From our results (presented in section
[\ref{sobs-results}])  it is  evident that  the field  evidently shows
signs of `freezing in' when we  stop our calculation. Of course, as we
have mentioned earlier, to  achieve millisecond period about 0.1 solar
mass  requires  to  be  accreted.  But  once  the  field  attains  its
`frozen-in'  value subsequent  accretion  does not  affect  it in  any
way. Therefore, the  final spin-period is determined by  this value of
the surface field in accordance with equation [\ref{ePeq}].

\section{results and discussions}
\label{sobs-results}

\subsection{solitary neutron stars}

\bef
\begin{center}{\mbox{\epsfig{file=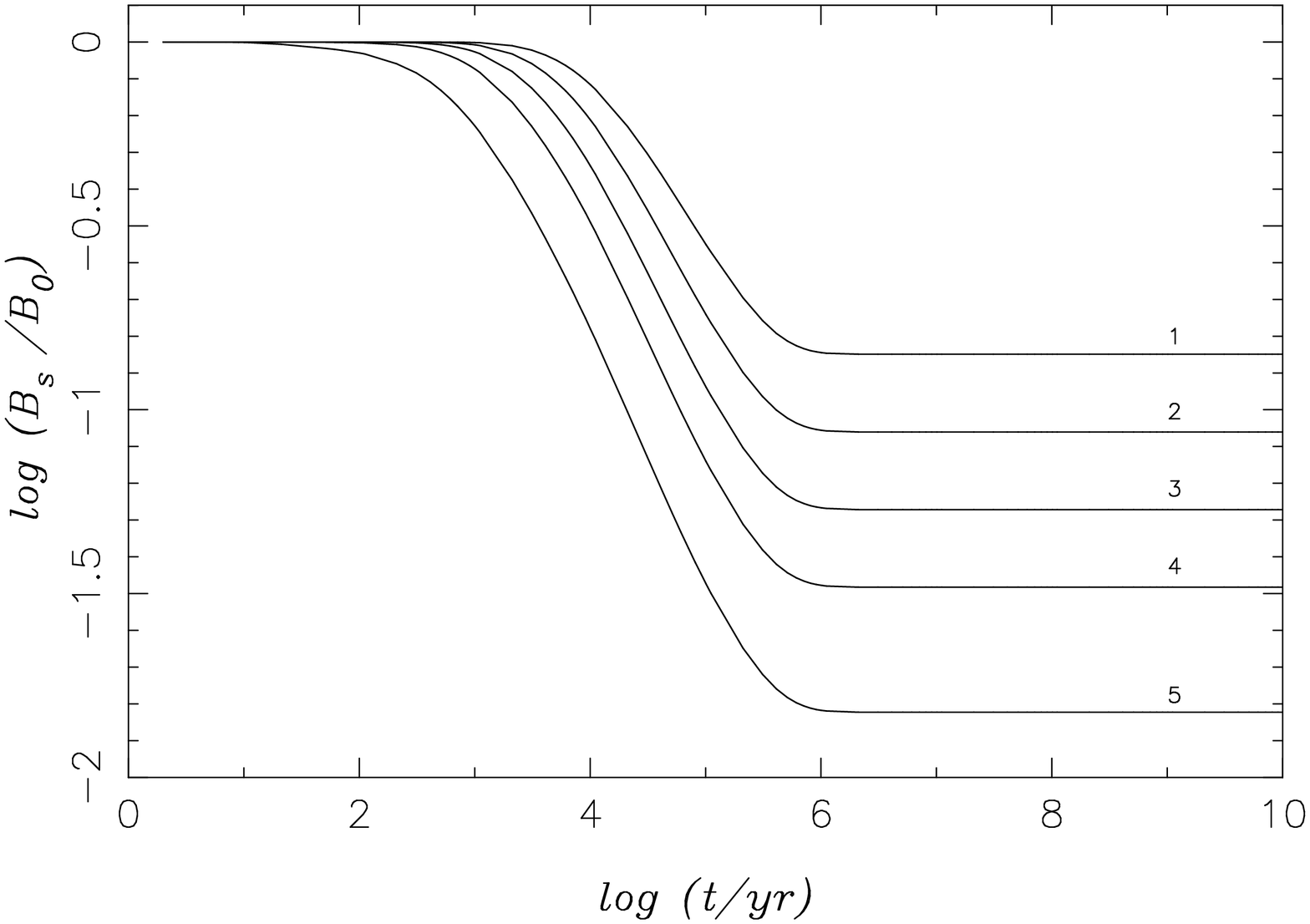,width=235pt,angle=-90}}}\end{center}
\caption[ohmic diffusion of the surface field Ia]{The evolution of the
surface  magnetic  field  due  to  pure  diffusion.   Curves  1  to  5
correspond  to densities of  $10^{13}, 10^{12.5},  10^{12}, 10^{11.5},
10^{11}~\gcc$ respectively, at which the $g$-profiles are centred. All
curves correspond to $Q$ = 0. Standard cooling has been assumed here.}
\label{fisolated_std_q1}
\eef

\bef
\begin{center}{\mbox{\epsfig{file=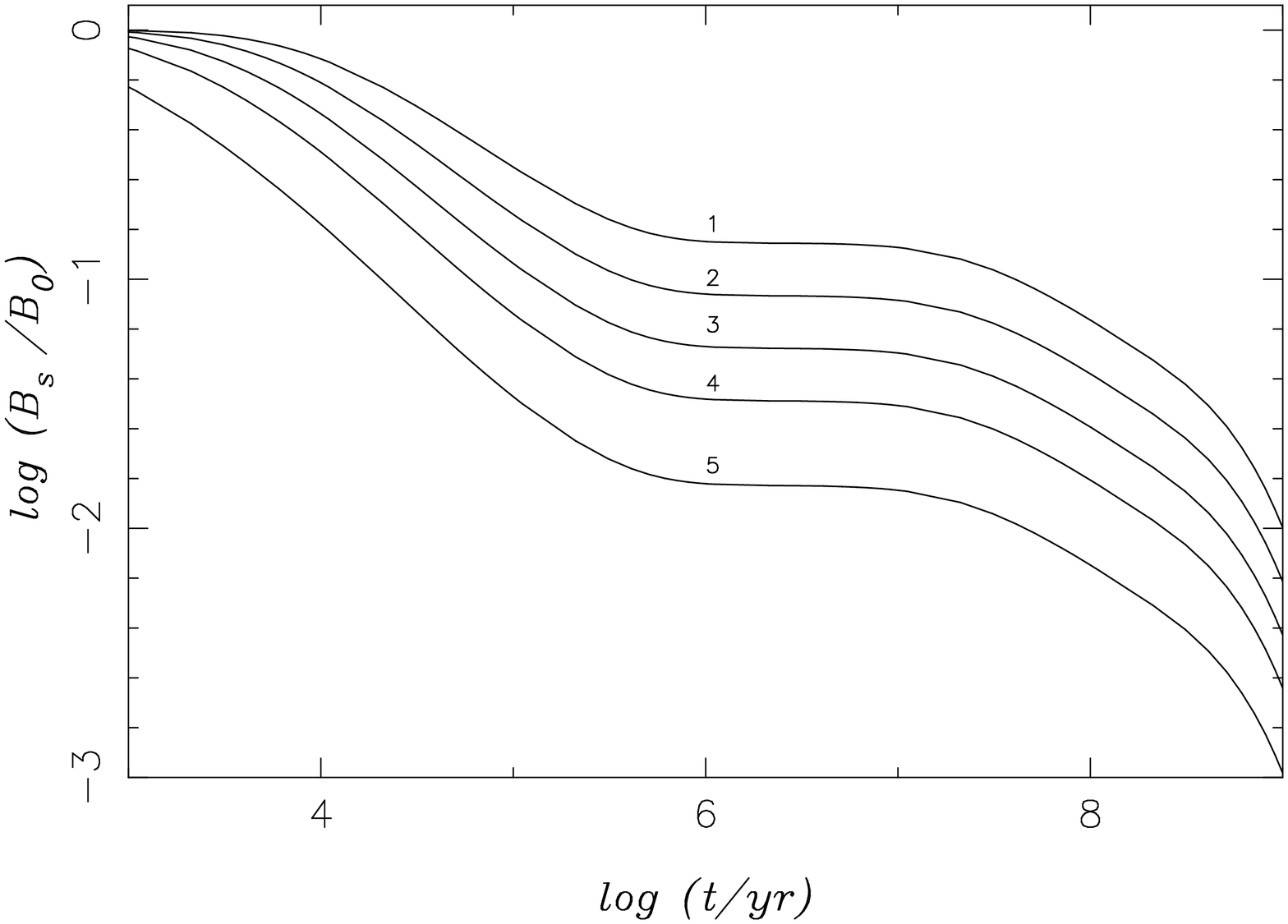,width=235pt,angle=-90}}}\end{center}
\caption[ohmic  diffusion  of the  surface  field  Ib]{Same as  figure
[\ref{fisolated_std_q1}] but with $Q$ = 0.05.}
\label{fisolated_std_q2}
\eef

\bef
\begin{center}{\mbox{\epsfig{file=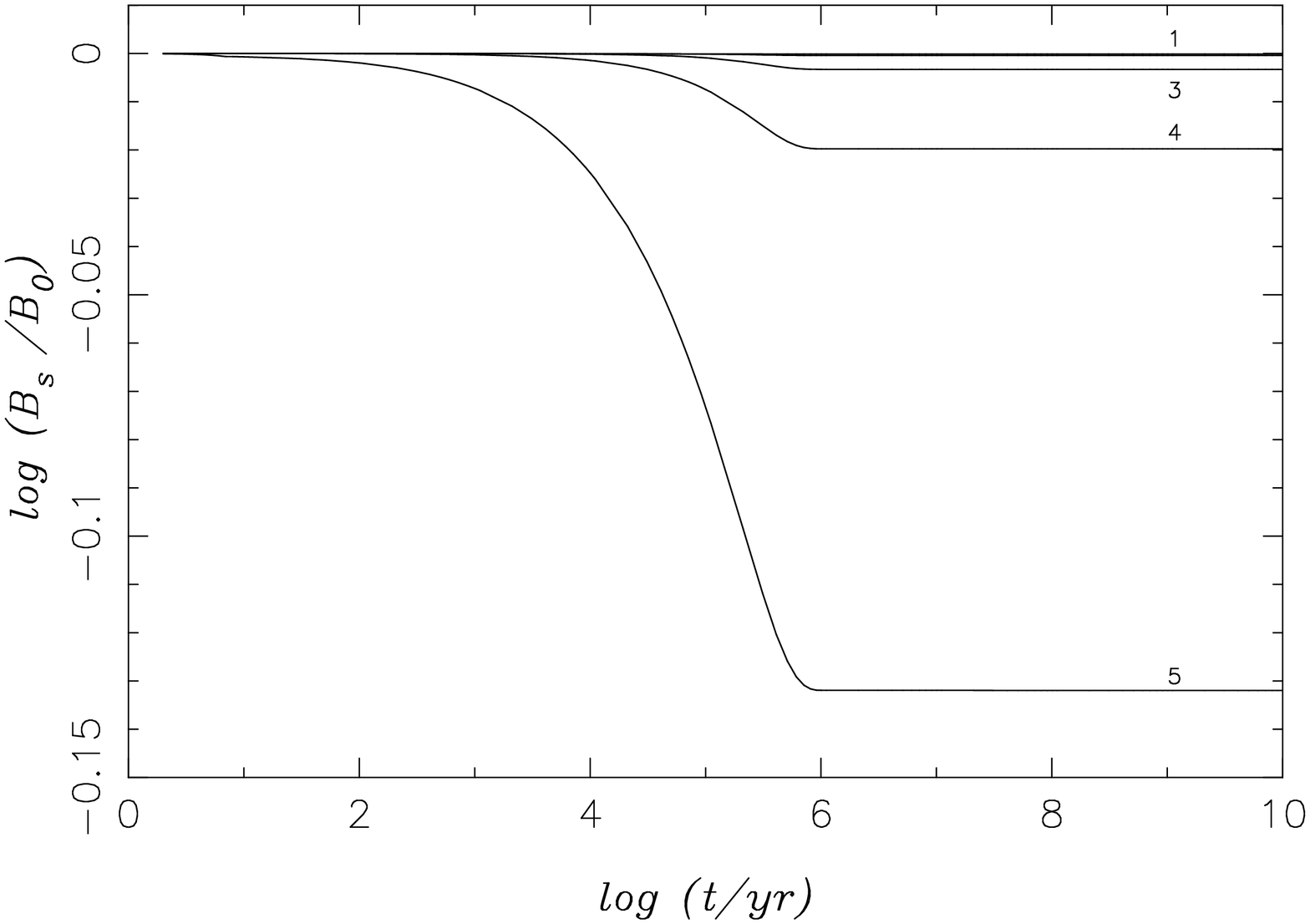,width=235pt,angle=-90}}}\end{center}
\caption[ohmic  diffusion of  the  surface field  IIa]{Same as  figure
[\ref{fisolated_std_q1}] but with accelerated cooling assumed.}
\label{fisolated_acc_q1}
\eef

\bef
\begin{center}{\mbox{\epsfig{file=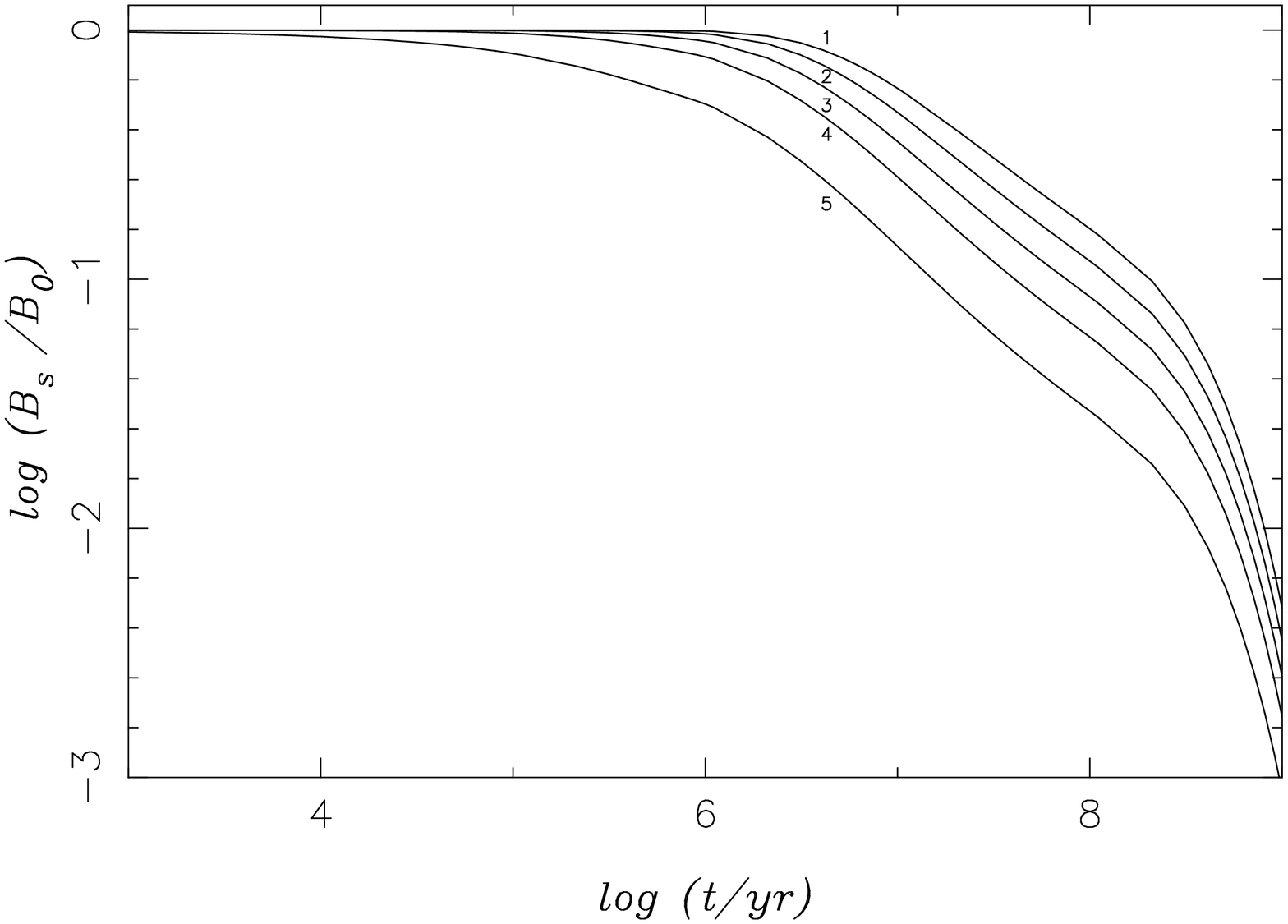,width=235pt,angle=-90}}}\end{center}
\caption[ohmic  diffusion of  the  surface field  IIb]{Same as  figure
[\ref{fisolated_acc_q1}] but with $Q$ = 0.1.}
\label{fisolated_acc_q2}
\eef

Figure  [\ref{fisolated_std_q1}] shows  the evolution  of  the surface
field  in  an isolated  neutron  star due  to  pure  ohmic decay,  for
different  densities  at which  the  initial  current distribution  is
concentrated. We have  seen it before too, that  the lower the density
of  current  concentration  the  more   rapid  is  the  decay  of  the
field. This figure  refers to a situation where  the impurity strength
has  been assumed to  be zero.  In figure  [\ref{fisolated_std_q2}] we
plot similar  curves for  the evolution of  the surface field  with an
assumed $Q$ = 0.05.  In this case the final field values are too small
to be  consistent with the  indication from the  statistical analyses.
We find  that the maximum  value of $Q$  that can be allowed  is about
0.01 if  allow for a maximum  decay by two orders  of magnitude within
solitary pulsar  active lifetime. In  figures [\ref{fisolated_acc_q1}]
and [\ref{fisolated_acc_q2}]  we plot  similar curves assuming  a fast
cooling process for the thermal behaviour of the star. It is seen that
the decay  is evidently  much less  than in the  case of  the standard
cooling. But  even in this case  the maximum $Q$  value permissible is
about 0.05. This we shall see  later that is in contradiction with the
requirements of a spin-down induced flux expulsion model.

\subsection{high mass binaries}

Figures [\ref{fcrust-hmxb_vRs1}]  to [\ref{fcrust-hmxb_vRf5}] show the
evolution of the  surface field in the HMXBs;  for different values of
the  density   at  which  the  initial  current   profile  is  centred
at.   Figures  [\ref{fcrust-hmxb_vRs1}]   to  [\ref{fcrust-hmxb_vRs5}]
correspond  to  computations with  standard  cooling  in the  isolated
phase,      whereas      figures      [\ref{fcrust-hmxb_vRf1}]      to
[\ref{fcrust-hmxb_vRf5}]   correspond    to   calculations   with   an
accelerated cooling in  the isolated phase. For all  of these cases we
have assumed  a Roche-contact phase  lasting for $10^4$ years  with an
uniform rate of accretion  of $10^{-8}$~\dmdt. Since the Roche-contact
phase is extremely short-lived, the  actual field decay takes place in
the wind phase. In fact, the decay attained in the Roche-contact phase
is   insignificant    and   is   not   quite    visible   in   figures
[\ref{fcrust-hmxb_vRs1}]   to   [\ref{fcrust-hmxb_vRf5}].  In   figure
[\ref{fcrust-hmxb_vRs1a}]  we  show  an  expanded  version  of  figure
[\ref{fcrust-hmxb_vRs1}] highlighting this final phase.

\bef
\begin{center}{\mbox{\epsfig{file=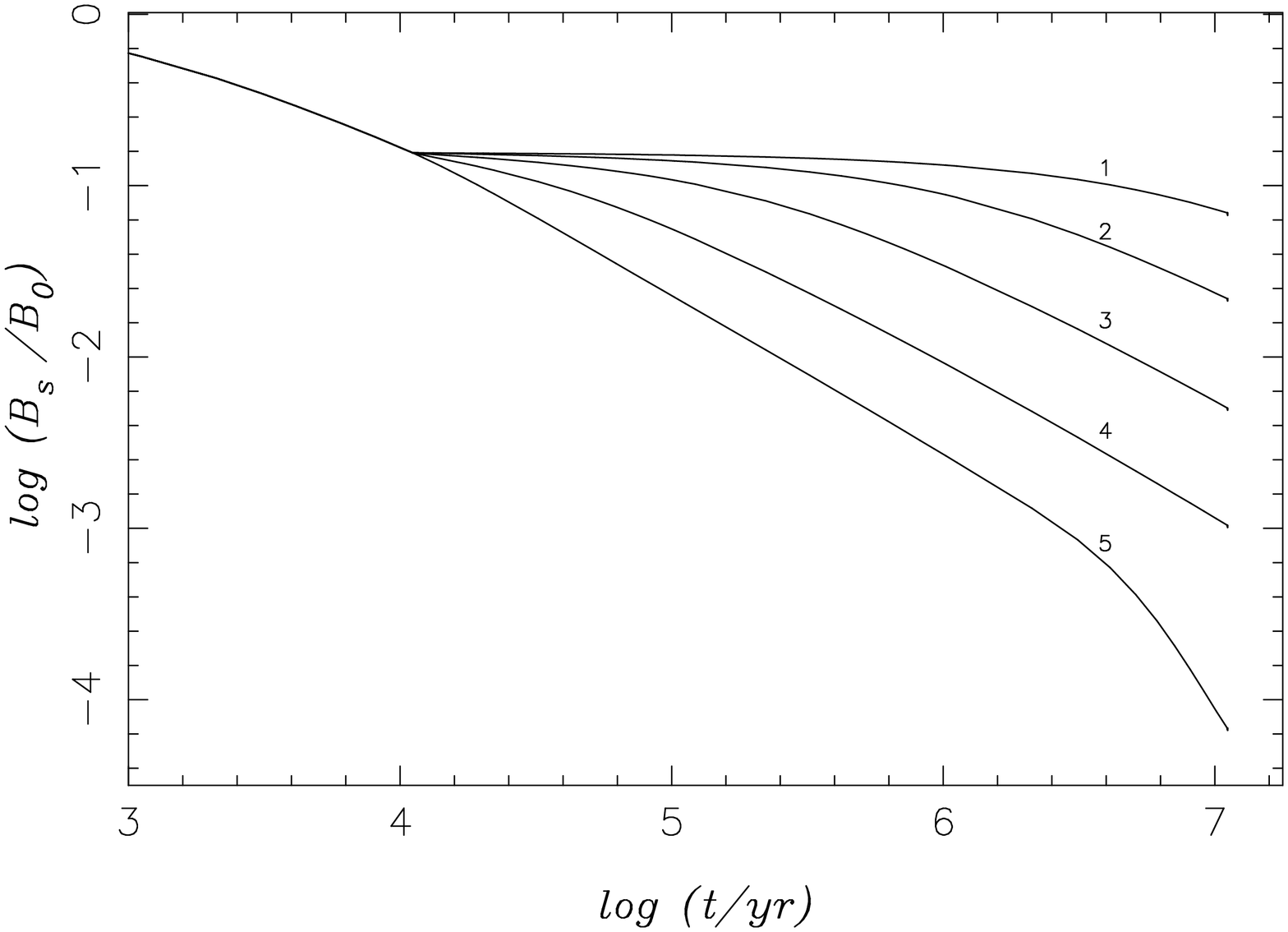,width=235pt,angle=-90}}}\end{center}
\caption[field  evolution  in   HMXBs  Ia]{Evolution  of  the  surface
magnetic field  in HMXBs for four  values of wind  accretion rate with
slow  cooling behaviour  in  the isolated  phase.  The curves  1 to  5
correspond  to  $\mdot   =  10^{-14},  10^{-13},  10^{-12},  10^{-11},
10^{-10}$~\dmdt.   All  curves  correspond   to  an   initial  current
configuration centred at  $\rho = 10^{11} \gcc$, an  accretion rate of
$\mdot = 10^{-8}$~\dmdt in the Roche-contact phase and $Q$ = 0.0.}
\label{fcrust-hmxb_vRs1}
\eef

\bef
\begin{center}{\mbox{\epsfig{file=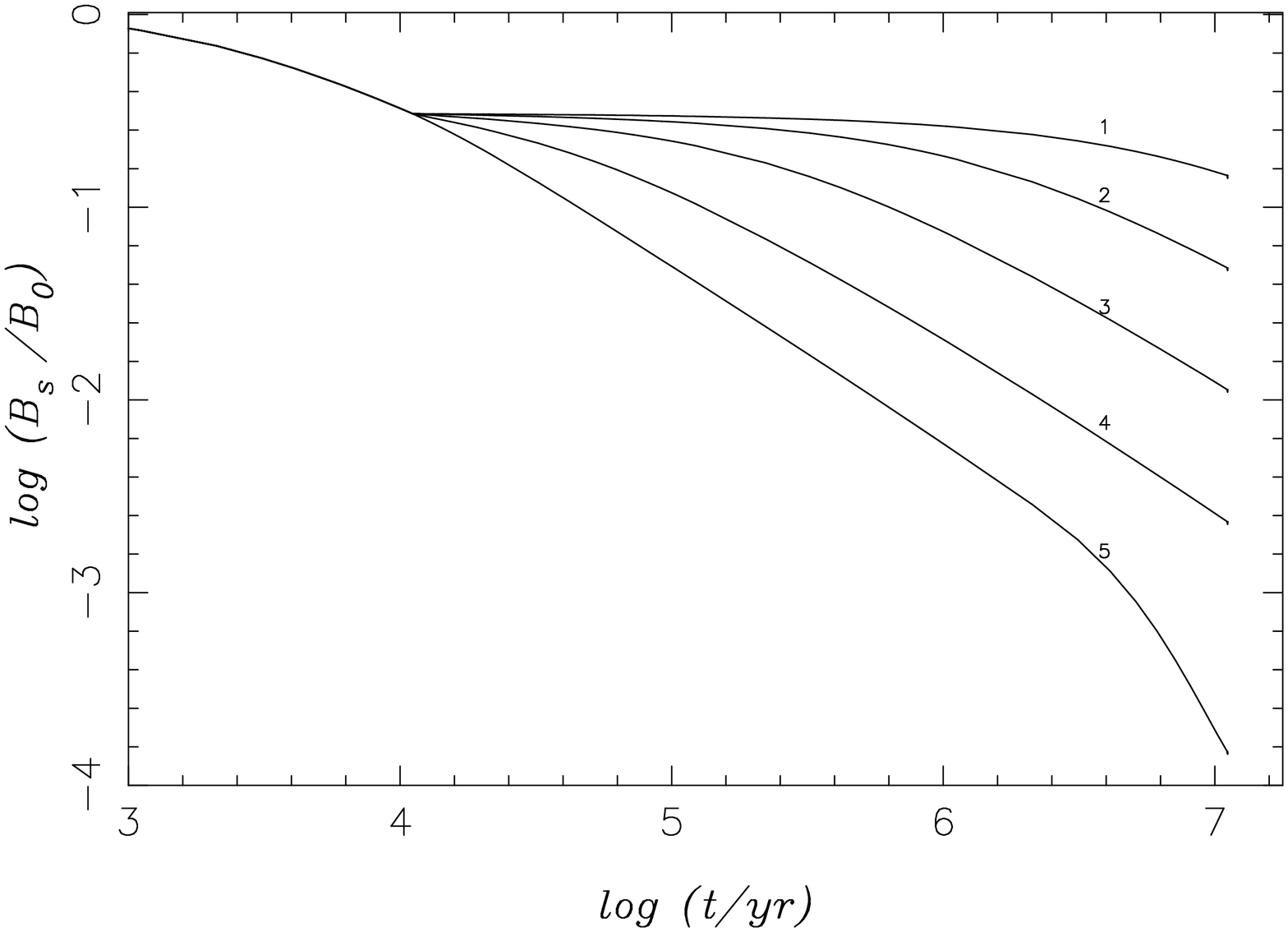,width=235pt,angle=-90}}}\end{center}
\caption[field    evolution    in    HMXBs    Ib]{Same    as    figure
[\ref{fcrust-hmxb_vRs1}] with an initial current configuration centred
at $\rho = 10^{11.5} \gcc$. }
\label{fcrust-hmxb_vRs2}
\eef
     
\bef
\begin{center}{\mbox{\epsfig{file=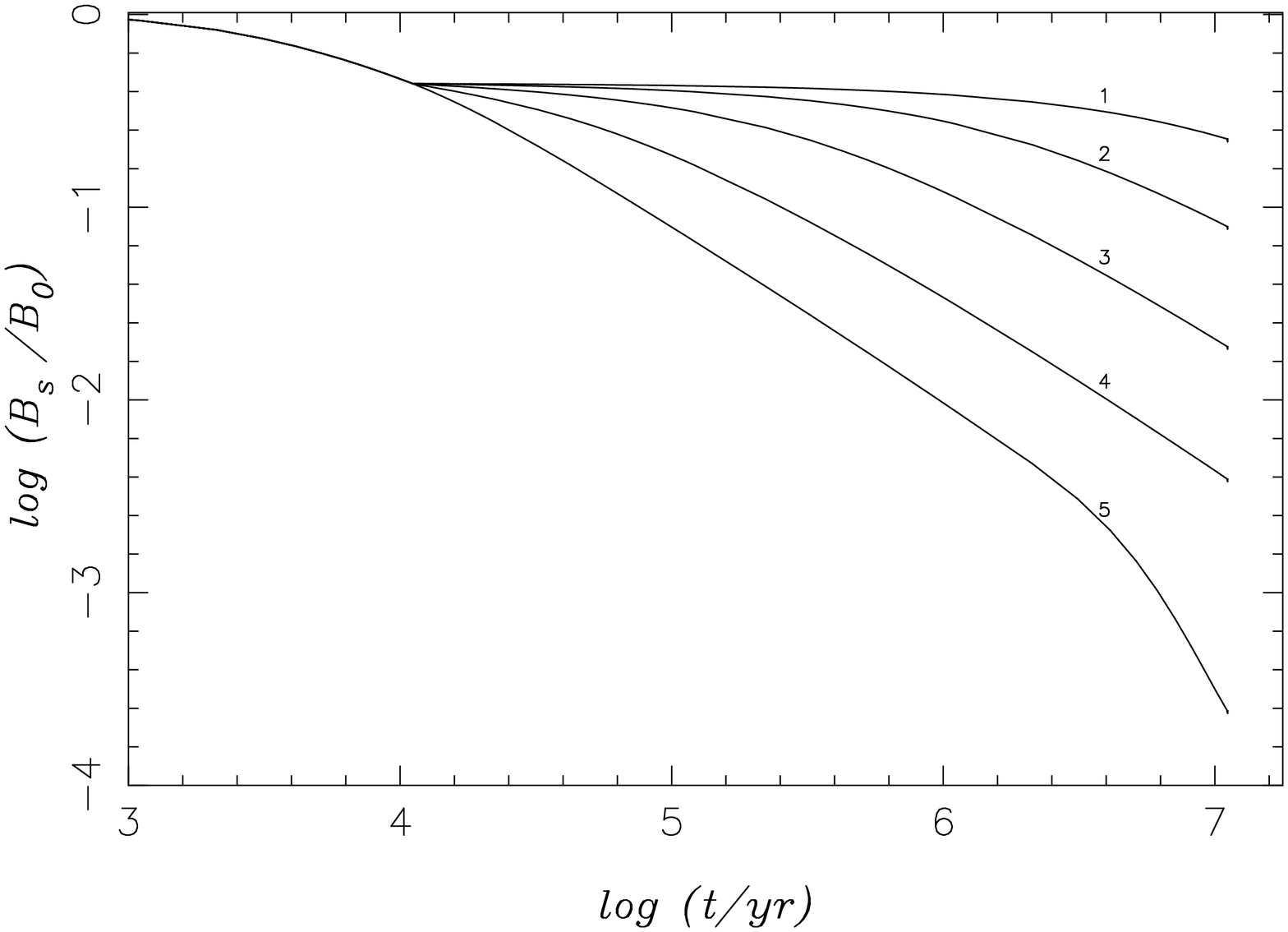,width=235pt,angle=-90}}}\end{center}
\caption[field    evolution    in    HMXBs    Ic]{Same    as    figure
[\ref{fcrust-hmxb_vRs1}] with an initial current configuration centred
at $\rho = 10^{12} \gcc$. }
\label{fcrust-hmxb_vRs3}
\eef
     
\bef
\begin{center}{\mbox{\epsfig{file=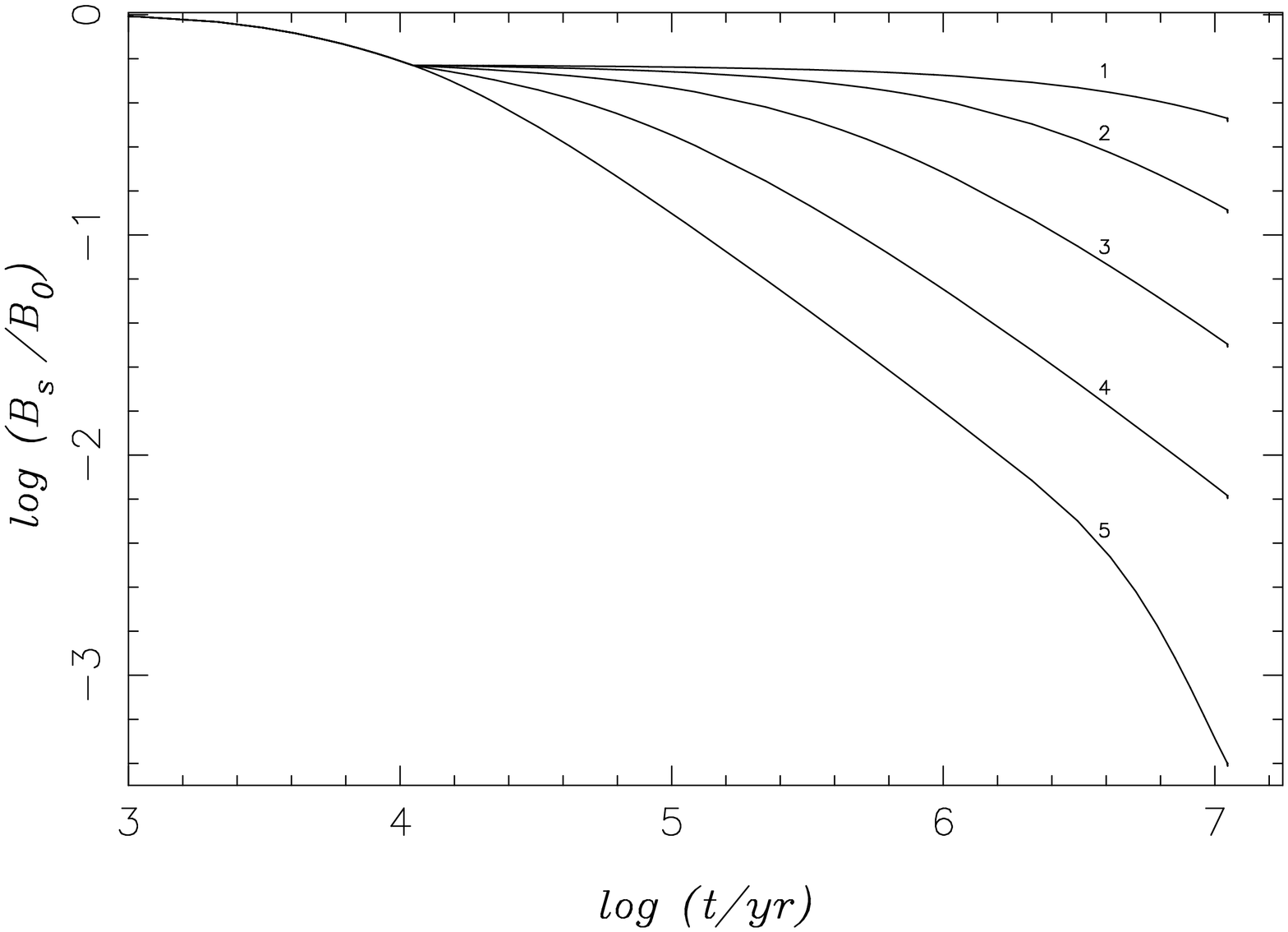,width=235pt,angle=-90}}}\end{center}
\caption[field    evolution    in    HMXBs    Id]{Same    as    figure
[\ref{fcrust-hmxb_vRs1}] with an initial current configuration centred
at $\rho = 10^{12.5} \gcc$. }
\label{fcrust-hmxb_vRs4}
\eef
     
\bef
\begin{center}{\mbox{\epsfig{file=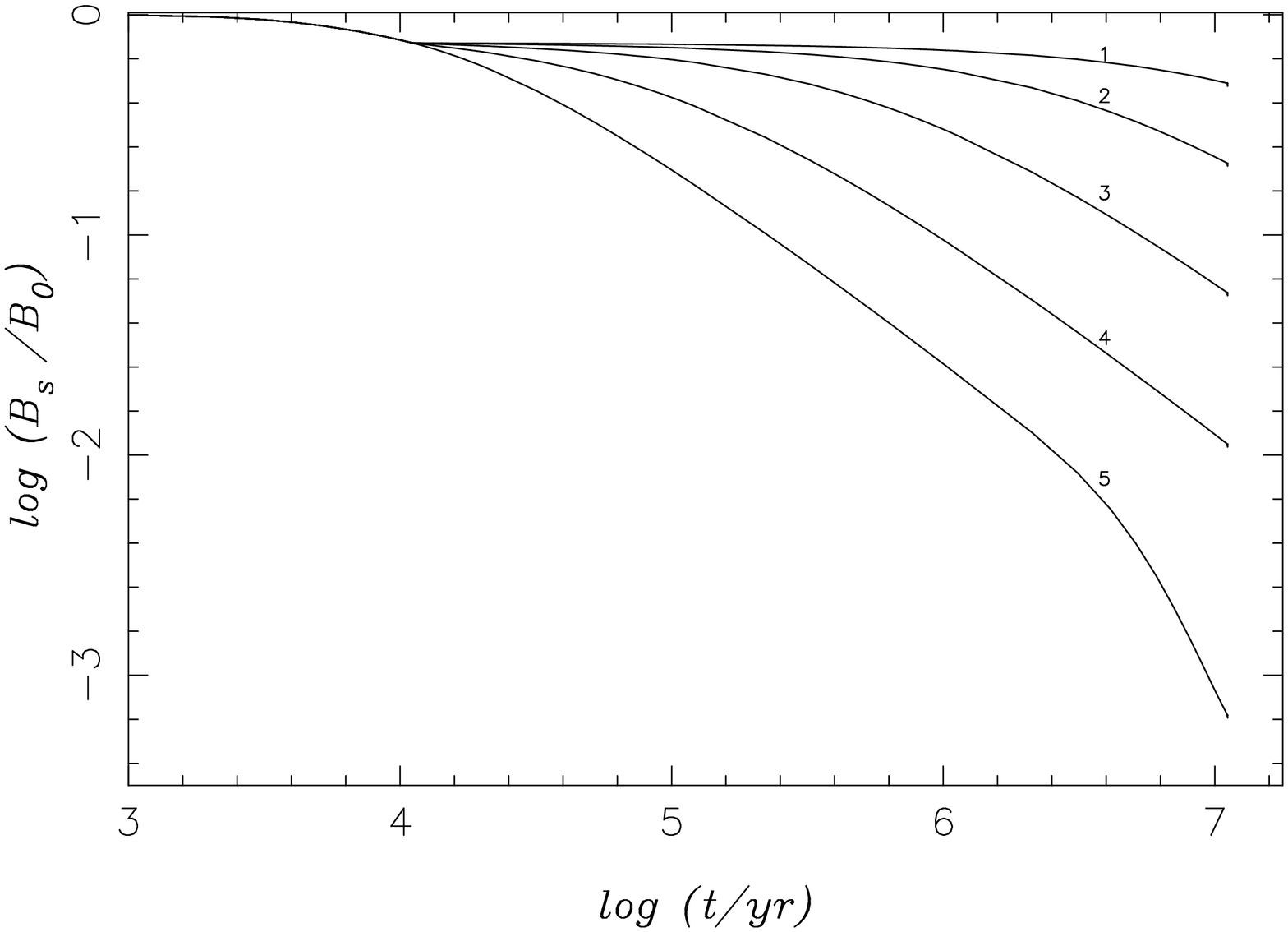,width=235pt,angle=-90}}}\end{center}
\caption[field    evolution    in    HMXBs    Ie]{Same    as    figure
[\ref{fcrust-hmxb_vRs1}] with an initial current configuration centred
at $\rho = 10^{13} \gcc$. }
\label{fcrust-hmxb_vRs5}
\eef
     
From  these  figures  it  is  evident  that  if  the  initial  current
distribution is  located at high densities the  objects from high-mass
systems will  retain fairly large  final fields. Even for  low density
current distributions,  if the duration of  the wind phase  is not too
long - again high-field objects are produced.  We expect these objects
to show  up as recycled pulsars  with relatively high  fields and long
periods like PSR  B1913+16 or PSR B1534+12. On the  other hand, if the
wind phase  lasts for about $10^7$  years, it is possible  to obtain a
significant  field  decay  for  higher  rates  of  accretion  in  that
phase.  But as the  total mass  and hence  the total  angular momentum
accreted is not sufficient to spin  the star up to very short periods,
these systems probably would not produce millisecond pulsars. In other
words  these so-called  `recycled' pulsars  would have  small magnetic
fields with relatively long spin-periods  and therefore may not at all
be active  as pulsars. We make  an estimate of the  actual spin-up for
objects processed in high-mass systems to check this fact.

The spin-up  of a neutron star, in  a binary system, is  caused by the
angular momentum brought in  by the accreted matter. In magnetospheric
accretion  matter  accretes  with  angular momentum  specific  to  the
Alfv\'{e}n radius. Therefore, the total angular momentum brought in by
accretion is
\beq
J_{\rm accreted} = \delta M R_A V_A
\eeq
where $\delta M$  is the total mass accreted. $R_A$  and $V_A$ are the
Alfv\'{e}n  radius and  Keplerian  velocity at  that  radius given  by
equations [\ref{ealfrad}] and [\ref{ealfvel}]. The final period of the
neutron star then is
\beq
P_{\rm final} = 2 \pi \frac{I_{\rm ns}}{J_{\rm accreted}},
\eeq
where $I_{\rm  ns}$ is the moment  of inertia of the  neutron star. In
figure [\ref{fhmxb_spinup}] we have indicated the possible location of
the  `recycled'  pulsars  originating  in  HMXBs.  We  find  that  for
sufficiently low  field strengths ($B \lsim  10^9$~Gauss) the recycled
pulsars indeed will be beyond  the death line. And active pulsars with
slightly   higher  fields   will   lie  very   close   to  the   death
line.  Comparing this  figure  with the  field-period  diagram of  the
observed  pulsars (figure  [\ref{fbp_pulsar}]) one  does  find pulsars
like PSR  0655+64 to  fall in this  category. The cases  of high-field
pulsars well above death line are also quite evident from this figure.

\bef
\begin{center}{\mbox{\epsfig{file=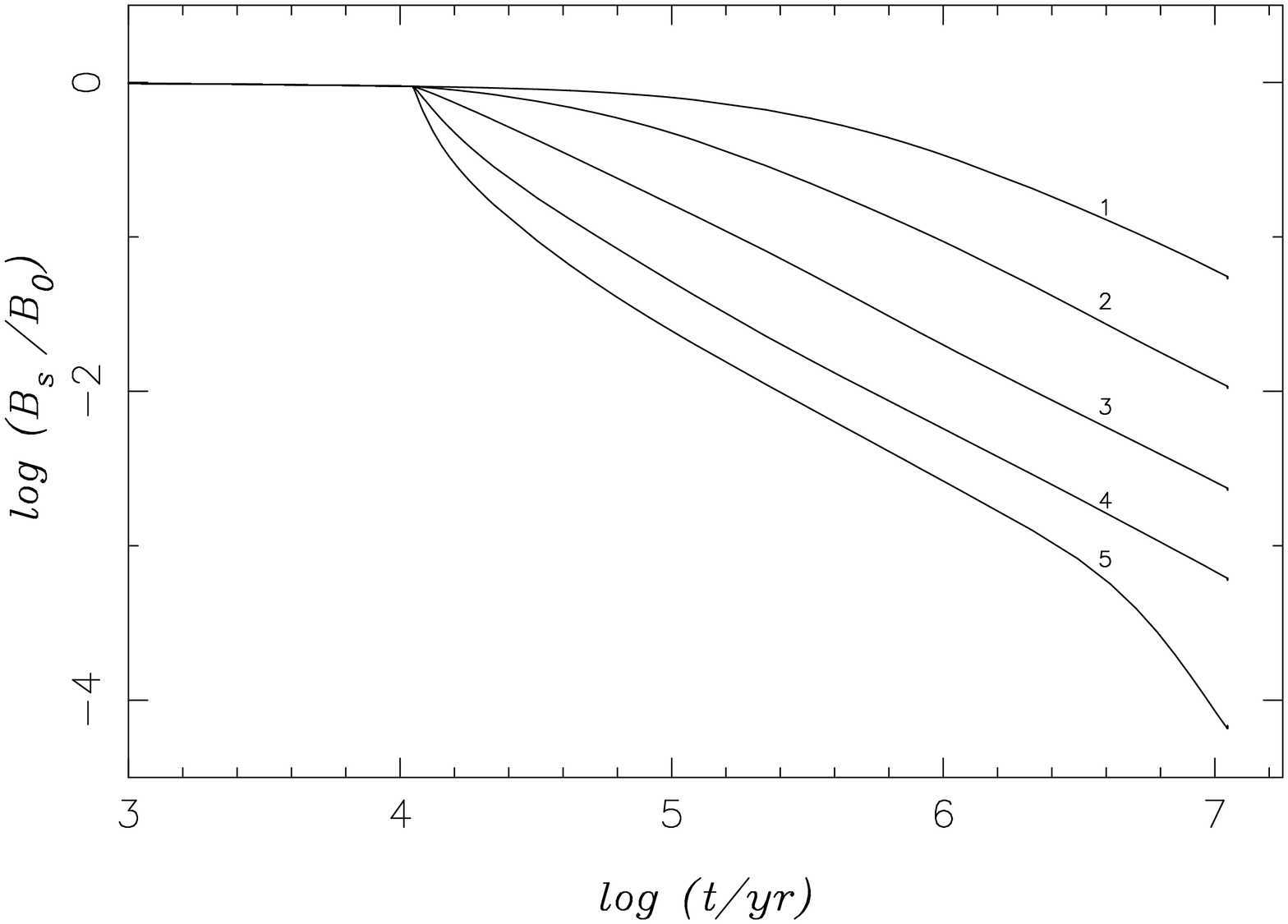,width=235pt,angle=-90}}}\end{center}
\caption[field  evolution  in  HMXBs  IIa]{Evolution  of  the  surface
magnetic field  in HMXBs for four  values of wind  accretion rate with
accelerated cooling behaviour in the isolated phase. The curves 1 to 5
correspond  to  $\mdot   =  10^{-14},  10^{-13},  10^{-12},  10^{-11},
10^{-10}$~\dmdt.   All  curves  correspond   to  an   initial  current
configuration centred at  $\rho = 10^{11} \gcc$, an  accretion rate of
$\mdot = 10^{-8}$~\dmdt in the Roche-contact phase and $Q$ = 0.0.}
\label{fcrust-hmxb_vRf1}
\eef

\bef
\begin{center}{\mbox{\epsfig{file=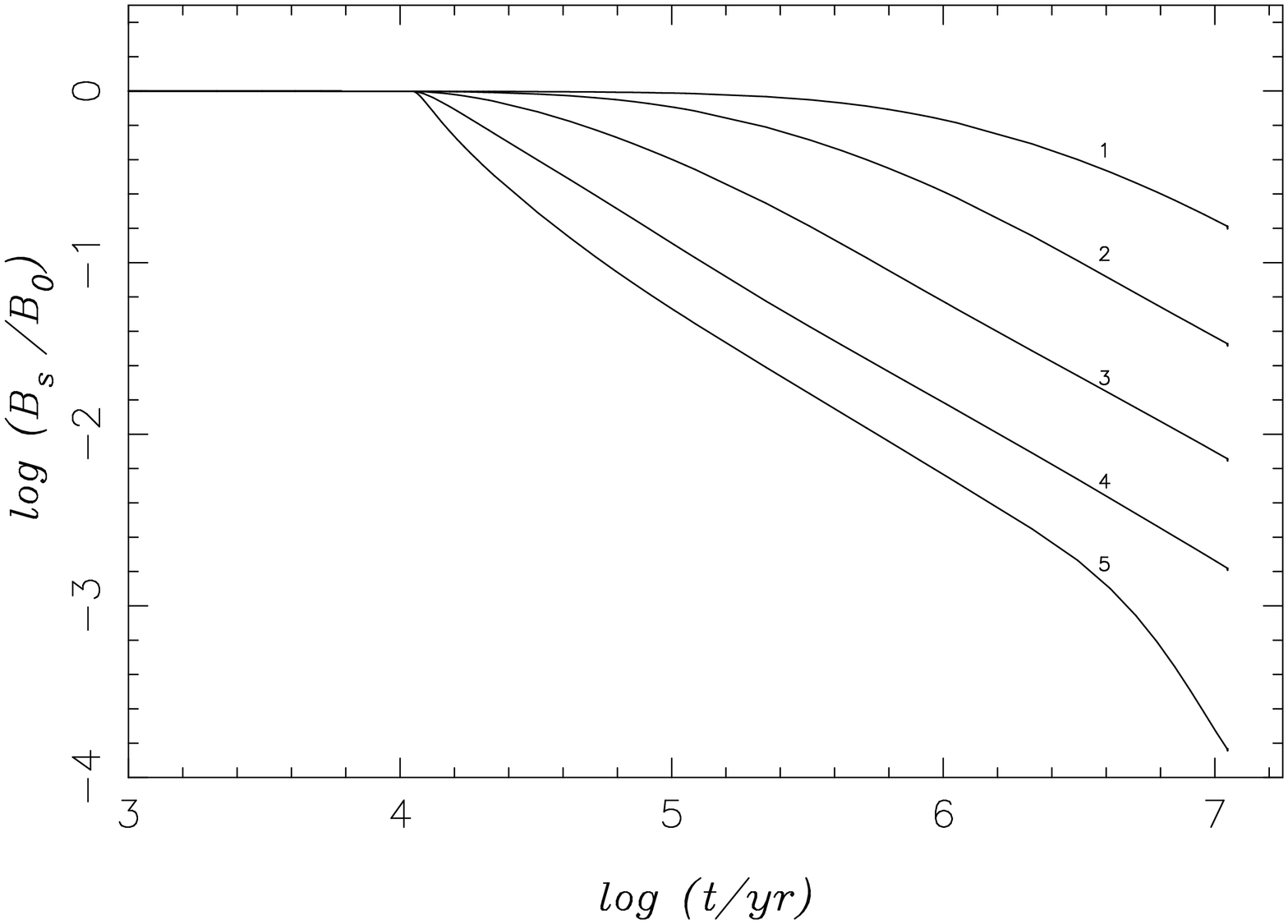,width=235pt,angle=-90}}}\end{center}
\caption[field    evolution    in    HMXBs   IIb]{Same    as    figure
[\ref{fcrust-hmxb_vRf1}] with an initial current configuration centred
at $\rho = 10^{11.5} \gcc$. }
\label{fcrust-hmxb_vRf2}
\eef
     
\bef
\begin{center}{\mbox{\epsfig{file=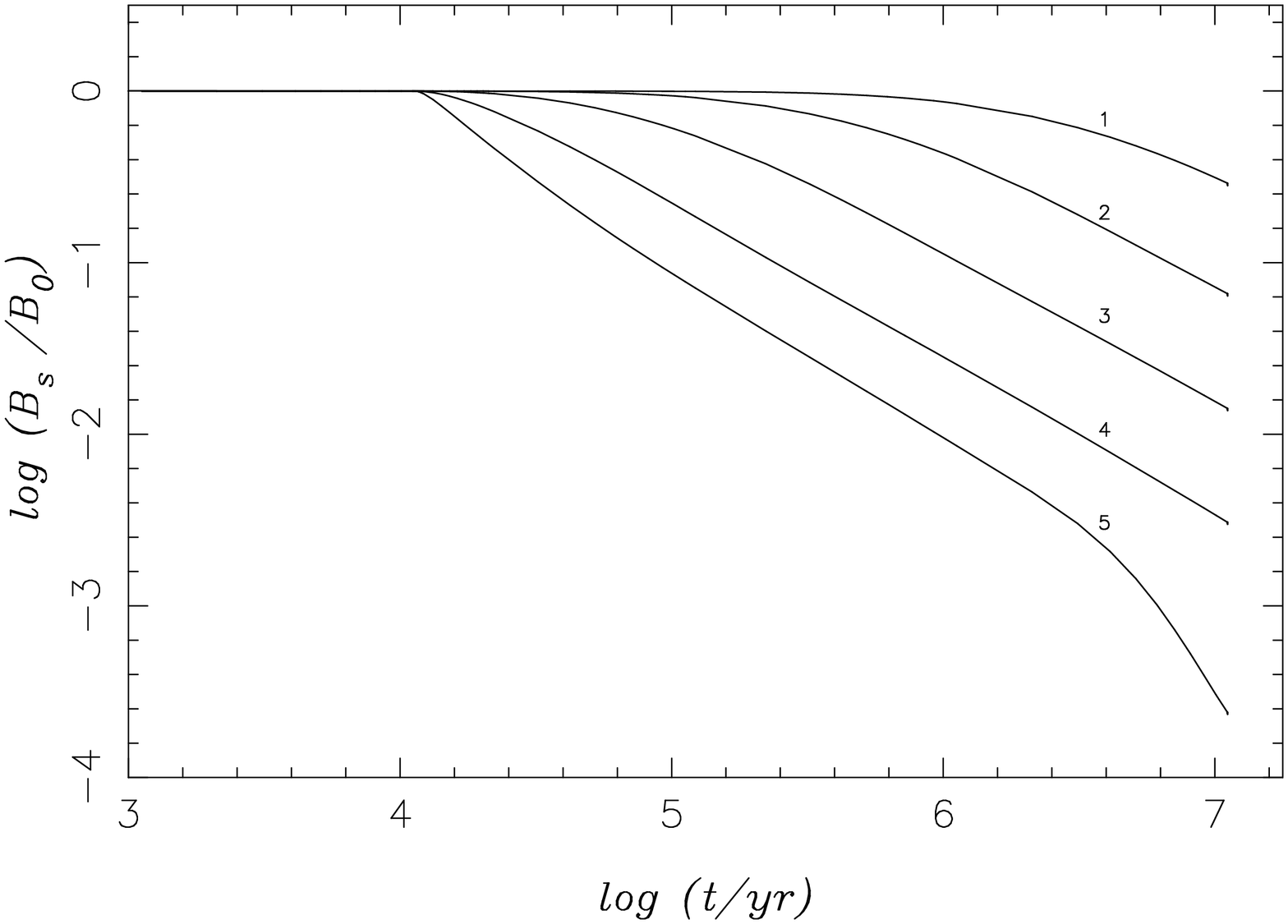,width=235pt,angle=-90}}}\end{center}
\caption[field    evolution    in    HMXBs   IIc]{Same    as    figure
[\ref{fcrust-hmxb_vRf1}] with an initial current configuration centred
at $\rho = 10^{12} \gcc$. }
\label{fcrust-hmxb_vRf3}
\eef
     
\bef
\begin{center}{\mbox{\epsfig{file=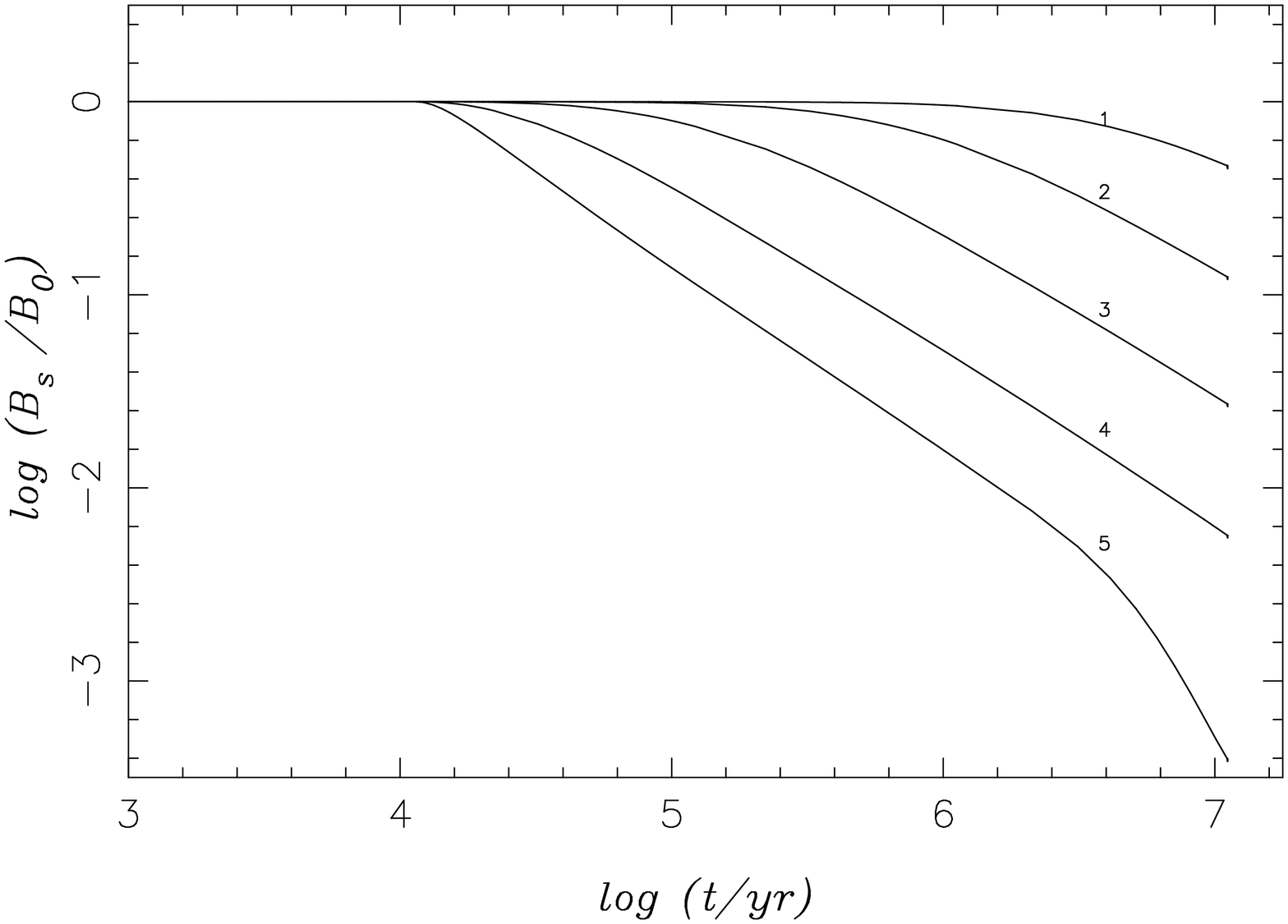,width=235pt,angle=-90}}}\end{center}
\caption[field    evolution    in    HMXBs   IId]{Same    as    figure
[\ref{fcrust-hmxb_vRf1}] with an initial current configuration centred
at $\rho = 10^{12.5} \gcc$. }
\label{fcrust-hmxb_vRf4}
\eef
     
\bef
\begin{center}{\mbox{\epsfig{file=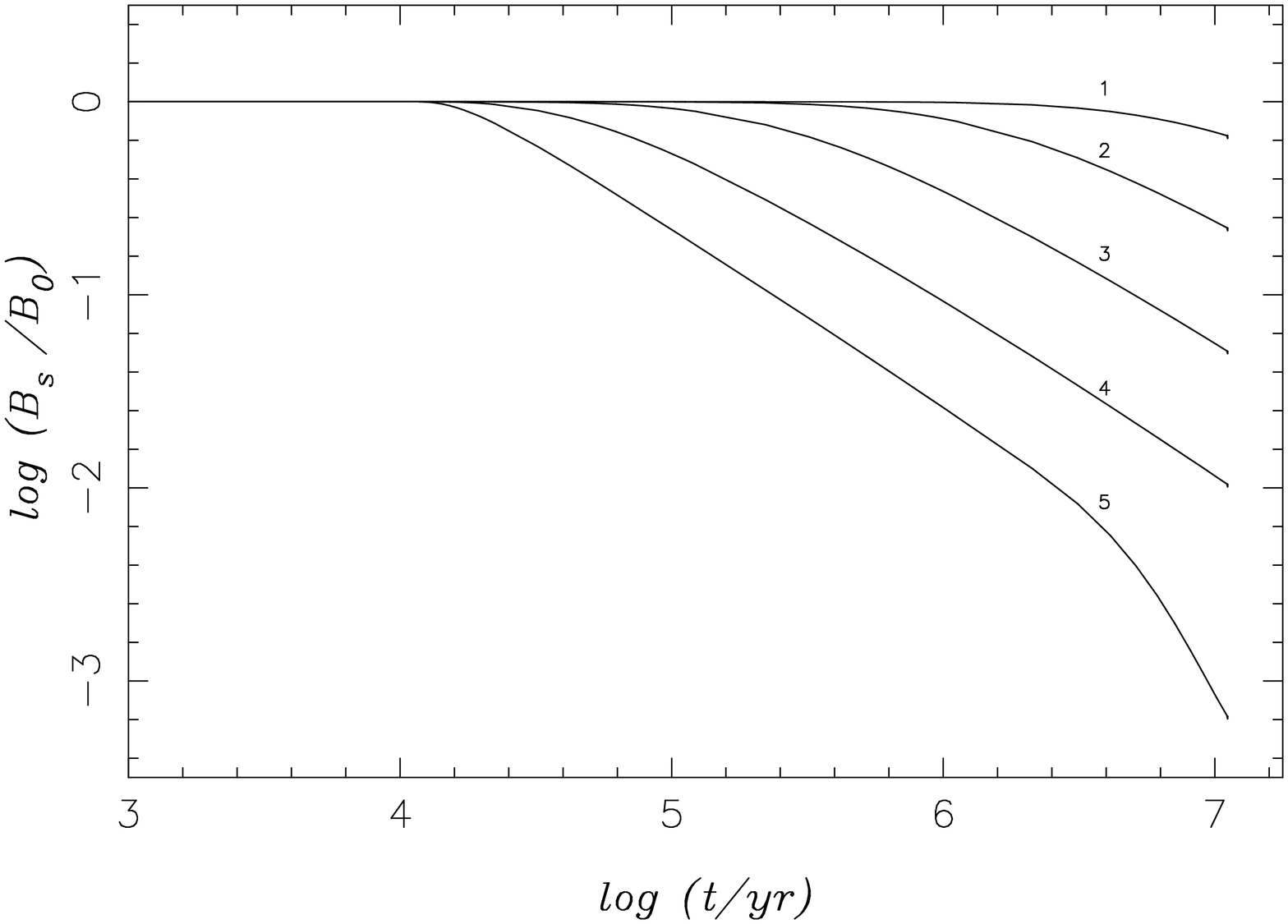,width=235pt,angle=-90}}}\end{center}
\caption[field    evolution    in    HMXBs   IIe]{Same    as    figure
[\ref{fcrust-hmxb_vRf1}] with an initial current configuration centred
at $\rho = 10^{13} \gcc$. }
\label{fcrust-hmxb_vRf5}
\eef

\bef
\begin{center}{\mbox{\epsfig{file=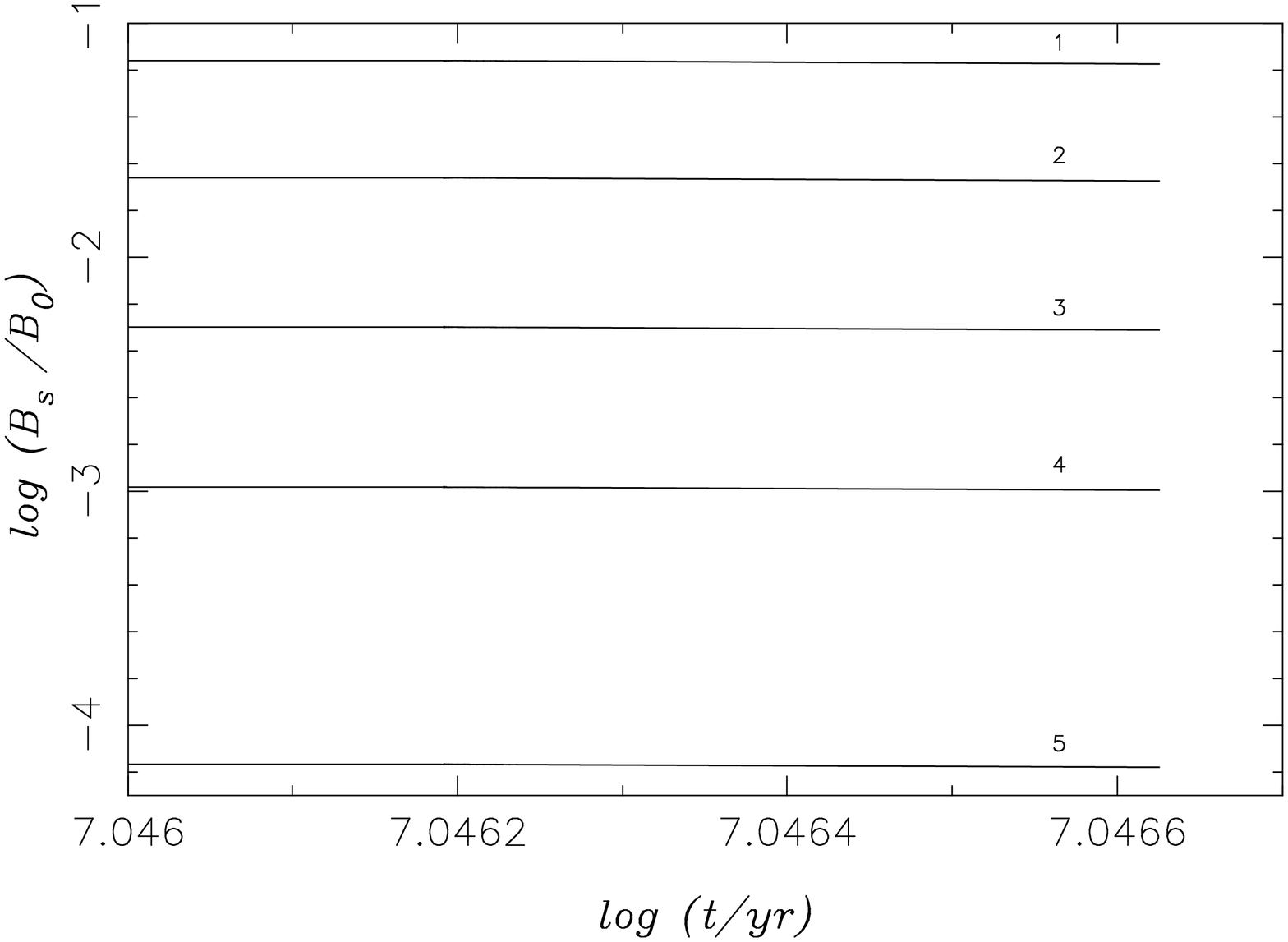,width=235pt,angle=-90}}}\end{center}
\caption[field    evolution    in    HMXBs   Ia2]{Same    as    figure
[\ref{fcrust-hmxb_vRs1}]   with    the   final   Roche-contact   phase
expanded. }
\label{fcrust-hmxb_vRs1a}
\eef
     
It should  also be  noted here  that even though  the nature  of field
evolution is  very different  in the isolated  phase for  standard and
accelerated cooling, the final surface  field values at the end of the
wind phase  are not very different. This  is due to the  fact that the
nature of the field  evolution is significantly influenced by previous
history.   We  have   already  seen   in  chapter   [5]   (see  figure
[\ref{fpre-acc}])  that subsequent decay  is slowed  down in  a system
with a history  of prior field decay than in  systems without any. The
decay in  the isolated phase is  less for accelerated  cooling, but in
the subsequent wind  accretion phase the field decays  more rapidly in
such systems  than in those  starting with standard cooling.  We shall
see  a more dramatic  manifestation of  this fact  in LMXBs  where the
duration of the isolated phase is much longer. 

\bef
\begin{center}{\mbox{\epsfig{file=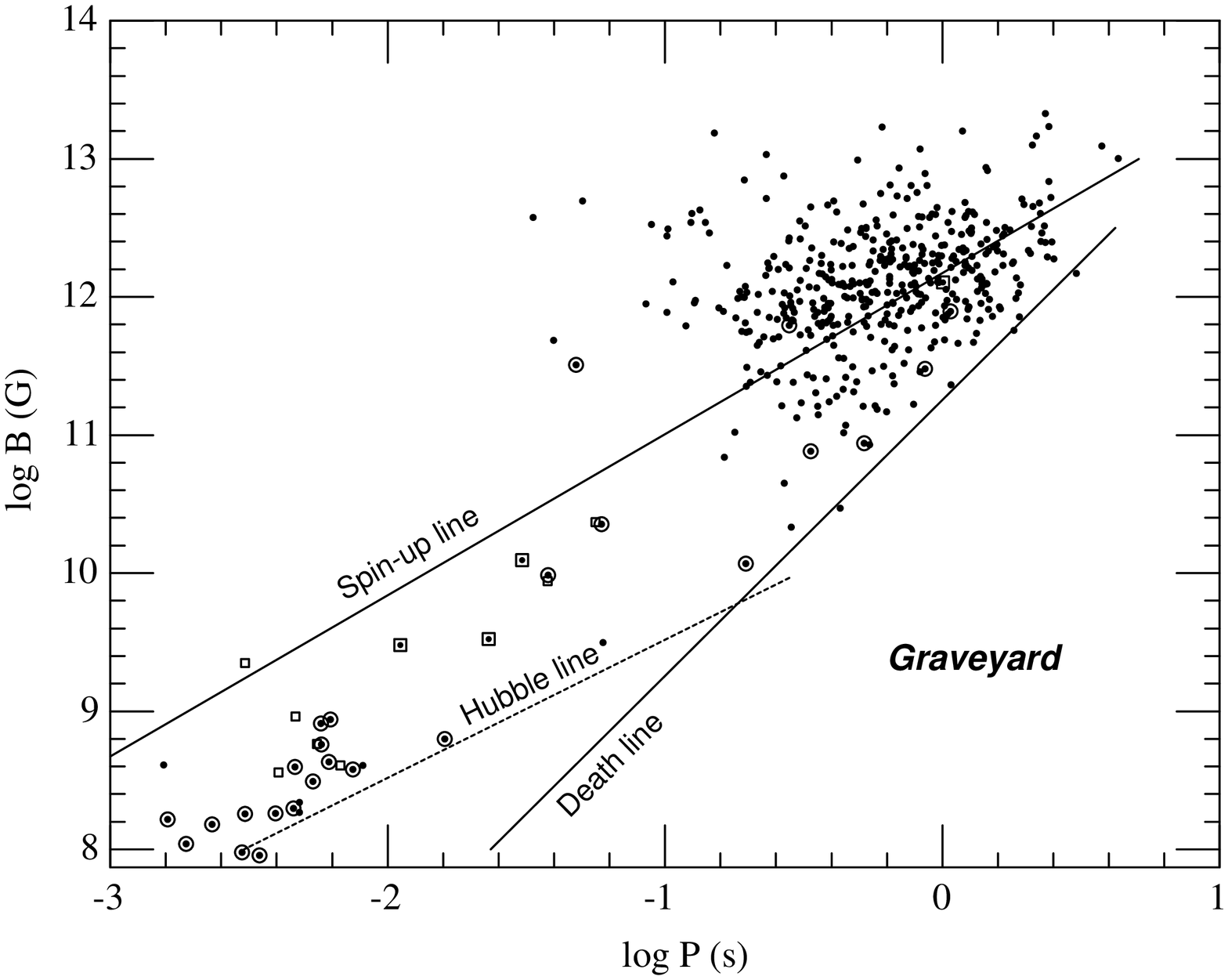 ,width=225pt,angle=90}}}\end{center}
\caption[radio  pulsar  field  period]{The  measured periods  and  the
derived surface dipole field  of observed pulsars.  The filled circles
and  the filled circles  within open  circles correspond  isolated and
binary pulsars in  the galactic disc. The open  squares and the filled
circles within open squares  correspond to isolated and binary pulsars
in globular clusters. On the right  hand side of the death line pulsar
activity stops. The spin-up line correspond to the minimum spin-period
that can  be achieved through  binary recycling assuming  an Eddington
accretion rate.  The Hubble line correspond to  the characteristic age
of the pulsars equal to the age of the universe.}
\label{fbp_pulsar}
\eef
     
\bef
\begin{center}{\mbox{\epsfig{file=hmxb_spinup.ps,width=225pt,angle=-90}}}\end{center}
\caption[HMXB  progenies]{The probable location  of HMXB  progenies in
the field-period  diagram. The dashed lines correspond  to the maximum
spin-up achievable,  the upper and  the lower lines being  for assumed
accreted masses of $10^{-4}$ and $10^{-3}$~\msun. The recycled pulsars
from HMXBs are expected to lie within the hatched region.}
\label{fhmxb_spinup}
\eef
     
\subsection{low mass binaries}

Figures               [\ref{fcrust_lmxb_whole_w1_vRs}]              to
[\ref{fcrust_lmxb_Roche_w2_vRf}]  show the  evolution  of the  surface
field  in  LMXBs.   In  figures  [\ref{fcrust_lmxb_whole_w1_vRs}]  and
[\ref{fcrust_lmxb_whole_w2_vRs}] we plot the complete evolution of the
surface field for  two values of the accretion rate  in the wind phase
and  five  values of  the  density  at  which currents  are  initially
concentrated, assuming standard cooling  in the isolated phase. But in
these  figures, the  wind  and Roche-contact  phases  are not  clearly
distinguishable.  Therefore  we have plotted the  expanded versions of
these   figures   to   highlight   the  individual   phases.   Figures
[\ref{fcrust_lmxb_wind_w1_vRs}]   and  [\ref{fcrust_lmxb_wind_w2_vRs}]
are        for       the       wind-phase        whereas       figures
[\ref{fcrust_lmxb_Roche_w1_vRs}]  and [\ref{fcrust_lmxb_Roche_w2_vRs}]
are  for the  Roche-contact  phase corresponding  to  those plots.  In
figures               [\ref{fcrust_lmxb_wind_w1_vRf}]               to
[\ref{fcrust_lmxb_Roche_w2_vRf}]  we  plot  the corresponding  figures
with an accelerated cooling in the isolated phase.

\bef
\begin{center}{\mbox{\epsfig{file=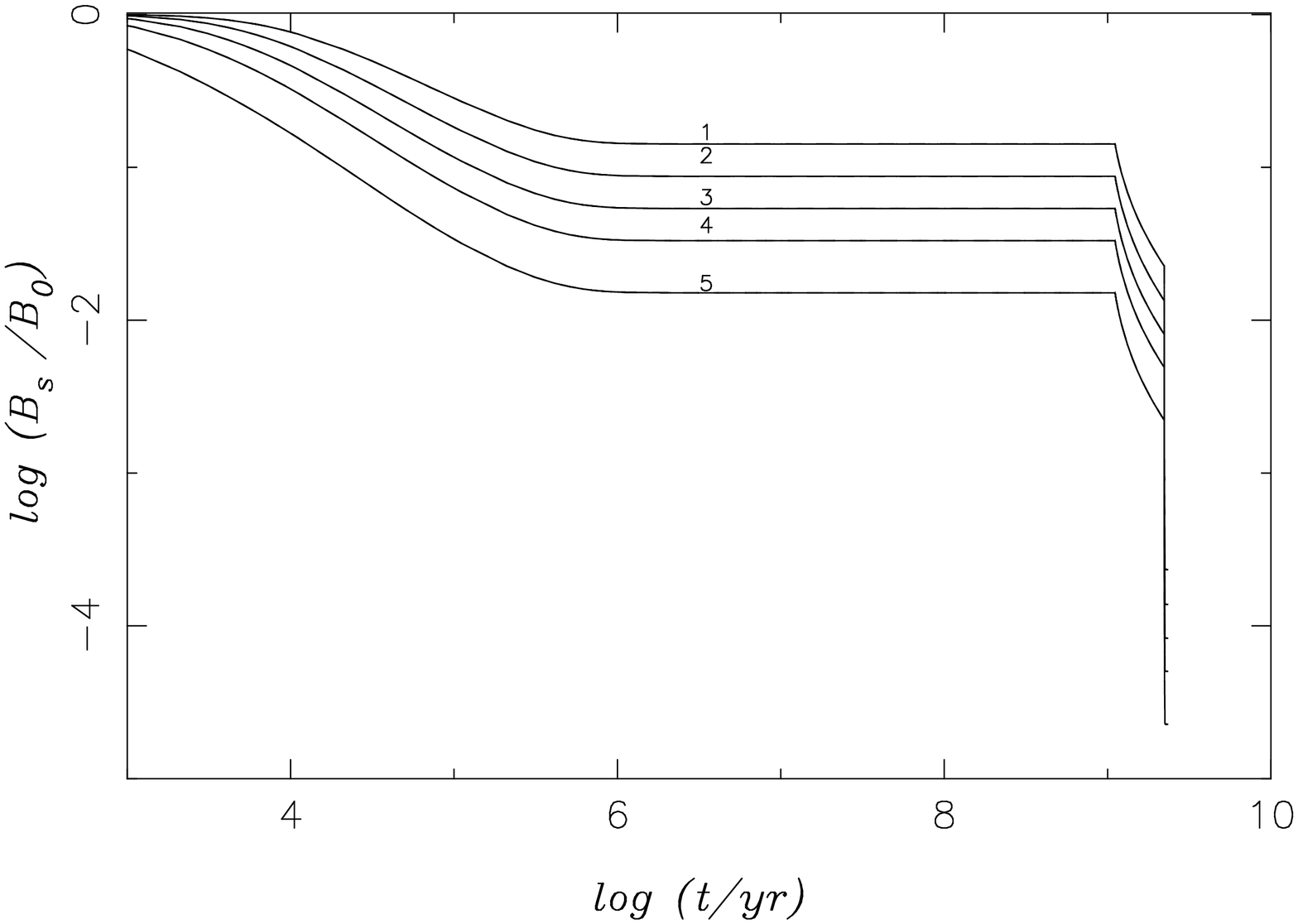,width=235pt,angle=-90}}}\end{center}
\caption[field  evolution  in   LMXBs  Ia]{Evolution  of  the  surface
magnetic  field in  LMXBs  with an  wind  accretion rate  of $\mdot  =
10^{-16}$~\dmdt, accretion rates  of $\mdot = 10^{-10}, 10^{-9}$~\dmdt
in  the Roche-contact  phase.  Curves  1 to  5  correspond to  initial
current configuration centred at  $\rho = 10^{11}, 10^{11.5}, 10^{12},
10^{12.5},  10^{13} \gcc$.  All  curves  correspond to  $Q$  = 0.0.  A
standard cooling has been assumed for the isolated phase here.}
\label{fcrust_lmxb_whole_w1_vRs}
\eef

\bef
\begin{center}{\mbox{\epsfig{file=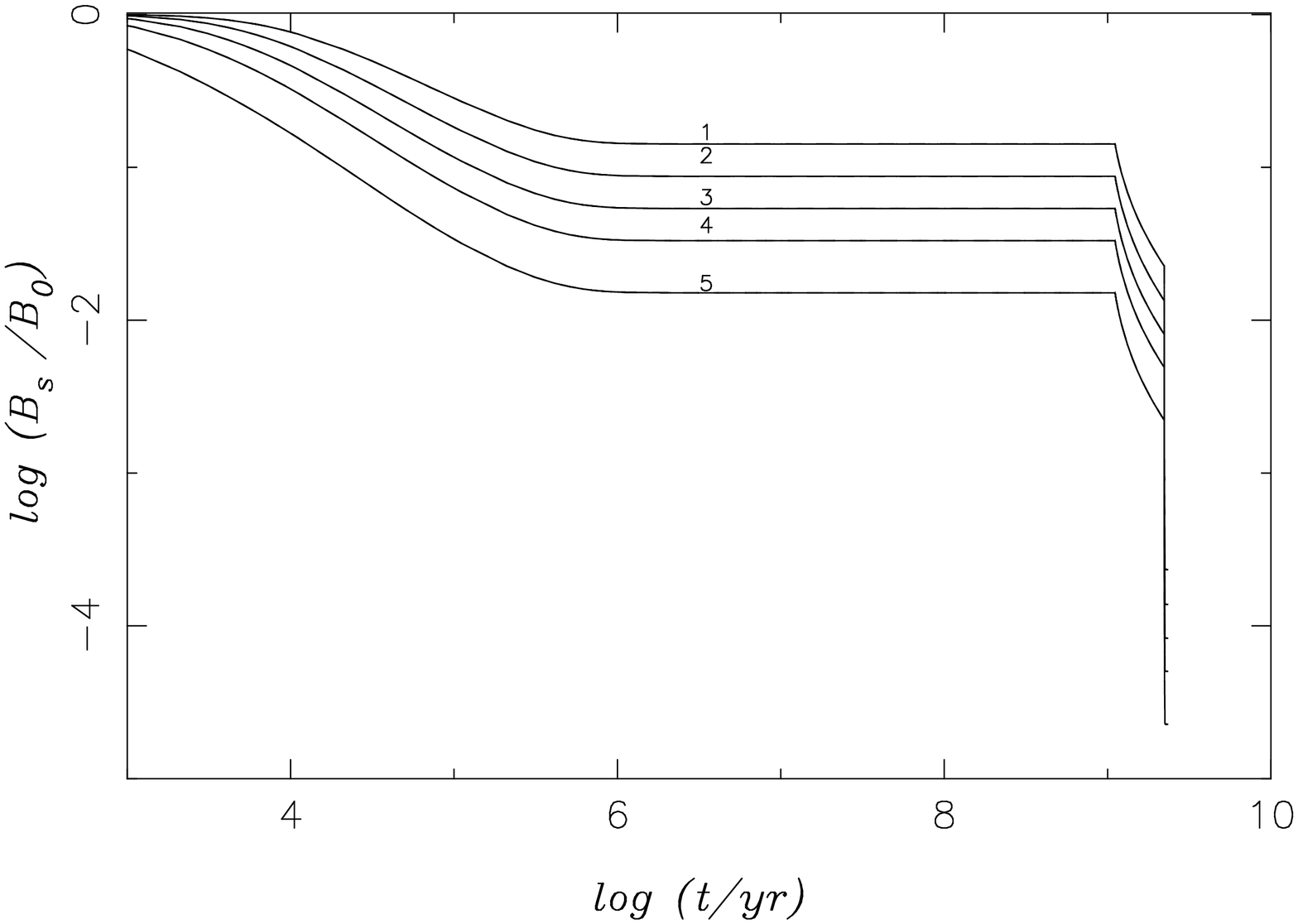,width=235pt,angle=-90}}}\end{center}
\caption[field    evolution    in    LMXBs   IIa]{Same    as    figure
[\ref{fcrust_lmxb_whole_w1_vRs}]  but with an  wind accretion  rate of
$\mdot = 10^{-14}$~\dmdt.}
\label{fcrust_lmxb_whole_w2_vRs}
\eef

\bef
\begin{center}{\mbox{\epsfig{file=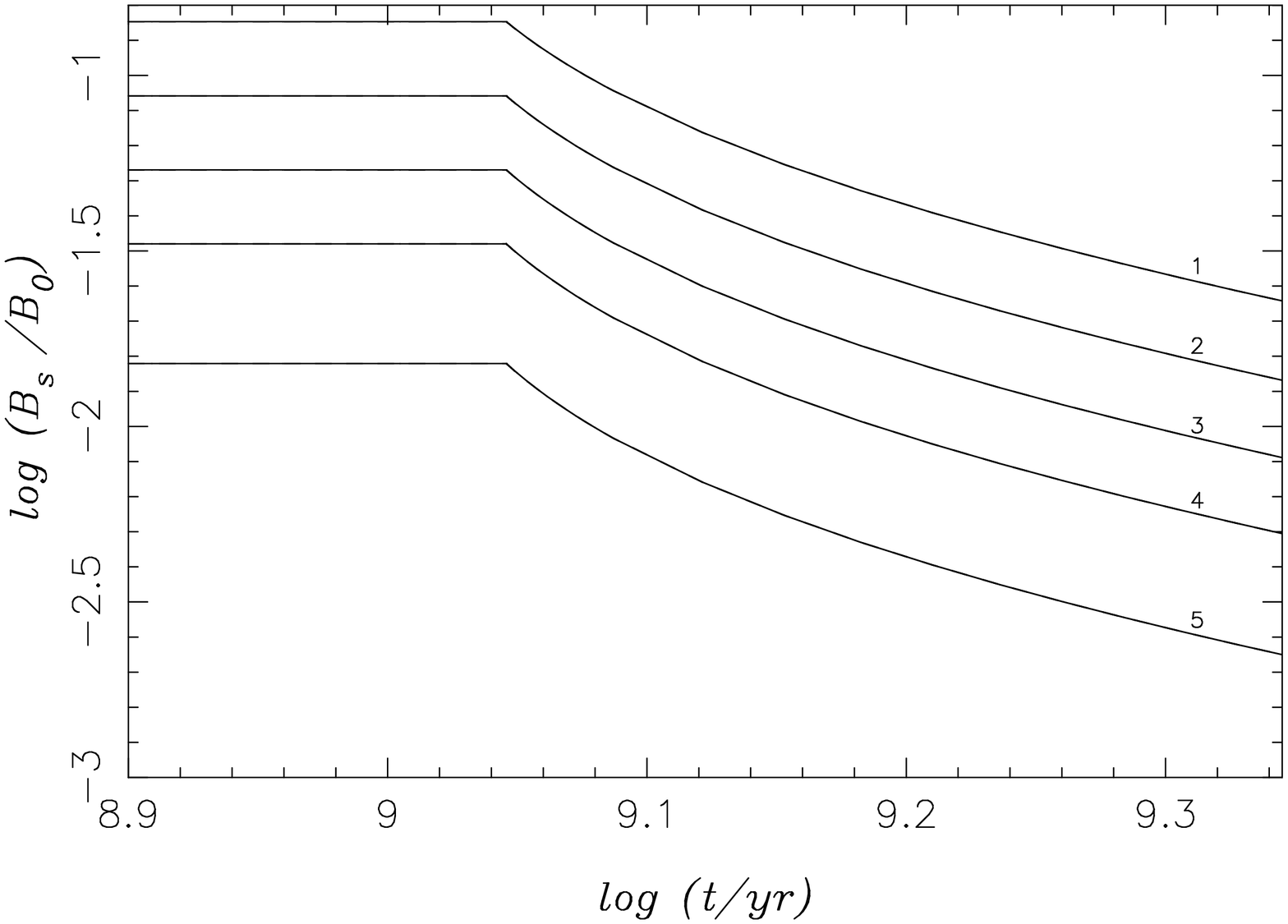,width=235pt,angle=-90}}}\end{center}
\caption[field    evolution    in    LMXBs    Ib]{Same    as    figure
[\ref{fcrust_lmxb_whole_w1_vRs}] with the wind phase expanded.}
\label{fcrust_lmxb_wind_w1_vRs}
\eef

\bef
\begin{center}{\mbox{\epsfig{file=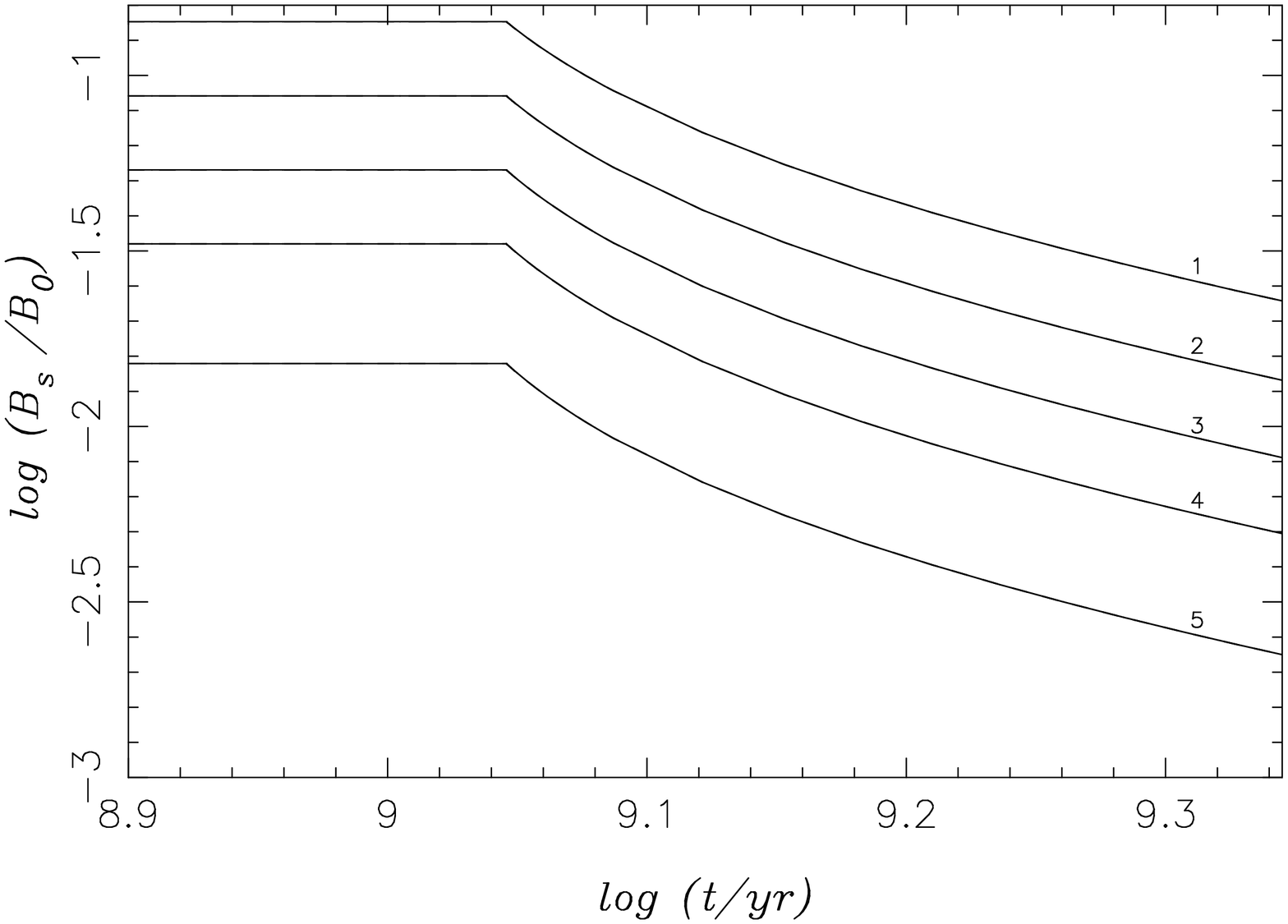,width=235pt,angle=-90}}}\end{center}
\caption[field    evolution    in    LMXBs   IIb]{Same    as    figure
[\ref{fcrust_lmxb_whole_w2_vRs}] with the wind phase expanded.}
\label{fcrust_lmxb_wind_w2_vRs}
\eef

\bef
\begin{center}{\mbox{\epsfig{file=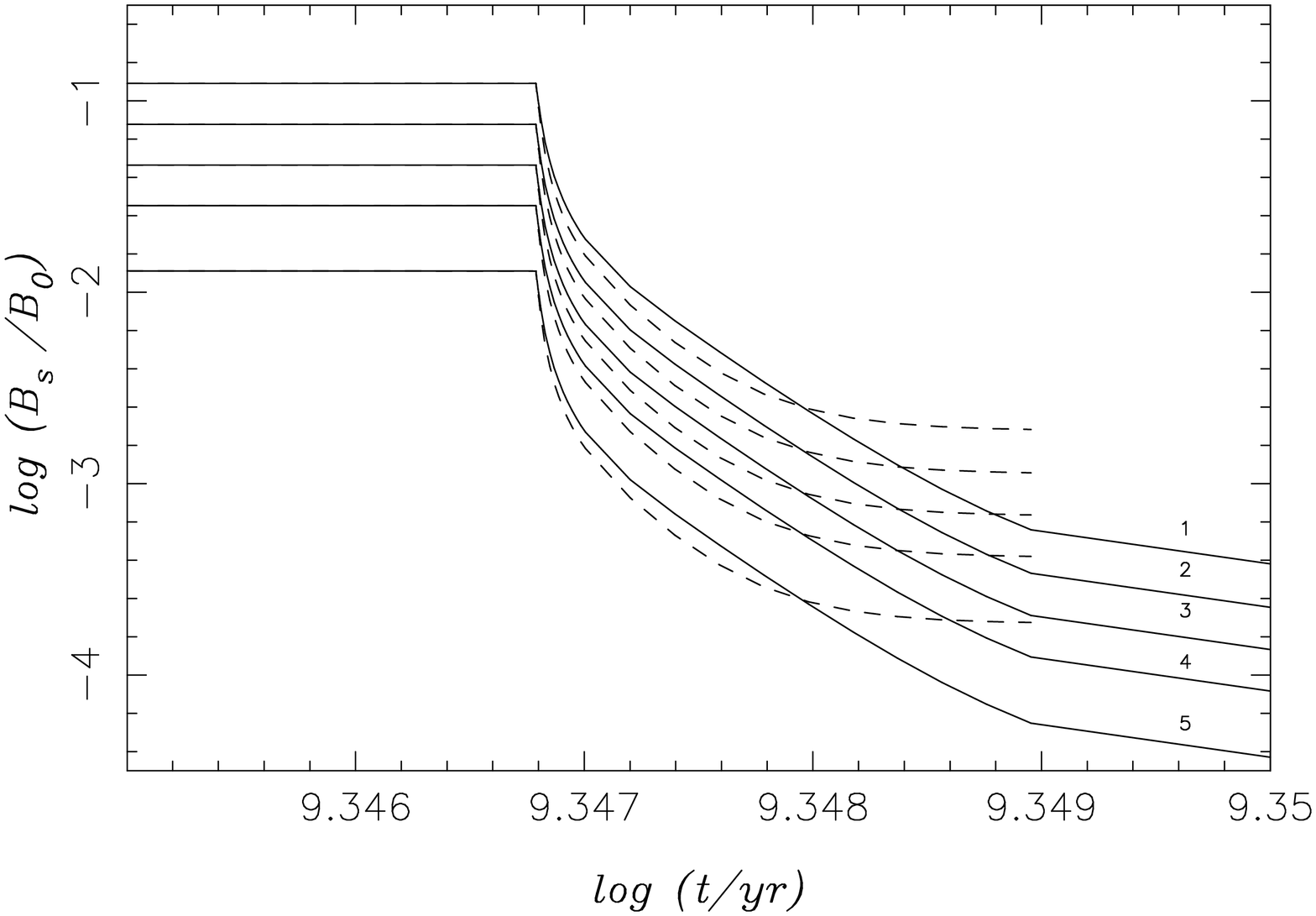,width=235pt,angle=-90}}}\end{center}
\caption[field  evolution in LMXBs  Ic]{Roche-contact phase  of figure
[\ref{fcrust_lmxb_whole_w1_vRs}] expanded.   The solid and  the dotted
lines  correspond  to  the  accretion rates  of  $10^{-10}$~\dmdt  and
$10^{-9}$~\dmdt  in  the  Roche-contact  phase corresponding  to  each
initial current concentration density.}
\label{fcrust_lmxb_Roche_w1_vRs}
\eef

\bef
\begin{center}{\mbox{\epsfig{file=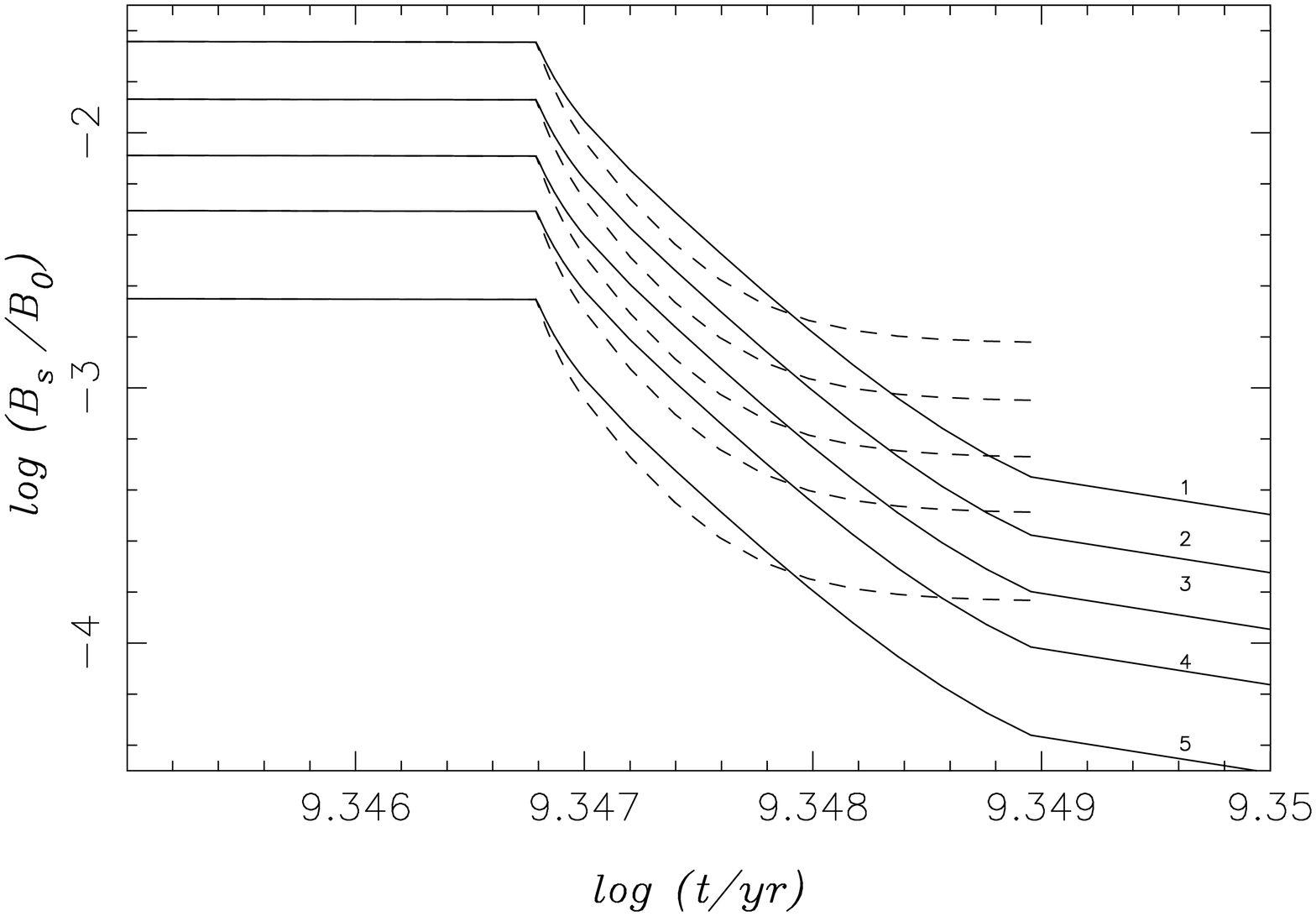,width=235pt,angle=-90}}}\end{center}
\caption[field evolution  in LMXBs IIc]{Roche-contact  phase of figure
[\ref{fcrust_lmxb_whole_w2_vRs}] expanded.   The solid and  the dotted
lines  correspond  to  the  accretion rates  of  $10^{-10}$~\dmdt  and
$10^{-9}$~\dmdt  in  the  Roche-contact  phase corresponding  to  each
initial current concentration density.}
\label{fcrust_lmxb_Roche_w2_vRs}
\eef

It is seen from these figures  that the surface field drops by half to
one order of magnitude in the wind phase of the binary evolution. When
the system is  in contact through Roche-lobe overflow  the field decay
depends  very much  on  the rate  of  accretion. A  difference in  the
accretion rate  in this phase  shows up as  a difference in  the final
value of the surface field, which freezes at a higher value for higher
rates of accretion. The total  decay in the Roche-contact phase may be
as  large as  two to  three orders  of magnitude  with respect  to the
magnitude of the field at the end of the wind phase.

\bef
\begin{center}{\mbox{\epsfig{file=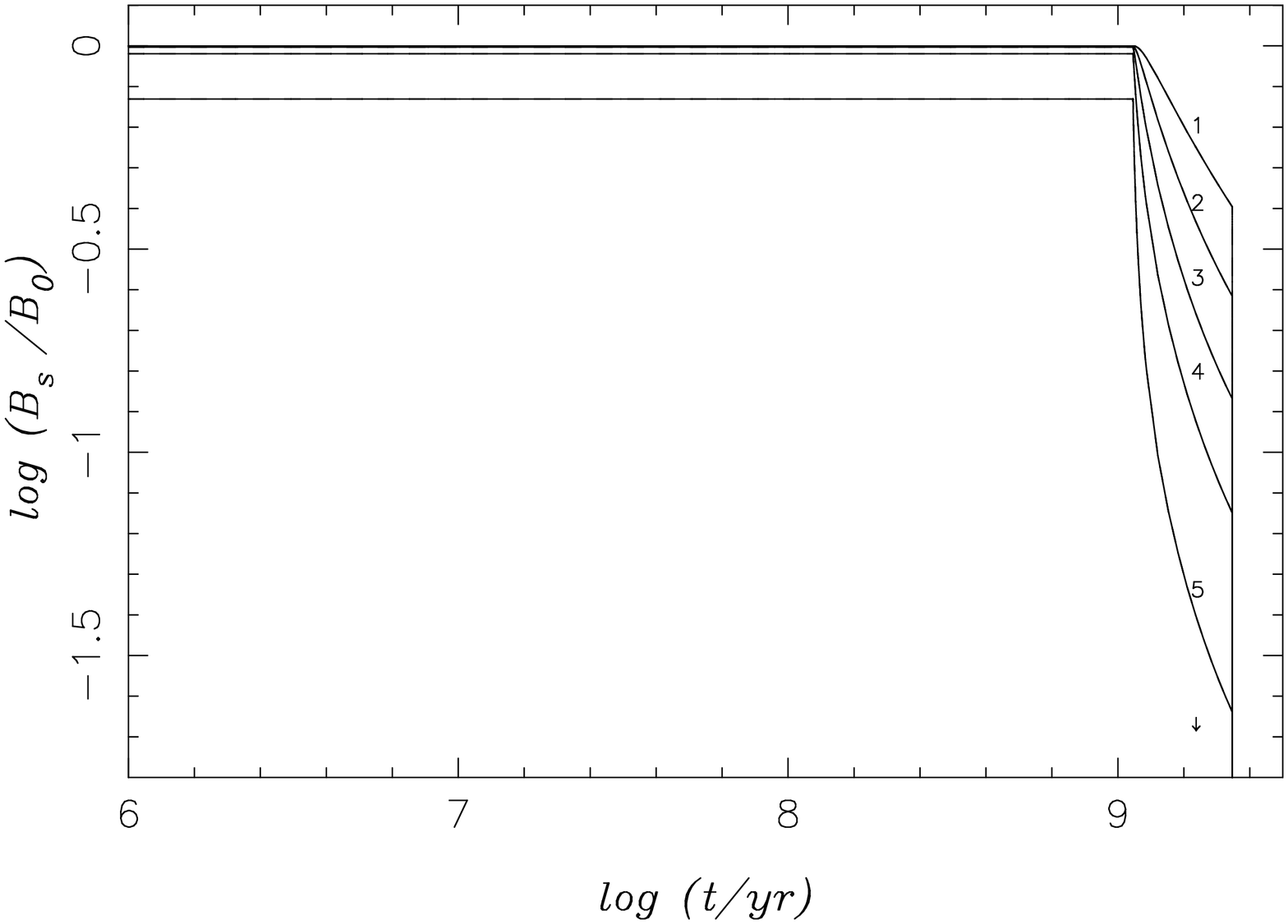,width=235pt,angle=-90}}}\end{center}
\caption[field    evolution    in    LMXBs   Ia']{Same    as    figure
[\ref{fcrust_lmxb_whole_w1_vRs}]  but with  an accelerated  cooling in
the isolated phase.} 
\label{fcrust_lmxb_whole_w1_vRf} 
\eef

\bef
\begin{center}{\mbox{\epsfig{file=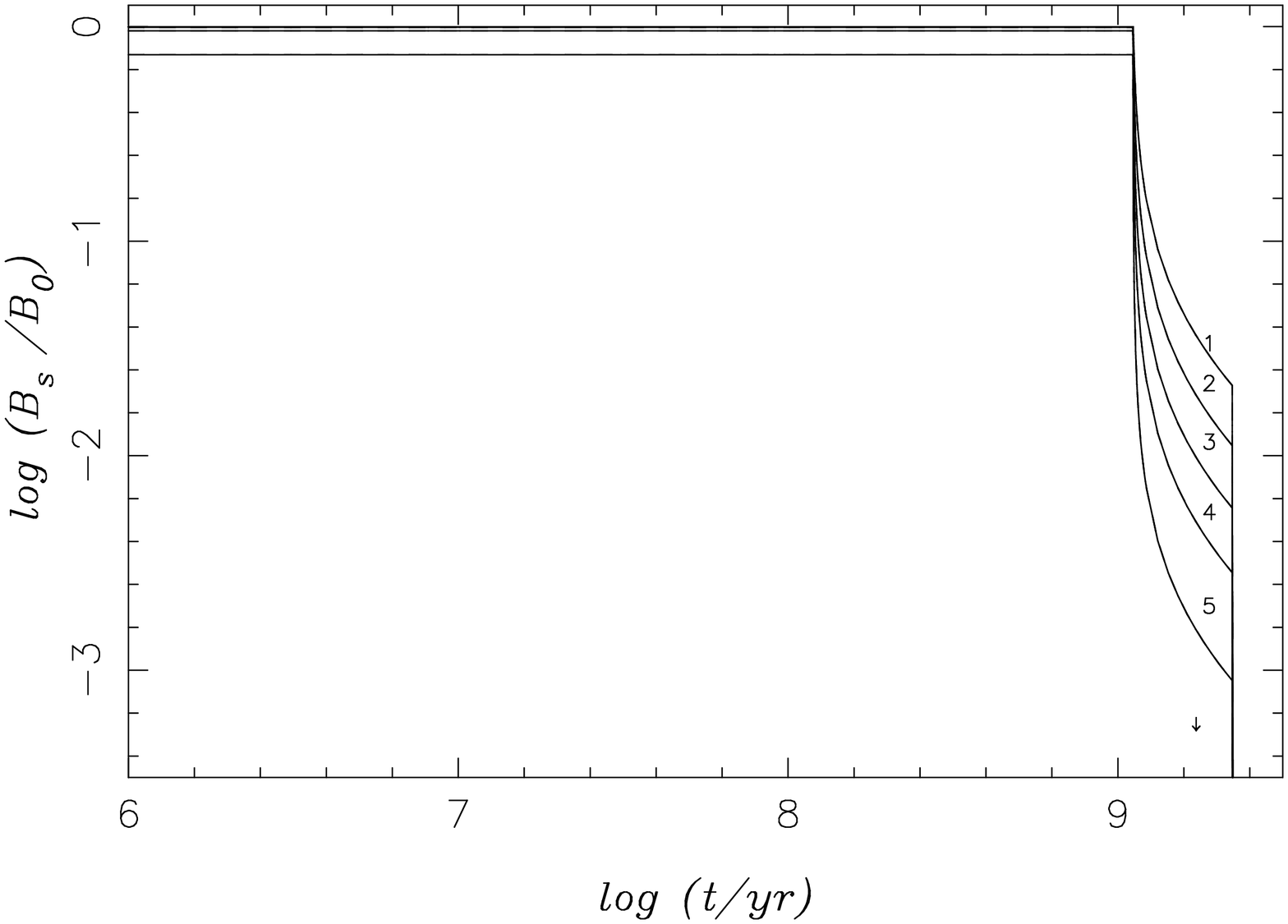,width=235pt,angle=-90}}}\end{center}
\caption[field    evolution   in    LMXBs    IIa']{Same   as    figure
[\ref{fcrust_lmxb_whole_w2_vRs}]  but with  an accelerated  cooling in
the isolated phase.}
\label{fcrust_lmxb_whole_w2_vRf}
\eef

\bef
\begin{center}{\mbox{\epsfig{file=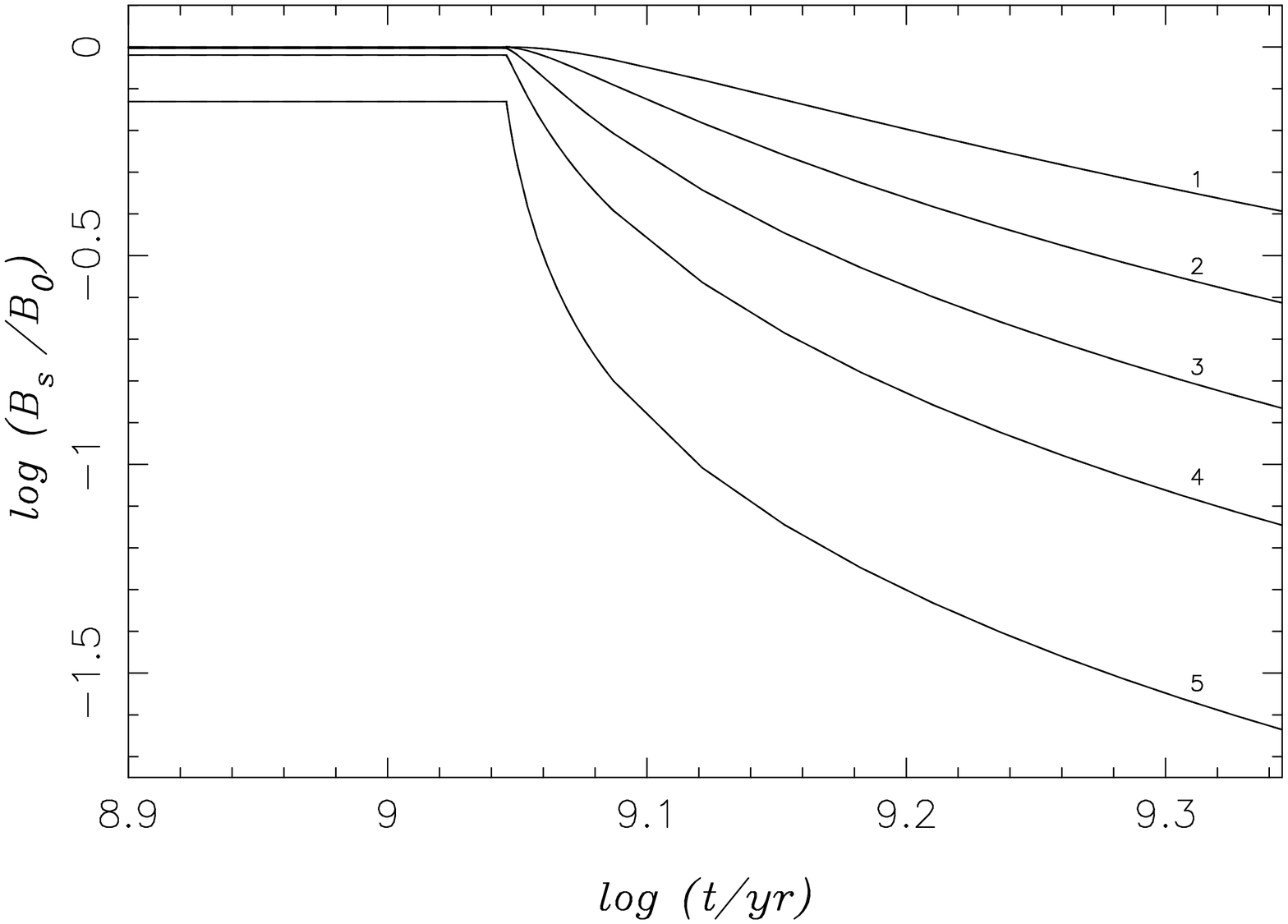,width=235pt,angle=-90}}}\end{center}
\caption[field  evolution in  LMXBs Ib']{The  wind accretion  phase of
figure [\ref{fcrust_lmxb_whole_w1_vRf}] expanded.}
\label{fcrust_lmxb_wind_w1_vRf}
\eef

\bef
\begin{center}{\mbox{\epsfig{file=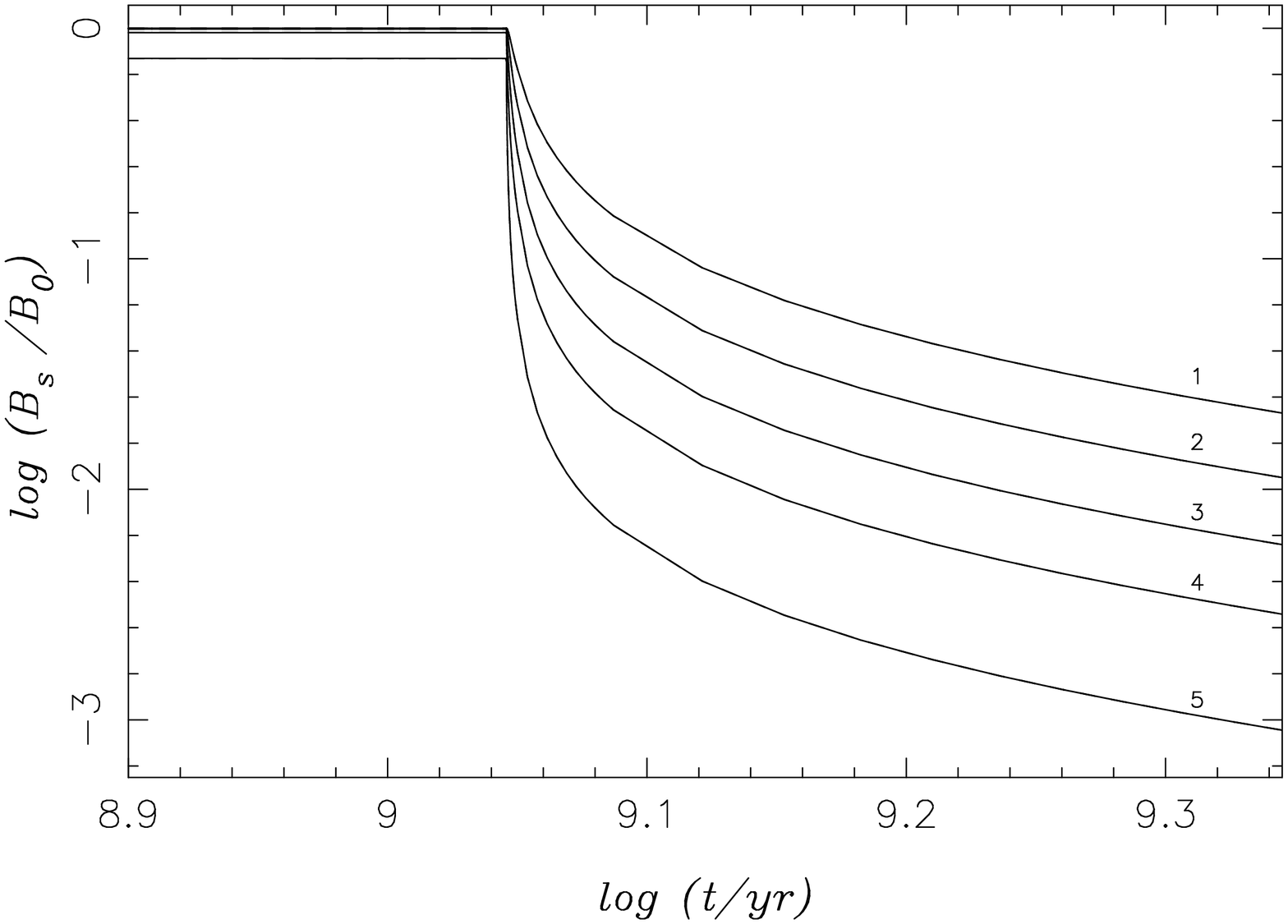,width=235pt,angle=-90}}}\end{center}
\caption[field evolution  in LMXBs  IIb']{The wind accretion  phase of
figure [\ref{fcrust_lmxb_whole_w2_vRf}] expanded.}
\label{fcrust_lmxb_wind_w2_vRf}
\eef

\bef
\begin{center}{\mbox{\epsfig{file=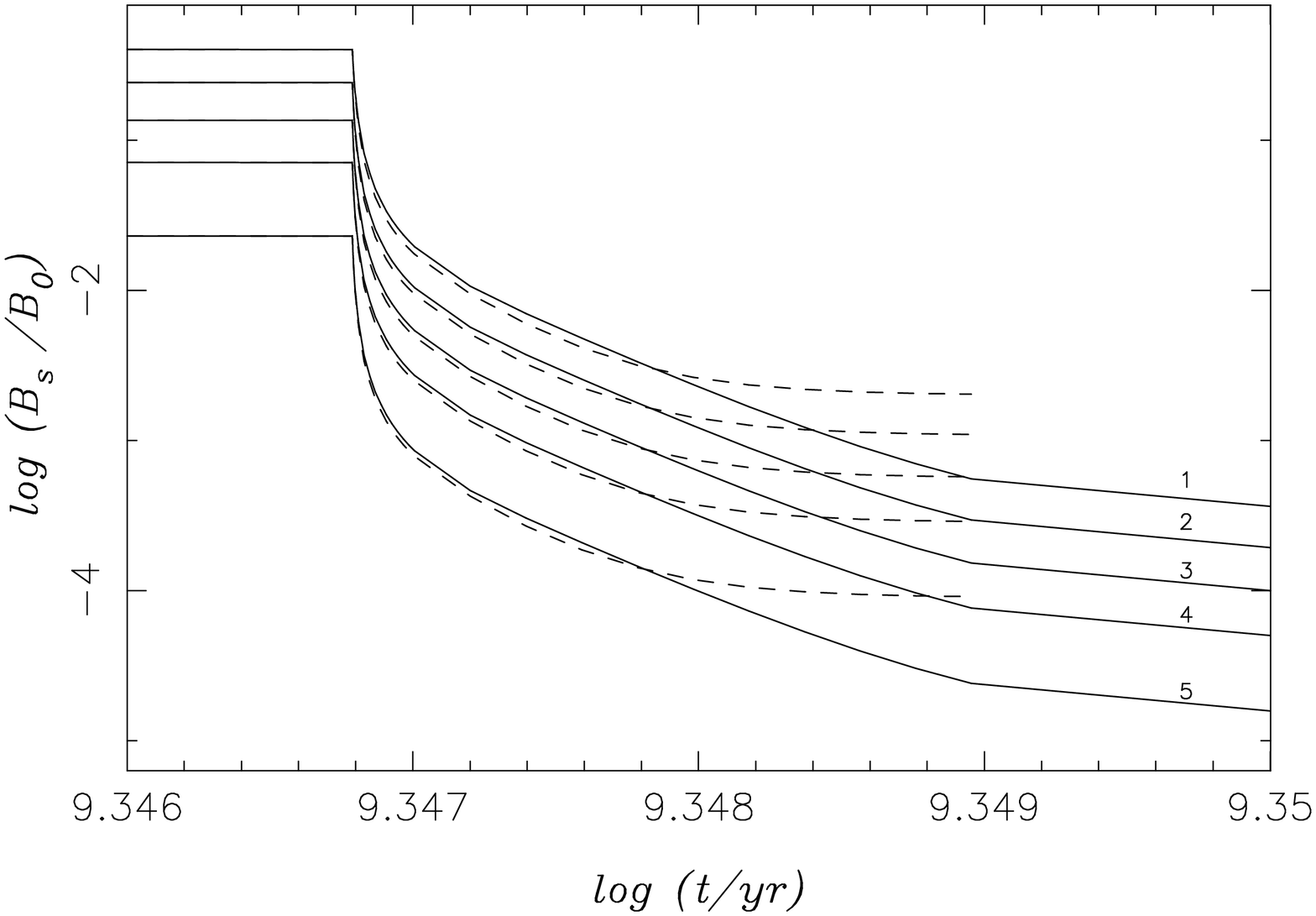,width=235pt,angle=-90}}}\end{center}
\caption[field evolution  in LMXBs Ic']{Roche contact  phase of figure
[\ref{fcrust_lmxb_whole_w1_vRf}] expanded.   The solid and  the dotted
lines  correspond  to  the  accretion rates  of  $10^{-10}$~\dmdt  and
$10^{-9}$~\dmdt  in  the  Roche-contact  phase corresponding  to  each
initial current concentration density.}
\label{fcrust_lmxb_Roche_w1_vRf}
\eef

\bef
\begin{center}{\mbox{\epsfig{file=crust_lmxb_Roche_w2_vRs.ps,width=235pt,angle=-90}}}\end{center}
\caption[field evolution in LMXBs  IIc']{Roche contact phase of figure
[\ref{fcrust_lmxb_whole_w2_vRf}] expanded.   The solid and  the dotted
lines  correspond  to  the  accretion rates  of  $10^{-10}$~\dmdt  and
$10^{-9}$~\dmdt  in  the  Roche-contact  phase corresponding  to  each
initial current concentration density.}
\label{fcrust_lmxb_Roche_w2_vRf}
\eef

We have mentioned  before that the phase of wind  accretion may not be
realized    in   some   of    the   cases    at   all.    In   figures
[\ref{fcrust_lmxb_whole_nw_vRs}]  to  [\ref{fcrust_lmxb_Roche_nw_vRf}]
we have  plotted the  evolution of the  surface field for  such cases,
both for  the standard and the  accelerated cooling. We  find that the
final field  strengths achieved without  a phase of wind  accretion is
not very different from the cases  where such a phase does exist. This
again is  indicative of  the fact  that a prior  phase of  field decay
slows down the decay in  the subsequent phase. And therefore the final
result from both the cases become similar. 

\bef
\begin{center}{\mbox{\epsfig{file=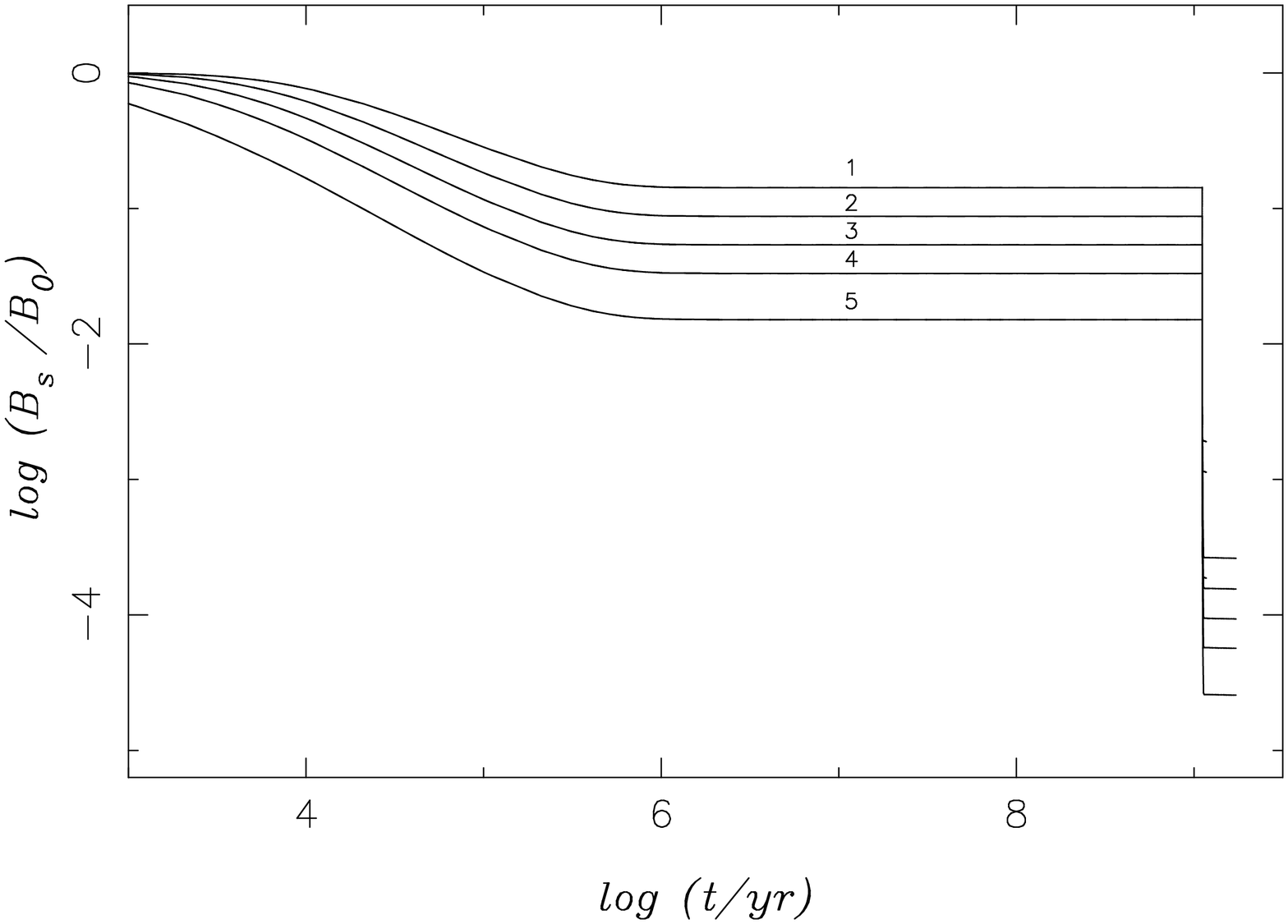,width=235pt,angle=-90}}}\end{center}
\caption[field  evolution  in  LMXBs  IIIa]{Evolution of  the  surface
magnetic  field  in  LMXBs  without  a phase  of  wind  accretion  and
accretion  rates   of  $\mdot   =  10^{-10},  10^{-9}$~\dmdt   in  the
Roche-contact  phase. Curves  1  to 5  correspond  to initial  current
configuration  centred   at  $\rho  =   10^{11},  10^{11.5},  10^{12},
10^{12.5},  10^{13} \gcc$.   All curves  correspond  to $Q$  = 0.0.  A
standard cooling has been assumed for the isolated phase here.}
\label{fcrust_lmxb_whole_nw_vRs}
\eef

\bef
\begin{center}{\mbox{\epsfig{file=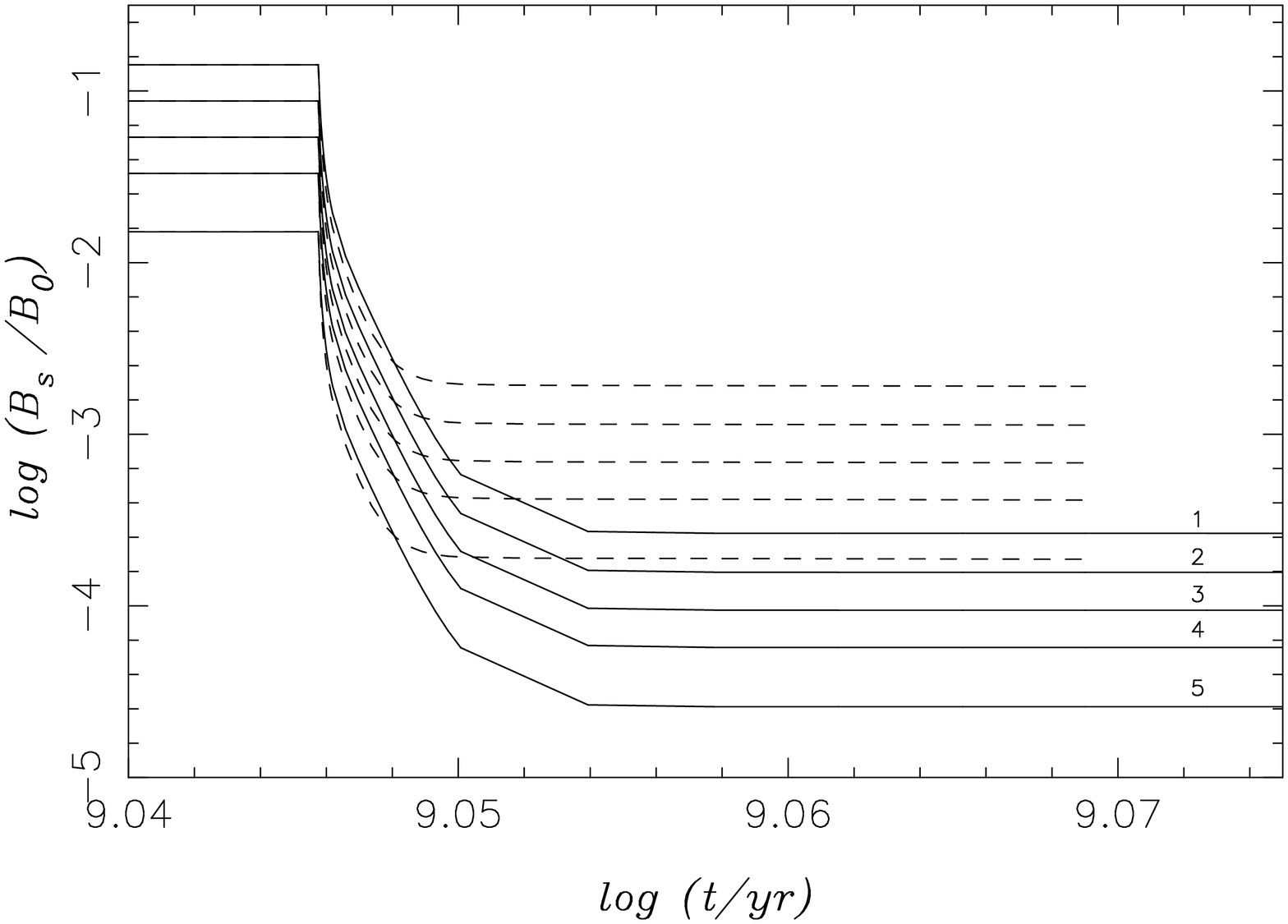,width=235pt,angle=-90}}}\end{center}
\caption[field  evolution in  LMXBs IIIb]{The  Roche-contact  phase of
figure [\ref{fcrust_lmxb_whole_nw_vRs}]  expanded.  The solid  and the
dotted lines correspond to the accretion rates of $10^{-10}$~\dmdt and
$10^{-9}$~\dmdt  in  the  Roche-contact  phase corresponding  to  each
initial current concentration density.}
\label{fcrust_lmxb_Roche_nw_vRs}
\eef

\bef
\begin{center}{\mbox{\epsfig{file=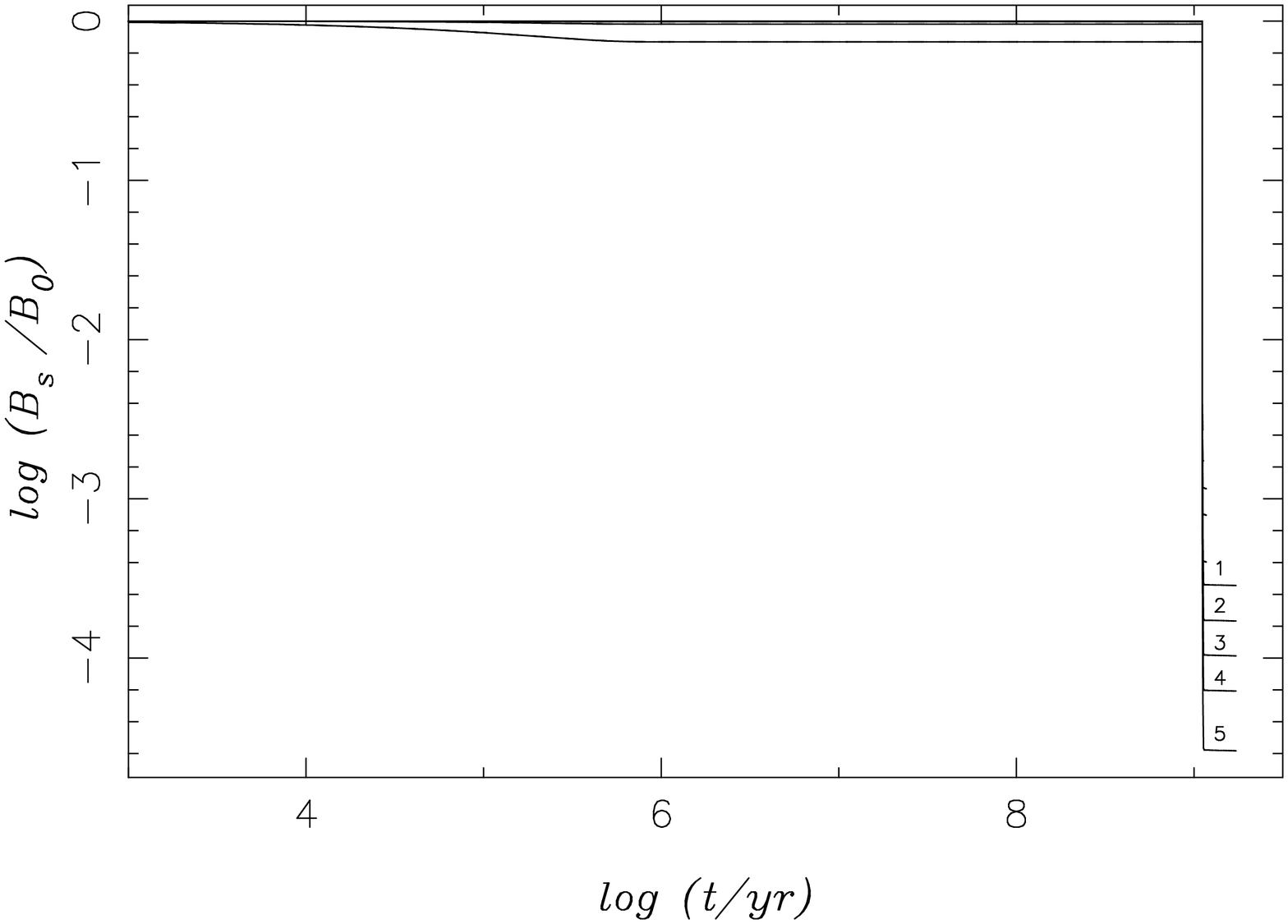,width=235pt,angle=-90}}}\end{center}
\caption[field    evolution   in    LMXBs   IIIa']{Same    as   figure
[\ref{fcrust_lmxb_whole_nw_vRs}] with  accelerated cooling assumed for
the isolated phase.}
\label{fcrust_lmxb_whole_nw_vRf}
\eef

\bef
\begin{center}{\mbox{\epsfig{file=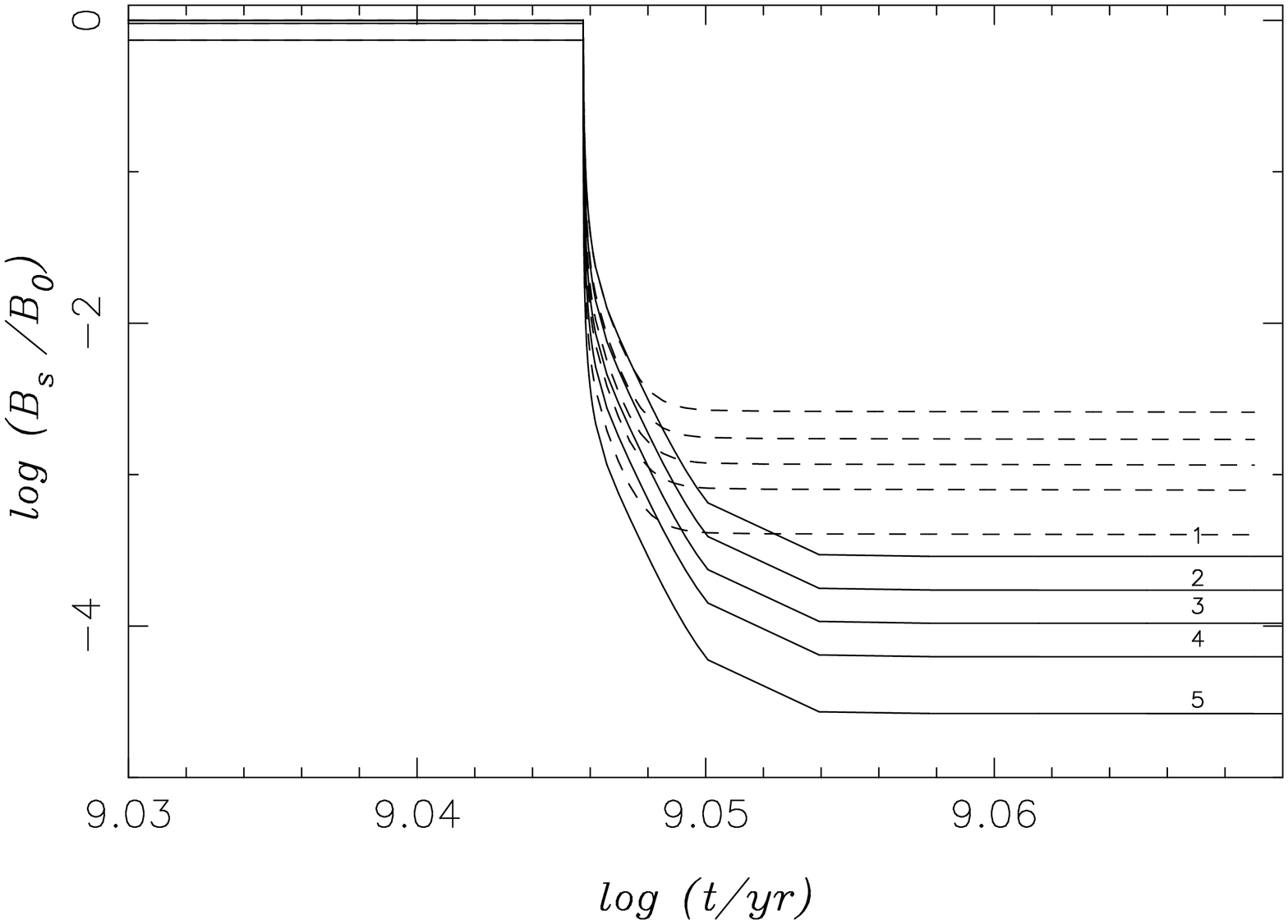,width=235pt,angle=-90}}}\end{center}
\caption[field    evolution   in    LMXBs   IIIb']{Same    as   figure
[\ref{fcrust_lmxb_Roche_nw_vRs}] with  accelerated cooling assumed for
the isolated phase.}
\label{fcrust_lmxb_Roche_nw_vRf}
\eef

There   are  several   interesting  points   to  note   here.  Figures
[\ref{fcrust_lmxb_Roche_w1_vRs}],
[\ref{fcrust_lmxb_Roche_w2_vRs}],[\ref{fcrust_lmxb_Roche_w1_vRf}], and
[\ref{fcrust_lmxb_Roche_w2_vRf}]  shows  that  for  higher  values  of
accretion rate in  the Roche-contact phase the final  field values are
higher.  We have not explored  the case of accretion with an Eddington
rate in this phase. From the  trends observed in our calculation it is
evident that  with such high rate  of accretion the  final field value
may remain fairly large. Under  such circumstances it will be possible
to  have  `recycled'  pulsars  of  high surface  magnetic  field  (and
therefore long  spin-period) from low  mass binaries and  pulsars like
PSR 0820+02 will fit in with the general scenario quite well. Then, of
course, we do find significant  amount of field decay with lower rates
of  accretion in  the  Roche-contact phase.  Such  low surface  fields
combined  with the  provision of  maximal spin-up  would  then produce
millisecond pulsars. 

Figures  [\ref{fcrust-hmxb_vRs1}]  to [\ref{fcrust_lmxb_Roche_w2_vRf}]
then indicate  that the model  of field evolution assuming  an initial
crustal  field  configuration is  quite  consistent  with the  present
scenario  of  field  evolution.  The LMXBs  will  produce  high-field,
long-period pulsars  in addition to  the expected crop  of millisecond
pulsars. Whereas the  only kind of recycled pulsars  that are expected
from  the HMXBs  would be  of the  relatively  high-field, long-period
variety. 

\bef
\begin{center}{\mbox{\epsfig{file=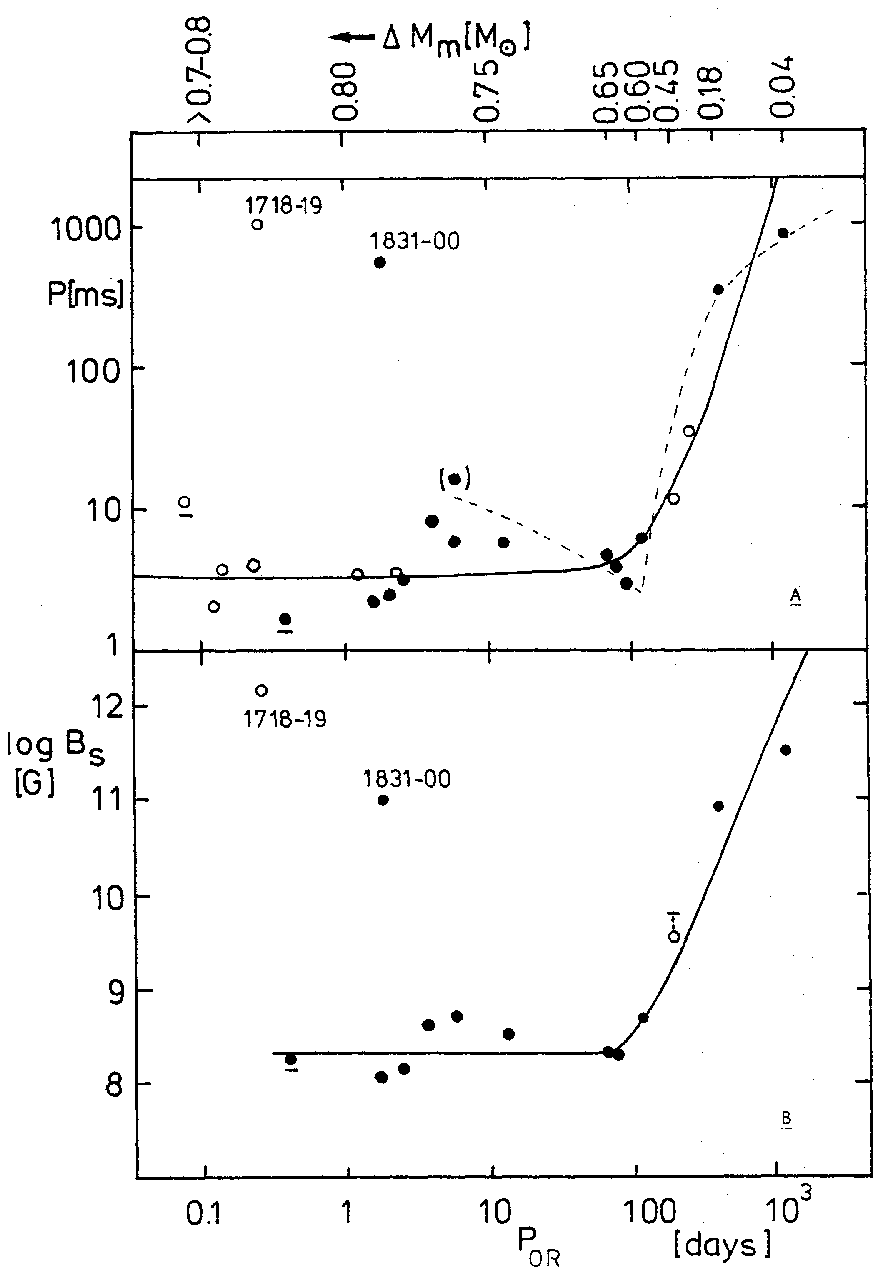,width=275pt}}}\end{center}
\vspace{-2.00cm}
\hspace{3.25cm}
\mbox{\epsfig{file=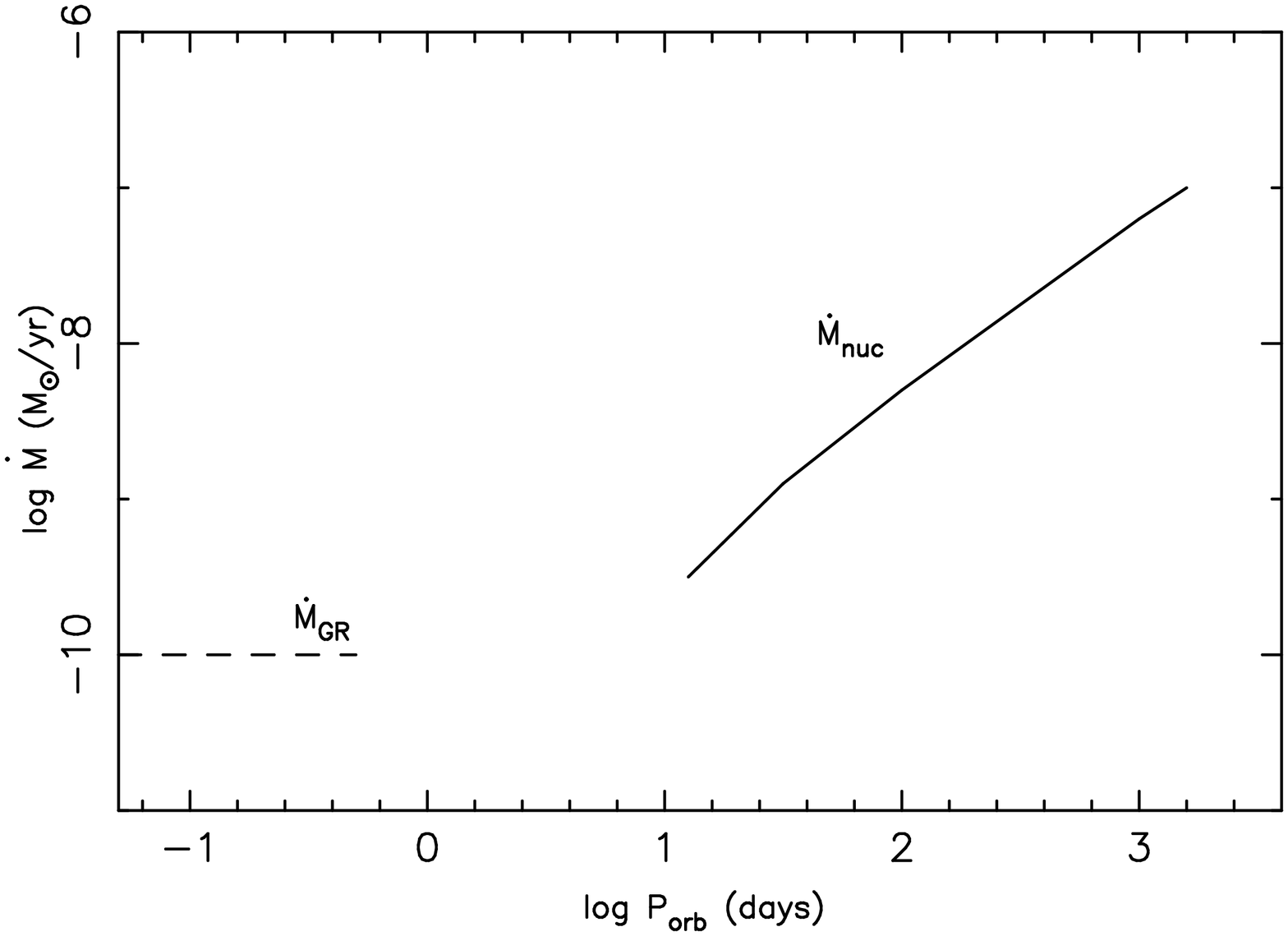,width=215pt,angle=-90}}
\caption[binary pulsar  parameter correlation]{The top  and the middle
panels correspond  to spin period,  magnetic field vs.  orbital period
for binary  pulsars with circular orbits and  low-mass companions. The
bottom panel depicts  rate of mass transfer vs.  the orbital period in
low-mass  X-ray  binaries.  $\mdot_{\rm  GR}$  and  $\mdot_{\rm  nuc}$
correspond to mass transfers  driven by gravitational radiation and by
the nuclear evolution of the companion to the neutron star.}
\label{fvdh+bitz+verb}
\eef

In the course of our investigations  in this chapter we have seen that
there is  a positive  correlation between the  rate of  accretion with
that  of the  final field  strength, namely,  the higher  the  rate of
accretion  the higher is  the final  field. There  has already  been a
mention  of such  a correlation  in connection  with the  Z  and Atoll
sources~\cite{hasn89}.  It  was suggested that  the difference between
these two classes of sources in regard to their fluctuation spectra is
not only due to  a difference in the accretion rate but  also due to a
difference in the  magnetic field. And there has  been indication from
the study of the radiation spectra of these sources that the accretion
rate  and  the  magnetic  field strength  are  positively  correlated.
Recently,   \citeN{lamb98}  on   the   basis  of   LMXB  spectra   and
\citeN{whit97} on  the basis of  the properties of the  kilohertz QPOs
have indicated  the existence of such a  correlation.  Consider figure
[\ref{fvdh+bitz+verb}] - the top and  the middle panels are taken from
\citeN{bitz95}  and the  bottom  panel  is made  using  the data  from
\citeN{verb95}.  The middle panel  shows the variation of the magnetic
field with the orbital period for  some of the low mass binary pulsars
whereas the  bottom panel  shows the rate  of mass accretion  with the
orbital period  in low-mass X-ray  binary systems. For  longer periods
the surface field increases with  period. In the bottom panel the rate
of   accretion   shows   a   similar   increase   with   the   orbital
period. Therefore we see that  for longer orbital periods higher rates
of  accretion  and  higher  values  of surface  field  are  positively
correlated. This  is another observational indication  that for higher
rates of accretion the magnetic field strengths tend to be higher.

\section{conclusions}
\label{sobs-concl}

In this chapter, we have looked  at one of the model of magnetic field
evolution assuming an initial crustal field configuration to check the
consistency with the overall  scenario of field evolution for isolated
as well as  binary pulsars. We find that the  model can explain almost
all the features that have  been observed to date. And our conclusions
can be summarized as follows :
\bei
\i  for this  model to  be  consistent with  the statistical  analyses
performed on the isolated pulsars at  the most a maximum value of 0.05
for the impurity strength can be allowed;
\i HMXBs produce high-field  long period pulsars provided the duration
of  the  wind  accretion  phase   is  short  or  the  initial  current
distribution is located at higher densities;
\i  Relatively   low-field  ($B  \sim   10^{10}$~Gauss)  objects  near
death-line (low-luminosity pulsars) are also predicted from HMXBs;
\i LMXBs will  produce both high-field long period  pulsars as well as
low-field short period pulsars inclusive of millisecond pulsars in the
later variety; and
\i a positive correlation between  the rate of accretion and the final
field  strength  is  indicated  that  is  supported  by  observational
evidence.
\eei

\chapter{spin-down induced flux expulsion and its consequences}
\label{csif}

\section{introduction}
\label{ssif-intro}

Almost all models of field  evolution depend on the mechanism of ohmic
decay of the underlying current  loops for a permanent decrease in the
field strength. It has been mentioned earlier (chapters [\ref{cfield}]
and [\ref{cmn}]) that  such ohmic dissipation is possible  only if the
current  loops  are  situated   in  the  crust  where  the  electrical
conductivity   is  finite.   Any   flux  that   may   reside  in   the
superconducting core of the star would remain unchanged forever unless
this flux is brought out to the crust. 

Models  that assume  an initial  core-field  configuration, therefore,
require a phase  of flux expulsion from the  core.  This expelled flux
then undergoes  ohmic dissipation in the crust  decreasing the surface
field strength. One such model uses the spin-evolution to achieve this
- that        of       the        `spin-down        induced       flux
expulsion'~\cite{srini89,srini90}.   The core of  the neutron  star is
believed to  contain two superfluids - the  neutral neutron superfluid
and  the charged proton  superconductor. Whereas  the rotation  of the
star is supported  by creation of vortices in  the neutron superfluid,
the magnetic  flux is  sustained by Abrikosov  fluxoids in  the proton
superconductor~\cite{bhat95}.   Spinning down of  the star  requires a
decrease in the  number of vortices in the  superfluid core. Therefore
as a result of spin-down  the vortices move out towards the core-crust
boundary of  the star. An  inter-pinning between the vortices  and the
magnetic fluxoids make the  fluxoids move outwards too, reducing their
number in the  core. A decrease in the number of  fluxoids in the core
reduces the field  strength there.  The nature of  such flux expulsion
as a result of spin-evolution has been investigated in detail for both
isolated  pulsars  (undergoing  pure  dipole spin-down)  and  for  the
neutron stars that are members of binaries (undergoing major spin-down
in the `propeller phase')~\cite{ding93,miri94,miri96}.

In this chapter  we look at the ohmic decay of  such expelled field in
the  crust of  an isolated  as well  as that  of an  accreting neutron
star. Some of the earlier  investigations in this direction assumed an
uniform  ohmic  decay time-scale  in  the  crust  irrespective of  the
accretion rate~\cite{miri94,miri96}.   The only detailed  work in this
context has been by \citeN{bhat96} where they incorporated the crustal
micro-physics into their calculation for the evolution of the expelled
field. However  this work did  not include the material  movement that
takes place in the crust as a result of accretion. In the present work
we incorporate the material movement  too and look at the evolution of
an expelled field in the crust  of an accreting neutron star using the
methodology developed in chapter [\ref{cmn}].

In the previous chapter we have discussed the overall scenario for the
evolution  of the  neutron  star magnetic  field  - encompassing  both
isolated neutron stars  and those that are members  of binary systems.
In  this chapter  we  investigate  the consequences  of  the model  of
`spin-down induced flux expulsion'  in the above mentioned systems. We
try to judge  whether this model, too, is  consistent with the overall
field    scenario   and    under   what    conditions.    In   section
[\ref{ssif-results}] we shall discuss  the results of our computations
and conclude in section [\ref{ssif-concl}].

\section{results and discussions}
\label{ssif-results}

Using the  methodology developed in  chapter [\ref{cmn}] we  solve the
equation  [\ref{edgdt}]  for an  initial  flux  just  expelled at  the
core-crust  boundary  due  to  spin-down.  Such an  expelled  flux  is
deposited at the bottom of the crust and we start our calculation with
such   a  field  configuration.   Figure  [\ref{fcore-g}]   shows  the
distribution  of  the  $g$-function  and  figure  [\ref{fcore-J}]  the
toroidal currents,  $J_{\phi}$, assumed at  the starting point  of our
evolution. 

\bef
\begin{center}{\mbox{\epsfig{file=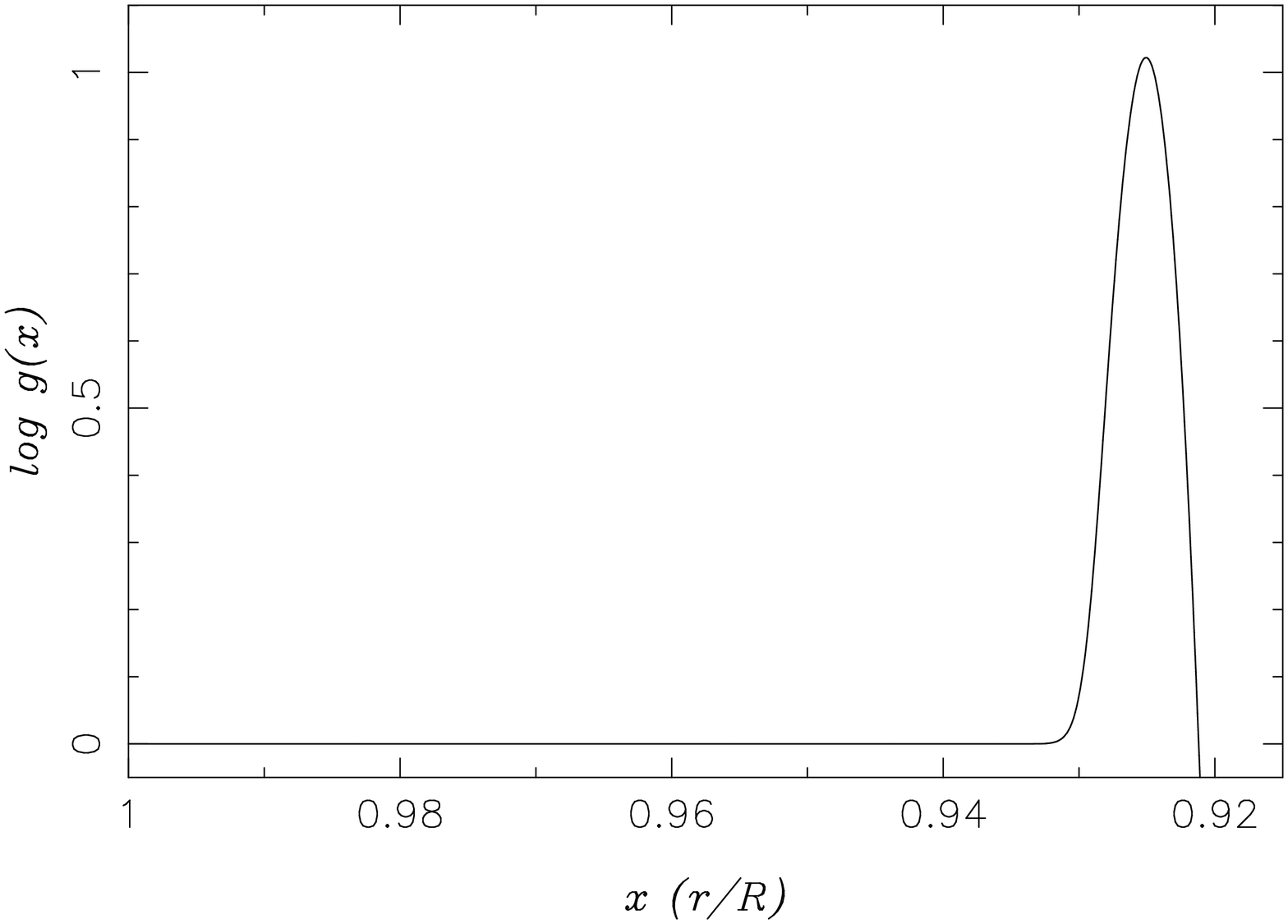,width=235pt,angle=-90}}}\end{center}
\caption[initial  $g$-profile]{The initial  radial  dependence of  the
$g$-profile, corresponding to an  expelled flux, centred at $x=0.925$,
with a  width $\delta x$ =  0.006; where $x$ is  the fractional radius
$r/R$.}
\label{fcore-g}
\eef
\bef
\begin{center}{\mbox{\epsfig{file=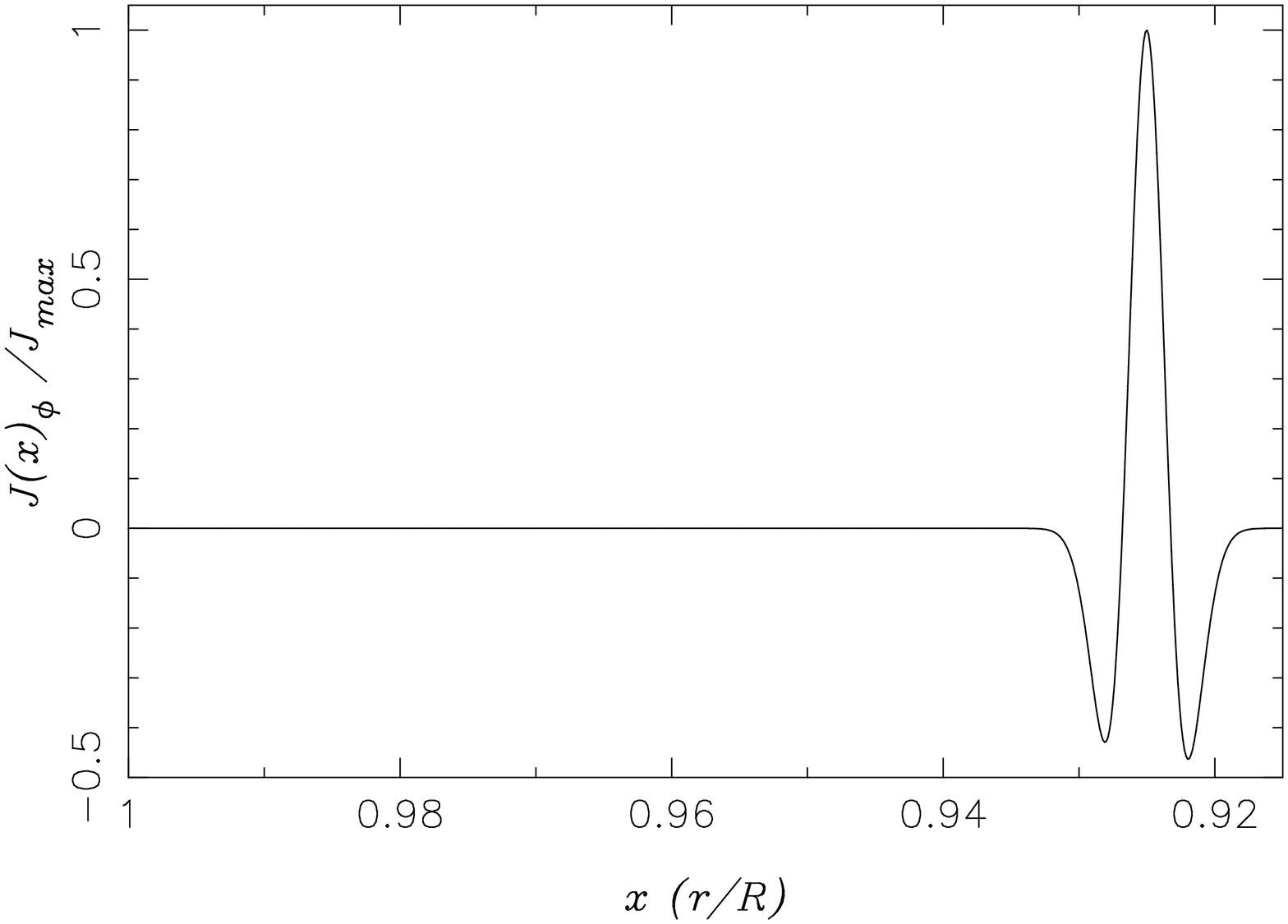,width=235pt,angle=-90}}}\end{center}
\caption[initial  $J$-profile]{The initial  radial  dependence of  the
$\phi$-component of the corresponding current configuration.}
\label{fcore-J}
\eef

\subsection{ohmic  decay  in  isolated  pulsars}  

The  cause of  flux  expulsion is  the  spinning down  of the  pulsar.
Therefore,  for  isolated  pulsars  too, experiencing  a  pure  dipole
slow-down, there would be some flux expulsion. The extent of such flux
expulsion has been worked out  by \citeN{miri96} and it has been shown
there that the flux expulsion is most effective for pulsars with large
values  of  magnetic  field  strength.  Before  discussing  the  field
evolution in  neutron stars  in binary systems,  we first look  at the
ohmic  decay  of such  an  expelled  field  in isolated  pulsars.   We
consider  the isolated neutron  star to  undergo standard  cooling, in
order to find the maximum extent of field reduction in this case.  But
by  the  time  significant  flux-expulsion is  achieved,  the  crustal
temperature goes down to very low values for an isolated neutron star.
In such a situation the conductivity would be mainly determined by the
scattering of the  conduction electrons by the impurities.  To see the
effect  of that  we have  considered here  several values  of impurity
strength $Q$ (see section [\ref{sns-transport}] for details).

\bef
\begin{center}{\mbox{\epsfig{file=gprofile_core_diffusion_1.ps,width=225pt,angle=-90}}}\end{center}
\caption[ohmic  diffusiono of $g$-profile  I]{Pure ohmic  diffusion of
the $g$-profile plotted in figure [\ref{fcore-g}] for $\tau \sim 10^9$
yrs, with  $Q$ =  0.1, in  a neutron star  with standard  cooling. The
curves are shown for intermediate  times with no discernible change in
the value at the surface ($x$ = 1).}
\label{fcore-cool-g1}
\eef
\bef
\begin{center}{\mbox{\epsfig{file=gprofile_core_diffusion_2.ps,width=225pt,angle=-90}}}\end{center}
\caption[ohmic   diffusiono   of   $g$-profile  II]{Same   as   figure
[\ref{fcore-cool-g1}]  with $Q$ =  0.2. The  profiles show  an initial
increase  and a  substantial  decrease  in the  surface  value over  a
time-scale of $10^9$ years. }
\label{fcore-cool-g2}
\eef
\bef
\begin{center}{\mbox{\epsfig{file=core_diffusion.ps,width=225pt,angle=-90}}}\end{center}
\caption[ohmic diffusion  of the  surface field]{The evolution  of the
surface  magnetic field  due to  pure  diffusion of  an expelled  flux
corresponding  to the  evolution  of the  $g$-profile  plotted in  the
previous figures. The dotted and  the solid curves correspond to $Q$ =
0.0, 0.1, 0.2, 0.3 and 0.4 respectively.}
\label{fcore-cool-b}
\eef

In  figures [\ref{fcore-cool-g1}]  and  [\ref{fcore-cool-g2}] we  have
plotted the  time evolution of the  $g$-profile for two  values of the
impurity  strength. In  figure [\ref{fcore-cool-b}]  the corresponding
evolution  of the  surface  fields  for five  values  of the  impurity
strength are plotted.  It is seen very clearly  that the surface field
actually  increases when  the expelled  flux diffuses  outwards before
finally decaying down  to smaller values. It is to  be noted that even
for  a very  large value  of  the impurity  strength ($Q$  = 0.4)  the
surface field shows significant decay  only over a time-scale of $\sim
10^9$  years.  The  active lifetime  of  an isolated  radio pulsar  is
$\lsim  10^8$  years.  Therefore,   an  isolated  radio  pulsar  would
experience  little field  decay over  its active  lifetime, consistent
with the  indications from statistical  analyses. So we  conclude that
with an  expelled flux there is  provision for large values  of $Q$ to
exist in  the crust of the neutron  star. In fact, we  shall see later
that  in  this  model  large   $Q$  is  a  necessary  requirement  for
millisecond pulsar generation. 

\subsection{field evolution with accretion}

\bef
\begin{center}{\mbox{\epsfig{file=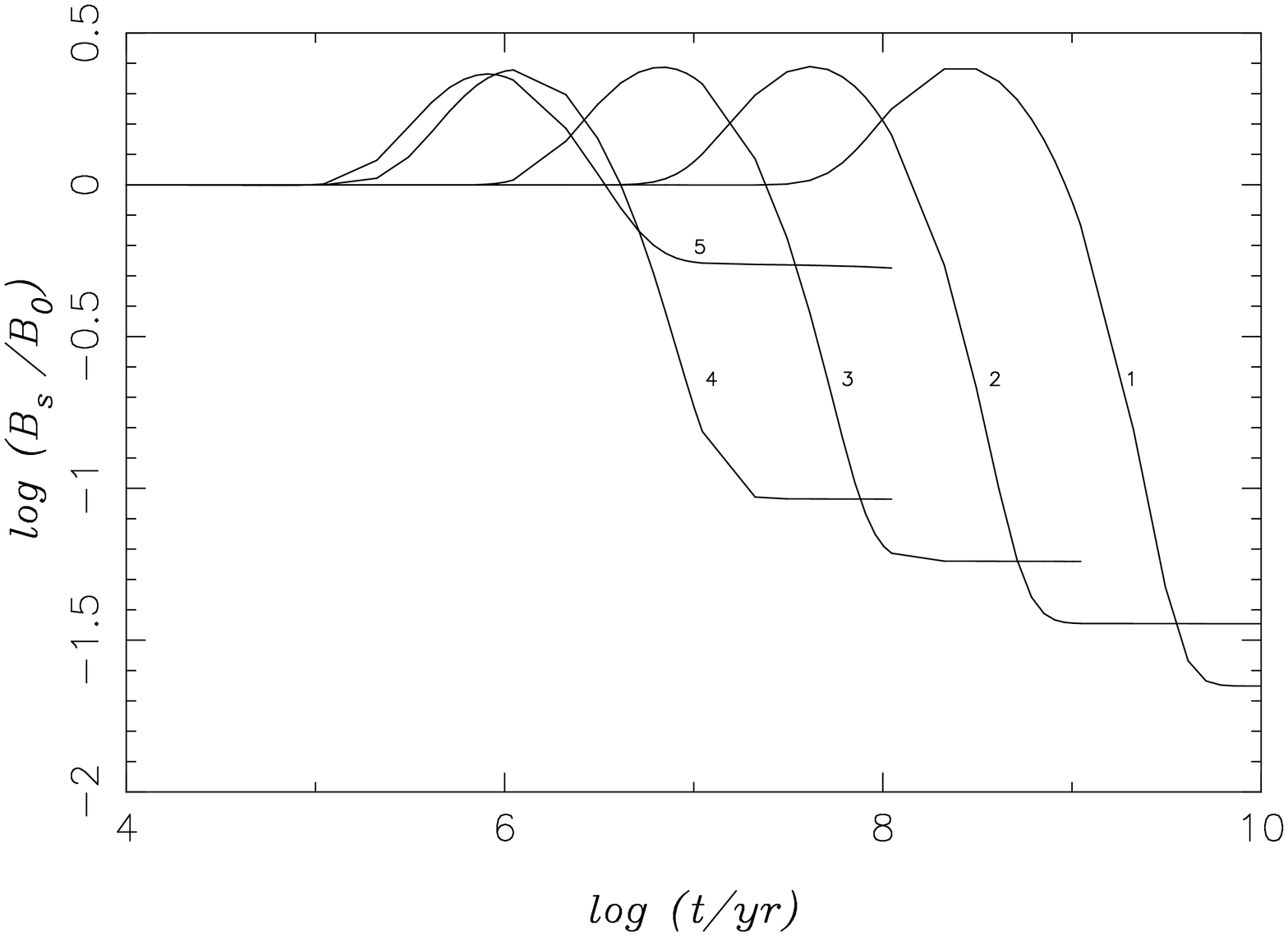,width=245pt,angle=-90}}}\end{center}
\caption[evolution of  the surface  field I]{Evolution of  the surface
magnetic field for  an expelled flux. The curves 1  to 5 correspond to
$\mdot =  10^{-13}, 10^{-12}, 10^{-11},  10^{-10}, 10^{-9}$~\dmdt. All
curves correspond to $Q$ = 0.0.}
\label{fcore_accretion_dm}
\eef
\bef
\begin{center}{\mbox{\epsfig{file=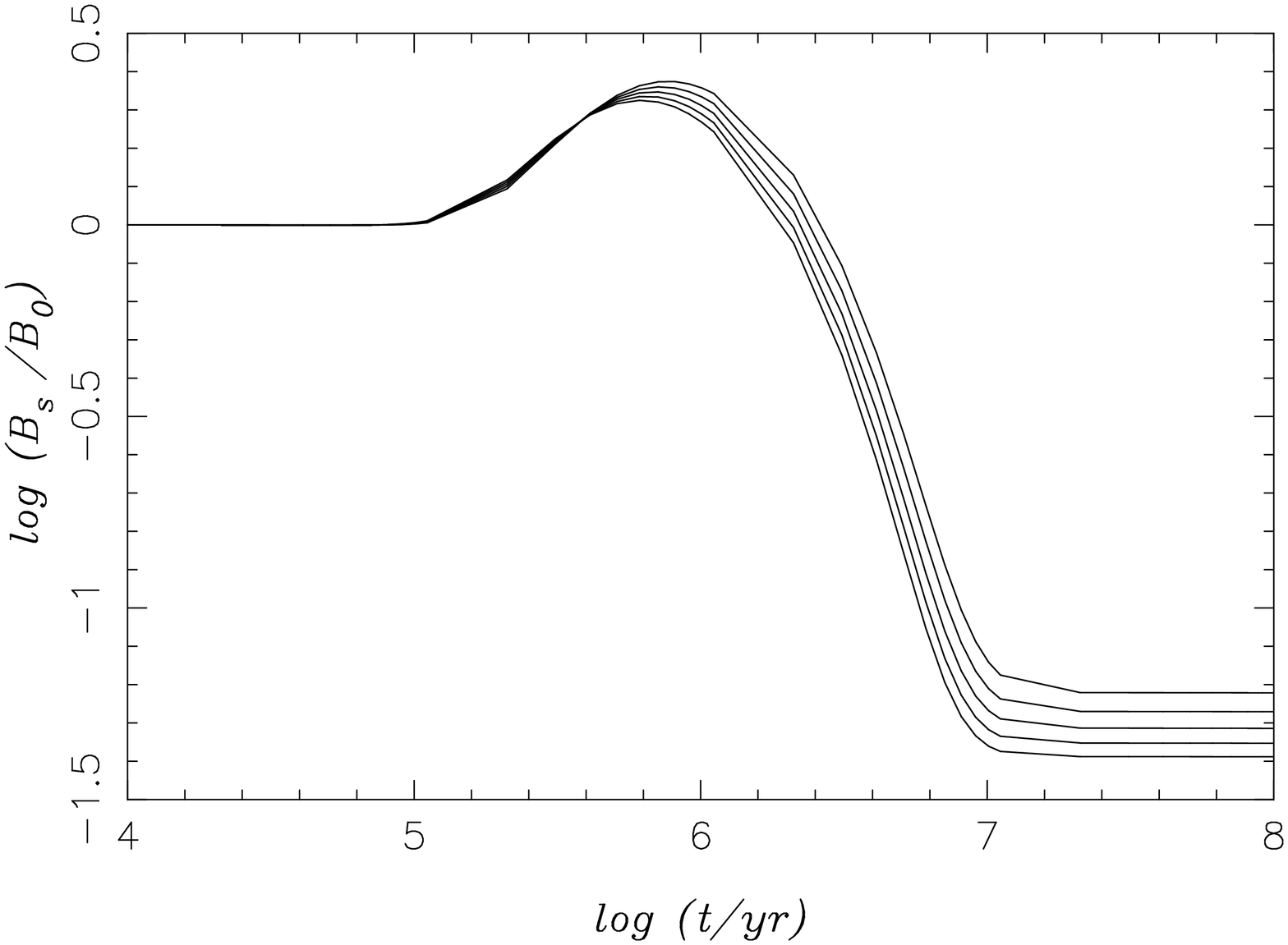,width=225pt,angle=-90}}}\end{center}
\caption[evolution of  the surface field II]{Evolution  of the surface
magnetic field for an expelled  flux. The different curves from top to
bottom correspond to $Q$ = 0,  0.1, 0.2, 0.3 and 0.4 respectively. All
curves correspond to $\mdot = 10^{-10}$~\dmdt.}
\label{fcore_accretion_q}
\eef

Before looking  into the nature  of field evolution in  various binary
systems, we first investigate the evolution of the surface field under
accretion, assuming the  $g$-profile plotted in figure [\ref{fcore-g}]
as the initial condition. In figure [\ref{fcore_accretion_dm}] we plot
the  evolution  of the  surface  field  for  different values  of  the
accretion  rate and  in figure  [\ref{fcore_accretion_q}] we  plot the
evolution for different values of  the impurity strength. We find that
the field  strengths go  down by about  only one  and a half  order of
magnitude   even  for   a   fairly  large   value   of  the   impurity
strength. Therefore,  even higher values of impurity  strength will be
required for  a larger reduction in  the field strength.  We shall see
later that  for certain cases  the required impurity strength  is much
larger than that  considered by us here (we  have considered a maximum
$Q$  value of  0.4) to  achieve millisecond  pulsar field  values. The
characteristic features of field evolution are then as follows.
\ben
\i An initial rapid decay  (neglecting the early increase) is followed
by a slow down and an eventual {\em freezing}.
\i The onset  of `freezing' is faster with  higher rates of accretion,
i.e., at higher values of crustal temperature.
\i  Lower  final `frozen'  fields  are  achieved  for lower  rates  of
accretion.
\i  To achieve  a significant  reduction in  the field  strength, very
large values of the impurity strength are required.
\een
It  is clear  from these  features that  the general  nature  of field
evolution in the case of  an expelled flux is qualitatively similar to
that in the  case of an initial crustal flux.  This indicates that the
behaviour of the  field evolution in different kinds  of binaries will
again  be similar to  what we  have found  in the  case of  an initial
crustal flux, discussed in chapter [6].

\subsection{field evolution in binaries}

In section  [\ref{sobs-binaries}] we have outlined the  three phase of
binary   evolution,  namely  -   the  isolated,   the  wind   and  the
Roche-contact  phase.  In  the  wind  phase  there  are  two  distinct
possibilities  of interaction between  the neutron  star and  its main
sequence companion.  If the  system is in  the `propeller  phase' then
there is no  mass accretion. But the importance of  this phase is that
the star  rapidly slows down  to very long  periods and as a  result a
significant flux-expulsion is achieved. From the point of view of flux
expulsion, therefore,  we assume the  flux to be  completely contained
within the  superconducting core (neglecting  the small flux-expulsion
caused by  the dipole spin-down in  the isolated phase)  prior to this
phase. Therefore,  the ohmic decay of  this flux will  take place only
after this phase is over -  that is in the phase of wind-accretion and
most-importantly in  the phase of  Roche-contact. It has  already been
noted that, in  case of low mass X-ray binaries, it  is not very clear
as to  how long the  phase of wind-accretion  lasts or whether  such a
phase   is  at   all   realized  after   the   `propeller  phase'   is
over. Therefore, in our calculations we have considered cases with and
without a phase of wind accretion, as before.

\subsubsection{high mass binaries}

\bef
\begin{center}{\mbox{\epsfig{file=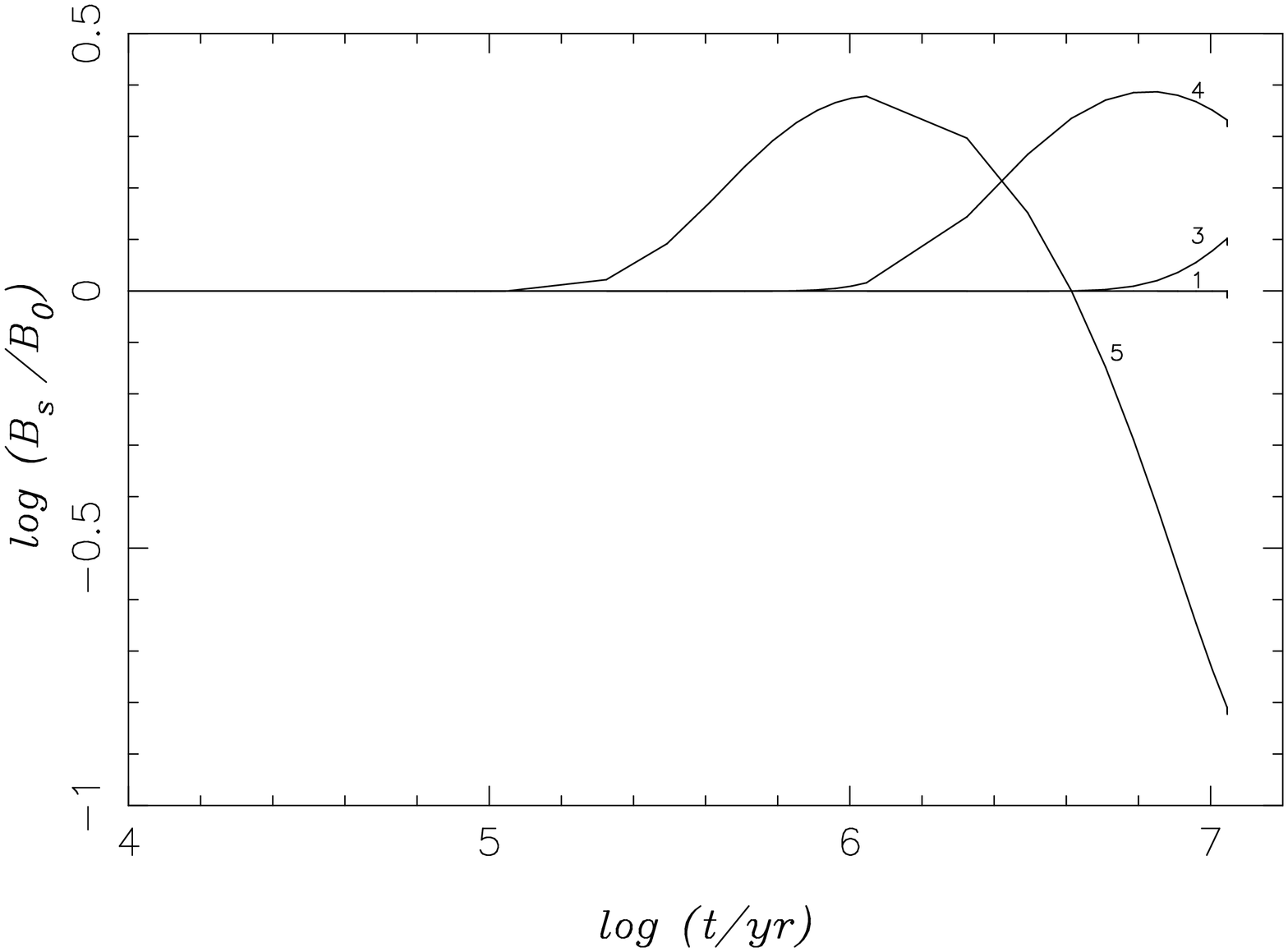,width=225pt,angle=-90}}}\end{center}
\caption[field evolution in HMXB Ia]{Evolution of the surface magnetic
field in  high mass X-ray binaries  for four values  of wind accretion
rate. The  curves 1  to 5 correspond  to $\mdot =  10^{-14}, 10^{-13},
10^{-12}, 10^{-11},  10^{-10}$~\dmdt.  Though here curves 1  and 2 are
indistinguishable. All curves correspond to $Q$ = 0.0.}
\label{fcore-hmxb1}
\eef
\bef
\begin{center}{\mbox{\epsfig{file=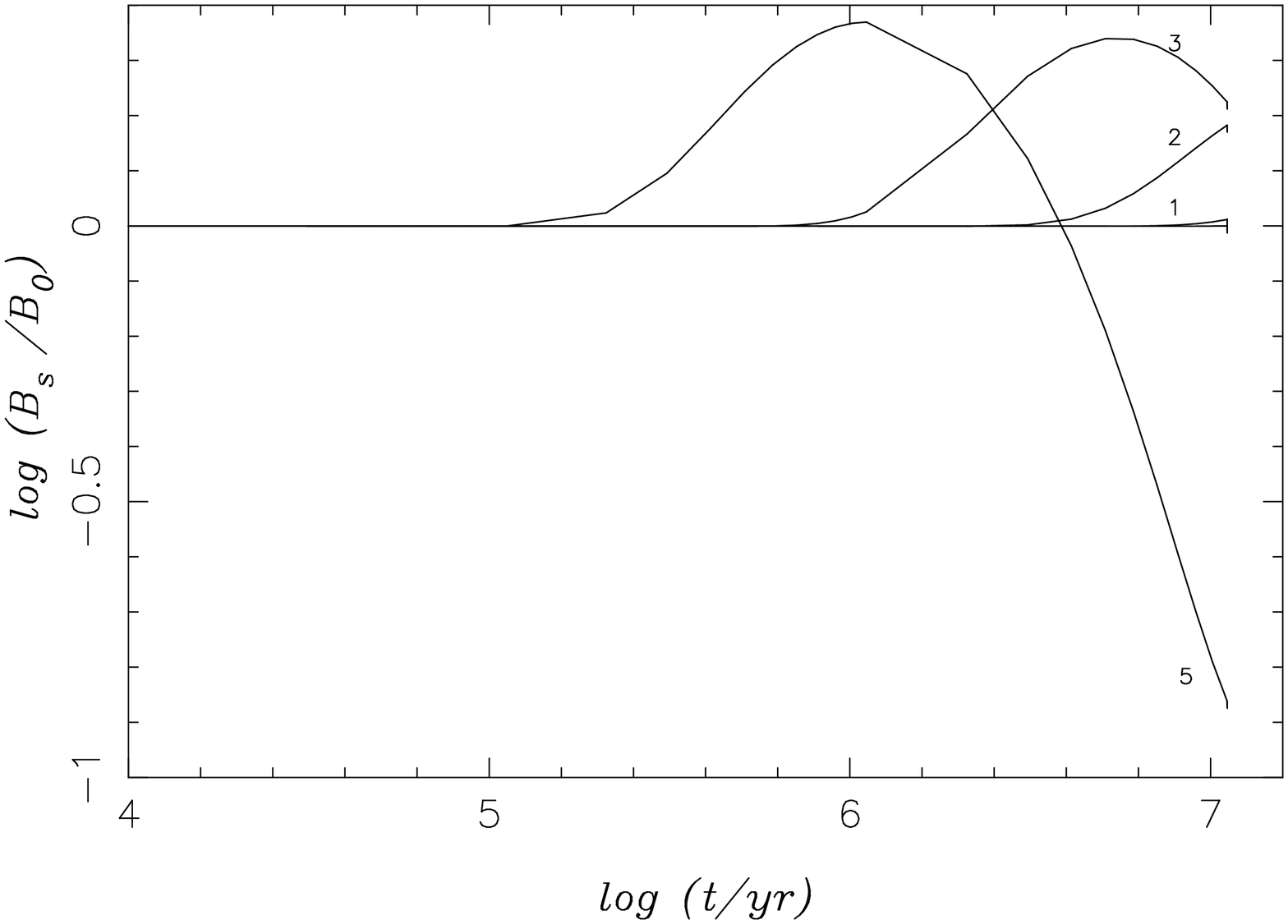,width=225pt,angle=-90}}}\end{center}
\caption[field    evolution    in     HMXB    II]{Same    as    figure
[\ref{fcore-hmxb1}] with $Q$ = 0.04.}
\label{fcore-hmxb5}
\eef
\bef
\begin{center}{\mbox{\epsfig{file=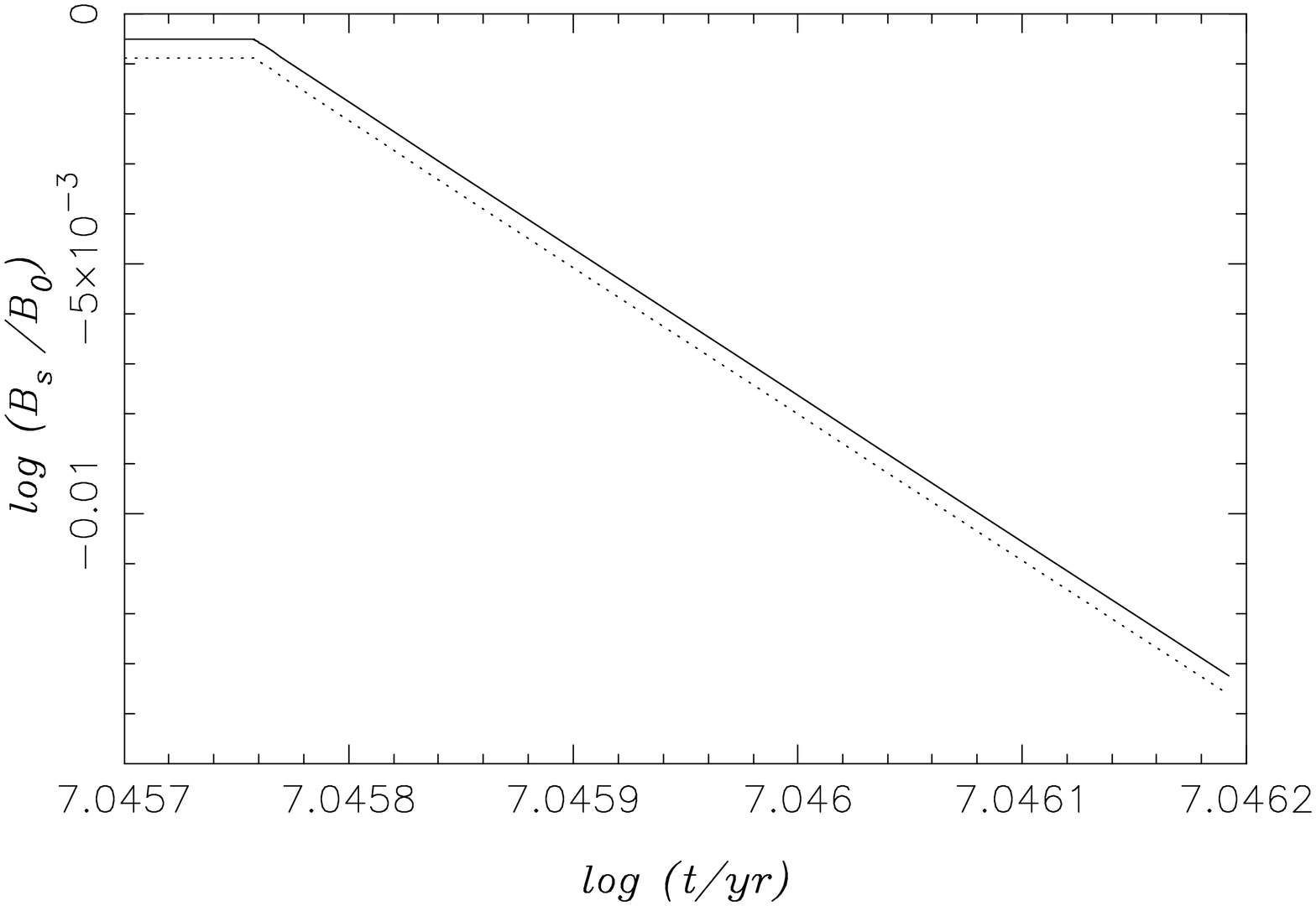,width=225pt,angle=-90}}}\end{center}
\caption[field    evolution    in     HMXB    Ib]{Same    as    figure
[\ref{fcore-hmxb1}] with  the Roche-contact phase  expanded. The solid
and  the  dotted  curves  correspond  to  curves 1  and  2  in  figure
[\ref{fcore-hmxb1}].}
\label{fcore-hmxb-Roche-q1}
\eef

Figures [\ref{fcore-hmxb1}] and [\ref{fcore-hmxb5}] show the evolution
of the surface field in  high mass X-ray binaries for different values
of the  impurity strength  in the crust.  The parameters for  the high
mass X-ray binary evolution are  as before.  The surface field shows a
clear initial  increase followed  by a sharp  decay. The decay  in the
Roche-contact  phase is  very small  and  is almost  invisible in  the
above-mentioned plots.  In figure [\ref{fcore-hmxb-Roche-q1}]  we have
expanded    the    Roche-phase    corresponding    to    the    figure
[\ref{fcore-hmxb1}]  to   highlight  this  phase.   For  the  impurity
strengths considered by  us, the field decreases by  about an order of
magnitude.

\subsubsection{low mass binaries}

\bef
\begin{center}{\mbox{\epsfig{file=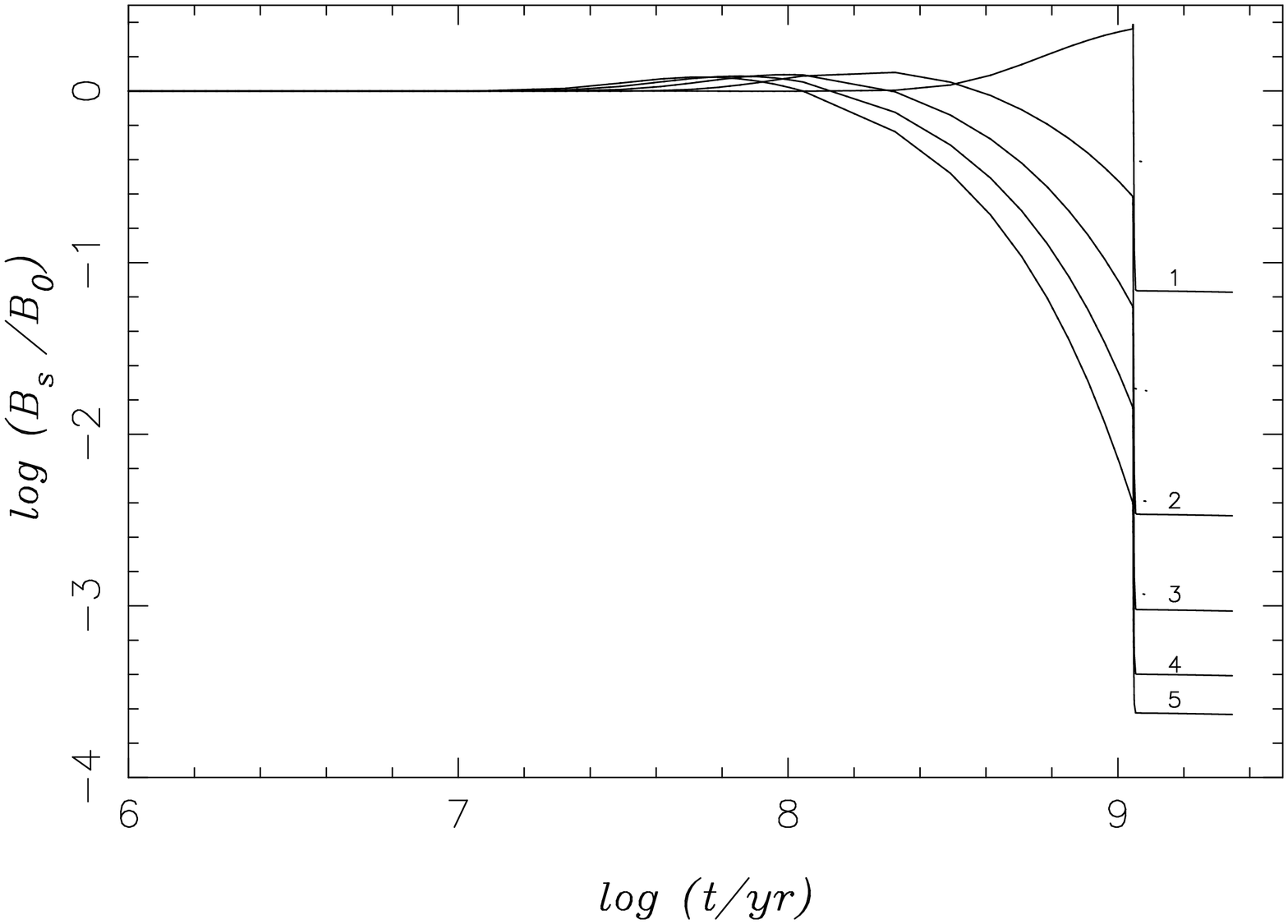,width=225pt,angle=-90}}}\end{center}
\caption[field evolution in LMXB Ia]{Evolution of the surface magnetic
field in low mass X-ray binaries  for two values of the wind accretion
rate. The curves 1 to 5 correspond  to $Q$ = 0.0, 0.01, 0.02, 0.03 and
0.04 respectively. All  curves correspond to a wind  accretion rate of
$\mdot = 10^{-16}$~\dmdt. }
\label{fcore-lmxb1}
\eef
\bef
\begin{center}{\mbox{\epsfig{file=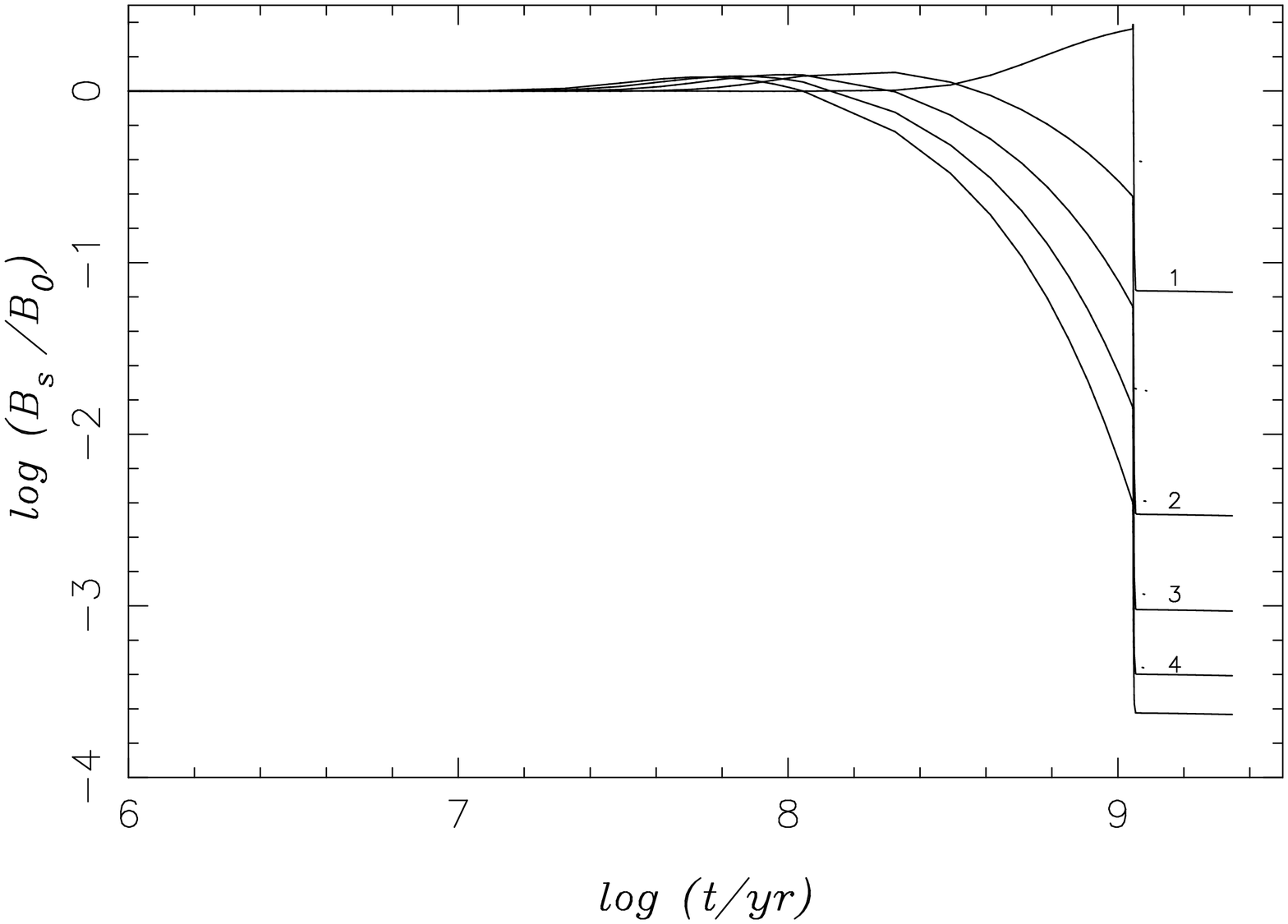,width=235pt,angle=-90}}}\end{center}
\caption[field    evolution    in     LMXB    II]{Same    as    figure
[\ref{fcore-lmxb1}]   with  a   wind  accretion   rate  of   $\mdot  =
10^{-14}$~\dmdt. }
\label{fcore-lmxb2}
\eef
\bef
\begin{center}{\mbox{\epsfig{file=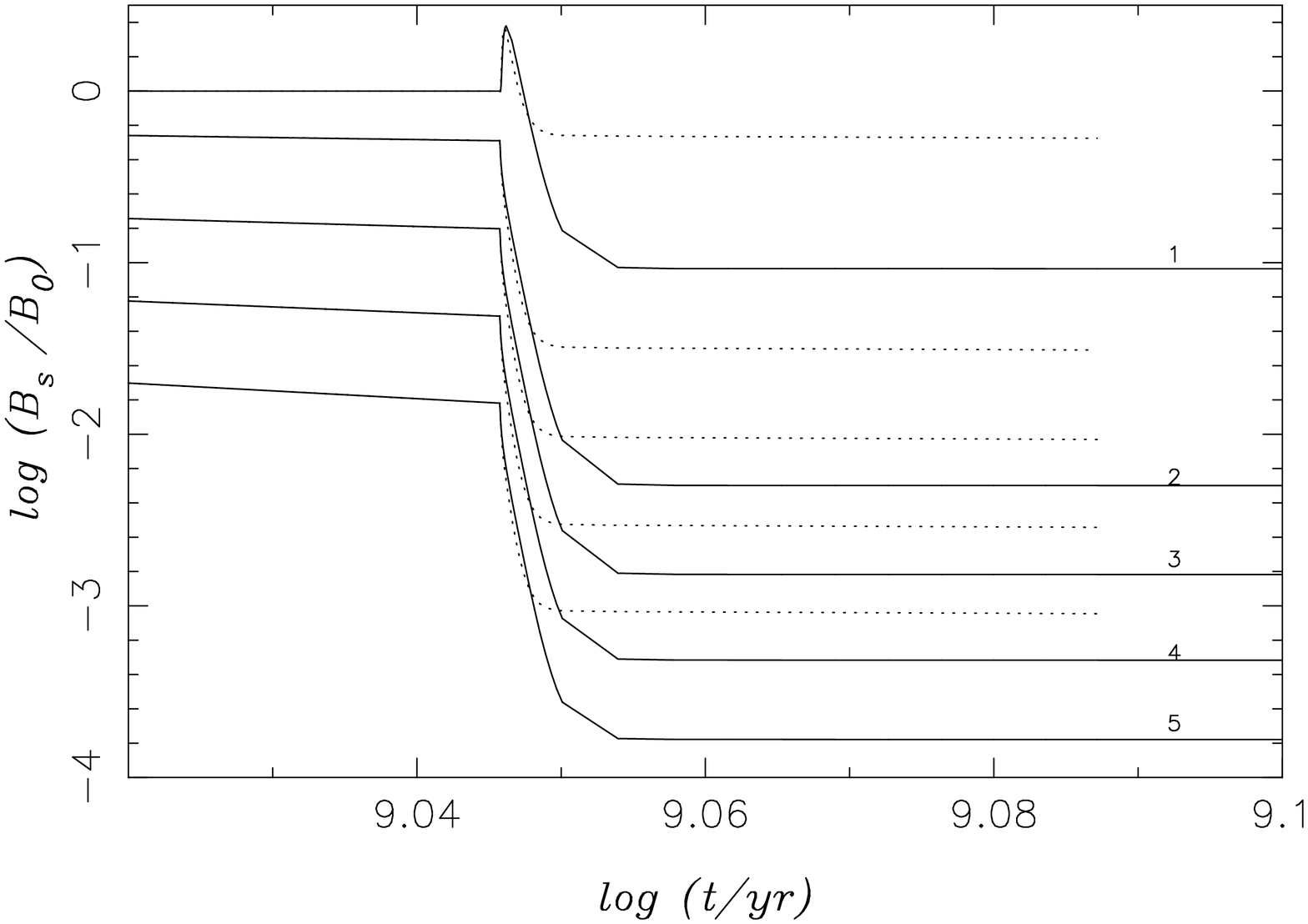,width=235pt,angle=-90}}}\end{center}
\caption[field    evolution    in     LMXB    Ib]{Same    as    figure
[\ref{fcore-lmxb1}] with  the Roche-contact phase  expanded. The solid
and  the dotted  curves  correspond  to accretion  rates  of $\mdot  =
10^{-10}, 10^{-9}$~\dmdt in the Roche contact phase.}
\label{fcore-lmxb-Roche1}
\eef
\bef
\begin{center}{\mbox{\epsfig{file=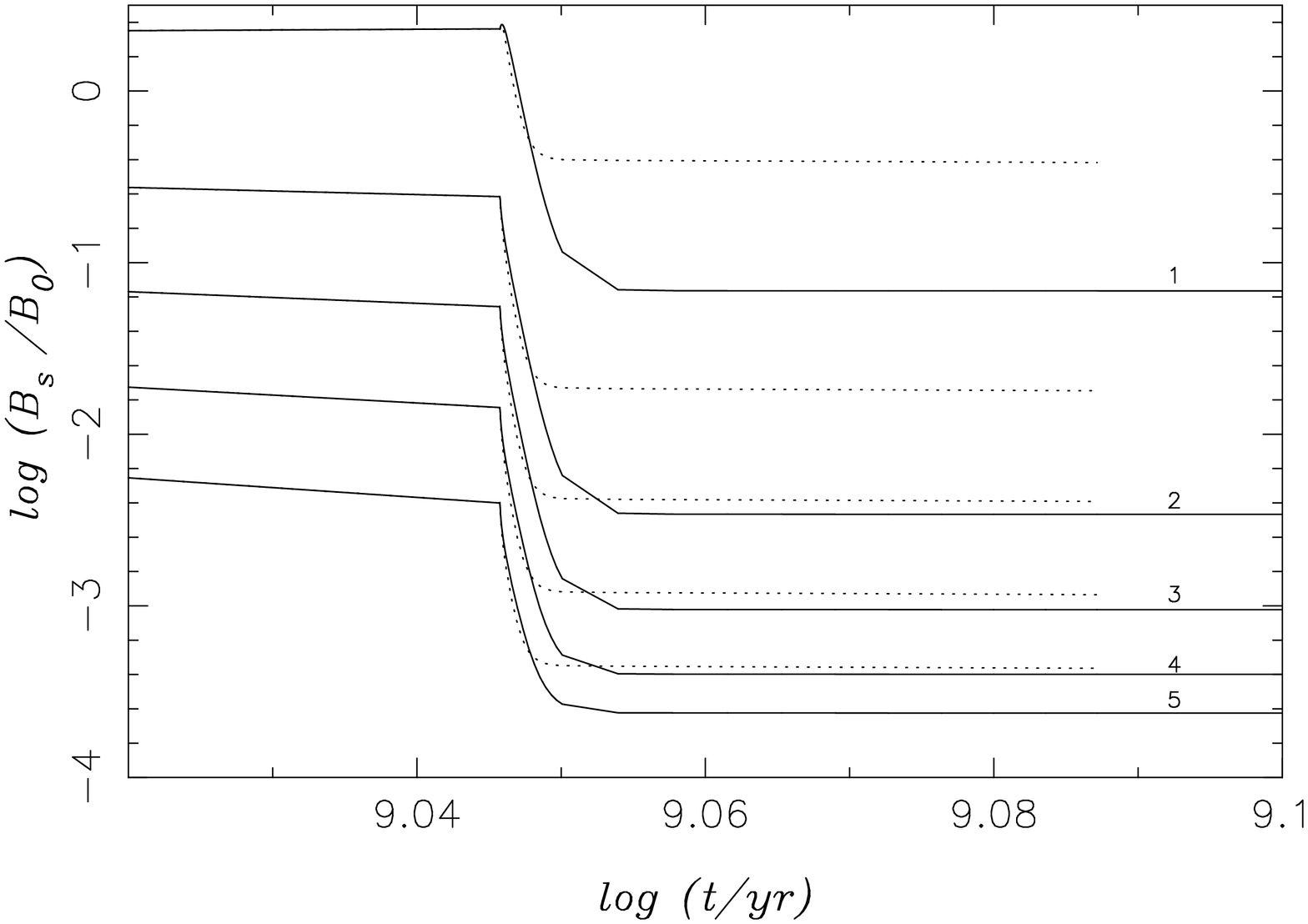,width=235pt,angle=-90}}}\end{center}
\caption[field    evolution    in    LMXB    IIb]{Same    as    figure
[\ref{fcore-lmxb2}]  with the  Roche-contact phase  expanded.The solid
and  the dotted  curves  correspond  to accretion  rates  of $\mdot  =
10^{-10}, 10^{-9}$~\dmdt in the Roche contact phase. }
\label{fcore-lmxb-Roche2}
\eef
\bef
%
%
\begin{center}{\mbox{\epsfig{file=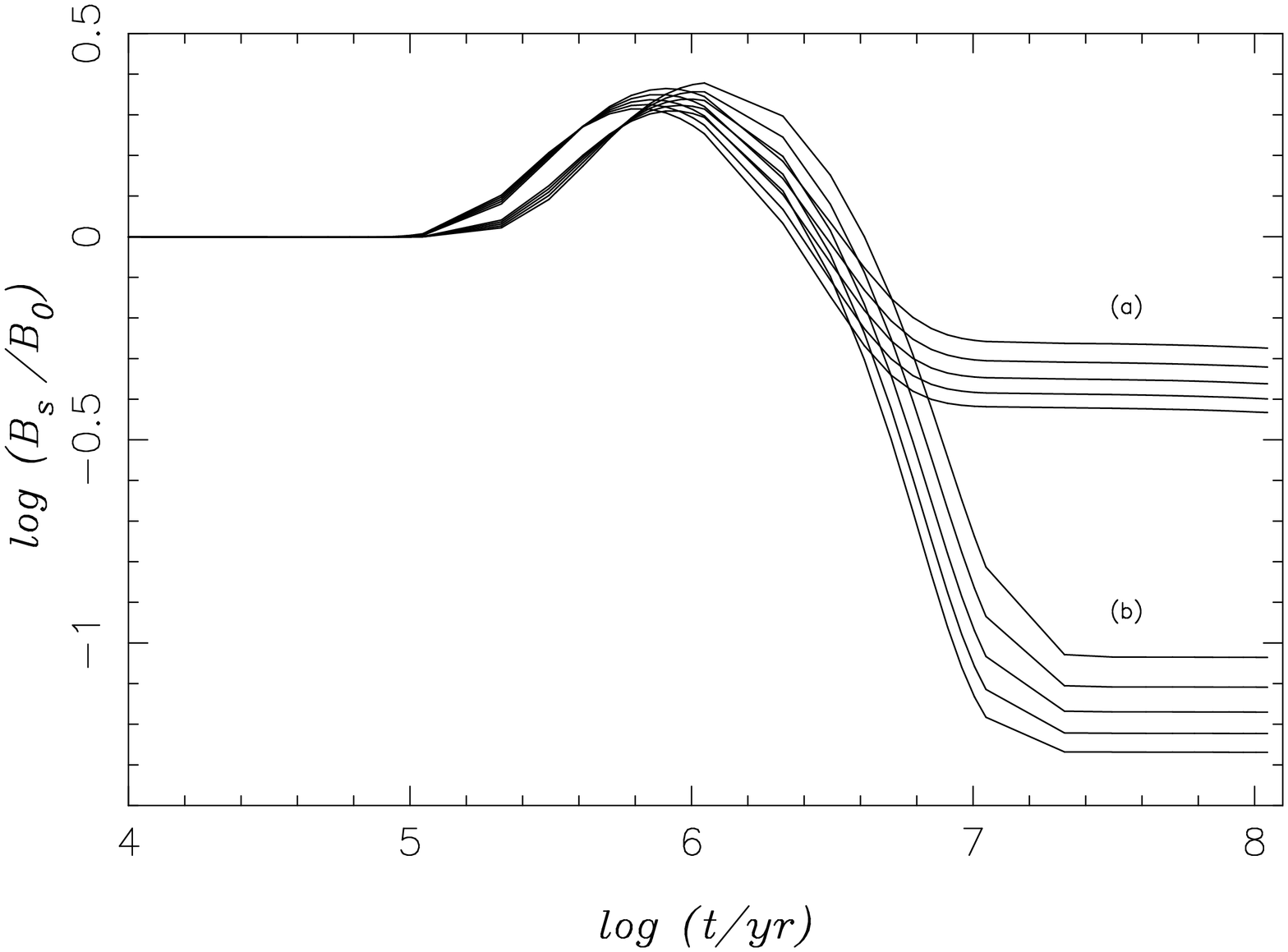,width=235pt,angle=-90}}}\end{center}
\caption[field  evolution  in   LMXB  III]{Evolution  of  the  surface
magnetic  field in low  mass X-ray  binaries without  a phase  of wind
accretion. The set of curves (a) and (b) correspond to accretion rates
of   $\mdot  =   10^{-9},   10^{-10}$~\dmdt  in   the  Roche   contact
phase. Individual  curves in  each set correspond  to $Q$ =  0.0, 0.1,
0.2, 0.3  and 0.4 respectively, the  upper curves being  for the lower
values of $Q$.}
\label{fcore-lmxb-nw}
\eef

Figures [\ref{fcore-lmxb1}] and [\ref{fcore-lmxb2}] show the evolution
of the  surface field  in the low  mass X-ray binaries;  for different
values of  the impurity  concentration in the  crust. The  two figures
correspond  to two  different values  of  accretion rate  in the  wind
phase. It should  be noted that this difference  in the wind accretion
rate  does not  manifest  itself in  either  the nature  of the  field
evolution or the  final field strengths. However, a  difference in the
accretion rate  in the  Roche-contact phase shows  up very  clearly in
figures [\ref{fcore-lmxb-Roche1}]  and [\ref{fcore-lmxb-Roche2}] where
the    Roche-contact   phases,    corresponding    to   the    figures
[\ref{fcore-lmxb1}]     and     [\ref{fcore-lmxb2}],     have     been
expanded. Comparing the different  curves (for different values of the
impurity parameter)  we see  that a large  value of  impurity strength
gives rise  to a rapid  decay and therefore  a low value of  the final
surface field. 

In figure  [\ref{fcore-lmxb-nw}] we have plotted the  evolution of the
surface field  assuming the  wind accretion phase  to be  absent. Once
again we  find that for higher  rates of accretion  higher final field
values are  obtained.  It  should be noted  here that the  final field
values obtained  now are only about  an order and a  half of magnitude
lower  than the  original  surface field  strengths.  Even though  the
impurity strengths assumed now are much higher than those assumed when
we investigated  the field evolution  in low mass X-ray  binaries {\em
with} a phase of wind-accretion.  In  absence of a prior phase of wind
accretion the  flux does not  have enough time  to diffuse out  to low
density regions when the  Roche-contact is established. Therefore here
the role of  accretion, in the Roche-contact phase,  is mainly to push
the currents in  rather than to enhance ohmic  decay rate.  Evidently,
much larger impurity strength is  required to be assumed in order that
the final  field values could  be reduced by  three to four  orders of
magnitude.  Unfortunately, due to  numerical instabilities  it becomes
very  difficult  to explore  situations  with  even  higher values  of
$Q$. But the above-mentioned figure  clearly establishes a trend as to
how the final  field values behave with $Q$ and it  is evident that we
need $Q$ values much larger than  those considered here to get down to
millisecond pulsar field strengths. 

The most important point to note  here is the fact that again, similar
to an initial  crustal field configuration, the amount  of field decay
is  much larger  than that  achieved in  the case  of high  mass X-ray
binaries. Although, in  low mass X-ray binaries with  higher values of
impurity  strength the surface  field does  go down  by three  to four
orders of magnitude from its  original value, the final field strength
could remain fairly  large if the impurity strength  is small.  But if
the wind-accretion  phase is absent  in these systems then  to achieve
large amount of reduction in  the field strengths much larger value of
the  impurity strength  will  be required.  Therefore, the  `spin-down
induced  flux expulsion  model' will  be consistent  with  the overall
scenario of field evolution  and in particular millisecond pulsars can
be produced in low mass  X-ray binaries provided the impurity strength
in the crust of the neutron stars is assumed to be extremely large. 

\section{conclusions}
\label{ssif-concl}

In this  chapter we have  investigated the consequences  of `spin-down
induced flux expulsion'. So far, the general nature of field evolution
seem to  fit the  overall scenario. The  nature of field  evolution is
quite  similar to  that in  case of  a purely  crustal model  of field
evolution though  the details  differ. Most significantly,  this model
has the  requirement of large values  of the impurity  strength $Q$ in
direct contrast to the crustal model. To summarize then :
\bei
\i The field in isolated  neutron stars do not undergo any significant
decay, over  the active  lifetime of the  pulsar, conforming  with the
statistical analyses.
\i The field values in the high mass X-ray binaries can reaming fairly
large for a moderate range of impurity strength.
\i  A reduction  of three  to four  orders of  magnitude in  the field
strength can be  achieved in the low mass  X-ray binaries provided the
impurity strength is as large as 0.5.
\i If the  wind accretion phase is absent  then to achieve millisecond
pulsar field values, impurity strength in excess of unity is required.
\eei

\chapter{conclusions}

In this final  chapter we shall summarize the  main conclusions of our
investigations. For our work we have drawn upon many results which are
not entirely free  of uncertainties.  We shall also  mention here such
uncertainties that  are likely  to affect our  results. And  lastly we
shall indicate the future directions of work in this context.

\ben
\i {\bf Generic  Features of Field Evolution in the  Crust --} In this
thesis we have mainly investigated  two models of field evolution that
of an  initial crustal current supporting  the field and  the model of
spin-down induced  flux expulsion.  The qualitative features  of field
evolution are  same for both  the models. We  have also looked  at the
effect  of diamagnetic  screening in  an accreting  neutron  star. The
nature  of this  screening  is  such that  an  assumption regarding  a
particular model of field configuration is unnecessary and the results
of this  investigation are  model-independent. Therefore, we  have the
following general conclusions regarding  the nature of field evolution
in the crust.
\bei
\i Pure Ohmic Decay in Isolated Neutron Stars :
\ben
\i  A  slow cooling  of  the  star gives  rise  to  a  fast decay  and
consequent  low  final field.  The  opposite  happens  in case  of  an
accelerated cooling.
\i  An  initial crustal  current  distribution  concentrated at  lower
densities  again gives  rise to  faster  decay and  low final  surface
field. Whereas if the current is located at higher densities the decay
is slow resulting in a higher final surface field.
\i A  large value of impurity  strength implies a rapid  decay and low
final field. If  the crust behaves more like a  pure crystal the decay
slows down considerably.
\een
\newpage
\i Accretion-Induced Field Decay in Accreting Neutron Stars :
\ben
\i In an  accreting neutron star the field  undergoes an initial rapid
decay, followed by slow down and an eventual {\em freezing}.
\i A positive correlation between  the rate of accretion and the final
field  strength is observed,  giving rise  to higher  final saturation
field strengths for higher rates of accretion.
\i  An expected  screening of  the  surface field  by the  diamagnetic
accreting  material   is  rendered  ineffective   by  the  interchange
instabilities in the liquid surface layers of the star.
\een
\eei
\i {\bf Nature of Field and Spin Evolution in Real Systems -- } In the
next phase  of our investigation we  have applied the  models of field
evolution to real systems - isolated neutron stars as well as to those
in binaries.  The paradigm of field evolution that have emerged out of
various    observations,   statistical   analyses    and   theoretical
expectations have  been summarized in the  following flow-diagram. The
arrows indicate the expected  evolutionary link between X-ray binaries
and binary  radio pulsars. The following table  summarizes the results
of our investigations.

\bef[h]
\hspace{1.0cm}
\epsfig{file=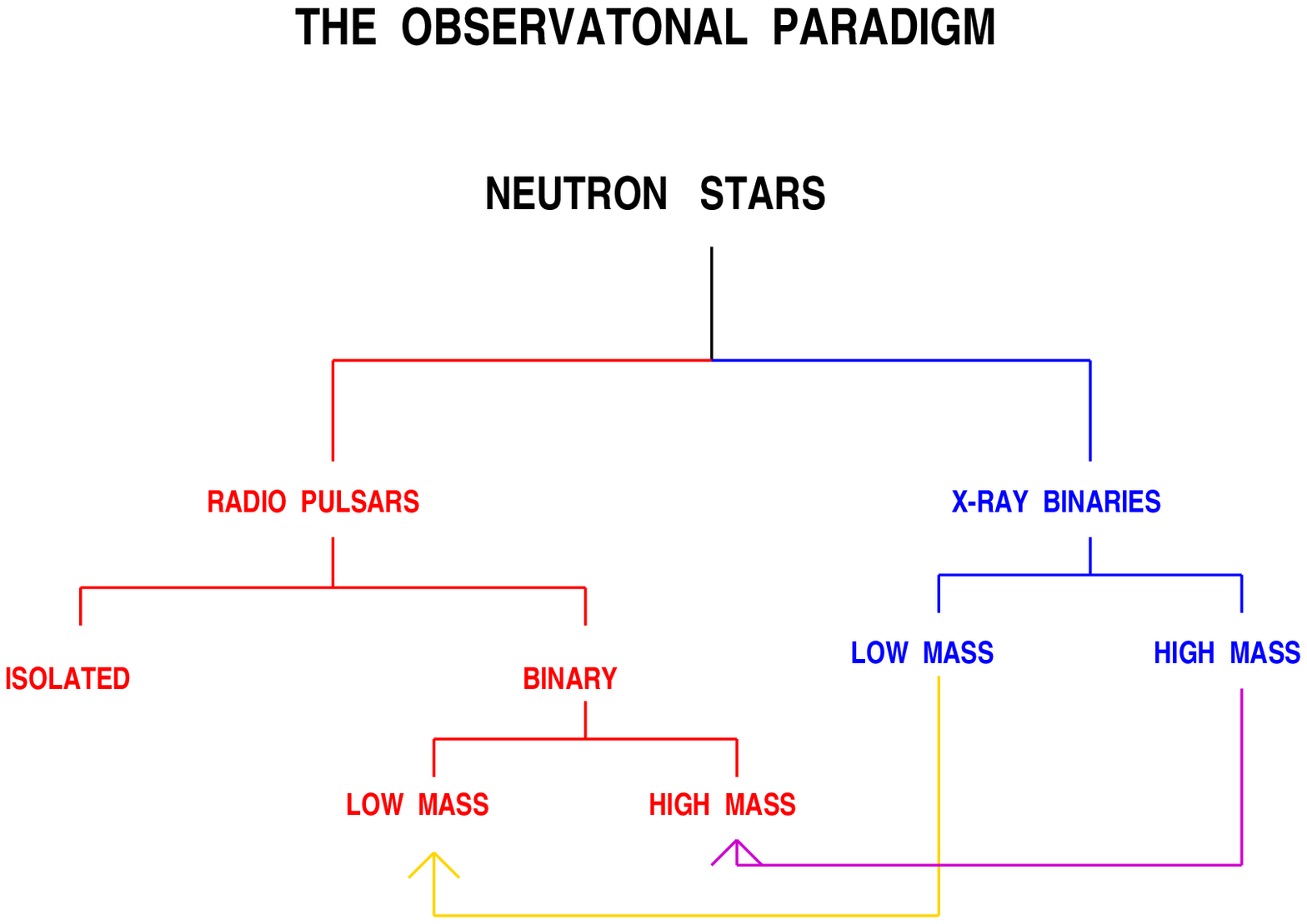,width=300pt,angle=-90}
\eef

\begin{tabular}{|l|l|r|} \hline
\multicolumn{3}{|c|}{} \\
\multicolumn{3}{|c|}{\bf Evolution of a Purely Crustal Field} \\
\multicolumn{3}{|c|}{\bf Our Results} \\
\multicolumn{3}{|c|}{} \\ \hline
&& \\
system & final field and period & comment \\
&& \\ \cline{1-3}
&& \\
isolated & high field, long period & no significant field decay \\
radio pulsars & & in $10^9$~ years  \\
&& \\ \cline{1-3}
&& \\
HMXB & high field, long period & high-mass binary pulsars \\
& & and solitary counterparts  \\
&& \\ \cline{2-3}
&& \\
& low field, long period & not active as pulsars  \\
&& \\ \cline{1-3}
&& \\
LMXB & high field, long period & high field low-mass binary pulsars \\
& & and solitary counterparts  \\
&& \\ \cline{2-3}
&& \\
& low field, short period & low field low-mass binary pulsars \\
& & and solitary counterparts, \\
& & millisecond pulsars  \\
&& \\ \cline{1-3}
\end{tabular}

\vspace{0.5cm}

It  is evident  that the  results  obtained from  the field  evolution
models agree well with the observational paradigm. The nature of field
evolution  is  similar  for   the  model  of  spin-down  induced  flux
expulsion.  Though   there  is   one  major  difference.   To  produce
millisecond pulsars in LMXBs in spin-down induced flux expulsion model
very  large values  of impurities  are required.  This makes  {\em the
surface field go  down to very low values  in $10^9$~years in isolated
pulsars} in contrast to a purely crustal model.

\i {\bf Ranges  of Physical Parameters -- } In  the following table we
summarize the constraints on  various physical parameters in the field
evolution  models   placed  by  the  requirement   to  match  observed
properties in a variety of  systems. The parameters discussed here are
- the  density at  which the initial  crustal current  distribution is
located  ($\rho_0$), the  impurity strength  in the  crust  ($Q$), the
duration of  wind-accretion phase in different binary  systems and the
rate of accretion in the Roche-contact phase for LMXBs.

\vspace{0.5cm}
\begin{tabular}{|l|l|l|r|r|} \hline
\multicolumn{5}{|c|}{} \\
\multicolumn{5}{|c|}{\bf Constraints on Physical Parameters} \\
\multicolumn{5}{|c|}{} \\ \hline
&&&& \\
parameter & model & system & requirement & parameter range \\
&&&& \\ \cline{1-5}
&&&& \\
$\rho_0$ & crustal & hmxb & high field & high $\rho_0$ \\
&&&& \\ \cline{1-5}
&&&& \\
$Q$ & crustal & isolated & no field decay & $Q \lsim 0.01$ for \\
& & radio pulsar & over active & standard cooling, \\
& & & pulsar life-time & $Q \lsim 0.05$ for \\
& & & & accelerated cooling \\
&&&& \\ \cline{2-5}
&&&& \\
& core & LMXB & millisecond pulsar & $Q \gsim 0.05$ with \\
& & & generation & wind accretion, \\
& & & &  $Q >> 1$ without \\
& & & & wind accretion \\
&&&& \\ \cline{1-5}
&&&& \\
duration of & crustal & HMXB & high field & short \\
wind accretion & & & & \\
&&&& \\ \cline{1-5}
&&&& \\
\mdot~in & crustal & LMXB & high field & Eddington rate \\
Roche-phase & & & & \\
&&&& \\ \cline{1-5}
\end{tabular} 

\i {\bf  Uncertainties -- }  The results and conclusions  stated above
suffer from  a number of uncertainties regarding  the micro-physics of
the neutron star, as listed below.
\bei
\i Thermal Behaviour -
\ben
\i Isolated Phase - The present date can be made to fit scenarios with
both a  {\em slow} or an  {\em accelerated} cooling.  Hence  it is not
clear which  is the correct  cooling behaviour of an  isolated neutron
star.
\i Accreting Phase - the  crustal temperature corresponding to a given
rate of  accretion has not  been determined with any  certainty. Also,
the  existing results  are  limited in  their  scope and  there is  no
agreement  between  various authors.   \i  Post-Accretion  Phase -  No
calculation exists for the thermal behaviour of this phase at all.
\een
\i Transport  Properties - There  are several factors, the  effects of
which have not been incorporated yet, namely those of
\ben
\i the change  in the chemical composition due to  a) accretion and b)
spin-down; and
\i the dislocations, defects etc. of the crustal lattice.
\een
\i Equation of  State - Apart from the  uncertainties already existent
for a cold  equation of state, the change  in the chemical composition
due to accretion introduces change  in the equation of state and hence
in a) the structure of the star and its b) transport properties.
\i All of our investigation has  been based on an assumption of a pure
dipolar model  for the magnetic  field. The validity of  these results
requires to be  checked by including higher order  multi-poles for the
field.
\eei
\i {\bf Future Directions -- }  One of the most important questions in
the context  of the evolution of  magnetic fields in  neutron stars is
regarding  the  models  of  field  generation (and  therefore  of  the
internal field configuration).  There is no consensus as  to which one
is  actually realized.  Our calculations  clearly point  out  that the
models have  very different requirements for the  impurity strength in
the crust. Possible  ways of resolving this dilemma  then could be the
following.
\bei
\i  Theoretical - A  better many-body  calculations may  determine the
state  of matter in  the crust  accurately in  regard to  its impurity
content.
\i Observational Prediction - An old solitary radio pulsar will have a
very low field if the impurity content of the crust is very large. The
detection  of such  a pulsar  accreting matter  from  the interstellar
matter will therefore immediately indicate the impurity content of the
crust. The results of our investigation clearly indicate that the both
the models  place very stringent  limits on the impurity  strength and
the limits from the different models are quite incompatible. Therefore
such an  observation will immediately  determine the viability  of one
model or the other - predictably providing an answer to this question.
\eei
\een

\bibliography{mnrasmnemonic,refs}
\bibliographystyle{mnras}

\end{document}